\documentclass[aps,prb,reprint,a4paper,longbibliography,superscriptaddress,citeautoscript,floatfix]{revtex4-2}

\usepackage{amsmath,amsfonts,amssymb}
\usepackage{newtxtext,newtxmath}
\usepackage[pdftex]{graphicx}
\usepackage[svgnames]{xcolor}
\usepackage{dcolumn}
\usepackage[expansion=false]{microtype}
\usepackage[
colorlinks=True,linkcolor=DarkRed,citecolor=ForestGreen,urlcolor=MediumBlue,
pdfstartview=FitH,bookmarks=False,pdfpagemode=UseNone
]{hyperref}
\usepackage{etoolbox}

\let\vec\mathbf
\let\vecg\boldsymbol
\def\ee{\mathrm{e}}
\def\ci{\mathrm{i}}
\def\dd{\mathrm{d}}

\begin{document}

\title{Extraction of the self energy and Eliashberg function from angle resolved photoemission spectroscopy using the \textsc{xARPES} code}

\homepage[This version of the article has been accepted for publication after peer review but is not the Version of Record and does not reflect post-acceptance improvements, or any corrections. The Version of Record is available online at: \vspace{0.2em}]{http://doi.org/10.1038/s41524-026-02026-9}

\author{Thomas P. van Waas}
\email[Corresponding author: ]{thomas@vanwaas.org}
\affiliation{European Theoretical Spectroscopy Facility, Institute of Condensed Matter and Nanosciences, Universit\'e catholique de Louvain, Chemin des \'{E}toiles 8, 1348 Louvain-la-Neuve, Belgium}

\author{Christophe Berthod}
\affiliation{Department of Quantum Matter Physics, University of Geneva, 1211 Geneva, Switzerland}

\author{Jan Berges}
\affiliation{U Bremen Excellence Chair, Bremen Center for Computational Materials Science, and MAPEX Center for Materials and Processes, University of Bremen, 28359 Bremen, Germany}

\author{Nicola Marzari}
\affiliation{Theory and Simulation of Materials (THEOS), and National Centre for Computational Design and Discovery of Novel Materials (MARVEL), \'{E}cole Polytechnique F\'{e}d\'{e}rale de Lausanne, 1015 Lausanne, Switzerland}
\affiliation{U Bremen Excellence Chair, Bremen Center for Computational Materials Science, and MAPEX Center for Materials and Processes, University of Bremen, 28359 Bremen, Germany}
\affiliation{Laboratory for Materials Simulations, Paul Scherrer Institut (PSI), 5232 Villigen, Switzerland}

\author{J. Hugo Dil}
\affiliation{Institut de Physique, \'{E}cole Polytechnique F\'{e}d\'{e}rale de Lausanne, 1015 Lausanne, Switzerland}
\affiliation{Center for Photon Science, Paul Scherrer Institut (PSI), 5232 Villigen, Switzerland}

\author{Samuel Ponc\'{e}}
\email[Corresponding author: ]{samuel.ponce@uclouvain.be}
\affiliation{European Theoretical Spectroscopy Facility, Institute of Condensed Matter and Nanosciences, Universit\'e catholique de Louvain, Chemin des \'{E}toiles 8, 1348 Louvain-la-Neuve, Belgium}
\affiliation{WEL Research Institute, Avenue Pasteur 6, 1300 Wavre, Belgium}

\begin{abstract}

Angle-resolved photoemission spectroscopy is a powerful experimental technique for studying anisotropic many-body interactions through the electron spectral function. Existing attempts to decompose the spectral function into non-interacting dispersions and electron-phonon, electron-electron, and electron-impurity self-energies rely on linearization of the bands and manual assignment of self-energy magnitudes. Here, we show how self-energies can be extracted consistently for curved dispersions. We extend the maximum-entropy method to Eliashberg-function extraction with Bayesian inference, optimizing the parameters describing the dispersions and the magnitudes of electron-electron and electron-impurity interactions. We compare these novel methodologies with state-of-the-art approaches on model data, then demonstrate their applicability with two high-quality experimental data sets. With the first set, we identify the phonon modes of a two-dimensional electron liquid on TiO$_2$-terminated SrTiO$_3$. With the second set, we obtain unprecedented agreement between two Eliashberg functions of Li-doped graphene extracted from separate dispersions. We release these functionalities in the novel Python code \textsc{xARPES}.

\end{abstract}

\maketitle

\section{Introduction}\label{sec:introduction}

The coupling of electrons with bosons is a central subject in condensed matter physics~\cite{bruus:2004}, governing many experimental phenomena~\cite{frohlich:1950, mahan:2000, bruus:2004, coleman:2015, allen:2015, berthod:2018a}. In solids, a commonly encountered boson is the phonon, where lattice vibrations affect electronic properties such as the resistivity in metals~\cite{ponce:2016}, Cooper pairing in conventional superconductors~\cite{allen:1983}, lifetimes of electron spins~\cite{park:2020}, and the formation of polarons~\cite{franchini:2021}. Depending on the material, the coupling of electrons with other types of bosons such as magnons~\cite{yu:2022a, mazzola:2022} and plasmons~\cite{lazzari:2018} may also show pronounced effects. In this work, we focus on systems where electron-phonon coupling (EPC) is the predominant type of electron-boson coupling, as it is intrinsic to all materials. In metals, EPC typically appears as a photoemission kink in the spectral function near the chemical potential ~\cite{johnson:2001}, which can be quantified in terms of the Eliashberg function~\cite{shi:2004}. The Eliashberg function directly affects the effective mass of the charge carriers in the metallic state, reveals the frequencies and coupling strength of the relevant phonons~\cite{grimvall:1981}, and enters into the Migdal-Eliashberg theory of superconductivity~\cite{marsiglio:2008}. In general, the Eliashberg function is an anisotropic function of the electron momentum~\cite{margine:2013}; for example, MgB$_2$ is an anisotropic superconductor with one of the highest ambient-pressure phonon-mediated critical temperatures of $T_{\mathrm c}=39$~K~\cite{nagamatsu:2001}. Its two superconducting gaps originate from the out-of-plane $\sigma$-state Fermi sheets and the two in-plane tubular structures arising from the $\pi$ states~\cite{choi:2002, ponce:2016}.

The Eliashberg function is accessible through experimental techniques, including optical-conductivity experiments~\cite{marsiglio:1999}, electron tunneling~\cite{mcmillan:1965}, Landau level spectroscopy~\cite{zeljkovic:2015}, and angle-resolved photoemission spectroscopy (ARPES)~\cite{damascelli:2003}. However, optical-conductivity experiments and electron tunneling only provide access to the isotropic Eliashberg function, making ARPES the experimental method of choice for accessing anisotropic EPC. In an ARPES experiment, photoelectrons are ejected out of a material via the photoelectric effect~\cite{einstein:1905}, after which their kinetic energies and emission angles are detected~\cite{moser:2017}. The kinetic energy is then related to the electronic binding energy and the angle to the momentum, via conservation equations~\cite{moser:2017}. Recent progress has made the spin degree of freedom also accessible via high-accuracy spin-ARPES~\cite{tusche:2011, dil:2019}. The effect of anisotropic EPC on an electron in the $n^{\mathrm{th}}$ band at wavevector $\vec{k}$ is described by a complex quantity known as the electron self-energy $\Sigma_{n}(E, \vec{k}) = \Sigma_{n}^{\prime}(E, \vec{k})+\ci\Sigma_{n}^{\prime\prime}(E, \vec{k})$~\cite{mahan:2000} where $\Sigma_{n}$ is an abbreviation for the band-diagonal $\Sigma_{nn}$~\cite{giustino:2017}. The real part $\Sigma_{n}^{\prime}(E, \vec{k})$ renormalizes the eigenenergy of an electron from the non-interacting dispersion $\varepsilon_n(\vec{k})$ into $E_n(\vec{k})=\varepsilon_n(\vec{k}) + \Sigma_{n}^{\prime}(E_n(\vec{k}), \vec{k})$, whereas the imaginary part $\Sigma_{n}^{\prime\prime}(E, \mathbf{k})$ defines a lifetime $\tau_n(\mathbf{k})$ via $\hbar/\tau_n(\mathbf{k})=-2 \Sigma_{n}^{\prime\prime}(E_n(\mathbf{k}),\mathbf{k})$~\cite{grimvall:1981}, where $\hbar$ is the reduced Planck constant, which after energy renormalization gives the quasiparticle lifetime. The self-energy and non-interacting dispersion enter into the electronic spectral function $A_{n}(E, \vec{k})$~\cite{abrikosov:1964}, which can be interpreted as a momentum- and band-resolved density of states (DOS).

To gain fundamental insight into the intrinsic EPC of materials, it is desirable to extract the electron self-energy from an ARPES band map. This extraction is often performed by fitting momentum-distribution curves (MDCs), during which the momentum dependence of the self-energy is neglected~\cite{veenstra:2011}, giving rise to an extracted self-energy $\Sigma_n(E)$ for a specific momentum cut and branch index, collectively labeled as $n$. Here, we call ``branch'' a dispersing feature of a band map that can be singled out during the MDC fitting, where it is jointly described by $\varepsilon_n(\mathbf{k})$ and $\Sigma_n(E)$. In this process, $\varepsilon_n(\vec{k})$ of such a branch is usually approximated by a polynomial dispersion and sometimes obtained from first-principles calculations~\cite{li:2024a}. The spectral function $A_n(E,\vec{k})$ is often approximated by a Lorentzian as a function of $\mathbf{k}$, which is exact only when $\Sigma_n(E,\vec{k})$ is momentum-independent and $\varepsilon_n(\vec{k})$ is linear in $\vec{k}$~\cite{veenstra:2011}. The angular/momentum resolution during the MDC fitting is usually incorporated by convolving each Lorentzian peak with a Gaussian, leading to a Voigt profile~\cite{iwasawa:2020}. Given that many non-interacting dispersions are non-linear, the use of Lorentzians leads to a biased extraction of $\Sigma_n(E)$~\cite{chien:2009a}. In this work, we propose a more general approach based on non-interacting dispersions described by polynomials, which improves the self-energy extraction when the dispersion relation is notably non-linear, as is the case for, e.g., Sr$_2$RuO$_4$~\cite{iwasawa:2012} or SrMoO$_3$~\cite{cappelli:2022}.

A common treatment of EPC expands $\Sigma_{n}^{\mathrm{ph}}(E, \vec{k})$ to lowest order in the reciprocal of the atomic mass, which includes the dynamical Fan-Migdal term $\Sigma_{n}^{\mathrm{FM}}(E, \vec{k})$~\cite{fan:1951, migdal:1958}, as well as the static Debye-Waller term $\Sigma_{n}^{\mathrm{DW}}(\vec{k})$~\cite{allen:1976, allen:1981} and the tadpole/polaron term $\Sigma_{n}^{\mathrm{P}}(\vec{k})$~\cite{marini:2015, lafuente-bartolome:2022a}. We argue here that for photoemission kinks, $\Sigma_n(E)$ is often dominated by the Fan-Migdal contribution $\Sigma_{n}^{\mathrm{FM}}(E)$, from which one can extract the Eliashberg function $\alpha^2F_n(\omega)$ with phonon energy $\omega$, which combines the phonon DOS $F(\omega)$ with a coupling function $\alpha_n^2(\omega)$~\cite{grimvall:1976}. Furthermore, we argue that purely static, real-valued self-energies such as $\Sigma_{n}^{\mathrm{DW}}(\vec{k})$ should be captured by $\varepsilon_n(\mathbf{k})$ during the fit. ARPES can therefore give access to anisotropic EPC through analysis of $\Sigma_{n}^{\mathrm{FM}}(E)$ along different momentum paths.

Some of the first determinations of $-\Sigma_n^{\prime\prime}(E)$ from ARPES band maps were performed for elemental metals in the mid-1990s~\cite{mcdougall:1995, balasubramanian:1998}, while one of the first extractions of $\Sigma_n^{\prime}(E)$ was performed for Bi$_{2}$Sr$_{2}$CaCu$_{2}$O$_{8+\delta}$ in 1999~\cite{valla:1999}. Furthermore, in 2004 Shi et al.~\cite{shi:2004} were among the first to obtain $\alpha^2F_n(\omega)$ by solving an inversion problem for $\Sigma_{n}^{\prime}(E)$ for a branch of the (10$\bar{1}$0) Be shallow $\bar{\mathrm{A}}$-point surface state using the maximum-entropy method (MEM)~\cite{jarrell:1996}, establishing ARPES as the method of choice to quantify anisotropic EPC. The MEM allows for the incorporation of prior knowledge in $\alpha^2F_n(\omega)$, such as positive semidefiniteness over a finite bandwidth~\cite{shi:2004}. In subsequent years, $\alpha^2F_n(\omega)$ was extracted for a wide range of materials, including a Be(0001) surface state~\cite{chien:2015}, MgB$_2$~\cite{ludbrook:2014}, doped graphene~\cite{haberer:2013, fedorov:2014, ludbrook:2015, verbitskiy:2016, usachov:2018, ehlen:2020, huempfner:2023}, the kagome metal CsV$_3$Sb$_5$~\cite{zhong:2023}, the two-dimensional electron liquid (2DEL) of Nb-doped SrTiO$_3$~\cite{sokolovic:2025}, a Ni(111) surface state also including electron-magnon coupling~\cite{rost:2024}, and several cuprates~\cite{bok:2010, zhao:2011, yun:2011}. In parallel to the experimental investigations into EPC, first-principles methods have grown into a powerful tool to simulate electronic band structures using density-functional theory (DFT) and beyond~\cite{marzari:2021}, as well as phonon energies and EPC from density-functional perturbation theory (DFPT)~\cite{gonze:1997b, gonze:1997, baroni:2001, giustino:2017}. Calculations of the latter have become computationally feasible by momentum extrapolation~\cite{ponce:2015}, Fourier interpolation of the perturbed potential~\cite{eiguren:2008, gonze:2020}, or Wannier interpolation~\cite{marzari:2012, pizzi:2020} of the EPC matrix elements~\cite{giustino:2007}, nowadays implemented in various first-principles codes~\cite{ponce:2016, lee:2023, gonze:2016, gonze:2020, cepellotti:2022, zhou:2021a, protik:2022, marini:2024}. State-of-the-art developments in the calculation of $\Sigma_n^{\mathrm{ph}}(E)$ include a description of the perturbation of the electronic potential at the GW level~\cite{li:2019a} and a description of the electron's Green's function at the level of dynamical mean-field theory (DMFT)~\cite{abramovitch:2023}. Thus, the time is ripe for the treatment of ARPES data in a many-body framework and to connect $\Sigma_n^{\mathrm{FM}}(E)$ and $\alpha^2F_n(\omega)$ from ARPES to their first-principles counterparts.

Matthiessen's rule states that $\Sigma_n(E)=\Sigma_n^{\mathrm{ph}}(E)+\Sigma_n^{\mathrm{el}}(E)+\Sigma_n^{\mathrm{imp}}(E)$~\cite{matthiessen:1864, kemper:2018}, with $\Sigma_n^{\mathrm{el}}(E)$ and $\Sigma_n^{\mathrm{imp}}(E)$ the respective electron-electron and electron-impurity contributions. A key problem in determining $\Sigma_{n}^{\mathrm{ph}}(E)$ from $A_n(E,\vec{k})$ is that $\varepsilon_n(\vec{k})$, $\Sigma_n^{\mathrm{el}}(E)$, and $\Sigma_n^{\mathrm{imp}}(E)$ are unknown. To distinguish between $\varepsilon_n(\vec{k})$ and $\Sigma_n(E)$, the current state of the art relies on the Kramers-Kronig relations between $\Sigma_{n}^{\prime}(E)$ and $\Sigma_{n}^{\prime\prime}(E)$~\cite{berthod:2018a}. Typically, one makes an initial guess of a linear $\varepsilon_n(\vec{k})$ leading to $\Sigma_n^{\prime}(E)$ and $-\Sigma_n^{\prime\prime}(E)$; the Kramers-Kronig relations are then used to obtain $\Sigma_{n}^{\prime}(E)$ as the Hilbert transform of $\Sigma_n^{\prime\prime}(E)$, which is then compared to the initial $\Sigma_n^{\prime}(E)$. The non-interacting parameters are then obtained by minimizing the difference between the two $\Sigma_n^{\prime}(E)$~\cite{kordyuk:2005, veenstra:2011, pletikosic:2012, usachov:2018}, a procedure sometimes referred to as the Kramers-Kronig bare-band fitting (KKBF)~\cite{veenstra:2011}. However, in practice, an ARPES band map yields a finite, noisy set $\{\Sigma_{n}(E)\}$, such that the transform of $\Sigma_{n}^{\prime}(E)$ will not perfectly reconstruct $\Sigma_{n}^{\prime\prime}(E)$, and some expression for $-\Sigma_n^{\prime\prime}(E)$ must be assumed outside the fitted range to perform the Hilbert integral. Several Python codes are readily available for the treatment of ARPES data, such as \textsc{PyARPES}~\cite{stansbury:2020}, \textsc{pesto}~\cite{polley:2024}, \textsc{NavARP}~\cite{navarp:website}, and \textsc{ERLabPy}~\cite{erlabpy:website}. While these codes offer advanced data visualization capabilities, their many-body functionality is generally limited to extracting $\Sigma_n(E)$ for a single linear dispersion relation without matrix-element correction, and no decomposition into $\Sigma_n^{\mathrm{ph}}(E)$, $\Sigma_n^{\mathrm{el}}(E)$, or $\Sigma_n^{\mathrm{imp}}(E)$ is available.

In this work, we describe a consistent way to extract $\Sigma_{n}(E)$ for parabolic non-interacting dispersions, which can be extended to bands described by polynomials of all orders. Next, we show how to extract $\alpha^2F_n(\omega)$ from $\Sigma_{n}(E)$ and demonstrate with an artificial example that this inversion monotonically converges towards the true result for a sufficient amount of unbiased data. Specifically, we extend the MEM with Bayes' rule~\cite{bryan:1990} to determine the non-interacting dispersion parameters and the magnitudes of $\Sigma_n^{\mathrm{el}}(E)$ and $\Sigma_n^{\mathrm{imp}}(E)$ from the most probable $\alpha^2F_n(\omega)$~\cite{bryan:1990}, eliminating human bias from their evaluation. We aim to describe these quantities in a terminology that unifies experimental and first-principles communities, and we show how the experimental quantities for $\varepsilon_n(\vec{k})$, $\Sigma_{n}(E)$, $\alpha^2F_n(\omega)$ are related to their first-principles counterparts. We distribute these novel functionalities in the first release of the GPL-v3-licensed code \textsc{xARPES} v1.0.0.

In Sec.~\ref{sec:results}, we introduce the photointensity containing $A_n(E,\vec{k})$, show how to obtain $\Sigma_{n}(E)$ in the presence of a parabolic $\varepsilon_n(\vec{k})$, and provide expressions for $\Sigma_{n}(E)$ from impurity, electron, and phonon interaction. We then show how a one-shot solution for $\alpha^2F_n(\omega)$ can be obtained from $\Sigma_{n}(E)$, or alternatively how $\Sigma_{n}(E)$, $\alpha^2F_n(\omega)$, and the model parameters can be iteratively obtained with Bayesian inference. We then illustrate the different capabilities of the code with three examples: a model system, and two case studies of TiO$_2$-terminated SrTiO$_3$ and of Li-doped graphene, which are also distributed as \textsc{xARPES} example \textsc{Jupyter} notebooks. In Sec.~\ref{sec:discussion}, we summarize our findings and provide an overview of future directions on the subject, including the use of approximated non-interacting dispersions from DFT, which may aid the first-principles community by offering controlled reference data to benchmark new and existing approaches.

\section{Results}\label{sec:results}

\subsection{Angle-resolved photoemission spectroscopy}\label{sec:arpes}

\begin{figure}[tb]
\includegraphics[width=0.99\columnwidth]{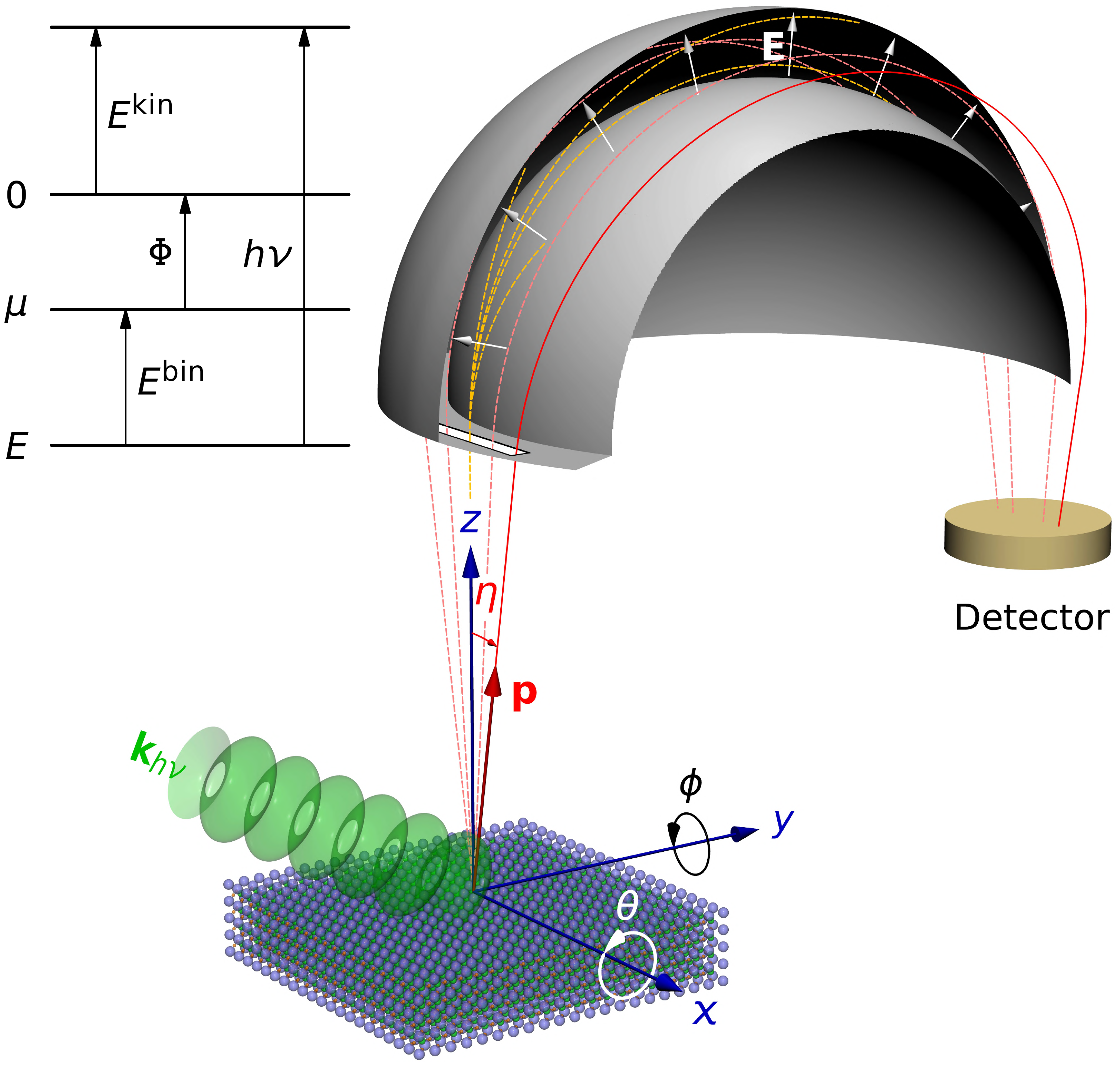}
\caption{\textbf{ARPES geometry and energy diagram}.
Light reaches the material, leading to emission of photoelectrons, where a selection passes through a hemispherical analyzer and gets detected. The inset shows an energy diagram for the one-step photoemission process. The displayed variables are defined in the main text.
\label{fig:ARPES}}
\end{figure}

First, we introduce the experimental geometry of a typical ARPES experiment -- as displayed in Fig.~\ref{fig:ARPES} -- with electrons collected along a detector slit characterized by an angle $\eta$. Photons of energy $h\nu$, with $h$ the Planck constant and $\nu$ the photon frequency, illuminate the material at an incident light wavevector $\vec{k}_{h\nu}$. The material contains electrons with a distribution of energies $E$, yielding a binding energy $E^{\mathrm{bin}}=\mu-E$ with respect to the chemical potential $\mu$, where we set the vacuum energy at the analyzer to zero. When $h\nu$ is sufficiently high for electrons to reach the vacuum energy at the sample, photoelectrons of rest mass $m_{\mathrm{e}}$ are detected with kinetic energies $E^{\mathrm{kin}}=h\nu - \Phi -E^{\mathrm{bin}}$, with $\Phi$ the work function. In a typical ARPES setup, the chemical potentials of the sample and the analyzer are aligned, while the vacuum level at the detector serves as the zero of energy, such that $\Phi$ can be identified as the work function of the analyzer~\cite{fero:2014}. Far from the sample, the wavefunctions of the photoemitted electrons can be approximated by plane waves with momenta $\hbar \vec{p}$ and energies $E^{\mathrm{kin}}=\hbar^2 |\mathbf{p}|^2/(2m_{\mathrm{e}})$. A radial electric field $\vec{E}$ is applied within the hemispherical analyzer, such that electrons arrive at the detector with $E^{\mathrm{kin}}$ within the energy resolution.

During the experiment, electrons enter a hemispherical analyzer through a detector slit characterized by an angle $\eta$. Fig.~\ref{fig:ARPES} shows the situation where the normal vector of the material and the normal vector of the detector lie along the same axis, with electrons collected at the detector angle $\eta$ in the $xz$-plane. Rotations of the sample along the $x$- and $y$-axes are described by the angles $\theta$ and $\phi$, respectively. Potentially, the set of accessible material rotations can be completed by including a rotation about the $z$ axis~\cite{ishida:2018}. Following these conventions, the components of $\mathbf{p}$ are given by:
\begin{equation}\label{eq:kinematic_relations}
	\vec{p} = |\mathbf{p}| \begin{pmatrix} \cos(\phi)\sin(\eta) + \sin(\phi) \cos(\theta) \cos(\eta) \\
		\sin(\theta) \cos(\eta) \\ \sin(\phi)\sin(\eta) + \cos(\phi) \cos(\theta) \cos(\eta) \end{pmatrix},
\end{equation}
where $|\vec{p}| = \sqrt{2m_{\mathrm{e}}E^{\mathrm{kin}}/\hbar^2}$. When $\theta=0$, $p_x =|\mathbf{p}|\sin(\eta + \phi)$, such that $p_x$ can directly be calculated after correcting with $\phi$, either to correct for misalignment or to cover a larger portion of reciprocal space. The in-plane momentum is conserved in the photoemission process, so that the in-plane components $k_{x}$ and $k_{y}$ can be inferred from $\eta$, $\theta$, and $\phi$.

The ARPES experiment can be described using photoemission theories at different levels of approximation. Early models based on a phenomenological three-step decomposition~\cite{berglund:1964} were superseded in the 1970s by a fundamental one-step theory that describes photoemission as a single quantum-mechanical process~\cite{schaich:1971, mahan:1970, keiter:1978, almbladh:2006}. Although formally rigorous, the fundamental theory is intractable when applied to real materials. Therefore, it is usually replaced by a simpler one-step description, where the steady-state photoelectron flux is regarded as a transition rate computed using Fermi's golden rule~\cite{hermeking:1975, hufner:1995, damascelli:2003}. Each possible final state of the transition is written as a product of a sample eigenstate and the photoelectron wavefunction. This factorization enables separating the spectroscopic properties of the sample -- represented by the spectral function -- from geometric factors related to the photoelectron wavefunction that are gathered in a matrix element. For a crystalline surface, the transition rate has the form~\cite{damascelli:2003}:
\begin{equation}\label{eq:totalrate}
	w(E,\mathbf{p})=\frac{2\pi}{\hbar}\sum_{n\vec{k}}
	|M_n(\vec{p},\vec{k})|^2 \delta_{\vec{k}_\parallel,\check{\vec{p}}_\parallel}
	A_n(E,\vec{k})f(E),
\end{equation}
where the sum runs over all Bloch waves with band index $n$ and wavevector $\vec{k}$ within the first Brillouin zone. In Eq.~\eqref{eq:totalrate}, $M_n(\vec{p},\vec{k})$ is the matrix element of the light-matter coupling between the Bloch wave and the photoelectron state identified by its wavevector $\vec{p}$. The notation of Eq.~(\ref{eq:totalrate}) emphasizes the conservation of in-plane crystal momentum at an ideal planar surface, where the in-plane wavevector $\vec{k}_\parallel$ of the Bloch state matches the vector corresponding to $\vec{p}_\parallel$ within the first Brillouin zone, here denoted $\check{\vec{p}}_\parallel$. As a consequence, changing $\vec{p}$ by an in-plane reciprocal lattice vector $\mathbf{G}_{\parallel}$, corresponding to a higher Brillouin zone, will affect the matrix element but not the spectral function, which has the periodicity of the lattice. The dependence of $M_n(\vec{p},\vec{k})$ on $\vec{k}_\perp$ and $\vec{p}_\perp$ is set by the material-specific wave functions and depends on $h\nu$ and the light polarization $\vecg{\epsilon}$. The photoemission matrix element (PME) is often assumed to be peaked at $\vec{k}_\perp=\check{\vec{p}}_\perp$~\cite{smith:1993, strocov:2023}, with a span of the order of $\ell_{\mathrm{e}}^{-1}$, where $\ell_{\mathrm{e}}$ is the photoelectron escape depth. $A_n(E,\vec{k})$ is the spectral function evaluated at the initial energy $E$ of the photoelectron (see inset of Fig.~\ref{fig:ARPES}), which includes the non-interacting dispersion and the self-energy discussed in Sec.~\ref{sec:introduction}: \begin{equation}\label{eq:AnEk}
	A_n(E,\vec{k})=\frac{1}{\pi}\frac{-\Sigma_n^{\prime\prime}(E,\vec{k})}
	{[E-\varepsilon_n(\vec{k})-\Sigma_n^{\prime}(E,\vec{k})]^2+[\Sigma_n^{\prime\prime}(E,\vec{k})]^2}.
\end{equation}
Non-interacting electrons have all the spectral weight at the band energy, corresponding to $\Sigma_n(E,\vec{k})=-\ci0^+$ and $A_n(E,\vec{k})=\delta\left(E-\varepsilon_n(\vec{k})\right)$. Sufficiently weak interactions preserve this peak structure, but broaden the $\delta$-function while shifting it to the quasiparticle energy $E_n(\vec{k})$. The broadening corresponds to the lifetime via $\hbar/\tau_n(\vec{k})=-2 \Sigma_{n}^{\prime\prime}(E_n(\vec{k}),\vec{k})$ and may be expressed in terms of a mean free path $\boldsymbol{\ell}_n (\mathbf{k}) \equiv \mathbf{v}_n(\mathbf{k}) \tau_n(\mathbf{k})$ via the group velocity $\vec{v}_n(\vec{k})=\hbar^{-1}\vec{\nabla}E_n(\vec{k})$. The last factor in Eq.~\eqref{eq:totalrate} is the Fermi-Dirac distribution $f(E) \equiv [\ee^{(E-\mu)/(k_{\mathrm{B}}T)}+1]^{-1}$ with $k_{\mathrm{B}}$ the Boltzmann constant and $T$ the temperature, representing the requirement that the photoelectron state is initially occupied.

The spectral function of surface states and of states in two-dimensional (2D) systems is independent of $\vec{k}_\perp$. As a result, $w$ factorizes into a part involving the matrix element and a part involving the spectral function. Such a factorization is in general impossible for three-dimensional (3D) states, such as the $\sigma$-state sheets of MgB$_2$~\cite{choi:2002}, since the sum over $\vec{k}_\perp$ in Eq.~\eqref{eq:totalrate} mixes the matrix element with the spectral function. The factorization is nevertheless possible if either $\ell_{\mathrm{e}}\gg|\boldsymbol{\ell}_n|$ or $|\boldsymbol{\ell}_n|\gg\ell_{\mathrm{e}}$. In the first regime, the matrix element is sharply peaked at $\vec{k}_\perp=\check{\vec{p}}_\perp$, while in comparison the spectral function varies slowly around $\vec{k}_\perp$ with a span of order $|\boldsymbol{\ell}_n|^{-1}$. Thus the matrix element enforces approximate conservation of perpendicular momentum and the spectral function is evaluated at $\vec{k}_\perp=\check{\vec{p}}_\perp$. If $|\boldsymbol{\ell}_n|\gg\ell_{\mathrm{e}}$, the spectral function is sharp and the matrix element is broad, such that $w$ is proportional to the surface-projected spectral function $\sum_{\vec{k}_{\perp}}A_n(E,\vec{k})$, leading to intrinsic broadening of the spectral features. In many 3D systems, neither of these two regimes is realized and the matrix element and spectral function are inextricably mixed. As this work focuses on the self-energy of 2D systems and surface states, we adopt the product ansatz and we treat the matrix element as a phenomenological ingredient.

The experimental band map is recorded through a detector slit characterized by an angle $\eta$ while selecting and counting photoelectrons according to their kinetic energy. As $E^{\mathrm{kin}}=h\nu - \Phi + E -\mu$, the photoelectron wavevector may be viewed as a function $\vec{p}(E,\eta)$ of the two variables $E$ and $\eta$. Due to finite energy and angular resolutions of the hemispherical analyzers~\cite{iwasawa:2017}, the measured intensity is convolved with resolution functions $R(E)$ and $Q(\eta)$ with full widths at half maximum (FWHM) $\Delta E$ and $\Delta \eta$, which are usually taken to be Gaussian distributions. Taking into account an energy- and possibly weakly angle-dependent background $B(E,\eta)$, we arrive at the photointensity at light polarization $\vecg{\epsilon}$ as the time-integrated transition rate:
\begin{multline}\label{eq:photointensity}
	P(E,\eta; h\nu, \vecg{\epsilon}) \propto
	\int_{-\infty}^{\infty}\dd E'\,R(E-E')\int_{-\infty}^{\infty}\dd\eta'\,Q(\eta-\eta')\\
	\kern-3em\times\Big[B(E',\eta')+f(E')\sum_n |M_n(E',\eta'; h\nu, \vecg{\epsilon})|^2 A_n(E',\eta')\Big].\kern-0.6em
\end{multline}
The photointensity at fixed $\eta$ yields an energy-distribution curve (EDC), and the photointensity at fixed $E$ yields an MDC. After specifying the rotation angles displayed in Fig.~\ref{fig:ARPES}, the $\eta$-dependent quantities in Eq.~\eqref{eq:photointensity} can be converted into $\mathbf{k}$-dependent quantities.

\subsection{xARPES workflow}

\begin{figure}[tb]
\includegraphics[width=0.9\columnwidth]{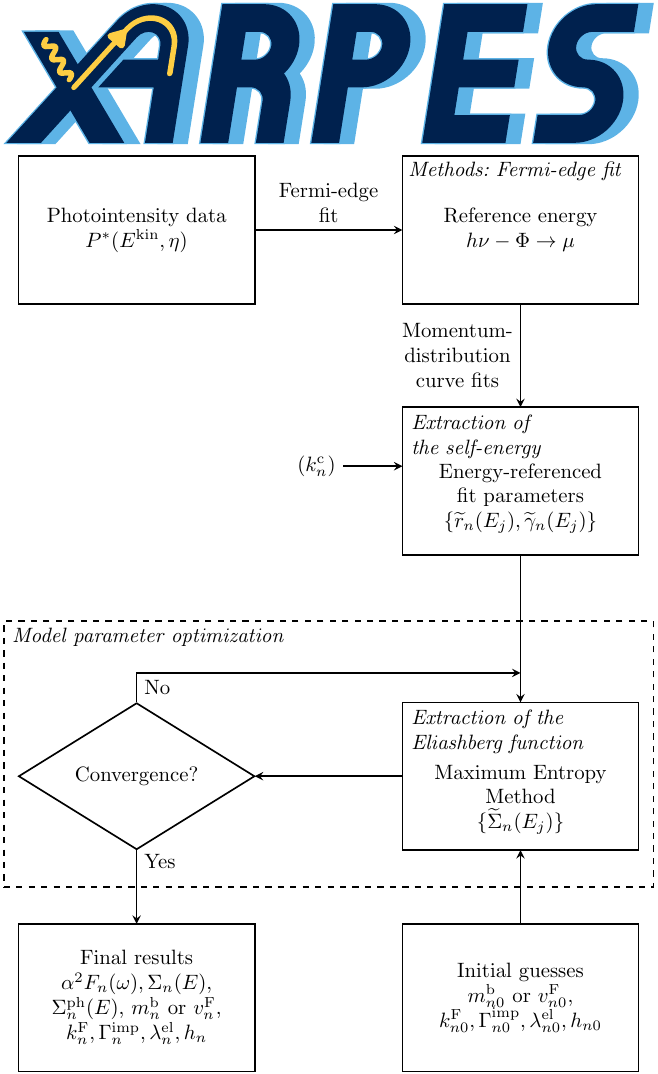}
\caption{\textbf{\textsc{xARPES} logo and flowchart}.
The \textsc{xARPES} logo and different steps of the workflow, as described in detail in the text. The workflow can be divided into three parts. First, the Fermi edge and the momentum-distribution curves from photointensity data $P^*(E^{\mathrm{kin}},\eta)$ are fitted to obtain the respective peak positions and widths $\{\widetilde{r}_n(E_j), \widetilde{\gamma}_n(E_j)\}$. Second, initial guesses for the model parameters are inserted into the maximum-entropy method to obtain the one-shot self-energy $\Sigma_{n0}(E)$ and Eliashberg function $\alpha^2F_{n0}(\omega)$. Third, the optimization loop can be called to obtain the optimized self-energy $\Sigma_{n}(E)$ and Eliashberg function $\alpha^2F_n(\omega)$. The boxes contain the variables and are tagged with the subsection in which the step is fully described. The subsections belong to the Results section unless described as a "Methods" section.
\label{fig:workflow}}
\end{figure}

In this section, we give an overview of the different steps that can be performed with the \textsc{xARPES} code. The workflow is displayed in Fig.~\ref{fig:workflow}, where boxes describe individual steps of the code and refer to the sections in which these steps are described in detail. The first step is to load raw photointensity data for a given photon energy $h\nu$ and polarization $\boldsymbol{\epsilon}$.

After loading the band map containing data in $E^{\mathrm{kin}}$, the user can either perform a Fermi-edge fit or provide a previous Fermi-edge fit result, which gives the electron energy $E$ in the photointensity $P(E,\eta)$. Next, the user selects a region from the band map to be used for MDC fitting -- with discrete energies indexed as $E_j$ -- and selects a set of linear/curved dispersions indexed with $n$ to be fitted in this region. When using parabolic non-interacting dispersions, the locations of the band extrema must also be provided, either as an angle $\eta_n^{\mathrm{c}}$ or wavevector $k_n^{\mathrm{c}}$. Next, the code fits the MDCs to capture the band map information in terms of the dimensionless peak positions $\widetilde{r}_n(E_j)$ and peak widths $\widetilde{\gamma}_n(E_j)$, where the tilde (\kern0.4em$\widetilde{}$\kern0.4em) refers to fitted quantities. Afterwards, initial guesses (denoted with $0$) for the real ($\widetilde{\Sigma}_{n0}^{\prime}(E_j)$) and minus imaginary parts ($-\widetilde{\Sigma}_{n0}^{\prime\prime}(E_j)$) of the self-energy are calculated by merging these fitting results with initial guesses for the non-interacting band parameters, such as the Fermi wavevector $k_{n0}^{\mathrm{F}}$. If a photoemission kink is present in the phonon contribution to $\widetilde{\Sigma}_{n0}^{\prime}(E)$, the user may wish to extract the Eliashberg function $\alpha^2F_{n0}(\omega)$ with the Maximum-Entropy Method (MEM) implemented in \textsc{xARPES}. In the one-shot mode, extraction of $\alpha^2F_{n0}(\omega)$ requires the user to assign the magnitude of the electron-electron coupling coefficient $\lambda_{n0}^{\mathrm{el}}$ and minus the imaginary part of the impurity contribution to the electron self-energy $\Gamma_{n0}^{\mathrm{imp}}$. After obtaining the one-shot results with the initial guesses for the model parameters -- including the non-interacting band parameters and the electron and impurity contributions -- the Bayesian inference loop can be called for a full optimization of the parameters. This full optimization also results in the final $\alpha^2F_n(\omega)$, from which $\Sigma_{n}(E)$ can be calculated at every $E$. Optionally, the results can be improved by repeating the above code steps for various $k_n^{\mathrm{c}}$, followed by choosing the most probable result from the individual optimizations. Finally, we remark that while $\alpha^2F_{n0}(\omega)$ is traditionally extracted using only $\widetilde{\Sigma}_{n0}^{\prime}(E_j)$, \textsc{xARPES} provides the possibility of employing $\widetilde{\Sigma}_{n0}^{\prime}(E_j)$, $-\widetilde{\Sigma}_{n0}^{\prime\prime}(E_j)$, or both.

\subsection{Extraction of the self-energy}\label{sec:se_extraction}

We consider photoemission kinks of electronic bands that are sufficiently far from other bands to ignore band mixing~\cite{forster:2016}, such that they can be described as individual branches $n$, starting out with Eqs.~\eqref{eq:AnEk} and~\eqref{eq:photointensity}. The MDC fitting is performed in $\eta$-space instead of wavevector space, because the angular resolution of the detector $\Delta \eta$ is approximately constant as a function of $\eta$. A wavevector-based fitting will be implemented in a future version of \textsc{xARPES}, which will be useful when the desired momentum-space path is poorly parameterized by a single angle, such as for cuts through 3D data sets. The photoelectron wavevector $\vec{p}(E,\eta)$ is mapped to the crystal wavevector $\vec{k}(E,\eta)$ in the first Brillouin zone. This is accurate for 2D states, which have no dispersion along $\vec{k}_{\perp}$, but ignores a broadening in $\vec{k}_{\perp}$ for 3D states. The MDCs are created as different slices of the band map, after which the branches labeled with $n$ are fitted on the selected angular range, leading to extraction of $\Sigma_n(E_j)$ from MDC fits at selected energies $E_j$, with $j$ the MDC index. The momentum dependence of the self-energy is ignored during the MDC fitting, resulting in a fitted self-energy $\Sigma_n(E)$. Therefore, the non-interacting band $\varepsilon_n(\vec{k})$ is the only momentum-dependent quantity during the fitting. As a consequence, any purely static, real-valued self-energy contributions $\Sigma_n^{\prime}(\vec{k})$ from the true self-energy are likely captured by $\varepsilon_n(\vec{k})$ instead of $\Sigma_n^{\prime}(E)$. If the MDC presents a sufficiently sharp single peak at wavevector $\vec{k}_n(E_j)$, the fitting is mostly sensitive to the self-energy in the vicinity of $\vec{k}_n(E_j)$. The solutions $\Sigma_n(E_j)$ should then be regarded as an experimental determination of the self-energies $\Sigma_n(E_j,\vec{k}_n(E_j))$~\cite{verga:2003, li:2021b}. Thus, when comparing to theoretical calculations (calc), the experimental $\varepsilon_n(\vec{k})$ should be compared to the sum of the non-interacting dispersion and static self-energy $\varepsilon_n^{\mathrm{calc}}(\vec{k})+\Sigma_n^{\mathrm{calc}\prime}(\vec{k})$, while $\Sigma_n(E)$ should be compared to the dynamical $\Sigma_n^{\mathrm{calc}}(E,\vec{k})$ evaluated at the momentum $\vec{k}$ corresponding to the quasiparticle energy $E_n^{\mathrm{calc}}(\mathbf{k})$ closest to $E$.

We start from the photointensity in Eq.~\eqref{eq:photointensity} with fixed $\vecg{\epsilon}$ and $h\nu$, where $n$ is now a branch index, and assume that the PME depends more strongly on $\eta$ than on $E$, which results in $|M_n(E,\eta; h\nu, \vecg{\epsilon})|^2 \rightarrow |M_n(\eta)|^2$, which is a dimensionless quantity that the user can set manually. We refer to the use of a non-constant $|M_n(\eta)|^2$ during the MDC fitting as the matrix-element correction (MEC). For the MDC $j$, we approximate the $E_{j}^{\mathrm{kin}}$-referenced photointensity $P^{*}(E_j^{\mathrm{kin}}, \eta)$ from Eq.~\eqref{eq:photointensity} with an expression $\widetilde{P}(E_j^{\mathrm{kin}}, \eta)$ for which the energy convolution is neglected:
\begin{multline}\label{eq:fit1}
	\widetilde{P}(E_j^{\mathrm{kin}}, \eta) = \int_{-\infty}^{\infty}\dd\eta'\,Q(\eta-\eta') \Bigg\{
	\widetilde{B}(E_j^{\mathrm{kin}}, \eta') +\sum_{n} |M_n(\eta')|^2 \\
	\times\frac{\widetilde{\mathcal{A}}_n^0(E_j)}{\pi}\frac{-\widetilde{\Sigma}_n^{\prime\prime}(E_j)E_{\mathrm{dim}}}
	{\big[E_j-\varepsilon_n(\eta')-\widetilde{\Sigma}_n^{\prime}(E_j)\big]^2
	+\big[\widetilde{\Sigma}_n^{\prime\prime}(E_j)\big]^2} \Bigg\},
\end{multline}
where $E_{\mathrm{dim}}$ is constant with dimensions of an energy and $\widetilde{B}(E_j^{\mathrm{kin}}, \eta)$ is a user-defined polynomial in $\eta$, which is fitted for each $E_j^{\mathrm{kin}}$. In Eq.~\eqref{eq:fit1}, $\widetilde{\mathcal{A}}_n^{0}(E_j)$ and $\widetilde{B}(E_j^{\mathrm{kin}},\eta)$ have units of photointensity, while the relation $E_j^{\mathrm{kin}}=h\nu - \Phi + E_j - \mu$ can be established after $h\nu-\Phi$ has been determined from the Fermi-edge fit. Interestingly, a convenient rewriting allows for eliminating two of the $l+1$ fitting parameters of a non-interacting band described by a polynomial of order $l$ during the fitting procedure. Consequently, a parabolic dispersion can be fitted with one parameter, which in \textsc{xARPES} is either the wavevector of the extremum $k_n^{\mathrm{c}}$ or the corresponding angle $\eta_n^{\mathrm{c}}(E)$ at $E=\mu$, with no parameters needed for a linear non-interacting dispersion. As an example, we perform the convenient rewriting for a parabolic dispersion, with example analyses provided in Secs.~\ref{sec:mock} and~\ref{sec:sto}. The implementation works for electron-like (positive mass) as well as hole-like (negative mass) bands. The code also supports the linear case, which is presented in Sec.~\ref{SI:linearized_dispersion}.

Here -- and in the first release of \textsc{xARPES} -- we assume that the parallel momentum $p_{\parallel}$ of the photoelectron is related to the detector angle $\eta$ as:
\begin{equation}\label{eq:path}
	p_{\parallel} = \sqrt{2m_{\mathrm{e}}E^{\mathrm{kin}}/\hbar^2}\sin(\eta + \phi).
\end{equation}
The parallel component in Eq.~\eqref{eq:path} can be obtained from the $x$-component in Eq.~\eqref{eq:kinematic_relations} by using a possible rotation around the $z$-axis, and by subtraction of the angle $\phi$, after which we denote $\eta + \phi \rightarrow \eta$ for brevity. Thus, Eq.~\eqref{eq:path} describes detection along all planes that contain the normal vector of the material, with more complicated expressions for $p_{\parallel}$ scheduled for future releases of \textsc{xARPES}. A generic parabolic dispersion along this path may be written as:
\begin{equation}\label{eq:curved}
	\varepsilon_n(\eta) =\mu+\frac{m_{\mathrm{e}}}{m_n^{\mathrm{b}}}E^{\mathrm{kin}}
	\left\{[\sin(\eta)-\sin(\eta_n^{\mathrm{c}})]^2-\sin^2(\eta_n^{\mathrm{F}}) \right\},
\end{equation}
where $m_n^{\mathrm{b}}$ is the mass of the non-interacting band, and the angles $\eta_n^{\mathrm{c}}$ and $\eta_n^{\mathrm{F}}$ are related to the projected centers $k_n^{\mathrm{c}}$ and Fermi wavevectors $k_n^{\mathrm{F}}$ of the parabola by $\sin^2(\eta_n^{\mathrm{c}})=({\hbar k_n^{\mathrm{c}}})^2/(2m_{\mathrm{e}}E^{\mathrm{kin}})$ and $\sin^2(\eta_n^{\mathrm{F}})=(\hbar {k_n^{\mathrm{F}}})^2/(2m_{\mathrm{e}}E^{\mathrm{kin}})$, respectively. The branch label $n$ takes into account the momentum path of Eq.~\eqref{eq:path} such that the band parameters $k_n^{\mathrm{F}}$, $m_n^{\mathrm{b}}$, and $k_n^{\mathrm{c}}$ are projected onto the momentum path, and the same underlying band may result in different projected parameters along a different path. With $\varepsilon_n(\eta)$ given by Eq.~\eqref{eq:curved}, Eq.~\eqref{eq:fit1} becomes:
\begin{multline}\label{eq:fit2}
	\widetilde{P}(E_j^{\mathrm{kin}}, \eta) = \int_{-\infty}^{\infty}\dd\eta'\,Q(\eta-\eta') \Bigg\{
	\widetilde{B}(E_j^{\mathrm{kin}}, \eta') + \sum_{n} |M_n(\eta')|^2 \\
	\times\frac{\widetilde{A}_n^{0}(E_j)}{\pi}\frac{\widetilde{\gamma}_n(E_j)}
	{\big\{\big[\sin(\eta')-\sin(\eta_n^{\mathrm{c}}(E_j))\big]^2-\widetilde{r\,}_n^2(E_j)\big\}^2
	+\widetilde{\gamma}_n^2(E_j)} \Bigg\},\\[-2em]
\end{multline}
after which the self-energy data can be computed via:
\begin{align}
	\label{eq:reobtain_real}
	\widetilde{\Sigma}_n^{\prime}(E_j) &= E_j-\mu-\frac{m_{\mathrm{e}}}
	{m_n^{\mathrm{b}}} E_j^{\mathrm{kin}} \left[\widetilde{r\,}_n^2(E_j) - \sin^2(\eta_n^{\mathrm{F}}) \right],\\
	\label{eq:reobtain_imag}
	-\widetilde{\Sigma}_n^{\prime\prime}(E_j) &= \frac{m_{\mathrm{e}}}{|m_n^{\mathrm{b}}|}
	E_j^{\mathrm{kin}} \,\widetilde{\gamma}_n(E_j),
\end{align}
where $\widetilde{\gamma}_n(E_j)$ is a dimensionless broadening parameter, and $\widetilde{r}_n(E_j)$ is a dimensionless peak maximum relative to $\sin(\eta_n^{\mathrm{c}}(E_j))$. Furthermore, the prefactors in Eqs.~\eqref{eq:fit1} and~\eqref{eq:fit2} are related by $\widetilde{\mathcal{A}}_n^0(E_j) = \widetilde{A}_n^0(E_j) m_{\mathrm{e}} E_j^{\mathrm{kin}}/(|m_n^{\mathrm{b}}| E_{\mathrm{dim}})$. The MDC maxima can then be recovered by assigning the fitted quantities to the left-hand or right-hand side of a parabola. Eq.~\eqref{eq:fit2} allows for fitting the MDCs with the dimensionless quantities $\widetilde{\gamma}_n(E_j)$ and $\widetilde{r}_n(E_j)$, while the fit no longer has to be performed with $m_n^{\mathrm{b}}$ and $k_n^{\mathrm{F}}$. The rewriting also simplifies finding a sufficiently good initial guess for the MDC fitting, as the angular distance between $\widetilde{r}_n(E_j)$ and $\sin(\eta_n^{\mathrm{c}}(E_j))$ can directly be visualized.

Once the fitting parameters $\{\widetilde{r}_{n}(E_j), \widetilde{\gamma}_{n}(E_j)\}$ have been obtained, they can be substituted into Eqs.~\eqref{eq:reobtain_real} and~\eqref{eq:reobtain_imag} to obtain the extracted self-energy $\widetilde{\Sigma}_n(E_j)$. In this process, $m_n^{\mathrm{b}}$ and $k_n^{\mathrm{F}}$ may either be provided by the user in the one-shot mode, or they can be optimized through the Bayesian inference feature of \textsc{xARPES} described in Sec.~\ref{sec:parameter_optimization}.

\subsection{Extraction of the Eliashberg function}\label{sec:eliashberg_extraction}

In this section, we describe how the Eliashberg function $\alpha^2F_n(\omega)$ can be extracted from $\widetilde{\Sigma}_n(E)$ for a given set of parameters, whose optimization is described in Sec.~\ref{sec:parameter_optimization}. Considering phonons as the only type of coupling bosons, we assume validity of Matthiessen's rule~\cite{matthiessen:1864, kemper:2018} which implies that $\Sigma_n(E)$ can be decomposed into the following contributions:
\begin{equation}\label{eq:matthiessen}
	\Sigma_n(E)=\Sigma_n^{\mathrm{ph}}(E) + \Sigma_n^{\mathrm{el}}(E) + \Sigma_n^{\mathrm{imp}}(E).
\end{equation}
Matthiessen's rule implies that mixed contributions, as for example an electron propagator renormalized by electron-electron interactions in the lowest-order electron-phonon coupling diagram~\cite{abramovitch:2023}, are negligible. The rule applies if all couplings are in the perturbative regime and if the self-energy is evaluated at leading order in each of them, since the mixed terms are of higher orders.

In \textsc{xARPES}, the impurity contribution to the electron self-energy $\Sigma_n^{\mathrm{imp}}(E)$ is fitted with an imaginary static decay rate term $-\ci \Gamma_n^{\mathrm{imp}}$. For $\Sigma_n^{\mathrm{el}}(E)$, it is desirable that it simultaneously satisfies (i) Fermi-liquid behavior: $\Sigma_n^{\mathrm{el}\prime}(E)=-\lambda_n^{\mathrm{el}}(E-\mu)$ and $-\Sigma_n^{\mathrm{el}\prime\prime}(E)\propto (E-\mu)^2+(\pi k_{\mathrm{B}}T)^2$ for small $E - \mu$~\cite{chubukov:2012}, (ii) particle-hole symmetry: $\Sigma(E+\mu)=-\overline{\Sigma}(-E+\mu)$, (iii) Kramers-Kronig consistency~\cite{berthod:2018a}, and (iv) $|E|^{-2}$-decay for large $|E|$ to avoid an ultraviolet divergence~\cite{chubukov:2012}. These considerations yield the expression:
\begin{multline}\label{eq:electron_power_4}
	\frac{\Sigma^{\mathrm{el}}_n(E)}{W_n} = \frac{\lambda_n^{\mathrm{el}}}{\left(1-\bar{T}_n^2\right) \left(1+\bar{E}_n^4\right)}\\ 
	\times\left\{\bar{E}_n\left[\bar{T}_n^2(1+\bar{E}_n^2)-(1-\bar{E}_n^2)\right]-\ci\sqrt{2}(\bar{E}_n^2+\bar{T}_n^2)\right\},
\end{multline}
where $\bar{E}_n=(E-\mu)/W_n$ and $\bar{T}_n=\pi k_{\mathrm{B}}T/W_n$ are energy ratios, with $W_n$ an ultraviolet scale that the user must specify (see Sec.~\ref{SI:eliashberg_details}). The derivation of Eq.~\eqref{eq:electron_power_4} is given in Sec.~\ref{SI:eliashberg_details}, where an alternative expression for $\Sigma^{\mathrm{el}}_n(E)$ (also implemented in \textsc{xARPES}) is provided. The Fermi-liquid behavior imposed on Eq.~\eqref{eq:electron_power_4} has relatively wide applicability, since it is obtained for various theoretical models, as well as observed experimentally in several systems, as detailed in Sec.~\ref{SI:eliashberg_details}.

Finally, we discuss $\Sigma_n^{\mathrm{ph}}(E)$ and its relation to $\alpha^2F_n(\omega)$. In the problem of Bloch electrons coupled to non-interacting phonons, the finite-temperature perturbation theory in the atomic displacements for the retarded self-energy yields at the lowest order in the reciprocal of the atomic mass, beside $E$-independent terms contributing to the static self-energy~\cite{allen:1976, lafuente-bartolome:2022a}, the dynamical Fan-Migdal term~\cite{fan:1951, migdal:1958, allen:1978}:
\begin{multline}\label{eq:self_one}
	\Sigma^{\mathrm{FM}}_n(E,\vec{k})=\int\frac{\dd\vec{q}}{\Omega^{\mathrm{BZ}}}
	\sum_{m\nu\pm}\left|g_{mn\nu}(\vec{k},\vec{q})\right|^2 \\
	\times \frac{f^{\pm}\left(\varepsilon_m(\vec{k+q})\right)+n\left(\omega_{\nu}(\vec{q})\right)}
	{E\pm \omega_{\nu}(\vec{q})-\varepsilon_m(\vec{k+q})+\ci 0^{+}},
\end{multline}
where $f^{+}(\varepsilon)\equiv f(\varepsilon)$, $f^{-}(\varepsilon) \equiv 1 - f(\varepsilon)$, $n(\omega) \equiv [\ee^{\omega/(k_{\mathrm{B}}T)}-1]^{-1}$ is the Bose-Einstein distribution, $\Omega^{\mathrm{BZ}}$ the Brillouin-zone volume, $g_{mn\nu}(\mathbf{k},\mathbf{q})$ is the EPC matrix element between electronic states $\varepsilon_n(\mathbf{k})$ and $\varepsilon_m(\mathbf{k}+\mathbf{q})$ coupled by a phonon of band $\nu$, wavevector $\mathbf{q}$, and energy $\omega_{\nu}(\vec{q})$, while $0^{+}$ is a positive infinitesimal. Eq.~\eqref{eq:self_one} can equivalently be written with the following two expressions:
\begin{align}
	\label{eq:self_two}
	\Sigma^{\mathrm{FM}}_n(E,\vec{k})&=
	\int_0^{\infty}\dd \omega\int_{-\infty}^{\infty}\dd\varepsilon\,
	\alpha^2F_n(\omega,\varepsilon,\vec{k}) \nonumber \\
	&\quad \times \sum_{\pm}\frac{f^{\pm}(\varepsilon) + n(\omega)}{E\pm \omega-\varepsilon+\ci 0^{+}},\\
	\label{eq:a2F_first}
	\alpha^2F_n(\omega,\varepsilon,\vec{k})&=
	\int\frac{\dd\vec{q}}{\Omega^{\mathrm{BZ}}}
	\sum_{m\nu}\left|g_{mn\nu}(\vec{k},\vec{q})\right|^2\delta(\omega-\omega_{\nu}(\vec{q})) \nonumber \\
	&\quad \times \delta(\varepsilon-\varepsilon_m(\vec{k+q})),
\end{align}
where we stress that $\omega$ is a vibrational energy, not a frequency. Eq.~\eqref{eq:self_two} can also be derived with a definition of $\alpha^2F_n(\omega,\varepsilon,\vec{k})$ that contains the full instead of the non-interacting phonon spectral function~\cite{allen:1983, allen:1974a, mahan:2000}, which includes a subset of the higher-order terms in the reciprocal atomic mass.

In ARPES experiments, we are mostly interested in $\Sigma^{\mathrm{FM}}_n(E,\vec{k})$ when $E$ remains close to $\mu$. The denominator in Eq.~\eqref{eq:self_two} shows that the relevant energies $\varepsilon$ are close to $E\pm \omega$. Phonon energies are typically a few tens of meV, which means that the $\varepsilon$ integral in Eq.~\eqref{eq:self_two} probes $\alpha^2F_n(\omega,\varepsilon,\vec{k})$ within no more than a few hundred meV around $\varepsilon=\mu$. If electronic energy scales are large compared to phonon energies, as occurs when studying photoemission kinks for bands $\varepsilon_n(\vec{k})$ that disperse much more steeply than the phonon dispersion $\omega_{\nu}(\vec{q})$, one expects that the variation of $\alpha^2F_n(\omega,\varepsilon,\vec{k})$ over the relevant $\varepsilon$ range is weak in comparison with the variation of the energy denominator in Eq.~\eqref{eq:self_two}. Under these conditions, one can retain the first term of the Taylor expansion of $\alpha^2F_n(\omega,\varepsilon,\vec{k})$ around $\varepsilon=\mu$~\cite{eiguren:2008}. Similarly, the wavevector dependence is weak if the electrons disperse much more steeply than the phonons, such that $\alpha^2F_n(\omega,\mu,\vec{k})$ can be evaluated at the Fermi wavevector $\vec{k}_n^{\mathrm{F}}$ of the photoemission kink being analyzed. Finally, contributions to $\Sigma_n^{\mathrm{ph}}(E)$ from higher-order dynamical terms are not captured independently with our formalism, while the $E$-independent terms should be captured in $\varepsilon_n(\mathbf{k})$ during the fitting, leading to $\Sigma_n^{\mathrm{ph}}(E)=\Sigma_n^{\mathrm{FM}}(E)$. These considerations lead to:
\begin{align}
	\label{eq:self_three}
	\Sigma_n^{\mathrm{FM}}(E) &= \int_0^{\infty}\dd\omega\,\alpha^2F_n(\omega)K(E,\omega), \\
	\nonumber
	K(E,\omega) &= \int_{-\infty}^{\infty}\dd\varepsilon\,\sum_{\pm}\frac{f^{\pm}(\varepsilon)+n(\omega)}
	{E\pm\omega-\varepsilon+\ci 0^{+}} \\
	\nonumber
	&= \Psi\left(\frac{1}{2}-\ci\frac{E-\mu-\omega}{2\pi k_{\mathrm{B}}T}\right)
	-\Psi\left(\frac{1}{2}-\ci\frac{E-\mu+\omega}{2\pi k_{\mathrm{B}}T}\right) \\
	\label{eq:kernel_function}
	&\quad -\ci\pi[2n(\omega)+1], 
\end{align}
where $K(E,\omega)$ is a kernel function~\cite{eiguren:2009, allen:1983} and $\Psi$ the digamma function. In the limit of infinitely large electronic energy scales over phonon energies and no higher-order terms, the extracted $\alpha^2F_n(\omega)$ may coincide with $\alpha^2F_n(\omega,\mu,\vec{k}_n^{\mathrm{F}})$, which we denote as the ``Fermi-surface Eliashberg function'' as all electronic scales are evaluated at the Fermi surface. When these conditions are not perfectly met, the extracted $\alpha^2F_n(\omega)$ will represent some mixture of $\alpha^2F_n(\omega,\varepsilon,\vec{k})$ for different $\varepsilon$ and $\mathbf{k}$, as well as higher-order contributions. The workflow uses the experimentally acquired $\{\widetilde{\Sigma}_n^{\mathrm{FM}}(E_j)\}$ to obtain $\alpha^2F_n(\omega)$ via kernel inversion of Eq.~\eqref{eq:self_three} as an estimate of the true $\alpha^2F_n^{\mathrm{true}}(\omega)$. Here, $\alpha^2F_n^{\mathrm{true}}(\omega)$ is the quantity that is recovered from Eq.~\eqref{eq:self_three} for of an infinite amount of unbiased data. Subsequently, $\Sigma_n^{\mathrm{FM}}(E)$ is obtained from $\alpha^2F_n(\omega)$ through Eq.~\eqref{eq:self_three}. However, direct inversion of Eq.~\eqref{eq:self_three} can result in negative values~\cite{wen:2018} or spurious behavior of $\alpha^2F_n(\omega)$ at low/high energies~\cite{rudin:1998}. The function $\alpha^2F_n(\omega)$ should be positive semidefinite for $\omega \in [0, \omega_n^{\mathrm{max}}]$ with $\omega_n^{\mathrm{max}}$ a maximum frequency, while $\alpha^2F_n(\omega)$ is zero outside this interval. This positive semidefiniteness can be encoded as prior knowledge in a regularization term $a S$, where $a$ is a Lagrange multiplier or hyperparameter, and $S$ is the generalized Shannon-Jaynes information entropy~\cite{gull:1989}:
\begin{multline}\label{eq:information_entropy}
	S(\alpha^2F_n, m_n) = \frac{1}{E_{\mathrm{dim}}} \int_{0}^{\infty}\dd\omega \\
	\times\left[\alpha^2 F_n(\omega)-m_n(\omega)-\alpha^2F_n(\omega)\ln\left(\frac{\alpha^2 F_n(\omega)}{m_n(\omega)} \right)\right],
\end{multline}
where $m_n(\omega)$ is a model function of maximum height $h_n$ that encodes the prior knowledge on $\alpha^2F_n(\omega)$. Additional details on $S$ and $m_n(\omega)$ are provided in Sec.~\ref{SI:eliashberg_details}, where we also provide the definition of the normalized Euclidean distance measure $M(\alpha^2F_{1}(\omega), \alpha^2F_{2}(\omega))$ for the comparison of two Eliashberg functions $\alpha^2F_{1}(\omega)$ and $\alpha^2F_{2}(\omega)$.

After inclusion of $aS$ as the log-prior, the most probable $\alpha^2F_n(\omega)$ is obtained in the MEM from the maximization of the log-posterior $L+aS$~\cite{jarrell:1996}:
\begin{equation}\label{eq:objective}
	\alpha^2F_n(\omega) = \mathop{\mathrm{arg\,max}}_{\alpha^2 F_n^{\bullet}(\omega)}(L + a S),
\end{equation} 
where $L$ is the log-likelihood after rendering the likelihood dimensionless, and $\alpha^2 F^{\bullet}_n(\omega)$ is the argument of the log-posterior optimization. In \textsc{xARPES}, the maximization of Eq.~\eqref{eq:objective} is performed using Bryan's algorithm~\cite{bryan:1990}. Furthermore, $\alpha^2F_n(\omega)$ can be extracted using $\widetilde{\Sigma}_n^{\prime}(E)$, $-\widetilde{\Sigma}_n^{\prime\prime}(E)$, or both. While $\alpha^2F_n(\omega)$ is commonly extracted using only $\widetilde{\Sigma}_n^{\prime}(E)$~\cite{shi:2004, bok:2010}, simultaneous incorporation of $-\widetilde{\Sigma}_n^{\prime\prime}(E)$ may lead to a better extraction. In that case, $L$ becomes:
\begin{multline}\label{eq:log-likelihood}
	L =-\frac{1}{2}\sum_{j=1}^{N_J}\Bigg[\left(\frac{\Sigma_n^{\prime}(E_j)-\widetilde{\Sigma}_n^{\prime}(E_j)}
	{\sigma_{n}^{\prime}(E_j)}\right)^2 + \ln\left(\frac{\sigma_n^{\prime}(E_j)}{E_{\mathrm{dim}}}\right) \\
	\kern-2em+\left(\frac{\Sigma_n^{\prime\prime}(E_j)-\widetilde{\Sigma}_n^{\prime\prime}(E_j)}
	{\sigma_{n}^{\prime\prime}(E_j)}\right)^2 
	+\ln\left(\frac{\sigma_n^{\prime\prime}(E_j)}{E_{\mathrm{dim}}}\right)\Bigg] - N_{J} \ln (2 \pi),
\end{multline}
where $\sigma_n^{\prime}(E_j)$ and $\sigma_n^{\prime\prime}(E_j)$ are the standard deviations from the MDC fitting of the respective $\widetilde{\Sigma}_n^{\prime}(E_j)$ and $\widetilde{\Sigma}_n^{\prime\prime}(E_j)$, whereas $\Sigma_n(E)$ is calculated from Eq.~\eqref{eq:matthiessen}, with $\Sigma_n^{\mathrm{ph}}(E)$ determined using Eq.~\eqref{eq:self_three}. Several approaches exist in the MEM community to determine $a$, including the historic method~\cite{jarrell:1996}, the classic method~\cite{jarrell:1996}, and Bryan's method~\cite{bryan:1990}. Recent approaches suggest determining $a$ by a transition from noise fitting to information fitting of the data upon increasing $a$~\cite{kaufmann:2023}.

Interestingly, there is a ``tail'' or ``upturn''~\cite{levy:2014, plumb:2010} in $\Sigma_n^{\prime}(E)$ near $E = \mu$ from the interplay of the FWHM energy resolution $\Delta E$ with a strong decrease in $P(E, \eta)$ through $f(E)$. Consequently, we exclude $\widetilde{\Sigma}_n(E)$ for $\mu - E < \Delta E$ during the inversion of Eq.~\eqref{eq:self_three}. Furthermore, optimization of $\Sigma_n^{\mathrm{el}}(E)$ and $\Sigma_n^{\mathrm{imp}}(E)$ as described in Sec.~\ref{sec:parameter_optimization} is only available when the full $\widetilde{\Sigma}_n(E)$ is used in $L$, as $\widetilde{\Sigma}_n^{\prime}(E)$ by itself was found to give insufficient information to optimize these terms.

Finally, the EPC strength $\lambda_n^{\mathrm{ph}} \equiv -\partial \Sigma_n^{\mathrm{ph}\prime}(E)/\partial E |_{E=\mu}$ is commonly determined by a linear fit through $\widetilde{\Sigma}_n^{\prime}(E)$ near the Fermi edge~\cite{jiang:2009}. However, this evaluation inconveniently coincides with the upturn for $E \approx \mu$. We observe that $\alpha^2F_n(\omega) K(E, \omega)$ is continuous over the domain of $\omega$-integration, so that Leibniz' integral rule can be applied. Combining $\lambda_n^{\mathrm{ph}}=-\partial \Sigma_n^{\mathrm{FM}\prime}(E)/\partial E |_{E=\mu}$ with Eq.~\eqref{eq:self_three} then yields:
\begin{multline}\label{eq:lambda_derivative}
	\lambda_n^{\mathrm{ph}} = \frac{1}{2\pi k_{\mathrm{B}}T}\int_0^{\infty}\dd\omega\,\alpha^2F_n(\omega) \\
	\times\mathrm{Im}\left[\Psi_1\left(\frac{1}{2}-\frac{\ci\omega}{2\pi k_{\mathrm{B}}T}\right)
	-\Psi_1\left(\frac{1}{2}+\frac{\ci\omega}{2\pi k_{\mathrm{B}}T}\right)\right],
\end{multline}
where $\Psi_1$ is the trigamma function. Instead of fitting near the upturn, $\lambda_n^{\mathrm{ph}}$ in \textsc{xARPES} is based on all the $\Sigma_n(E_j)$ for which $\mu - E_j > \Delta E$ through Eq.~\eqref{eq:lambda_derivative}. We remark that $\lambda_n \equiv - \partial \Sigma_n^{\prime}(E)/ \partial E |_{E=\mu}=\lambda_n^{\mathrm{ph}}+\lambda_n^{\mathrm{el}}$ because $\Sigma_n^{\mathrm{imp}}=-\ci \Gamma_n^{\mathrm{imp}}$ is used.

\subsection{Model parameter optimization}\label{sec:parameter_optimization}

A key problem in the parameterization of $A_n(E,\mathbf{k})$ is determining the magnitude of $\Sigma_n^{\mathrm{ph}}(E)$, $\Sigma_n^{\mathrm{el}}(E)$, $\Sigma_n^{\mathrm{imp}}(E)$, and the coefficients that represent $\varepsilon_n(\vec{k})$. While the KKBF can be used to distinguish between $\varepsilon_n(\vec{k})$ and $\Sigma_n(E)$, the decomposition of the latter into its constituents is usually still performed by visual inspection~\cite{zhou:2005} instead of using a quantitative approach. Here, we extend the probabilistic procedure for extracting $\alpha^2F_n(\omega)$ with Bayesian inference to obtain a quantitative procedure for obtaining the model parameters, whose set we denote with $V$. Therefore, $V$ includes $m_{n}^{\mathrm{b}}$ or $v_{n}^{\mathrm{F}}$, $k_{n}^{\mathrm{F}}$, $\Gamma_{n}^{\mathrm{imp}}$, $\lambda_{n}^{\mathrm{el}}$, and $h_{n}$, where inclusion of the latter implies that the shape of $m_n(\omega)$ is considered known, but its height is not.

We denote with $D$ a set of self-energy data, comprising $\{\widetilde{\Sigma}_n^{\prime}(E)\}$, $\{-\widetilde{\Sigma}_n^{\prime\prime}(E)\}$, or $\{\widetilde{\Sigma}_n^{\prime}(E), -\widetilde{\Sigma}_n^{\prime\prime}(E)\}$. The posterior probability density $p(\alpha^2F_n(\omega) | D, a, m_n(\omega))$ over $\alpha^2F_n(\omega)$ for given $D$, $a$, and model function $m_n(\omega)$ can be expressed as~\cite{gull:1989, bryan:1990}:
\begin{equation}\label{eq:a2f_prob}
	p\left(\alpha^2F_n(\omega) | D, a, m_n(\omega) \right) = \frac{\ee^{L+aS}}{Z^S(a)Z^L(D)},
\end{equation}
where $Z^S(a)$ and $Z^L(D)$ are normalization factors over the respective $S$ and $L$, and where Eq.~\eqref{eq:objective} is equivalent to maximizing the logarithm of Eq.~\eqref{eq:a2f_prob}. It may be recognized in Eq.~\eqref{eq:log-likelihood} that $L$ is not just a function of $D$, but also a function of $V$. Thus, after realizing that $Z^{L}(D)\rightarrow Z^{L}(D, V)$, the expression in Eq.~\eqref{eq:a2f_prob} may be recognized as $p\left(\alpha^2F_n(\omega) | V, D, a, m_n(\omega)\right)$. Applying Bayes' rule~\cite{gull:1989} to Eq.~\eqref{eq:a2f_prob} then yields:
\begin{multline}\label{eq:parameter_prob}
	p\left(V | \alpha^2F_n(\omega), D, a, m_n(\omega) \right) = \\
	\frac{p\left(\alpha^2F_n(\omega)| V, D, a, m_n(\omega) \right)}{p\left(D, \alpha^2F_n(\omega) \right)} p(V, D),
\end{multline}
where the evidence $p\left(D, \alpha^2F_n(\omega) \right)$ is a normalization factor that is constant during determination of $p\left(V | \alpha^2F_n(\omega), D, a, m_n(\omega) \right)$, and $p(V, D)$ contains the prior probabilities over $V$ and $D$. Thus, Eq.~\eqref{eq:parameter_prob} provides a quantitative criterion to determine the most probable $V$ for a given $\alpha^2F_n(\omega)$. The \textsc{xARPES} code supports uniform probability distributions over $V$ in $p(V, D)$, although different expressions may be implemented in the future, such that previous experimental/theoretical knowledge can be incorporated for subsequent data sets. Iterative optimization of Eqs.~\eqref{eq:a2f_prob} and~\eqref{eq:parameter_prob} constitutes the outer loop in Fig.~\ref{fig:workflow}. First, $\alpha^2F_n(\omega)$ is determined for a given $V$, after which $V$ is determined for the updated $\alpha^2F_n(\omega)$, until the change of $p\left(V | \alpha^2F_n(\omega), D, a, m_n(\omega) \right)$ is below a given threshold, see Sec.~\ref{sec:methods}. The Bayesian procedure described here can be generalized to other inversion problems in which the data $D$ themselves depend on unknown parameters $V$.

\subsection{Introduction to the model system and use cases}\label{sec:examples}

In the following three sections, we showcase the capabilities of \textsc{xARPES} by studying a model system and two use cases. In subsection~\ref{sec:mock}, we use artificial data to demonstrate that \textsc{xARPES} recovers 95\% overlap of the 95\% confidence intervals of $\widetilde{\Sigma}(E)$ with the true $\Sigma(E)$, for an energy resolution $\Delta E \rightarrow 0$. We then compare our approach with a frequently used Lorentzian fitting approach~\cite{shi:2004,king:2014, ludbrook:2015, huempfner:2023}, which performs increasingly poorly towards higher binding energies. Subsequently, we show that the sharpness of recovered phonon modes is limited by $\Delta E$, and that the optimization loop recovers $k^{\mathrm{F,true}}$, $m^{\mathrm{b,true}}$, $\Gamma^{\mathrm{imp,true}}$, and $\lambda^{\mathrm{true}}=\lambda^{\mathrm{el,true}}+\lambda^{\mathrm{ph,true}}$ within about 5\% for realistic values. Finally, we show that $\alpha^2F(\omega)$ converges towards $\alpha^2F^{\mathrm{true}}(\omega)$ with increasing amounts of unbiased data.

The first use case concerns the 2DEL of the $d_{xy}$-derived bands of Nb-doped TiO$_2$-terminated SrTiO$_3$, showcasing extraction with a parabolic $\varepsilon_n(\mathbf{k})$ from experimental data, displaying high similarity between the Eliashberg functions from left/right branches of the inner $d_{xy}$-derived band. We additionally show that omission of $|M_n(\eta)|^2$ matrix elements can change the extracted $\Sigma(E)$ by over a factor of two. The second use case is on Li-doped graphene, showcasing the implementation for a linear $\varepsilon_n(\mathbf{k})$. We provide \textsc{Jupyter} notebooks for the three examples as described in Sec.~\ref{sec:code}. The SrTiO$_3$ and Li-doped graphene examples are remarkable due to PMEs $|M_n(\eta)|^2$ for which theoretical expressions exist and notably improve on the self-energy fitting. Their importance depends primarily on interplay between experimental geometry, light polarization, and orbital character~\cite{moser:2017}, and has to be evaluated on a case-by-case basis. Users interested in specifying $|M_n(\eta)|^2$ for fitting their data with \textsc{xARPES} may estimate the PMEs from their data with a heuristic approach~\cite{sokolovic:2025}, from tight-binding calculations separately~\cite{day:2019} or combined with Wannierization~\cite{schuler:2022}, from the so-called scattered-wave approximation~\cite{kern:2023}, or from Lippmann-Schwinger-based simulations~\cite{ryoo:2025}. PMEs are also naturally incorporated in one-step photoemission simulations~\cite{ebert:2011}, although separating their effects from $A_n(E,\mathbf{k})$ is not always straightforward from such calculations.

\subsection{Verification using model data}\label{sec:mock}

\begin{figure*}[t]
\includegraphics[width=\textwidth]{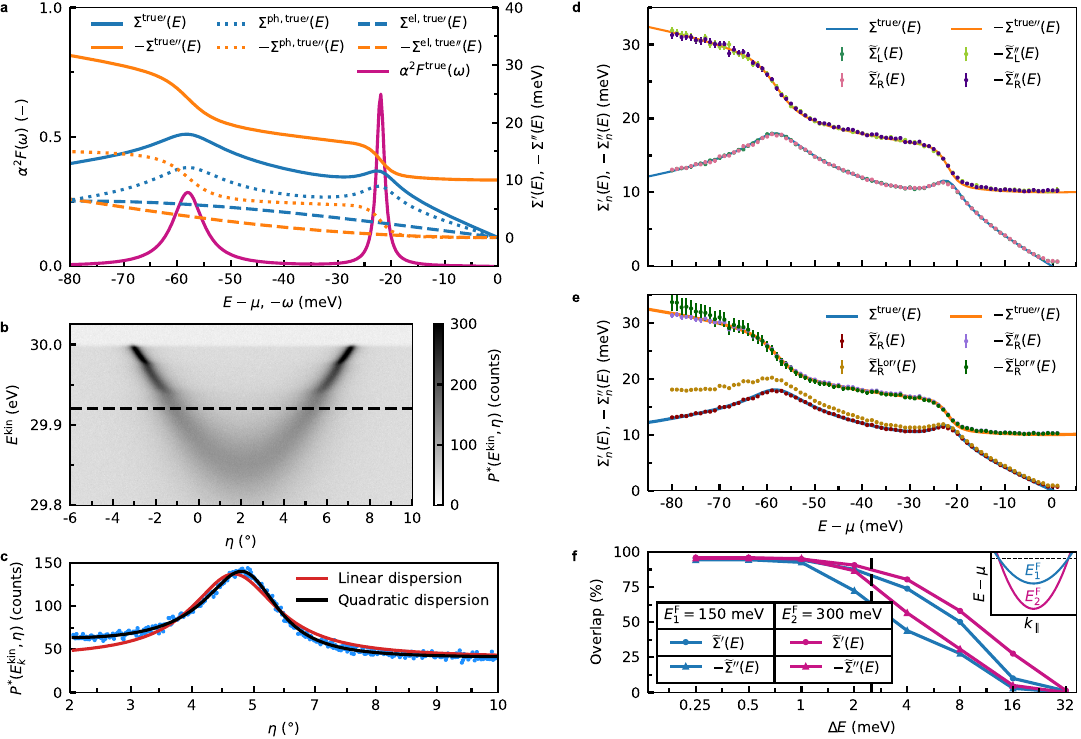}
\caption{\textbf{Verification band map and MDC analysis}.
\textbf{a} The total real $\Sigma^{\mathrm{true}\prime}(E)$ (blue) and minus imaginary parts $-\Sigma^{\mathrm{true}\prime\prime}(E)$ (orange), comprising $\Sigma^{\mathrm{ph,true}\prime}$ (dotted blue) and $-\Sigma^{\mathrm{ph,true}\prime\prime}$ (dotted orange) based on $\alpha^2F^{\mathrm{true}}(\omega)$ (magenta), the electron real $\Sigma^{\mathrm{el,true}\prime}$ (dashed blue) and minus imaginary parts $-\Sigma^{\mathrm{el,true}\prime\prime}$ (dashed orange), and the impurity term $\Sigma^{\mathrm{imp,true}}(E)=-\ci \Gamma^{\mathrm{imp,true}}$ (not shown). \textbf{b} Artificial photointensity $P^*(E^{\mathrm{kin}},\eta)$ for a parabolic dispersion $\varepsilon(\vec{k})$ displaced by $k^{\mathrm{c}}=0.1~\Angstrom^{-1}$, after adding the energy and angle convolutions, Gaussian noise, and $\Sigma^{\mathrm{true}}(E)$ from \textbf{a}. The dashed line corresponds to the selected binding energy $E^{\mathrm{bin}}=80$~meV in \textbf{c}. \textbf{c} MDC photointensity $P^{*}(E_{k}^{\mathrm{kin}},\eta)$ (blue dots), fitted with quadratic $\varepsilon(\vec{k})$ (black) and linear $\varepsilon(\vec{k})$ (red) non-interacting dispersions. \textbf{d} Reconstruction of $\Sigma^{\mathrm{true}\prime}(E)$ (blue) and $-\Sigma^{\mathrm{true}\prime\prime}(E)$ (orange) for the left-hand $\widetilde{\Sigma}_{\mathrm{L}}^{\prime}(E)$ (dark green) and $-\widetilde{\Sigma}_{\mathrm{L}}^{\prime\prime}(E)$ (lime) and the right-hand $\widetilde{\Sigma}_{\mathrm{R}}^{\prime}(E)$ (pink) and $-\widetilde{\Sigma}_{\mathrm{R}}^{\prime\prime}(E)$ (indigo). \textbf{e} Reconstruction at $\Delta E=0$ with $\widetilde{\Sigma}_{\mathrm{R}}^{\prime}(E)$ (maroon) and $-\widetilde{\Sigma}_{\mathrm{R}}^{\prime\prime}(E)$ (violet) from \textsc{xARPES}, compared to $\widetilde{\Sigma}_{\mathrm{R}}^{\mathrm{Lor}\prime}(E)$ (gold) and $-\widetilde{\Sigma}_{\mathrm{R}}^{\mathrm{Lor\prime\prime}}(E)$ (green) from the method described in the text. The confidence intervals of some $\widetilde{\Sigma}_{\mathrm{L,R}}^{\prime}(E)$ and $\widetilde{\Sigma}_{\mathrm{L,R}}^{\mathrm{Lor}\prime}(E)$ in \textbf{d}--\textbf{e} are smaller than the markers. \textbf{f} Bands with $E_{1}^{\mathrm{F}}=150$~meV (blue) and $E_{2}^{\mathrm{F}}=300$~meV (magenta) shown in the inset (different $m^{\mathrm{b}}$ but identical $k^{\mathrm{F}}$), which for a range of energy resolutions $\Delta E$ leads to the displayed overlap of 95\% confidence intervals against the energy resolution of $\widetilde{\Sigma}^{\prime}(E)$ with $\Sigma^{\mathrm{true}\prime}(E)$ (blue/magenta circles) and $-\widetilde{\Sigma}^{\prime\prime}(E)$ with $-\Sigma^{\mathrm{true}\prime\prime}(E)$ (blue/magenta triangles), with the dashed vertical line corresponding to $\Delta E$ used in \textbf{a}--\textbf{d}.
\label{fig:model}}
\end{figure*}

We analyze an artificial band map with Gaussian noise, $|M_n(\eta)|^2=1$, and a parabolic $\varepsilon_n(\vec{k})$ to verify our implementation and investigate how noise and energy resolution affect the extraction of $\Sigma_n(E)$ and the reconstruction of $\alpha^2F_n(\omega)$. We perform the self-energy analysis on the left-hand (L) and right-hand (R) branches of a parabolic dispersion, follow up with the extraction of $\alpha^2F_{\mathrm{R}}(\omega)$, and omit the branch index for symmetric quantities. The added noise is the sole source of difference for the extracted quantities. We generate $A(E,\vec{k})$ from a single parabolic $\varepsilon(\vec{k})$ at $T=10$~K with a known $\alpha^2F^{\mathrm{true}}(\omega)$ composed of a sum over two peaks with phonon energies $\omega_k \in \{22, 58\}$~meV, broadenings $\rho_k \in \{0.9, 3.5\}$~meV, and matrix elements $g_k^2 \in \{0.6, 1.0\}$~meV computed from:
\begin{equation}\label{eq:a2f_input}
	\alpha^2F^{\mathrm{true}}(\omega) = \sum_{k,\pm} g^2_k\mathrm{Im}\left(\frac{\pm1}{\omega\pm\omega_k+\ci\rho_k}\right).
\end{equation}
From $\alpha^2F^{\mathrm{true}}(\omega)$, we generate $\Sigma^{\mathrm{ph,true}}(E)=\Sigma^{\mathrm{FM,true}}(E)$ using Eq.~\eqref{eq:self_three}, to which we add $\Sigma^{\mathrm{el,true}}(E)$ using Eq.~\eqref{eq:electron_power_4} while setting $W=E^{\mathrm{F}}$ (the default for parabolic bands), where we define the Fermi energy as $E^{\mathrm{F}}\equiv (\hbar^2k^{\mathrm{F}})^2/(2m^{\mathrm{b}})$, and we add $\Sigma^{\mathrm{imp,true}}=-\ci \Gamma^{\mathrm{imp,true}}$. The resulting $\alpha^2F^{\mathrm{true}}(\omega)$ and $\Sigma^{\mathrm{true}}(E)$ are displayed in Fig.~\ref{fig:model}\textbf{a}. The parabolic dispersion is displaced by $k^{\mathrm{c}}=0.1~\Angstrom^{-1}$, while the values completing the description of $A(E,\vec{k})$ constitute the first row of Table~\ref{tab:parameters} and result in $E^{\mathrm{F}}=150$~meV.

\begin{table}[tb]
\vspace{-0.7em}\caption{True, optimized, and relative differences of parameters for the verification example.
\label{tab:parameters}}
	\renewcommand{\arraystretch}{1.1}
	\begin{tabular*}{\columnwidth}{@{\extracolsep{\fill}}lddddd}
		\hline\hline
		Parameter & 
		\multicolumn{1}{r}{$m_{\mathrm{R}}^{\mathrm{b}}$ ($m_{\mathrm{e}}$)} & 
		\multicolumn{1}{c}{$k_{\mathrm{R}}^{\mathrm{F}}$ ($\Angstrom^{-1}$)} & 
		\multicolumn{1}{c}{$\Gamma_{\mathrm{R}}^{\mathrm{imp}}$ (meV)} & 
		\multicolumn{1}{c}{$\lambda_{\mathrm{R}}^{\mathrm{el}}$ (--)} & 
		\multicolumn{1}{c}{$\lambda_{\mathrm{R}}^{\mathrm{ph}}$ (--)}\\[0.2em]
		\hline
		True      & 1.5875 & 0.25     & 10.0   & 0.12  & 0.28 \\
		Opt.      & 1.596  & 0.25001  & 10.04  & 0.113 & 0.30 \\
		Diff. \%  & 0.5    & 0.004    & 0.4    & -6    & 7 \\
		\hline\hline
	\end{tabular*}
\end{table}

We discretize the resulting $A(E,\vec{k})$ with $\vec{p}(E^{\mathrm{kin}},\eta)$ parameterized according to Eq.~\eqref{eq:path} with $N_J=80$ pairs of data ($\widetilde{\Sigma}_{\mathrm{R}}^{\prime}(E)$, $-\widetilde{\Sigma}_{\mathrm{R}}^{\prime\prime}(E)$) for realistic $E^{\mathrm{kin}} \in [29.8, 30.25]$~eV with the photon energy minus the work function $h\nu-\Phi=30$~eV and $0.02$\textdegree~angular steps for $\eta \in [-6, 10]$\textdegree, multiply with an energy-independent $\widetilde{\mathcal{A}}^{0}=10^{4}$ counts, and set the polynomial terms $\widetilde{B}_0/\widetilde{\mathcal{A}}^{0}=0.2$\% and $\widetilde{B}_1=0$ in Eq~\eqref{eq:fit1}. Next, we convolve with Gaussian distributions with FWHMs $\Delta E=2.5$~meV and $\Delta \eta=0.1$\textdegree~and add Gaussian noise $\mathcal{N}(\mu^{\mathrm{noise}}, (\sigma^{\mathrm{noise}})^2)$ at each $(E,\eta)$ with $\mu^{\mathrm{noise}}/\widetilde{\mathcal{A}}^{0}=0.2$\%, and noise-to-intensity ratio $\sigma^{\mathrm{noise}}/\widetilde{\mathcal{A}}^{0}=0.025$\%, obtaining the photointensity $P^*(E^{\mathrm{kin}}, \eta)$ displayed in Fig.~\ref{fig:model}\textbf{b}. We obtain the estimate $h\widehat{\nu}-\widehat{\Phi}$ from the Fermi-edge fitting and compute $P(E, \eta)$. We find $h\widehat{\nu}-\widehat{\Phi} =h\nu-\Phi - (0.06 \pm 0.03)$~meV. Next, we perform the MDC fits using a parabolic $\varepsilon(\vec{k})$, showing the fits at a selected binding energy $E^{\mathrm{bin}}=80$~meV in Fig.~\ref{fig:model}\textbf{c}, which corresponds to the dashed line in Fig.~\ref{fig:model}\textbf{b}, and compared to a fit with a linear $\varepsilon(\vec{k})$. We find that the quadratic dispersion fits the data better than the linear dispersion.

We remark that the Gaussian noise results in $\chi^2(E_{j}^{\mathrm{kin}})=\sum_{l}^{N_L} (P^{*}(E_{j}^{\mathrm{kin}}, \eta_l)-\widetilde{P}^{*}(E_j^{\mathrm{kin}}, \eta_l))^2/(\sigma^{\mathrm{noise}})^2$ with $N_L$ the number of angular data points. For $\Delta E \rightarrow 0$, we find the expected $\chi^2(E_j^{\mathrm{kin}}) \approx N_L$ for the parabolic $\varepsilon(\vec{k})$, versus $\chi^2(E_j^{\mathrm{kin}}) \approx 7.7$~$N_L$ for the linear dispersion, demonstrating that a Lorentzian MDC fitting of $A(E,\vec{k})$ yields biased results when the underlying $\varepsilon(\vec{k})$ is curved. We temporarily use the true $m^{\mathrm{b}}$ and $k^{\mathrm{F}}$ for the self-energy extraction step to highlight the best possible recovery of $\widetilde{\Sigma}_{\mathrm{L,R}}(E)$ in presence of a finite energy resolution (Fig.~\ref{fig:model}\textbf{d}) and to evaluate the performance of \textsc{xARPES} versus another, commonly used approach (Fig.~\ref{fig:model}\textbf{e}). At $\Delta E=2.5$~meV, the left- and right-hand $\widetilde{\Sigma}_{\mathrm{L,R}}(E)$ differ from each other only due to noise, as shown with 95\% confidence intervals in Fig.~\ref{fig:model}\textbf{d} on top of the model $\Sigma(E)$. The finite $\Delta E$ leads to three distinct deviations from $\Sigma^{\mathrm{true}}(E)$. First, $-\widetilde{\Sigma}^{\prime\prime}(E)$ is overestimated everywhere because a finite $\Delta E$ combined with a dispersive band broadens the MDCs at every $E$. Second, the relatively sharp peak in $\Sigma^{\mathrm{true}\prime}(E)$ near $E-\mu=-22$~meV cannot be fully recovered. Third, we observe the tail/upturn discussed in Sec.~\ref{sec:eliashberg_extraction} for $\widetilde{\Sigma}^{\prime}(E\rightarrow \mu)$.

In Fig.~\ref{fig:model}\textbf{e}, we compare the performance at $\Delta E=0$ with a commonly used approach that we will call the Lorentzian-based (Lor) method, found in Refs.~\cite{shi:2004,king:2014, ludbrook:2015, huempfner:2023}. We will refer to the self-energy extracted with this method as $\widetilde{\Sigma}_n^{\mathrm{Lor}}(E)$. Strictly speaking, a linear non-interacting dispersion $\varepsilon_n(\mathbf{k})$ leads to a Lorentzian lineshape in MDC fitting at a given $E_j$, with peak maxima $\widetilde{k}_n(E_j) = \widetilde{r}_n(E_j)\sqrt{2m_{\mathrm{e}}E^{\mathrm{kin}}_j}/\hbar$ and FWHMs $\Delta \widetilde{k}_n(E_j)$, as described in Sec.~\ref{SI:linearized_dispersion}. By contrast, if the non-interacting dispersion is curved, fitting MDCs with Lorentzian lineshapes results in biased lineshapes, as was demonstrated in Fig.~\ref{fig:model}\textbf{c}. The Lorentzian approach inconsistently inserts Lorentzian-based fit parameters into formulae corresponding to a curved $\varepsilon_n(\mathbf{k})$, commonly $\widetilde{\Sigma}_n^{\mathrm{\mathrm{Lor}\prime}}(E_j)=E_j-\mu-\varepsilon_n(\widetilde{k}_n(E_j))$ and $-\widetilde{\Sigma}_n^{\mathrm{Lor}\prime\prime}(E_j) = (\Delta \widetilde{k}_{n}(E_j)/2) \partial \varepsilon_n(k_n(E_j))/\partial k$. We compare this approach in Fig.~\ref{fig:model}\textbf{e}, obtaining 95\% overlap of the 95\% confidence intervals for the \textsc{xARPES} approach, whereas $\widetilde{\Sigma}_n^{\mathrm{Lor}}(E)$ deviates increasingly from $\widetilde{\Sigma}_n(E)$ towards higher binding energies.

Finally, we quantify the bias originating from a finite energy resolution $\Delta E$ in the extraction of the real and minus imaginary parts of the self-energy data $\widetilde{\Sigma}^{\prime}(E)$ and $-\widetilde{\Sigma}^{\prime\prime}(E)$. To do so, we calculate the overlap of the 95\% confidence intervals of these data with the respective $\Sigma^{\mathrm{true}\prime}(E)$ and $-\Sigma^{\mathrm{true}\prime\prime}(E)$ for energy resolutions ranging from 0.25~meV to 32~meV, and repeat this analysis while changing the Fermi energy (band bottom below $E=\mu$) $E^{\mathrm{F}}$ from 150~meV to 300~meV, based on a different effective mass $m^{\mathrm{b}}$ but the same Fermi wavevector $k^{\mathrm{F}}$. To single out the bias effect, we use the true chemical potential $\mu$ instead of its estimate from the Fermi edge fit described in Sec.~\ref{sec:fermi_edge}. The calculated overlaps are shown in Fig.~\ref{fig:model}\textbf{f} for bands with Fermi energies $E_{1}^{\mathrm{F}}=150$~meV and $E_{2}^{\mathrm{F}}=300$~meV shown in the inset. The overlaps converge to 95\% for $\Delta E \rightarrow 0$, demonstrating that the recovery of $\widetilde{\Sigma}(E)$ is unbiased for the correct choice of the non-interacting dispersion $\varepsilon(\vec{k})$. The overlap decreases more rapidly with increasing $\Delta E$ for $-\widetilde{\Sigma}^{\prime\prime}(E)$ than for $\widetilde{\Sigma}^{\prime}(E)$ because the energy convolution broadens the MDCs with intensity from adjacent MDCs, an effect that largely cancels out for the peak centers. This broadening effect is further illustrated with the steeper non-interacting band having $E_{2}^{\mathrm{F}}=300$~meV, which experiences less broadening and peak shifting upon increasing $\Delta E$. The results in Fig.~\ref{fig:model}\textbf{f} are invariant to the photointensity noise $\sigma^{\mathrm{noise}}$ as the self-energy confidence intervals increase accordingly until the MDC fitting fails, occurring at $\sigma^{\mathrm{noise}}/\widetilde{\mathcal{A}}^0=1.5$\% for this example.

\begin{figure}[tb]
\includegraphics[width=\columnwidth]{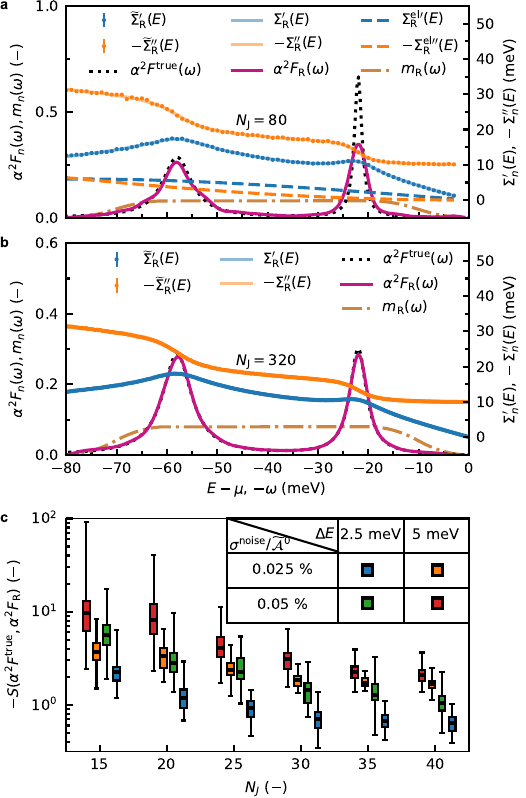}
\caption{\textbf{Quantitative analysis of the extraction of $\alpha^2F(\omega)$}.
\textbf{a} The optimized $\alpha^2F_{\mathrm{R}}(\omega)$ (magenta) versus $\alpha^2F^{\mathrm{true}}(\omega)$ (black dotted) and model function $m_{\mathrm{R}}(\omega)$ (dash-dotted gold), with the real part of the self-energy data $\widetilde{\Sigma}_{\mathrm{R}}^{\prime}(E)$ (blue bars), the reconstruction $\Sigma_{\mathrm{R}}^{\prime}(E)$ (blue translucent) and electron contribution $\Sigma_{\mathrm{R}}^{\mathrm{el}\prime}(E)$ (blue dashed); minus imaginary part of the self-energy data $-\widetilde{\Sigma}_{\mathrm{R}}^{\prime\prime}(E)$ (orange bars), reconstruction $-\Sigma_{\mathrm{R}}^{\prime\prime}(E)$ (orange translucent) and electron contribution $-\Sigma_{\mathrm{R}}^{\mathrm{el}\prime\prime}(E)$ (orange dashed) for the right-hand branch with $N_J=80$ pairs of data ($\widetilde{\Sigma}_{\mathrm{R}}^{\prime}(E)$, $-\widetilde{\Sigma}_{\mathrm{R}}^{\prime\prime}(E)$). \textbf{b} The same quantities as in \textbf{a} for $N_J=320$ pairs of data, while using the true model parameters and ideal experimental conditions, as described in the main text. \textbf{c} Box plot for minus the information entropy $-S(\alpha^2F^{\mathrm{true}}, \alpha^2F_{\mathrm{R}})$ for combinations of $\Delta E \in \{2.5, 5\}$~meV and noise-to-intensity ratio $\sigma^{\mathrm{noise}}/\mathcal{A}^{0} \in \{0.025, 0.05\}$\%, based on 50 code executions at each combination of $(N_J, \Delta E, \sigma^{\mathrm{noise}}/\mathcal{A}^0)$.
\label{fig:a2F}}
\end{figure}

We now extract the Eliashberg function from the right-hand branch $\alpha^2F_{\mathrm{R}}(\omega)$ and determine the model parameters in the parameter optimization loop: the non-interacting band mass $m_{\mathrm{R}}^{\mathrm{b}}$ and Fermi wavevector $k_{\mathrm{R}}^{\mathrm{F}}$, which together describe the non-interacting dispersion $\varepsilon_{\mathrm{R}}(\mathbf{k})$, the electron-electron coupling coefficient $\lambda_{\mathrm{R}}^{\mathrm{el}}$, the electron-impurity coupling magnitude $\Gamma^{\mathrm{imp}}$, and the height $h_{\mathrm{R}}$ of the model function $m_{\mathrm{R}}(\omega)$. For this artificial example, the code converges to the same maximum for a sufficiently consistent set of initial parameters. As a consequence, we obtain similar results for $\alpha^2F_{\mathrm{L}}(\omega)$, which the user can verify through Sec.~\ref{sec:data}. We set $W=E^{\mathrm{F}}$ during the optimization, and we initialize $m_{\mathrm{R}}(\omega)$ discretized on 250 evenly spaced energies for $\omega \in [\omega_{\mathrm{R}}^{\mathrm{min}},~\omega_{\mathrm{R}}^{\mathrm{max}}]$ with $\omega_{\mathrm{R}}^{\mathrm{min}}=1$~meV, $\omega_{\mathrm{R}}^{\mathrm{max}}=80$~meV, and $h_{\mathrm{R}0}=0.08$. The parameter ranges provided in Tables~\ref{tab:sup1}, \ref{tab:sup2}, and \ref{tab:sup3} indicate for this example how much the parameters can deviate in the respective one-shot, inference loop, and tightly converged cases, without causing the code to fail, while a typical convergence scenario is shown in Fig.~\ref{fig:convergence}. The resulting $\alpha^2F_{\mathrm{R}}(\omega)$ is shown in Fig.~\ref{fig:a2F}\textbf{a} against the known $\alpha^2F^{\mathrm{true}}(\omega)$ together with the extracted $\widetilde{\Sigma}_{\mathrm{R}}(E)$, the reconstructions $\Sigma(E)$ and $\Sigma^{\mathrm{el}}(E)$ calculated from $\alpha^2F_{\mathrm{R}}(\omega)$ via Eq.~\eqref{eq:a2f_input}, and $m_{\mathrm{R}}(\omega)$ based on the optimized $h_{\mathrm{R}}$. The reconstructions closely follow the extracted self-energies. In addition, $\Gamma_{\mathrm{R}}^{\mathrm{imp}}$, $k_{\mathrm{R}}^{\mathrm{F}}$, and $m_{\mathrm{R}}^{\mathrm{b}}$ are obtained with high accuracy, as listed in Table~\ref{tab:parameters}. By contrast, $\lambda_{\mathrm{R}}^{\mathrm{el}}$ is systematically underestimated, originating from the imposed positive semidefiniteness of $\alpha^2F_{\mathrm{R}}(\omega)$. An unconstrained $\alpha^2F_{\mathrm{R}}(\omega)$ would otherwise contain negative values due to the noise in $\widetilde{\Sigma}_{\mathrm{R}}(E)$. Due to the positive semidefiniteness, $\alpha^2F_{\mathrm{R}}(\omega)$ and $\lambda_{\mathrm{R}}$ are overestimated with respect to $\alpha^2F^{\mathrm{true}}(\omega)$ and $\lambda^{\mathrm{true}}$. We find an increase of 7\% from $\lambda^{\mathrm{ph,true}}$ to $\lambda_{\mathrm{R}}^{\mathrm{ph}}$, while $\lambda_{\mathrm{R}}^{\mathrm{tot}}$ only differs from $\lambda^{\mathrm{true}}$ by 3\%, emphasizing that the decrease of 6\% from $\lambda^{\mathrm{el,true}}$ to $\lambda_{\mathrm{R}}^{\mathrm{el}}$ is a compensation mechanism. Furthermore, $\alpha^2F_{\mathrm{R}}(\omega)$ captures the two peaks from $\alpha^2F^{\mathrm{true}}(\omega)$, although the peak at $\omega=22$~meV is broadened due to the energy resolution and the somewhat sparsely distributed $\widetilde{\Sigma}(E)$ over the energy range. The peak at $\omega=58$~meV is slightly broadened as the MEM has less confidence in $\widetilde{\Sigma}_{\mathrm{R}}(E)$ at higher binding energies due to its larger confidence intervals.

To show that the extraction of $\alpha^2F_{\mathrm{R}}(\omega)$ can in principle be unbiased, we repeat the extraction for an ideal scenario. In this scenario, we use $N_J=320$ pairs of data ($\widetilde{\Sigma}_{\mathrm{R}}^{\prime}(E)$, $-\widetilde{\Sigma}_{\mathrm{R}}^{\prime\prime}(E)$), energy resolution $\Delta E=0$, and noise-to-intensity ratio $\sigma^{\mathrm{noise}}/\widetilde{\mathcal{A}}^{0}=0.025$\%. Furthermore, we broaden the first peak of $\alpha^2F^{\mathrm{true}}(\omega)$ modeled with Eq.~\eqref{eq:a2f_input} from $\rho_1=0.9~\text{meV}$ to $2.0~\text{meV}$, use $h_{\mathrm{R}}=0.08$ as well as the true model parameters $k^{\mathrm{F,true}}$, $m^{\mathrm{b,true}}$, $\lambda^{\mathrm{el,true}}$, and $\Gamma^{\mathrm{imp,true}}$. The resulting $\alpha^2F_{\mathrm{R}}(\omega)$ and related quantities are displayed in Fig.~\ref{fig:a2F}\textbf{b}, showing a complete recovery of $\alpha^2F^{\mathrm{true}}(\omega)$ for a large amount of unbiased data. While these ideal conditions might never be obtained experimentally, the agreement demonstrates convergence toward $\alpha^2F^{\mathrm{true}}(\omega)$ with increasing amounts of unbiased data.

Lastly, we quantify the difference between the input $\alpha^2F_{\mathrm{R}}(\omega)$ and $\alpha^2F^{\mathrm{true}}(\omega)$ for representative laser ($\Delta E=2.5$~meV) and synchrotron ($\Delta E=5$~meV) resolutions~\cite{iwasawa:2020}, with noise-to-intensity ratios $\sigma^{\mathrm{noise}}/\widetilde{\mathcal{A}}^0=0.025$\% (used previously for Fig.~\ref{fig:model}\textbf{c}, with $\widetilde{\mathcal{A}}_n^0$ defined in Eq.~\eqref{eq:fit1}) and $\sigma^{\mathrm{noise}}/\widetilde{\mathcal{A}}^0=0.05$\%. We calculate minus the information entropy $-S(\alpha^2F^{\mathrm{true}}, \alpha^2F_{\mathrm{R}})$ according to Eq.~\eqref{eq:information_entropy}, which represents the extent of surprise in learning $\alpha^2F^{\mathrm{true}}(\omega)$ when already possessing $\alpha^2F_{\mathrm{R}}(\omega)$. Box plots with extrema, quartiles, and medians are shown in Fig.~\ref{fig:a2F}\textbf{c} for 50 code executions with random noise for each combination of $(\Delta E, \sigma^{\mathrm{noise}}/\widetilde{\mathcal{A}}^0)$ and $N_J$ pairs of $(\widetilde{\Sigma}_{\mathrm{R}}^{\prime}(E), -\widetilde{\Sigma}_{\mathrm{R}}^{\prime\prime}(E))$. We find several trends in the generated Eliashberg functions: spurious peaks and peak-shoulder splittings usually originate from a small amount of $N_J$ for a given phonon energy range, with their effect on $S$ depending on the magnitude of the Eliashberg function. Furthermore, peak magnitudes are governed by the standard deviations $\sigma_{n}^{\prime}(E_j)$ and $\sigma_{n}^{\prime\prime}(E_j)$ of the respective $\widetilde{\Sigma}_{\mathrm{R}}^{\prime}(E)$ and $-\widetilde{\Sigma}_{\mathrm{R}}^{\prime\prime}(E)$. At the smallest $N_J$, the large $-S$ is mostly due to overconfidence of the ``chi2kink'' method, resulting in overfitting. The curves for $\Delta E=2.5$~meV at high $N_J$ are illustrative of Fig.~\ref{fig:model}\textbf{b}, reiterating that $\alpha^2F^{\mathrm{true}}(\omega)$ can be fully recovered for a sufficiently small $\Delta E$ and sufficient $N_J$. For the $\Delta E=5$~meV cases, $-S$ remains large at higher $N_J$ because the sharp peak in $\alpha^2F^{\mathrm{true}}(\omega)$ is broadened during the MDC fitting. We conclude that inversion of $\alpha^2F_n(\omega)$ is successful for sufficiently large amounts of unbiased data, although phonon peaks generally broaden with increasing $\Delta E$, while $\sigma^{\mathrm{noise}}$ increases the variance.

\subsection{Photoemission matrix elements in TiO$_2$-terminated SrTiO$_3$}\label{sec:sto}	

In this first use case, we extract $\Sigma_n(E)$ and $\alpha^2F_n(\omega)$ for a 2DEL of two non-degenerate $d_{xy}$ bands on the TiO$_2$-terminated surface of Nb-doped SrTiO$_3$ along the $k_{[110]}$ direction. In cubic semiconducting SrTiO$_3$, DFPT calculations for the LO$_3$ (longitudinal-optical) and LO$_4$ modes show large long-wavelength coupling ($|g_{mn\nu}(\vec{k},\vec{q\rightarrow 0})|$)~\cite{zhou:2018a}, while cumulant expansion calculations for this system exhibit sidebands displaced from the quasiparticle peak by the LO$_3$ and LO$_4$ energies~\cite{zhou:2019}. Sidebands displaced by the LO$_4$ mode were experimentally observed in several Nb-doped SrTiO$_3$ 2DELs with the Fermi energy $E^{\mathrm{F}}$ comparable to the phonon bandwidth~\cite{chen:2015, wang:2016, guedes:2020} (the satellite case~\cite{giustino:2017}), whereas sidebands of the FeSe dispersion displaced by the LO$_3$ and LO$_4$ modes were observed in monolayer FeSe/SrTiO$_3$~\cite{faeth:2021}. Here, we quantify the EPC with $E^{\mathrm{F}}$ larger than the phonon bandwidth (the photoemission kink case~\cite{giustino:2017}), where $\Sigma_n^{\mathrm{ph}}(E)$ shows kinks at $\mu-E$ corresponding to several phonon modes~\cite{sokolovic:2025}.

We consider a 0.7\% Nb-doped SrTiO$_3$ sample (SrTi$_{0.993}$Nb$_{0.007}$O$_3$) at $T=20$~K. The other experimental conditions are identical to those listed in the Methods section of Ref.~\cite{sokolovic:2025}, and we repeat the PME correction discussed therein. We extract the Eliashberg functions from the inner left (IL) and inner right (IR) branches.

\begin{figure}[tb]
\includegraphics[width=\columnwidth]{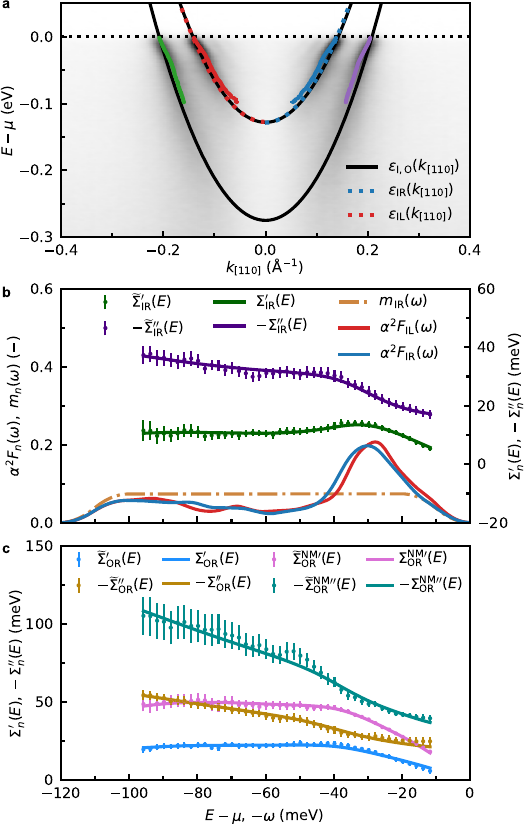}
\caption{\textbf{Self-energy and Eliashberg function of SrTiO$_3$.}
\textbf{a} Band map of the 2DEL on the TiO$_2$-terminated surface of SrTiO$_3$ along $k_{[110]}$, showing MDC maxima with 95\% confidence intervals for the outer left (green), inner left (red), inner right (blue), and outer right (purple) branches, with inner/outer bands (black) from initial guesses and the optimized inner left/right branches (dashed red/blue). \textbf{b} Comparison of the inner left $\alpha^2F_{\mathrm{IL}}(\omega)$ (red) with the inner right $\alpha^2F_{\mathrm{IR}}(\omega)$ (blue) with the optimized model function $m_{\mathrm{IR}}(\omega)$ (dash-dotted gold), based on the inner right real $\widetilde{\Sigma}^{\prime}_{\mathrm{IR}}(E)$ (green bars) and minus imaginary data $-\widetilde{\Sigma}^{\prime\prime}_{\mathrm{IR}}(E)$ (indigo bars), with reconstructions $\Sigma_{\mathrm{IR}}^{\prime}(E)$ (blue) and $-\Sigma_{\mathrm{IR}}^{\prime\prime}(E)$ (indigo). \textbf{c} Self-energies of the outer right branch obtained with and without (NM) photoemission matrix elements, comprising the corrected real $\Sigma_{\mathrm{OR}}^{\prime}(E)$ (light blue) and imaginary parts $-\Sigma_{\mathrm{OR}}^{\prime\prime}(E)$ (gold), and the real $\Sigma_{\mathrm{OR}}^{\mathrm{NM}\prime}(E)$ (pink) and imaginary parts $-\Sigma_{\mathrm{OR}}^{\mathrm{NM}\prime\prime}(E)$ (teal) without the correction. Corresponding data are denoted using a tilde.
\label{fig:STO}}
\end{figure}

The PMEs of the $d_{xy}$ orbitals of the 2DEL are dependent on the polar emission angle, which we assume to be equal to the detector angle $\eta$ after correcting for the small misalignment in $\phi$ inside Eq.~\eqref{eq:path}. Fortunately, performing the MEC with the physically motivated~\cite{sokolovic:2025, goldberg:1981} photoemission matrix elements $|M_{d_{xy}}(\eta)|^2 \propto \sin^2(\eta - \phi)$ surrounding the $\bar{\Gamma}_{00}$-point for the $d_{xy}$ bands improves the accuracy of extracted quantities. As in Ref.~\cite{sokolovic:2025}, we fit a $\sin^2(\eta-\phi)$ distribution to the energy-integrated photointensity $P(\eta)$, after which we subtract the fitted $\phi=0.57$\textdegree~from the band map to account for detector misalignment. We then determine $\widetilde{\gamma}_n(E)$ and $\widetilde{r}_n(E)$ for the four branches, while manually iterating over $k_n^{\mathrm{c}}$, finding that the zone center displacements and inner/outer (I/O) Fermi vectors, namely $k_n^{\mathrm{c}}=-0.0014~\Angstrom^{-1}$, $k_{\mathrm{I}}^{\mathrm{F}}=0.142~\Angstrom^{-1}$, and $k_{\mathrm{O}}^{\mathrm{F}}=0.208~\Angstrom^{-1}$, yield the desired particle-hole symmetry for $\Sigma_{n}^{\prime}(E)$ calculated with Eq.~\eqref{eq:reobtain_real} simultaneously for all four branches. We preliminarily assign $m_{\mathrm{I,O}}^{\mathrm{b}}=0.6$~$m_{\mathrm{e}}$ from visual inspection for the inner (I) and outer (O) bands. The $d_{xy}$ bands together with the MDC maxima along the $k_{[110]}$ direction are shown in Fig.~\ref{fig:STO}\textbf{a}, while $\widetilde{\Sigma}_{n}^{\prime}(E)$ and $-\widetilde{\Sigma}_{n}^{\prime\prime}(E)$ are shown for all four branches in Fig.~\ref{fig:STO-SM}. Next, we perform the optimization loop with an electron-electron scale parameter $W_{\mathrm{IR}}=E_{\mathrm{IR}}^{\mathrm{F}}=(\hbar k^{\mathrm{F}}_{\mathrm{IR}})^2/(2m_{\mathrm{IR}}^{\mathrm{b}})$ in Eq.~\eqref{eq:electron_power_4}. The optimization loop yields $m_{\mathrm{IR}}^{\mathrm{b}}=0.59$~$m_{\mathrm{e}}$, $k_{\mathrm{IR}}^{\mathrm{F}}=0.141~\Angstrom^{-1}$, $\Gamma_{\mathrm{IR}}^{\mathrm{imp}}=16.5$~meV, and $\lambda_{\mathrm{IR}}^{\mathrm{el}}=0$, such that $\lambda_{\mathrm{IR}}^{\mathrm{el,true}}$ might be too small to be resolved. We calculate $\lambda_{\mathrm{IR}}^{\mathrm{ph}}=0.46$ using Eq.~\eqref{eq:electron_power_4}, appreciably smaller than the outer branch values $\lambda_{\mathrm{OR}}^{\mathrm{ph}}=0.63$ and $\lambda_{\mathrm{OL}}^{\mathrm{ph}}=0.68$ from Ref.~\cite{sokolovic:2025}. The TO$_4$ and LO$_4$ modes~\cite{vogt:1988} found previously in $\alpha^2F_{\mathrm{IR}}(\omega)$~\cite{sokolovic:2025} at $\omega_{\mathrm{TO}_4}=69$~meV and $\omega_{\mathrm{LO}_4}=84$~meV are also observed in our $\alpha^2F_{\mathrm{IR}}(\omega)$ within approximately 2~meV. By contrast, the previously observed $\omega_{\mathrm{LO}_3}=56$~meV now appears as a shoulder, while the large feature near 40~meV has been red-shifted towards 30~meV. Furthermore, $\alpha^2F_{\mathrm{IR}}(\omega)$ is resolved with less detail compared to those in Ref.~\cite{sokolovic:2025}, primarily because the current data have been sampled for fewer $\eta$-values. Our LO$_4$ mode is red-shifted by approximately 8~meV compared to the DFPT calculations of Ref.~\cite{zhou:2018}, potentially due to accumulated charges~\cite{cancellieri:2016} from the in-gap states~\cite{alarab:2024} and from the 2DEL.

We calculate the Euclidean distance measure provided in Sec.~\ref{SI:eliashberg_details} as $M(\alpha^2F_{\mathrm{L}}(\omega), \alpha^2F_{\mathrm{R}}(\omega))=0.0099$. The high similarity raises the question whether $\alpha^2F_{\mathrm{IL}}^{\mathrm{true}}(\omega)$ and $\alpha^2F_{\mathrm{IR}}^{\mathrm{true}}(\omega)$ should be identical. The Fermi surface of the TiO$_2$-termination suggests that the band structure still has the $C_{4}$-symmetry obtained by terminating tetragonal SrTiO$_3$ along the $[001]$-direction. However, previously measured spin components for the $d_{xy}$-derived bands find a small difference in the $z$-component~\cite{santander-syro:2014}, potentially affecting the spin dependence of the Eliashberg function~\cite{fabian:1999}. Future first-principles calculations and experiments on the Eliashberg functions and the spin texture of the 2DEL on SrTiO$_3$ might elucidate if the underlying quantities are identical indeed.

Finally, we investigate the effect of the PMEs by comparing the self-energies after the optimization loop for the \textsc{xARPES} approach for the outer right (OR) branch versus omitting the PMEs (NM for ``no MEC'') during the MDC fitting. We choose the OR branch as $-\widetilde{\Sigma}_{\mathrm{IL}}^{\mathrm{ph,NM}\prime\prime}(E)$ and $-\widetilde{\Sigma}_{\mathrm{IR}}^{\mathrm{ph,NM}\prime\prime}(E)$ contain negative values after running the optimization loop, which is unphysical. Fig.~\ref{fig:STO}\textbf{c} shows the self-energies and reconstructions with and without the MEC, with $\Sigma_{\mathrm{OR}}^{\mathrm{NM}}(E)$ being approximately twice as large as $\Sigma_{\mathrm{OR}}(E)$, primarily because the optimization loop assigns a small $m_{\mathrm{OR}}^{\mathrm{b,NM}}=0.46$~$m_{\mathrm{e}}$ versus $m_{\mathrm{OR}}^{\mathrm{b}}=0.59$~$m_{\mathrm{e}}$ for a good simultaneous fitting of $\Sigma_{\mathrm{OR}}^{\mathrm{NM}\prime}(E)$ and $-\Sigma_{\mathrm{OR}}^{\mathrm{NM}\prime\prime}(E)$ (compare Eqs.~\eqref{eq:reobtain_real}--\eqref{eq:reobtain_imag}). By contrast, $\Sigma_{\mathrm{OR}}^{\prime}(E)$ and $-\Sigma_{\mathrm{OR}}^{\prime\prime}(E)$ from the fit with the MEC agree well with the respective data $\widetilde{\Sigma}_{\mathrm{OR}}^{\prime}(E)$ and $-\widetilde{\Sigma}_{\mathrm{OR}}^{\prime\prime}(E)$, and closely resemble those of Ref.~\cite{sokolovic:2025}. In agreement with Ref.~\cite{sokolovic:2025} we also obtain $\lambda_{\mathrm{OR}}^{\mathrm{el}}=0=\lambda_{\mathrm{OR}}^{\mathrm{el,NM}}$. Finding that $\Sigma(E)$ can be over twice as large when neglecting the MEC, we conclude that the MEC can be essential for reducing bias in extracted quantities.

\subsection{Eliashberg-function similarity in quasi-freestanding graphene}\label{sec:graphene}

In this second use case, we extract $\Sigma_n(E)$ and $\alpha^2F_n(\omega)$ for two photoemission kinks denoted as the L and R branches with $k_x \approx \textnormal{K}_{x} - 0.13~\Angstrom^{-1}$, perpendicular to the $\Gamma-\mathrm{K}$ high-symmetry line of Li-doped, Si-intercalated graphene on a Co(0001) substrate. The data were provided by the authors of Ref.~\cite{usachov:2018} and are shown as cut~\#2 in Fig.~7\textbf{b} of Ref.~\cite{usachov:2018}. Bands of the Dirac cones surrounding K-points are approximately linear~\cite{hwang:2007}, such that we fit them with the linear $\varepsilon_{\mathrm{L,R}}(\mathbf{k})$ fitting procedure of \textsc{xARPES} described in Sec.~\ref{SI:linearized_dispersion}. The geometry of the setup is obtained by performing a $90$\textdegree~rotation of the detector around the $z$-axis -- aligning the detector slit along the $y$-axis -- followed by a positive $\phi$-rotation that results in the finite $k_x$. We follow Ref.~\cite{usachov:2018} in employing a ``small-angle approximation''~\cite{zhang:2022}, resulting in $k_y=p_y \approx |\mathbf{p}| \sin(\theta + \eta)$, followed by correcting for misalignment with $\theta=-2.28$\textdegree. In Ref.~\cite{usachov:2018}, the data from the two kinks were combined into one pair of real and minus imaginary parts by a method that we describe in Sec.~\ref{SI:additional_graphene}, followed by the extraction of a single Eliashberg function. Instead, we separately extract $\Sigma_{\mathrm{L}}(E)$, $\alpha^2F_{\mathrm{L}}(\omega)$, $\Sigma_{\mathrm{R}}(E)$, and $\alpha^2F_{\mathrm{R}}(\omega)$, and discuss the symmetries that relate them to one another. A linear dispersion can be described by two non-interacting band parameters: the Fermi velocity $v_n^{\mathrm{F}}$ and Fermi wavevector $k_n^{\mathrm{F}}$. Since two parameters can be eliminated during the MDC fitting, all model parameters with a linearized band can be quantified in the optimization loop.

Prior DFPT calculations of the decay rate in graphene display two large peaks near $\omega=160$~meV and $\omega=195$~meV~\cite{ponce:2023}, arising largely from the in-plane A$^{\prime}_1$ mode at the K-point and the in-plane E$_{2\mathrm{g}}$ mode at the $\Gamma$-point~\cite{margine:2014}, respectively. Charge dopants red-shift these modes, bring them closer together, increase the relative intensity of the A$^{\prime}_1$-based mode, and introduce additional spectral weight in $\alpha^2F_n(\omega)$ below $\omega=100$~meV from out-of-plane coupling of the C-atoms with dopants~\cite{novko:2017}.

\begin{figure}[tb]
\includegraphics[width=\columnwidth]{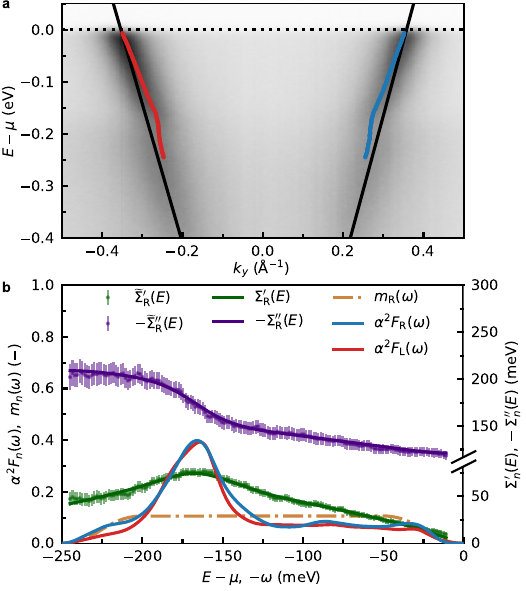}
\caption{\textbf{Self-energy and Eliashberg function of Li-doped graphene}.
\textbf{a} Band map of Li-doped graphene along cut~\#2 from Ref.~\cite{usachov:2018} with two optimized linear bands, as well as the left-hand (L, blue) and right-hand (R, red) MDC maxima with 95\% confidence intervals. \textbf{b} Comparison of $\alpha^2F_{\mathrm{L}}(\omega)$ (red) against $\alpha^2F_{\mathrm{R}}(\omega)$ (blue), with optimized model function $m_{\textnormal{R}}(\omega)$ (dash-dotted gold), based on the right-hand real $\widetilde{\Sigma}^{\prime}_{\mathrm{R}}(E)$ (green bars) and minus imaginary data $-\widetilde{\Sigma}^{\prime\prime}_{\mathrm{R}}(E)$ (indigo bars), with reconstructions $\Sigma^{\prime}_{\mathrm{R}}(E)$ (green) and $-\Sigma^{\prime}_{\mathrm{R}}(E)$ (indigo).
\label{fig:Li-gr}}
\end{figure}

Using $\Delta E=10$~meV~\cite{usachov:2018}, we perform the Fermi-edge correction described in Sec.~\ref{sec:fermi_edge}, finding the best fit for $T=50$~K, which we assign as the sample temperature. Next, we fit the MDCs on an interval $\mu-E \in [0, 246]$~meV to capture the E$_{2\mathrm{g}}$ and A$^{\prime}_1$ contributions. We use the MEC from Ref.~\cite{usachov:2018} with its details discussed in Sec.~\ref{SI:additional_graphene}. Afterwards, we perform the optimization loop for the left-hand and right-hand sides using $W_{\mathrm{L,R}}=1.5$~eV in Eq.~\eqref{eq:electron_power_4}, equal to the binding energy of the Dirac crossing. The MDC maxima and optimized non-interacting dispersions are displayed in Fig.~\ref{fig:Li-gr}\textbf{a}, based on $v_{\mathrm{L}}^{\mathrm{F}}=-2.67$~eV~$\Angstrom/\hbar$, and $k_{\mathrm{L}}^{\mathrm{F}}=-0.354~\Angstrom^{-1}$, $v_{\mathrm{R}}^{\mathrm{F}}=2.85$~eV~$\Angstrom/\hbar$, and $k_{\mathrm{R}}^{\mathrm{F}}=0.358~\Angstrom^{-1}$.

Fig.~\ref{fig:Li-gr}\textbf{b} displays the extracted $\widetilde{\Sigma}_{\mathrm{R}}(E)$ and reconstructed $\Sigma_{\mathrm{R}}(E)$, as well as $\alpha^2F_{\mathrm{L}}(\omega)$ compared to $\alpha^2F_{\mathrm{R}}(\omega)$ displayed together with its optimized $m_{\mathrm{R}}(\omega)$. The complete set of $\widetilde{\Sigma}_{\mathrm{L}}(E)$ and $\Sigma_{\mathrm{L}}(E)$ are provided in Sec.~\ref{SI:additional_graphene}. We calculate $\lambda_{\mathrm{L}}^{\mathrm{ph}}=0.48$ and $\lambda_{\mathrm{R}}^{\mathrm{ph}}=0.53$, with most of the discrepancy potentially originating from experimental asymmetry. A large peak at $\omega=166$~meV can be observed in $\alpha^2F_{\mathrm{R}}(\omega)$, likely incorporating the E$_{2\mathrm{g}}$ and A$^{\prime}_1$ contributions. These peaks are red-shifted by $\omega=10$~meV in DFPT calculations of the isotropic $\alpha^2F(\omega)$ in Li-doped graphene~\cite{novko:2017}, where absence of the substrate and intercalation compounds might contribute to the discrepancy. Some spectral weight can be observed in $\alpha^2F_{\mathrm{R}}(\omega)$ near $\omega=35$~meV and $\omega=85$~meV, reflecting the low-frequency spectrum found in the DFPT calculations~\cite{novko:2017}. The optimization also yields $\Gamma_{\mathrm{L}}^{\mathrm{imp}}=115.6$~meV, $\Gamma_{\mathrm{R}}^{\mathrm{imp}}=120.9$~meV, and $\lambda_{\mathrm{L}}^{\mathrm{el}}=0=\lambda_{\mathrm{R}}^{\mathrm{el}}$. Introduction of curvature in the non-interacting dispersion -- with the center wavevector $k_n^{\mathrm{c}}$ determined by the KKBF -- did not lead to a finite $\lambda_n^{\mathrm{el}}$ or other appreciably different results. While calculations for doped graphene in the random-phase approximation (RPA) predict a plasmon contribution in $-\Sigma_n^{\mathrm{el}\prime\prime}(E)$ of the order of 100~meV as $E-\mu$ approaches the Dirac point~\cite{hwang:2008, park:2009, gruneis:2009a}, our results suggest that $\Sigma_n^{\mathrm{el}}(E)$ is weak for binding energies up to the phonon bandwidth. However, while $\Sigma_n^{\mathrm{el}}(E)$ in Eq.~\eqref{eq:electron_power_4} may be related to coupling with particle-hole excitations in the RPA for a parabolic $\varepsilon_n(\mathbf{k})$ as described in Sec.~\ref{SI:eliashberg_details}, its functional form should be different for approximately linear $\varepsilon_n(\mathbf{k})$~\cite{hwang:2008}. Moreover, Eq.~\eqref{eq:electron_power_4} does not include coupling with plasmons, rendering it increasingly unsuitable to describe electron-electron interactions in doped graphene towards higher binding energies.

We observe good agreement between $\alpha^2F_{\mathrm{L}}(\omega)$ and $\alpha^2F_{\mathrm{R}}(\omega)$, obtaining the large peak in the latter at $\omega=164$~meV, consistent with verification for large $N_J$ found in Sec.~\ref{sec:mock}. The weaker resolution of the phonon modes in $\widetilde{\Sigma}_{\mathrm{L}}(E)$ reflects lower confidence of the MEM in the left-hand data. The Euclidean distance measure between the two Eliashberg functions is $M(\alpha^2F_{\mathrm{L}}(\omega), \alpha^2F_{\mathrm{R}}(\omega))=0.0067$. This Euclidean distance is lower than $M(\alpha^2F_{\mathrm{L}}(\omega), \alpha^2F_{\mathrm{R}}(\omega))=0.0099$ obtained for the SrTiO$_3$ in Sec.~\ref{sec:sto}, and much lower than the distance between two previously compared experimental Eliashberg functions~\cite{tang:2004}, suggesting that the agreement for the Li-doped graphene is unprecedented. As with the SrTiO$_3$, this high similarity between $\alpha^2F_{\mathrm{L}}(\omega)$ and $\alpha^2F_{\mathrm{R}}(\omega)$ suggests that the underlying underlying $\alpha^2F_{\mathrm{R}}^{\mathrm{true}}(\omega)$ and $\alpha^2F_{\mathrm{L}}^{\mathrm{true}}(\omega)$ may be identical. The two kinks are given by two $\pm k_y$ components perpendicular to $\Gamma$--$\mathrm{K}$ high-symmetry line, which in monolayer graphene are related by a $C_2$ axis and a $\sigma_d$ plane~\cite{fajardo:2019}. The Si/Co(0001) subsystem (see Fig.~3 of Ref.~\cite{usachov:2018}) breaks the $\sigma_d$ plane symmetry, whereas the introduction of Si/Co(0001) and Li intercalation both break the $C_2$ symmetry. However, the Li/graphene subsystem is often regarded as quasi-freestanding~\cite{usachov:2018}, which would still preserve the $\sigma_d$ symmetry. Although a spin splitting between the sublattices of doped graphene could also break the symmetry along the $\Gamma$--$\mathrm{K}$ path via spin dependence of the Eliashberg function~\cite{fabian:1999}, this degree of freedom should not be important as the magnetization in the graphene sheet was found to be negligible~\cite{usachov:2018}. Therefore, we expect that the asymmetry of the experimental setup might be the main cause of the observed difference between $\alpha^2F_{\mathrm{L}}(\omega)$ and $\alpha^2F_{\mathrm{R}}(\omega)$, as the underlying functions should be identical based on the symmetry. Interestingly, the omission of the Fermi-edge correction for linearized bands leads to a much larger $M(\alpha^2F_{\mathrm{L}}(\omega), \alpha^2F_{\mathrm{R}}(\omega))=0.022$, with $\alpha^2F_{\mathrm{L}}(\omega)$ and $\alpha^2F_{\mathrm{R}}(\omega)$ obtained without the edge correction shown in Fig.~\ref{fig:Li-gr-SM}. This increase of $M(\alpha^2F_{\mathrm{L}}(\omega), \alpha^2F_{\mathrm{R}}(\omega))$ of over a factor of three shows that experimental factors can lead to large differences in extracted self-energies and Eliashberg functions.

\section{Discussion}\label{sec:discussion}

We have introduced rigorous and quantitative approaches for extracting many-body properties from angle-resolved photoemission band maps. We make our methodologies available to the scientific community through the first release of \textsc{xARPES}, a GPL-v3-licensed Python package. This initial release focuses on the analysis of dispersion kinks arising from electron-phonon interactions. Two of its building blocks are based on the current state of the art. Starting from raw experimental data, the code first quantifies the band map data via non-interacting dispersions and causal self-energies, with the user specifying the number of dispersions, the self-energies, and the values for parameterizing them, optionally complemented with a photoemission matrix element. Second, it decomposes the self-energy into impurity, electron-electron, and electron-phonon contributions, concurrently extracting the electron-phonon Eliashberg coupling function (Fig.~\ref{fig:model}).

\textsc{xARPES} complements existing codes for angle-resolved photoemission analysis -- which are usually tied to a beamline and excel at data visualization -- with novel and powerful functionalities for many-body analysis. Concerning the self-energy extraction, we show that traditional Lorentzian fitting of momentum-distribution curves, suitable for linear dispersions~\cite{veenstra:2011}, introduces bias when applied to curved dispersions. The present approach improves on the extraction by fitting curved dispersions, which has been implemented in \textsc{xARPES} for parabolic bands. Furthermore, we introduce the possibility of specifying the photoemission matrix element during the fitting process. Existing methods for extraction of the Eliashberg function with the maximum-entropy method usually rely on manual assignment of the parameters governing the non-interacting dispersion and individual scattering channels. By contrast, the extraction is wrapped inside a bilevel optimization, using a newly derived criterion based on Bayesian inference for determining the most probable model parameters. This framework is designed to eliminate tedious and arbitrary manual parameter optimization, resulting in a robust and reproducible platform which is well-suited for objective comparisons between different datasets. We illustrate this by extracting and comparing Eliashberg functions from two symmetry-equivalent kinks within the same photoemission band map (Fig.~\ref{fig:Li-gr}).

We have extensively verified and validated the performance to demonstrate the potential of the present analysis. The robustness of the maximum-entropy-based Bayesian inference is confirmed using synthetic data with added noise and energy resolution as input. We show that the model parameters are accurately recovered, with precision converging monotonically to the exact result as energy resolution improves, the signal-to-noise ratio increases, and the data sampling becomes finer (Figs.~\ref{fig:model}--\ref{fig:a2F}). 

Using a case where the quadratic dispersion is indispensable, we analyze photoemission kinks from the 2-dimensional electron liquid on TiO$_2$-terminated SrTiO$_3$ (Fig.~\ref{fig:STO}\textbf{a}). We achieve good agreement between the Eliashberg functions extracted from both sides of the parabolic band. Furthermore, we demonstrate that omission of the photoemission matrix element can modify the results by over a factor of two (Fig.~\ref{fig:STO}\textbf{b}). We also apply our approach to published data on Li-doped graphene~\cite{usachov:2018}, showcasing the applicability of linearized non-interacting dispersions in \textsc{xARPES}. Here, a comparison of the Eliashberg functions extracted from two different kinks of a band map shows that Li-doped graphene is an excellent platform for future benchmarking of the extraction procedure (Fig.~\ref{fig:Li-gr}). With the advent of ever-improving experimental equipment, we introduce an Euclidean distance measure to compare Eliashberg functions that are believed to be identical, either via symmetry or from one experiment to the next.

The development of the present formalism was driven by the need for a platform to accurately compare experimentally extracted electron spectral functions with first-principles calculations in the presence of many-body interactions, particularly electron-phonon coupling. Achieving this long-term objective requires a standardized method for extracting the Eliashberg function from photoemission measurements. Like other ARPES-based approaches, \textsc{xARPES} extracts an Eliashberg function that depends only on the band, its location in reciprocal space, and on the exchanged phonon energy. However, in fundamental theory, this function also depends on the initial energy and momentum of the electron~\cite{mahan:2000}. We have shown under which conditions the experimentally extracted Eliashberg function can be compared with its theoretical counterpart, usually evaluated at the chemical potential and Fermi momentum relevant to the experimental conditions~\cite{eiguren:2008,eiguren:2009}. In the future, \textsc{xARPES} may be expanded to compare the non-interacting dispersion calculated from theoretical methods with its experimental counterpart, as well as for comparing the self-energy evaluated on the renormalized dispersion. As a first step in this direction, in Fig.~\ref{fig:Al} we have fitted the mass of the non-interacting dispersion to Kohn-Sham eigenvalues of the bottom of the zone-center surface state on the surface of Al(001) calculated with \textsc{Abinit}~\cite{gonze:2020, romero:2020, gonze:2016}, followed by extracting the self-energies and the Eliashberg function with this fixed non-interacting band mass for a band map published in Ref.~\cite{jiang:2011}. The relevant parameters of this fitting procedure are provided in Table~\ref{tab:sup4}. The modular structure of \textsc{xARPES} supports future development, extension, and potential interfacing with existing codes for analyzing angle-resolved photoemission spectroscopy. With respect to the underlying physical model, we foresee three main avenues for evolution. First, implementing higher-order polynomial forms of the dispersion would improve accuracy in cases where linear or quadratic models fall short. Second, incorporating higher-order phonon processes could be pursued, for example through self-consistent calculations at the self-energy level~\cite{lihm:2025} or via the cumulant expansion applied directly to the spectral function~\cite{aryasetiawan:1996}. Third, the library of Kramers-Kronig-consistent phenomenological models for the electron-electron self-energy, currently limited to Fermi-liquid forms, could be expanded to capture a wider range of behaviors found in quantum materials, such as marginal Fermi liquids~\cite{varma:1989} or systems with strong plasmon coupling~\cite{bostwick:2007a}. On the implementation side, the current method relies on fitting isolated momentum-distribution curves, which does not account for the finite energy resolution. A natural next step would be a code that analyzes the full two-dimensional photoemission intensity while incorporating the effects of energy resolution, by directly extracting the Eliashberg function from the photointensity~\cite{bryan:1990}. Additional enhancements could broaden the user base and improve usability. These include support for multiple experimental file formats, support for different experimental geometries, and interfaces with existing codes that specialize in data visualization.

The integration of machine learning could accelerate various tasks of the formalism. One such option is the automatic selection of the energy and momentum windows for the fitting. Ultimately, the current optimization loop could be replaced by an artificial neural network that directly optimizes all the model parameters. Meanwhile, the standardized extraction approach presented here can be used to procure harmonized training data for the supervised learning of these future machine learning models.

With the release of \textsc{xARPES}, we aim to foster a collaborative effort between the photoemission and first-principles communities to establish shared standards that enable unbiased comparisons between first-principles many-body theories and experimental data, thereby driving progress on both theoretical and experimental fronts.

\section{Methods}\label{sec:methods}

\subsection{Technical details of the xARPES code}\label{sec:technical}

The code is object-oriented, allowing the user to manipulate data through objects such as the \texttt{BandMap} class. It can be executed in \textsc{Jupyter} notebooks or with regular \textsc{Python} files, requiring \textsc{Python} version 3.10 or higher. Fitting of momentum-distribution curves is wrapped around the Python fitting package \textsc{LMFIT}~\cite{lmfit:website}, allowing the specification of the number of bands. The Bayesian inference loop employs \texttt{scipy.optimize.minimize}~\cite{virtanen:2020}, so that the user can choose different optimization algorithms. On a 12$^{\rm th}$ Gen Intel(R) Core(TM) i7-1255U 2.60~GHz CPU, individual code blocks from Fig.~\ref{fig:workflow} typically complete within one second. The Bayesian inference loop wrapping the MEM can take several minutes to complete, depending on the initial guess and the size of the singular space controlled with the cutoff value $\sigma_n^{\mathrm{cut}}$ in Bryan's algorithm~\cite{bryan:1990}. \textsc{xARPES} can currently load band maps from \texttt{NumPy} arrays and from the \textsc{IGOR Pro} binary wave format (\texttt{.ibw}). The band map needs to have the detector angle on the abscissa and the kinetic energy on the ordinate, while the user must provide the angular resolution and the sample temperature. The energy resolution must be provided by the user if the chemical potential has to be determined from the Fermi edge fitting.

\subsection{Maximum-entropy method in xARPES}\label{sec:maxent}

Within the MEM, \textsc{xARPES} uses the so called ``chi2kink'' method~\cite{kaufmann:2023} to determine the hyperparameter/Lagrange multiplier $a$. With this method, $\chi^2$ is defined as the sum of the squared terms in Eq.~\eqref{eq:log-likelihood}, followed by fitting $\mathrm{log}_{10}[\chi^2(\mathrm{log}_{10}(a))]$ with the following function:
\begin{equation}\label{eq:chi2kink_fit}
	\phi(x; g, b, c, d) = g + \frac{b}{1+\ee^{-d(x-c)}}.
\end{equation}
Once the fitting has been completed, $a$ is obtained from $a=10^{c-f^{\chi^2} /d}$ with a tuning parameter $f^{\chi^2} \in [2, 2.5]$~\cite{kaufmann:2023}. The default values in \textsc{xARPES} are $f^{\chi^2}=2.5$ when fitting with both $\widetilde{\Sigma}_{n}^{\prime}(E)$ and $-\widetilde{\Sigma}_{n}^{\prime\prime}(E)$, contrasting $f^{\chi^2}=2.0$ when fitting just one of them (commonly $\widetilde{\Sigma}_{n}^{\prime}(E)$). In practice, users must provide a chi2kink range minimum $\log_{10}(a_n^{\mathrm{min}})$ and a maximum $\log_{10}(a_n^{\mathrm{max}})$ and check that the kink in $\chi^2$ is correctly fitted within this interval, after which $\alpha^2F_n(\omega)$ is determined in a final iteration from the fitted $a$. The default parameters are $\sigma_n^{\mathrm{cut}}=1$ for the initial one-shot calculation and the optimization loop versus $\sigma_n^{\mathrm{cut}}=10^{-4}$ for a final extraction; $\log_{10}(a_{n}^{\rm{min}})=3$ for the initial calculation and optimization, versus $\log_{10}(a_{n}^{\rm{min}})=1$ during the final extraction extraction, with $\log_{10}(a_{n}^{\rm{max}})=\log_{10}(a_{n}^{\rm{min}})+8$ in all cases. The Eliashberg function extraction employs Bryan's algorithm~\cite{bryan:1990} with a default convergence threshold $t = 10^{-8}$.

\subsection{Fermi-edge fit}\label{sec:fermi_edge}

\begin{figure}[tb]
\includegraphics[width=\columnwidth]{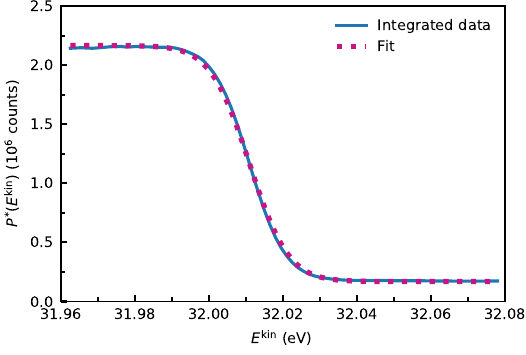}
\caption{\textbf{Fermi-edge fit of graphene}.
Photointensity $P^{*}(E^{\rm{kin}})$ for Li-doped graphene versus the kinetic energy $E^{\rm{kin}}$. The angle-integrated data (blue) and its corresponding fit (magenta).
\label{fig:Fermi}}
\end{figure}

During the band map analysis for a fixed photon energy $h\nu$ and polarization $\boldsymbol{\epsilon}$, it is beneficial to have the photointensity as a function of the electron energy $E-\mu$, which is written as $P(E,\eta)$ in Eq.~\eqref{eq:photointensity}. Here, we outline how $P(E,\eta)$ can be obtained, and highlight possible sources of bias in establishing the relation between $E$ and $E^{\mathrm{kin}}$. Integration over the momentum results in the angle-integrated photointensity:
\begin{equation}\label{eq:integrated_photointensity}
	P(E) \propto \int_{-\infty}^{\infty} \dd E' R(E-E')D(E'),
\end{equation}
where we define the shorthand notation $D(E) \equiv B(E) + f(E) \sum_{n} |M_n(E)|^2 A_n(E)$. However, for a given work function $\Phi$, the energy measured by the detector is $E^{\rm{kin}}$, yielding:
\begin{multline}\label{eq:shifted_photointensity}
	P^{*}(E^{\rm{kin}}) \propto \int_{-\infty}^{\infty} \dd E^{\mathrm{kin}\prime} R(E^{\mathrm{kin}}-E^{\mathrm{kin}\prime}) \\
	\times D(E^{\mathrm{kin}\prime} - h\nu + \Phi + \mu).
\end{multline}
Thus, to go from the experimentally detected $E^{\mathrm{kin}}=h\nu-\Phi+E-\mu$ to the binding energy coordinate $E-\mu$ requires estimating $h\nu-\Phi$. This can be done through the so-called \textit{Fermi-edge fit}~\cite{hellbruck:2024}, where a profile containing the Fermi-Dirac distribution is fitted to $P^{*}(E^{\mathrm{kin}})$ to yield the estimate $h\widehat{\nu}-\widehat{\Phi}$, with the hat (\kern0.4em$\widehat{}$\kern0.4em) denoting estimates. Furthermore, the Fermi-edge fit can be used to determine the energy resolution of an experimental setup for a known electronic temperature, or vice versa. In \textsc{xARPES}, the user has to provide these two quantities, while $P^{*}(E^{\mathrm{kin}})$ is fitted with parameters $A$, $B$, and $C$ for a function of the following form:
\begin{multline}\label{eq:fit_fermi}
	P^{*}(E^{\mathrm{kin}}) = \int_{-\infty}^{\infty}\dd E^{\mathrm{kin}\prime}\,R(E^{\mathrm{kin}}-E^{\mathrm{kin}\prime}) \\
	\times \left[ \frac{A}{\ee^{(E^{\mathrm{kin}\prime}-C)/(k_{\mathrm{B}}T)}+1}+B \right],
\end{multline}
resulting in $C=h\widehat{\nu}-\widehat{\Phi}$. As an example, we show in Fig.~\ref{fig:Fermi} the Fermi-edge fit for the Li-doped graphene from Sec.~\ref{sec:graphene}.

Afterwards, $h\widehat{\nu}-\widehat{\Phi}$ can be subtracted from $E^{\rm{kin}}$ to obtain $\widehat{E}-\widehat{\mu}=E^{\rm{kin}}-h\widehat{\nu}-\widehat{\Phi}$, yielding $P$ as a function of $\widehat{E}-\widehat{\mu}$ from Eq.~\eqref{eq:integrated_photointensity}. We subsequently write these quantities as $E-\mu$ and $P(E)$, although it should be kept in mind that this coordinate transformation involves an estimate.

We note that $h\widehat{\nu}-\widehat{\Phi}$ is often established before an experiment, for instance from Cu or Au references~\cite{hellbruck:2024}, especially for samples with a poorly resolved Fermi edge or a Fermi edge lying inside an energy gap, which can be the case for insulators, superconductors, and sometimes charge-density waves. Notably, neglecting the dispersion $\varepsilon_{n}(\vec{k})$, the many-body features in $A_n(E,\vec{k})$, or $|M_n(E,\eta; h\nu, \vecg{\epsilon})|^2$ may result in a biased estimate $h\widehat{\nu}-\widehat{\Phi}$. The DOS from 1D, 2D, or 3D parabolic dispersion relations projected onto the single momentum dimension of a band map is proportional to $1/\sqrt{E_n^{\mathrm{F}}-(\mu-E)}$, with $E_n^{\mathrm{F}}$ the Fermi energy of a branch $n$. The absence of this factor in Fermi-edge fits leads to bias in $h\widehat{\nu}-\widehat{\Phi}$, even with the commonly used amorphous Au and Cu standards. Furthermore, it is the largest source of bias in the fit from the artificial example in Sec.~\ref{sec:mock}. Different photoelectric responses between a reference material and the experimental sample can lead to a different $\Phi$, acting as another potential source of bias in $\widehat{E}-\widehat{\mu}$. The case of graphene from Fig.~\ref{fig:Fermi} is ideal as its 1D-projected DOS is constant, while its many-body and matrix-element features are weak, leading to a precise and accurate fit. Aside from yielding $h\widehat{\nu}-\widehat{\Phi}$, the Fermi-edge fitting can be used to determine $\widehat{\mu}=\widehat{\Phi}$ within the uncertainty of $h\nu$.

Lastly, we describe the so-called Fermi-edge correction~\cite{erlabpy:website}. Acquisition of a band map while using a straight detection slit on a hemispherical analyzer may result in detection of the photocurrent at shifted $(E^{\mathrm{kin}}, \eta)$. This distortion can be partially corrected with the following steps. First, Fermi-Dirac distributions are fitted for each angle $\eta$, leading to edge estimates $h\widehat{\nu}(\eta)-\widehat{\Phi}(\eta)$. Next, the resulting $h\widehat{\nu}(\eta)-\widehat{\Phi}(\eta)$ are fitted with a polynomial in $\eta$, and a reference angle $\eta^{\mathrm{ref}}$ is assigned for which $h\widehat{\nu}(\eta^{\mathrm{ref}})-\widehat{\Phi}(\eta^{\mathrm{ref}})$ is assumed to be undistorted. Afterwards, the data at each angle $\eta_j$ with angle index $j$ are shifted according to the difference between the polynomial fit result $h\widehat{\nu}(\eta_j)-\widehat{\Phi}(\eta_j)$ and $h\widehat{\nu}(\eta^{\mathrm{ref}})-\widehat{\Phi}(\eta^{\mathrm{ref}})$. In \textsc{xARPES}, the edges can be fitted with a polynomial up to second order, while an example with Li-doped graphene is provided in Fig.~\ref{fig:Li-gr-SM}.

\section{Data availability}\label{sec:data}

The raw data for the examples in Secs.~\ref{sec:sto} and~\ref{sec:graphene} are available in the examples directory of the \textsc{xARPES} \href{https://github.com/xARPES/xARPES}{GitHub repository}, with further download instructions available in the \href{https://xarpes.readthedocs.io}{xARPES documentation}. All data for reproducing the figures and the analysis are provided on the \href{https://doi.org/10.24435/materialscloud:2n-ap}{Materials Cloud archive}.

\section{Code availability}\label{sec:code}

The documentation of \textsc{xARPES} is available at \href{https://xarpes.readthedocs.io}{Read the Docs}, the source code is available on the \href{https://github.com/xARPES/xARPES}{GitHub repository}, while the public version of \textsc{xARPES} can be downloaded from \href{https://pypi.org/project/xarpes}{PyPI} and \href{https://anaconda.org/conda-forge/xarpes}{Conda Forge}.

\section{Acknowledgments}\label{sec:acknowledgements}

We gratefully acknowledge the data for Li-doped graphene shared by Dmitry Yu. Usachov and Denis V. Vyalikh~\cite{usachov:2018}, as well as the SrTiO$_3$ data shared by Igor Sokolovi{\'c} (not previously published), and the Al(100) data provided by Kenya Shimada~\cite{jiang:2011}. The authors would like to thank Eduardo Bonini Guedes, TeYu Chien, Kenya Shimada, Dmitry Yu. Usachov, Simon Moser, Cheol-Hwan Park, Jae-Mo Lihm, and Feliciano Giustino for useful discussions. T.P.v.W. is a Research Fellow of the F.R.S.-FNRS. S.P. is a Research Associate of the Fonds de la Recherche Scientifique - FNRS. This publication was supported by the Walloon Region in the strategic axe FRFS-WEL-T. J.B. and N.M. acknowledge funding by the Deutsche Forschungsgemeinschaft (DFG, German Research Foundation) under Germany's Excellence Strategy (University Allowance, EXC 2077, University of Bremen). N.M. acknowledges support by the NCCR MARVEL, a National Centre of Competence in Research, funded by the Swiss National Science Foundation (grant number 205602). Computational resources have been provided by the EuroHPC JU award granting access to MareNostrum5 at Barcelona Supercomputing Center (BSC), Spain (Project ID: EHPC-EXT-2023E02-050), and by the Consortium des \'Equipements de Calcul Intensif (C\'ECI), funded by the FRS-FNRS under Grant No. 2.5020.11 and computational resources on Lucia, the Tier-1 supercomputer of the Walloon Region with infrastructure funded by the Walloon Region under the grant agreement No. 1910247.

\section{Author contributions}\label{sec:contributions}

T.P.v.W. wrote the first draft with contributions from C.B.; T.P.v.W. wrote the code with contributions from J.B.; S.P., J.H.D., and N.M. conceived and supervised the project. All authors reviewed and approved the final version of the manuscript.

%

%

\onecolumngrid
\newpage

\setcounter{section}{0}
\renewcommand{\thesection}{S\arabic{section}}

\newcounter{equationSM}
\pretocmd{\refstepcounter}{%
  \ifstrequal{#1}{equation}{\stepcounter{equationSM}}{}%
}{}{}
\renewcommand{\theequation}{S\arabic{equationSM}}

\newcounter{figureSM}
\newcounter{tableSM}
\pretocmd{\stepcounter}{%
  \edef\@temparg{#1}%
  \def\@tempfig{figure}%
  \def\@temptab{table}%
  \ifx\@temparg\@tempfig
    \addtocounter{figureSM}{1}%
  \fi
  \ifx\@temparg\@temptab
    \addtocounter{tableSM}{1}%
  \fi
}{}{}
\renewcommand{\thefigure}{S\arabic{figureSM}}
\renewcommand{\thetable}{S\Roman{tableSM}}

\begin{center}\textbf{\Huge Supplemental Material\vspace{1em}}

{\large\textbf{
Extraction of the self energy and Eliashberg function from angle resolved photoemission\\[0.2em] spectroscopy using the \textsc{xARPES} code}}\\[1.5em]

Thomas P. van Waas,$^1$ Christophe Berthod,$^2$ Jan Berges,$^3$ Nicola Marzari,$^{4, 3, 5}$ J. Hugo Dil,$^{6, 7}$ and Samuel Ponc\'{e}$^{1, 8}$\\[0.5em]

\textit{\footnotesize
$^1$European Theoretical Spectroscopy Facility, Institute of Condensed Matter and Nanosciences,\\ Universit\'e catholique de Louvain, Chemin des \'{E}toiles 8, 1348 Louvain-la-Neuve, Belgium\\
$^2$Department of Quantum Matter Physics, University of Geneva, 1211 Geneva, Switzerland\\
$^3$U Bremen Excellence Chair, Bremen Center for Computational Materials Science,\\ and MAPEX Center for Materials and Processes, University of Bremen, 28359 Bremen, Germany\\
$^4$Theory and Simulation of Materials (THEOS), and National Centre for Computational Design and Discovery of Novel Materials (MARVEL),\\ \'{E}cole Polytechnique F\'{e}d\'{e}rale de Lausanne, 1015 Lausanne, Switzerland\\
$^5$Laboratory for Materials Simulations, Paul Scherrer Institut (PSI), 5232 Villigen, Switzerland\\
$^6$Institut de Physique, \'{E}cole Polytechnique F\'{e}d\'{e}rale de Lausanne, 1015 Lausanne, Switzerland\\
$^7$Center for Photon Science, Paul Scherrer Institut (PSI), 5232 Villigen, Switzerland\\
$^8$WEL Research Institute, Avenue Pasteur 6, 1300 Wavre, Belgium
}

\vspace{2em}
\end{center}

\twocolumngrid

\section{Linear dispersion relation}\label{SI:linearized_dispersion}

In this section, we outline the MDC fitting approach for a linear band $\varepsilon_n(\vec{k})$. In the linear case, one can eliminate the two parameters that characterize the non-interacting dispersion for a branch $n$ -- the projected Fermi wavevector $k_n^{\rm{F}}$ and the projected Fermi velocity $v_n^{\rm{F}}$ -- thus allowing for extraction of all the dispersion information with a single set of MDC fits. A non-interacting dispersion linear in the wavevector can be described by:
\begin{equation}\label{eq:linear_dispersion}
	\varepsilon_n(\eta) = \mu + v_n^{\rm{F}} \sqrt{2m_{\rm{e}} E^{\mathrm{kin}}}
	[\sin(\eta)-\sin(\eta_n^{\rm{F}})].
\end{equation}
Subsequently, the MDC fitting expression in Eq.~\eqref{eq:fit2} of the main text becomes:
\begin{multline}\label{eq:fit_linear}
	\widetilde{P}(E_j^{\rm{kin}}, \eta) = \int_{-\infty}^{\infty}\dd\eta'\,Q(\eta-\eta')\Bigg\{
	\widetilde{B}(E_j^{\rm{kin}}, \eta') +\sum_{n} |M_n(\eta')|^2\\
	\times\frac{\widetilde{A}_n^{0}(E_j)}{\pi}\frac{\widetilde{\gamma}_n(E_j)}
	{\bigl[\sin(\eta')-\widetilde{r}_n(E_j)\bigr]^2 + \widetilde{\gamma}_n^2(E_j)}\Bigg\},
\end{multline}
where the dimensionless $\widetilde{r}_n(E_j)$ and $\widetilde{\gamma}_n(E_j)$ define the fitted self-energy in the linear case via:
\begin{align}
	\label{eq:linear_real}
	\widetilde{\Sigma}_n^{\prime}(E_j) &= E_j-\mu-v_n^{\rm{F}}\sqrt{2m_{\rm{e}}E_j^{\mathrm{kin}}}
	\bigl[\widetilde{r}_n(E_j)-\sin(\eta_n^{\rm{F}})\bigr]\\
	\label{eq:linear_imaginary}
	-\widetilde{\Sigma}_n^{\prime\prime}(E_j) &= |v_n^{\rm{F}}|\sqrt{2m_{\rm{e}}E_j^{\mathrm{kin}}}
	 \widetilde{\gamma}_n(E_j),
\end{align}
with $\widetilde{\mathcal{A}}_n^0(E_j) = |v_n^{\rm{F}}|\sqrt{2 m_{\rm{e}} E_j^{\rm{kin}}} \widetilde{A}_n^{0}(E_j)/E_{\rm{dim}}$ relating the prefactors between Eqs.~\eqref{eq:fit1} and~\eqref{eq:fit2} of the main text. The MDC maxima can subsequently be calculated as $\widetilde{k}_n(E_j)=\widetilde{r}_n(E_j)\sqrt{2m_{\rm{e}} E_j^{\rm{kin}}}/\hbar$.

\section{Details on the Eliashberg-function extraction}\label{SI:eliashberg_details}

In this section, we provide details on the description of $\Sigma_n(E)$ and extraction of $\alpha^2F_n(\omega)$. First, we explain how Eq.~\eqref{eq:electron_power_4} of the main text is obtained. Fermi liquids are characterized by a sharp Fermi surface, owing to the existence of long-lived quasiparticles at $T=0$~\cite{luttinger:1961}. The self-energy in such systems presents a universal energy dependence for small $E-\mu$, given by $\Sigma_n^{\mathrm{el}}(E)=-\lambda_n^{\mathrm{el}}(E-\mu)-\ci C[(E-\mu)^2+(\pi k_{\mathrm{B}}T)^2]$, where the system-dependent constants $\lambda_n^{\mathrm{el}}$ and $C$ are related by the Kramers-Kronig relation (see Eq.~\eqref{eq:KK} below). The $(E-\mu)^2$ dependence of $\Sigma_n^{\mathrm{el}\prime\prime}(E)$ ensures that the width of the quasiparticle peak vanishes faster than the quasiparticle energy above the Fermi surface upon approaching it, which is the necessary condition for long-lived quasiparticles to exist. The Fermi-liquid behavior is predicted by a variety of theories, ranging from dynamical mean-field theory (DMFT) for the paramagnetic phase in the Hubbard model with a single-site Anderson impurity in infinite dimensions~\cite{georges:1992}, to the three-dimensional random-phase approximation for the long-range Coulomb interaction~\cite{berthod:2018a-SM}, while logarithmic corrections appear in some 2D models~\cite{bloom:1975, jungwirth:1996}. Two examples where experimental observation of Fermi liquid behavior is captured by DMFT calculations are SrVO$_3$~\cite{abramovitch:2024} and SrMoO$_3$~\cite{cappelli:2022-SM}. Experimental signatures of Fermi-liquid behavior also include a resistivity vanishing as $T^2$ at low $T$~\cite{behnia:2022}, an optical relaxation rate scaling as $(h\nu)^2+(2\pi k_{\mathrm{B}}T)^2$~\cite{stricker:2014}, and a Kadowaki-Woods relation between the resistivity and the electronic specific heat~\cite{jacko:2009}.

The $(E-\mu)^2$ increase of $-\Sigma_n^{\mathrm{el}\prime\prime}(E)$ away from the Fermi surface crosses over to a different behavior as the energy increases. Indeed, any retarded self-energy $\Sigma(E)$ must satisfy the Kramers-Kronig relation:
\begin{equation}\label{eq:KK}
	\Sigma(E)=-\frac{1}{\pi}\int_{-\infty}^{\infty}\dd E'\,\frac{\Sigma^{\prime\prime}(E')}{E-E'+\ci 0^+},
\end{equation}
which requires that $-\Sigma^{\prime\prime}(E\to\pm\infty)$ behaves as $|E|^{-\nu}$ with $\nu\geq 0$. Fermi-liquid theory does not describe the ultraviolet properties of Fermi liquids~\cite{chubukov:2012-SM}. To model the entire spectrum with a minimal number of parameters, we impose a soft cutoff $W_n$ to the quadratic increase of $-\Sigma_n^{\mathrm{el}\prime\prime}(E)$. We consider two models with different asymptotic behaviors above the cutoff. In the first model $-\Sigma_n^{\mathrm{el}\prime\prime}(E)$ saturates, while in the second model it vanishes as $|E|^{-2}$ for $|E|\to \infty$. These two cases can be described by the expression $\Sigma_n^{\mathrm{el}\prime\prime}(E)\propto[(E-\mu)^2+(\pi k_{\mathrm{B}}T)^2]/[1+(E-\mu)^{\nu}/W_n^{\nu}]$, where $\nu=2$ in the first model and $\nu=4$ in the second. We then insert this expression in Eq.~\eqref{eq:KK}, perform the integral using Cauchy's residue theorem, and fix the constant $C$ such that $\lambda_n^{\mathrm{el}}\equiv - \partial\Sigma_n^{\mathrm{el}\prime}(E)/\partial E|_{E=\mu}$ holds. The resulting model of a Kramers-Kronig consistent self-energy in the first case is:
\begin{equation}\label{eq:electron_power_2}
	\frac{\Sigma^{\mathrm{el}}_n(E)}{W_n}=\lambda_n^{\mathrm{el}}
	\frac{\bar{E}_n(\bar{T}_n^2-1)-\ci(\bar{E}_n^2+\bar{T}_n^2)}
	{(1-\bar{T}_n^2)(1+\bar{E}_n^2)},
\end{equation}
where we use the dimensionless ratios $\bar{E}_n\equiv(E-\mu)/W_n$ and $\bar{T}_n\equiv\pi k_{\mathrm{B}}T/W_n$. In the second case, we arrive at Eq.~\eqref{eq:electron_power_4} of the main text.

A key novelty of this work is the embedding of the MEM within a Bayesian inference loop for obtaining the most probable parameters, which represents a bilevel optimization problem. For bilevel optimization problems, the inner solution must be sufficiently precise, or the outer optimization may not observe the correct loss function, resulting in an incorrect solution. Fortunately, it appears that the default parameter of the MEM optimization with convergence parameter $t = 10^{-8}$ in Bryan's algorithm~\cite{bryan:1990-SM} results in complete convergence of the inner loop. By contrast, performing the parameter optimization with the inner loop and the MEM optimization with the outer loop would require complete convergence of the parameter optimization, resulting in many more iterations for the entire optimization.

In the definition of $L$ in Eq.~\eqref{eq:log-likelihood} of the main text, incorporation of $\widetilde{\Sigma}_n^{\prime}(E)$ entails $\ln(\sigma_n'(E_j))$, incorporation of $-\widetilde{\Sigma}_n^{\prime\prime}(E)$ entails $\ln(\sigma_n''(E_j))$, while both sets of data each entail a factor $N_{J} \ln (2 \pi)/ 2$. The logarithmic terms in Eq.~\eqref{eq:log-likelihood} of the main text are often omitted~\cite{shi:2004-SM, bok:2010-SM}, but they are relevant when optimizing the model parameters described in Sec.~\ref{sec:parameter_optimization} of the main text, since they depend on the model parameters through propagation of uncertainty. Furthermore, the likelihood has dimensions of an energy raised to the power $N_J$ for each set of $\{\widetilde{\Sigma}_n^{\prime}(E)\}$ and $\{-\widetilde{\Sigma}_n^{\prime\prime}(E)\}$. Therefore, we multiply the likelihood with $E_{\mathrm{dim}}^{N_J}$ per set before taking the logarithm, leading to the dimensionless form of the log-likelihood $L$ provided in Eq.~\eqref{eq:log-likelihood} of the main text. In a similar vein, compared to Ref.~\cite{shi:2004-SM}, the information entropy $S$ in our definition is divided by $E_{\rm{dim}}$ to ensure that $S$ and consequently $a$ are dimensionless. Therefore, $S$ will be dependent on the choice of $E_{\mathrm{dim}}$, and a different tuning parameter $f^{\chi^2}$ from Sec.~\ref{sec:maxent} of the main text will be needed if $E_{\mathrm{dim}}$ is changed in the ``chi2kink'' formalism. In \textsc{xARPES}, $E_{\mathrm{dim}}=1$~meV for the calculation of $S$, which we suggest as the standard for self-energy-based Eliashberg extraction, as it leads to values of $S$ and $f^{\chi^2}$ close to unity. The expression in Eq.~\eqref{eq:information_entropy} of the main text can then be approximated by a discrete sum, with the user setting the number of discrete values $N_n^{\omega}$:
\begin{multline}\label{eq:discrete_energy}
	S(\alpha^2F_n, m_n) \approx \frac{\Delta \omega_n}{E_{\rm{dim}}} \sum_{k=1}^{N_n^{\omega}}
	\bigg[ \alpha^2 F_n(\omega_k) - m_n(\omega_k) \\
	-\alpha^2F_n(\omega_k) \ln \left( \frac{\alpha^2 F_n(\omega_k)}{m_n(\omega_k)} \right) \bigg], 
\end{multline}
where the uniform energy steps are $\Delta \omega_n = (\omega_n^{\rm{max}}-\omega_n^{\rm{min}})/ N_n^{\omega}$ and with discrete energies $\omega_k$. Decreasing $N_n^{\omega}$ may speed up the iterations of Bryan's algorithm~\cite{bryan:1990-SM}, although the sampling should be sufficiently dense for the convergence of $\alpha^2F_n(\omega)$ following Eq.~\eqref{eq:objective} of the main text. Furthermore, the likelihood $\exp({L})$ is a probability density over an energy, and consequently has units of a reciprocal energy. We find that $E_{\rm{dim}}=1$~meV in Eq.~\eqref{eq:discrete_energy} gives a good bias-variance trade-off for $f^{\chi^2} \in [2.0, 2.5]$. Eq.~\eqref{eq:discrete_energy} is a summation over $\Delta \omega_n \alpha^2F_n(\omega_k)$, which also enters into the discretized form of Eq.~\eqref{eq:self_three}:
\begin{equation}\label{eq:self_discrete}
	\Sigma_n^{\rm{FM}}(E_j) \approx \sum_{k=1}^{N_n^{\omega}} \Delta \omega_n \alpha^2F_n(\omega_k)
	K(E_j,\omega_k).
\end{equation}
Thus, the numerical quantities used during the maximum-entropy method (MEM) inversion are $\Delta \omega_n \alpha^2F_n(\omega_k)$ and $\Delta \omega_n m_n(\omega_k)$. In the MEM formalism, the user has to provide the model function $m_n(\omega)$ that encodes the prior knowledge on $\alpha^2F_n(\omega)$.

In \textsc{xARPES}, the user determines the interval $\omega \in [\omega_n^{\rm{min}}, \omega_n^{\rm{max}}]$ and the parameters governing the shape of $m_n(\omega)$, where the height $h_n$ can be adjusted in the optimization loop of Sec.~\ref{sec:parameter_optimization} of the main text. These parameters result in a smooth $m_n(\omega)$ with the following domains:
\begin{align}\label{eq:model_function}
	&\frac{m_n(\omega)}{2h_n}= \nonumber \\ 
	& \begin{cases}
		\left(\frac{\omega}{\omega_{n}^{\rm{I}}}\right)^2
		& \: \omega_n^{\rm{min}} \leq \omega < \frac{\omega_{n}^{\rm{I}}}{2}, \\
		\frac{1}{2} - \left(\frac{\omega}{\omega_{n}^{\rm{I}}}-1\right)^2
		& \: \frac{\omega_{n}^{\rm{I}}}{2} \leq \omega < \omega_{n}^{\rm{I}}, \\
		\frac{1}{2}
		& \: \omega_n^{\rm{I}} \leq \omega < \omega_n^{\rm{M}}, \\
		\frac{1}{2} - \left(\frac{\omega-\omega_{n}^{\rm{M}}}{\omega_{n}^{\rm{max}} + \omega_{n}^{\rm{S}}
			- \omega_{n}^{\rm{M}}} \right)^2
		& \: \omega_{n}^{\rm{M}} \leq \omega < \frac{\omega_{n}^{\rm{max}}
			+ \omega_{n}^{\rm{S}} + \omega_{n}^{\rm{M}}}{2}, \\
		\left(\frac{\omega-\omega_{n}^{\rm{M}}}{\omega_{n}^{\rm{max}}
			+\omega_{n}^{\rm{S}} - \omega_{n}^{\rm{M}}}-1\right)^2
		& \: \frac{\omega_{n}^{\rm{max}} + \omega_{n}^{\rm{S}}
			+ \omega_{n}^{\rm{M}}}{2} \leq \omega \leq \omega_{n}^{\rm{max}},
	\end{cases}
\end{align}
where $\omega_n^{\rm{I}}$ governs the onset of $m_n(\omega)$, such that it has quadratic concave upwards shape corresponding to a Debye spectrum~\cite{grimvall:1981-SM}, after which it has a quadratic concave downwards shape up to its maximum height $h_n$ at $\omega=\omega_n^{\rm{I}}$. The second parameter $\omega_n^{\rm{M}}$ governs the decay of $m_n(\omega)$, as from $\omega=\omega_n^{\rm{I}}$ onward it has a concave downwards quadratic shape until $\omega=(\omega_n^{\rm{max}}+\omega_n^{\rm{S}}+\omega_n^{\rm{M}})/2$, turning into concave upwards quadratic decay until reaching $\omega=\omega_n^{\rm{max}}+\omega_n^{\rm{S}}$. For a given $\omega_n^{\rm{S}}$ (ideally less than 1~meV), $m_n(\omega)$ returns to zero at $\omega=\omega_n^{\rm{max}}+\omega_n^{\rm{S}}$, as large ratios of $\alpha^2F_n(\omega)/m_n(\omega)$ for $\omega \in [\omega_n^{\rm{min}}, \omega_n^{\rm{max}}]$ are found to be numerically unstable. An example shape for $m_n(\omega)$ is found in Fig.~\ref{fig:STO-SM}\textbf{d} (Sec.~\ref{SI:additional_STO}) for SrTiO$_3$ with a dash-dotted line.

For the comparison of two extracted Eliashberg functions $\alpha^2F_{1}(\omega)$ and $\alpha^2F_{2}(\omega)$, we use the Euclidean distance measure $M(\alpha^2F_{1}(\omega), \alpha^2F_{2}(\omega))$:
\begin{equation}\label{eq:euclidean_measure}
	M(\alpha^2F_{1}(\omega), \alpha^2F_{2}(\omega)) =
	\frac{\|\alpha^2F_1 - \alpha^2F_2\|^2}{\|\alpha^2F_1\|^2 + \|\alpha^2F_2\|^2}, 
\end{equation}
with $\|\alpha^2F_n\|^{m}=\int_0^{\infty}\dd\omega\,[\alpha^2F_n(\omega)]^{m}$. In the context of comparing Eliashberg functions, this function can be used as follows. The denominator in Eq.~\eqref{eq:euclidean_measure} ensures that the measure is intensive in $\omega$, such that one can compare Eliashberg functions with different $\omega$-samplings, although interpolation might be necessary to evaluate the numerator of Eq.~\eqref{eq:euclidean_measure}. Furthermore, the denominator also normalizes $M(\alpha^2F_{1}(\omega),\alpha^2F_{2}(\omega))$ via the magnitude of the Eliashberg functions, such that the distance caused by the bias from an experimental setup is relative to the magnitude of the Eliashberg function. Considering the magnitude, $M(\alpha^2F_{1}(\omega), \alpha^2F_{2}(\omega))=0$ when $\alpha^2F_{1}(\omega)=\alpha^2F_{2}(\omega)$, $M(\alpha^2F_{1}(\omega), \alpha^2F_{2}(\omega))=1/4$ when $\alpha^2F_{1}(\omega)$ and $\alpha^2F_{2}(\omega)$ are two uniformly random functions of equal height and width, and $M(\alpha^2F_{1}(\omega), \alpha^2F_{2}(\omega))=1$ when at most one of the two functions is everywhere 0.

\section{Supplement on verification example}\label{SI:additional_mock}

In this section, the parameter optimization of the artificial example from Sec.~\ref{sec:mock} of the main text is outlined in more detail. Concerning the metaparameters, we start the optimization with an initial model function $m_n(\omega)$ and a value for $a_{\mathrm{n}}^{\mathrm{min}}$ that allow for a wide range of initial conditions, as the MEM may otherwise fail to perform the extraction of $\alpha^2F_n(\omega)$. For the model function, Eq.~\eqref{eq:model_function}, large values of $\omega_n^{\mathrm{min}}$ and $\omega_n^{\mathrm{S}}$ generally result in a stabler MEM, although we suggest to decrease both of them below 1~meV in the final calculation for a physically meaningful $\alpha^2F_n(\omega)$. For the final calculation, we modify $a_{\mathrm{n}}^{\mathrm{min}}$ and $a_{\mathrm{n}}^{\mathrm{max}}$ such that the full kink in $\chi^2(\log_{10}(a))$ is visible, although a smaller $a_{\mathrm{n}}^{\mathrm{min}}$ usually requires better parameter choices for the $\alpha^2F_n(\omega)$-extraction step. Furthermore, the code speed versus accuracy can be controlled with the cutoff value $\sigma_n^{\mathrm{cut}}$ in Bryan's algorithm~\cite{bryan:1990-SM}. A smaller $\sigma_n^{\mathrm{cut}}$ leads to a larger number of search directions in the singular space generated by the singular value decomposition of the kernel matrix $K(E,\omega)$, denoted as $T$ in Ref.~\cite{bryan:1990-SM}, at the cost of slower iterations. Thus, we run the optimization loop with a large $\sigma_n^{\mathrm{cut}}=1$, while using $\sigma_n^{\mathrm{cut}}=10^{-4}$ for the final calculation. If there are appreciably different final parameter sets with different log-posteriors $L+aS$, for example from using a smaller $\sigma_n^{\mathrm{cut}}$ or after a calculation with a different initial guess, Eqs.\eqref{eq:a2f_prob}--\eqref{eq:parameter_prob} of the main text suggest selecting the result with the highest log-posterior.

We now explain the steps that we follow for the optimization loop of \textsc{xARPES}. We first visualize $\Sigma_n(E)$ followed by manually fine-tuning the model parameters to perform a one-shot extraction with large $\log_{10}(a_{\mathrm{R}}^{\rm{min}})$ and $\sigma_{\mathrm{R}}^{\mathrm{cut}}$, followed by calling the optimization loop. After loading $\Sigma_n(E)$ -- but before a one-shot extraction of $\alpha^2F_{n0}(\omega)$ -- we visually inspect the resulting $\Sigma_{n0}^{\mathrm{el}}(E)$, $\Sigma_{n0}^{\mathrm{imp}}(E)$, and $\Sigma_{n0}^{\mathrm{ph}}(E)$ individually, to see if $\Sigma_{n0}^{\mathrm{ph}}(E)$ has a typical shape from which the Eliashberg function can be extracted. In particular, $\Sigma_{n0}^{\mathrm{ph}\prime}(E)$ and $-\Sigma_{n0}^{\mathrm{ph}\prime\prime}(E)$ should be of approximately the same magnitude, while the former should decay approximately as $1/(\mu-E)$ for $|\mu-E|>\omega_n^{\mathrm{max}}$, while the latter should become constant.

We use the default $\log_{10}(a_{\mathrm{R}}^{\rm{min}})=3$ to determine how much the parameters can vary individually from a set of initial conditions. Denoting an element of the set of model parameters $V$ at iteration $j$ as $v_j$, we initialize the one-shot parameters $v_0$ with their true values $v^{\mathrm{true}}$, and set $h_{\mathrm{R}0}=0.08$, corresponding to $||m_{n0}||\approx||\alpha^2F_n^{\mathrm{true}}||$. These initial values and their possible percentage deviations are listed in Table~\ref{tab:sup1}.

\begin{table}[tb]
\vspace{-0.7em}\caption{Initial values and upper/lower allowed percentage deviations during the one-shot extraction step with $\log_{10}(a_{\mathrm{R}}^{\rm{min}})=3$ and $\sigma_{\mathrm{R}}^{\mathrm{cut}}=1$.
\label{tab:sup1}}
	\renewcommand{\arraystretch}{1.1}
	\begin{tabular*}{\columnwidth}{@{\extracolsep{\fill}}lddddd}
		\hline\hline
		Parameter & 
		\multicolumn{1}{r}{$m_{\mathrm{R}}^{\mathrm{b}}$ ($m_{\mathrm{e}}$)} & 
		\multicolumn{1}{c}{$k_{\mathrm{R}}^{\mathrm{F}}$ ($\Angstrom^{-1}$)} & 
		\multicolumn{1}{c}{$\Gamma_{\mathrm{R}}^{\mathrm{imp}}$ (meV)} & 
		\multicolumn{1}{c}{$\lambda_{\mathrm{R}}^{\mathrm{el}}$ (--)} & 
		\multicolumn{1}{c}{$h_{\mathrm{R}}$ (--)}\\[0.2em]
		\hline
		Upper (\%)     & 20     & 8      & 300  & 350    & \multicolumn{1}{c}{$5\times10^4$} \\
		Initial        & 1.5875 & 0.25   & 10   & 0.12   & 0.08   \\
		Lower (\%)     & -50    & -0.03  & -2   & -100   & -95    \\
		\hline \hline
	\end{tabular*}
\end{table}

The code is more lenient towards overestimation than underestimation, with 8\% or more overestimation allowed for all of the parameters. We further fine-tune the model parameters by visual inspection of the extracted $\alpha^2F_n(\omega)$ until the solution is sufficiently good for running the optimization loop. In particular, we verify that $\Sigma_{n0}^{\mathrm{\prime}}(E)$ closely follows $\widetilde{\Sigma}_{n0}^{\mathrm{\prime}}(E)$ for $E\rightarrow \mu$, which allowing for a precise determination of $k_{n0}^{\mathrm{F}}$.

We initialize the optimization loop with the same initial parameters, while using the default parameters for the optimization described in Sec.~\ref{sec:technical} of the main text. The Nelder-Mead algorithm~\cite{nelder:1965} is the default in \textsc{xARPES} as we find that it always arrives at the same objective function minimum for the artificial data set, while rarely taking problematically large steps in parameter space. The minimization algorithm is executed with a callback feature that reverts the parameters by several steps if (i) a new set of parameters causes the MEM extraction to fail, or (ii) the search algorithm attempts to restart with the initial conditions, such as when the simplex in the Nelder-Mead algorithm collapses back to the start~\cite{nelder:1965}. Whenever the callback feature has been used too many times, the code exits while warning that the obtained solution may not yet be optimal. The identical initial parameters, as well as the allowed parameter deviations during the optimization loop are listed in Table~\ref{tab:sup2}.

\begin{table}[tb]
\caption{Initial values and upper/lower allowed percentage deviations for the optimization loop with $\log_{10}(a_{\mathrm{R}}^{\rm{min}})=3$ and $\sigma_{\mathrm{R}}^{\mathrm{cut}}=1$.
\label{tab:sup2}}
	\renewcommand{\arraystretch}{1.1}
	\begin{tabular*}{\columnwidth}{@{\extracolsep{\fill}}lddddd}
		\hline\hline
		Parameter & 
		\multicolumn{1}{r}{$m_{\mathrm{R}}^{\mathrm{b}}$ ($m_{\mathrm{e}}$)} & 
		\multicolumn{1}{c}{$k_{\mathrm{R}}^{\mathrm{F}}$ ($\Angstrom^{-1}$)} & 
		\multicolumn{1}{c}{$\Gamma_{\mathrm{R}}^{\mathrm{imp}}$ (meV)} & 
		\multicolumn{1}{c}{$\lambda_{\mathrm{R}}^{\mathrm{el}}$ (--)} & 
		\multicolumn{1}{c}{$h_{\mathrm{R}}$ (--)}\\[0.2em]
		\hline
		Upper (\%)    & 2      & 0.4     & 5      & 20     & 500    \\
		Initial       & 1.5875 & 0.25    & 10     & 0.12   & 0.08   \\
		Lower (\%)    & -0.3   & -0.02   & -0.4   & -90    & -20    \\
		\hline \hline
	\end{tabular*}
\end{table}

The allowed deviations are smaller for the loop by approximately one order of magnitude compared to the one-shot initial conditions, as the algorithm may not be able to find a solution for certain starting configurations. If the loop fails due to an error or a sub-optimal solution, the user should initialize the loop with a different initial guess. The number of significant figures in Table~\ref{tab:sup2} indicates the precision in the parameters in the obtained solution. A slightly higher precision in $h_{\mathrm{R}}$ can sometimes rapidly be found by re-initializing the optimization loop with the new output parameters, while using a slightly modified $h_0$. This potential precision increase will likely be automated in a future version of \textsc{xARPES}. The parameter bias is greatly reduced during the optimization, allowing for a decrease of $\log_{10}(a_{\mathrm{R}}^{\rm{min}})$ and $\sigma_{\mathrm{R}}^{\mathrm{cut}}$ towards the default values of $1.5$ and $10^{-4}$, respectively. We perform a final optimization with these parameters, leading to $\alpha^2F_{\mathrm{R}}(\omega)$ displayed in Fig.~\ref{fig:a2F}\textbf{a} of the main text, with the parameters provided in Table~\ref{tab:sup3}.

\begin{table}[tb]
\vspace{-0.7em}\caption{Optimized values and upper/lower allowed percentage boundary deviations for the final extraction with $\log_{10}(a_{\mathrm{R}}^{\rm{min}})=1.5$ and $\sigma_{\mathrm{R}}^{\mathrm{cut}}=10^{-4}$.
\label{tab:sup3}}
	\renewcommand{\arraystretch}{1.1}
	\begin{tabular*}{\columnwidth}{@{\extracolsep{-0.05em}}lddddd}
		\hline\hline
		Parameter & 
		\multicolumn{1}{r}{$m_{\mathrm{R}}^{\mathrm{b}}$ ($m_{\mathrm{e}}$)} & 
		\multicolumn{1}{r}{$k_{\mathrm{R}}^{\mathrm{F}}$ ($\Angstrom^{-1}$)} & 
		\multicolumn{1}{c}{$\Gamma_{\mathrm{R}}^{\mathrm{imp}}$ (meV)} & 
		\multicolumn{1}{c}{$\lambda_{\mathrm{R}}^{\mathrm{el}}$ (--)} & 
		\multicolumn{1}{c}{$h_{\mathrm{R}}$ (--)}\\[0.2em]
		\hline
		Upper (\%)     & 3       & 0.2     & 2       & 40      & \multicolumn{1}{c}{$4\times10^{4}$} \\
		Opt.           & 1.596   & 0.25001 & 10.04   & 0.113   & 0.08 \\
		Lower (\%)     & -4      & -0.04   & -2      & -100    & -90 \\
		\hline \hline
	\end{tabular*}
\end{table}

Next, we individually vary these parameters with respect to the final solution to determine the allowed deviation from the final solution. We find more stringent conditions compared to the previous two sets of conditions in most but not all cases, including a widened lower deviation of $m_{\mathrm{R}}^{\mathrm{b}}$ from $-2$\% to $-4$\%, and a maximum allowed variation of $k_{\mathrm{R}}^{\mathrm{F}}$ of only $0.2$\%. The conditions are more symmetric around the optimized results than around the initial guesses, as the final result represents a maximum in the probability density surface.

Finally, we show that the initial guesses for most of the parameters of the optimization loop are allowed to deviate appreciably from the final/true values, as long as the underlying data are still well-described by these parameters. For example, setting $m_n^{\mathrm{b}}$ twice as large while setting $\Gamma_n^{\mathrm{imp}}$ half as large in Eq.~\eqref{eq:reobtain_imag} of the main text yields identical $\widetilde{\gamma}_n(E)$, showing that these two effects on the broadening cancel out. For the optimization parameters, we define their values $\widehat{v}_j$ during function evaluation $j$, the estimate after the optimization $\widehat{v}$, and the true value $v^{\mathrm{true}}$. The relative difference at the function evaluation $j$ can then be defined as:
\begin{equation}\label{eq:relative_difference}
	q_j = \frac{\widehat{v}_j - v^{\mathrm{true}}}{\widehat{v} - v^{\mathrm{true}}}.
\end{equation}
The relative difference is the ratio between the error $\widehat{v}_j - v^{\mathrm{true}}$ at function call $j$ and at convergence $\widehat{v} - v^{\mathrm{true}}$, quantifying the ability of the code to improve on an initial guess. To minimize the bias in the final parameters -- which will serve as a reference -- we set the estimate of the chemical potential from the Fermi-edge fit result from Sec.~\ref{sec:fermi_edge} equal to the true result.

\begin{figure}[tb]
\includegraphics[width=\columnwidth]{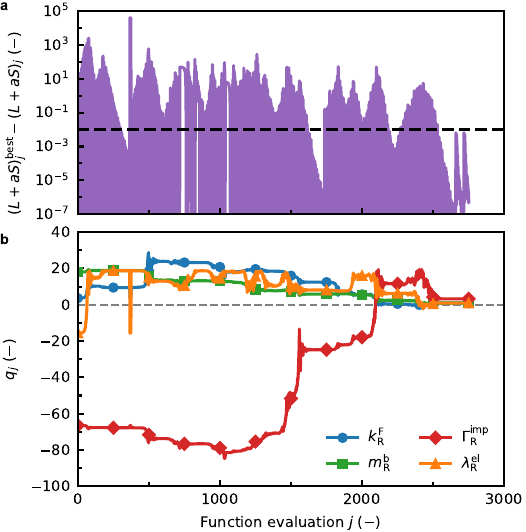}
\caption{\textbf{Artificial example code convergence for the right-hand (R) branch}.
\textbf{a} Difference between the preceding function evaluation $(L+aS)^{\mathrm{best}}_j$ at function evaluation $j$ and the objective function $(L+aS)_j$. \textbf{b} Relative difference $q_j$ defined in Eq.~\eqref{eq:relative_difference} for the four model parameters $m_{\mathrm{R}}^{\mathrm{b}}$, $k_{\mathrm{R}}^{\mathrm{F}}$, $\lambda_{\mathrm{R}}^{\mathrm{el}}$, and $\Gamma_{\mathrm{R}}^{\mathrm{imp}}$ during the optimization.
\label{fig:convergence}}
\end{figure}

\begin{figure*}[tb]
\includegraphics[width=\textwidth]{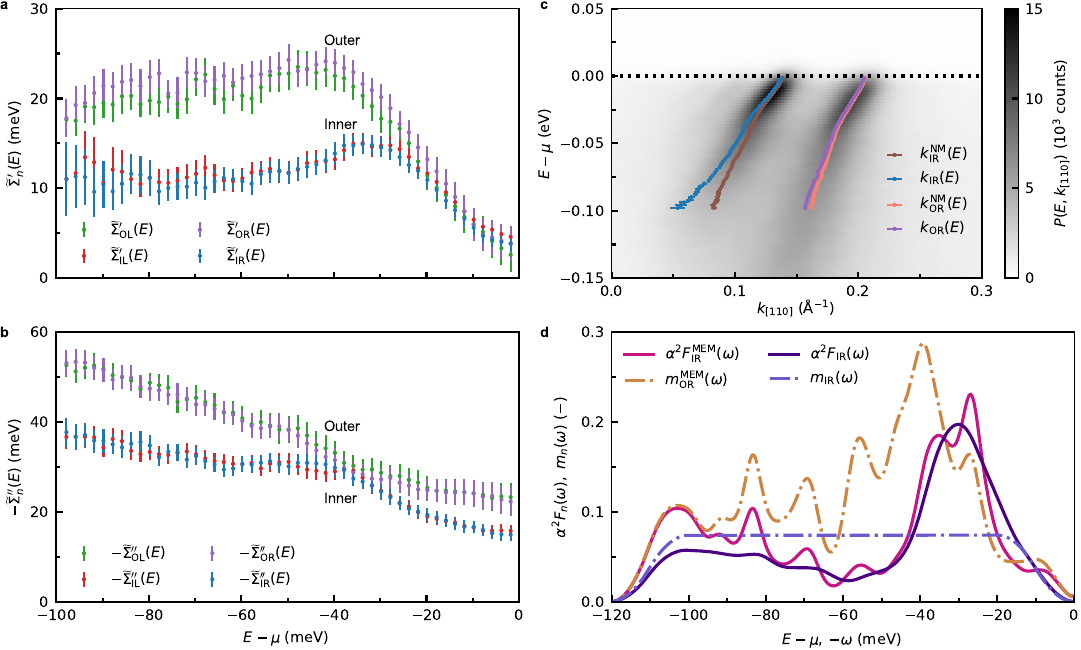}
\caption{\textbf{Supplemental analysis for TiO$_2$-terminated SrTiO$_3$}.
\textbf{a} Self-energy data of the real part $\widetilde{\Sigma}^{\prime}_n(E)$ for the outer left (OL, green), inner left (IL, red), inner right (IR, blue), and outer right (OR, purple) of the SrTiO$_3$ branches shown in Fig.~\ref{fig:STO}\textbf{a} of the main text. \textbf{b} Minus imaginary parts $-\widetilde{\Sigma}^{\prime\prime}_n(E)$ for the same branches and with the same color coding as in \textbf{a}. \textbf{c} The MDC maxima for the IR branch with the matrix-element correction (blue, $k_{\mathrm{IR}}(E)$) and without it (brown, $k_{\mathrm{IR}}^{\mathrm{NM}}(E)$), as well as for the outer right branch with (violet, $k_{\mathrm{OR}}(E)$) and without (pink, $k_{\mathrm{OR}}^{\mathrm{NM}}(E)$) the matrix-element correction. \textbf{d} $\alpha^2F_{\mathrm{IR}}^{\mathrm{MEM}}(\omega)$ (magenta) based on $m_{\mathrm{OR}}^{\mathrm{MEM}}(\omega)=\alpha^2F_{\mathrm{OR}}(\omega)$ -- with the latter extracted from the OR branch in Ref.~\cite{sokolovic:2025-SM} (dash-dotted brown) -- compared to $\alpha^2F_{\mathrm{IR}}(\omega)$ (dark purple) based on its optimized default-shaped $m_{\mathrm{IR}}(\omega)$ (light purple), with the latter two reproduced from Fig.~\ref{fig:STO}\textbf{a} of the main text.
\label{fig:STO-SM}}
\end{figure*}

We construct a set of initial parameters with $q_0$ ranging from a factor of 4 to a factor of 66 and use the optimization loop. The callback is also used for the optimization criterion: the difference between $-(L+aS)_{j}$ at iteration $j$ and the best previous value $-(L+aS)^{\mathrm{best}}_j\equiv\mathrm{min} \{-(L+aS)_i \}_{i=0}^{j-1}$ has to be below a threshold for a fixed number of iterations. Fig.~\ref{fig:convergence}\textbf{a} shows $(L+aS)_j^{\mathrm{best}}-(L+aS)_j$ during the optimization versus the convergence threshold, where the optimization terminates at approximately 2750 function evaluations. Certain optimization algorithms may perform multiple objective function evaluations per iteration, so that the number of function evaluations below the threshold may exceed the number of iterations. For Nelder-Mead, the number of evaluations per iteration depends on whether the iteration concerns reflection, expansion, contraction, or shrinkage~\cite{nelder:1965}, although the number of evaluations below the threshold should not strongly exceed the number of iterations below the threshold, with the majority of final iterations being reflection and contraction. The ratios $q_j$ are displayed in Fig.~\ref{fig:convergence}\textbf{b}, where use of the callback feature is seen in large, reversible jumps in $q_j$ for small $j$-intervals. The code greatly improves on the guesses for $m_{\mathrm{R}}^{\mathrm{b}}$, $\lambda_{\mathrm{R}}^{\mathrm{el}}$, and $\Gamma_{\mathrm{R}}^{\mathrm{imp}}$, while a good guess for $k_{\mathrm{R}}^{\mathrm{F}}$ can already be obtained during the initial one-shot extraction.

\section{Supplement on TiO$_2$-terminated SrTiO$_3$}\label{SI:additional_STO}

In this section, we provide additional details on the extraction of the self-energy and the Eliashberg function from the TiO$_2$-terminated surface of SrTiO$_3$ discussed in Sec.~\ref{sec:sto} of the main text. The real $\widetilde{\Sigma}_n^{\prime}(E)$ and minus imaginary parts $-\widetilde{\Sigma}_n^{\prime\prime}(E)$ of the self-energy are displayed in Figs.~\ref{fig:STO-SM}\textbf{a} and \ref{fig:STO-SM}\textbf{b} respectively, using the same color coding as for the MDC maxima in Fig.~\ref{fig:STO}\textbf{a} of the main text. These self-energies are based on $m_{\mathrm{I}}^{\mathrm{b}}=0.6$~$m_{\mathrm{e}}=m_{\mathrm{O}}^{\mathrm{b}}$, $k_{\mathrm{I}}^{\mathrm{F}}=0.142~\Angstrom^{-1}$, and $k_{\mathrm{O}}^{\mathrm{F}}=0.208~\Angstrom^{-1}$ based on visually inspecting that $\widetilde{\Sigma}_{n}^{\prime}(E)$ (i) satisfies particle-hole symmetry and (ii) decays as $1/(\mu-E)$ for $\mu-E>\omega_n^{\mathrm{max}}$. Comparison of the dotted red and blue lines with the black lines in Fig.~\ref{fig:STO}\textbf{a} shows that the visually inspected results are in good agreement with the output of the optimization loop. As remarked previously in Ref.~\cite{sokolovic:2025-SM}, there is a high similarity for the self-energies of branches originating from the same band, with lower similarity between self-energies originating from different bands. The pronounced dip near $E-\mu=-50$~meV in $-\widetilde{\Sigma}_{\mathrm{IL}}^{\prime\prime}(E)$ and $-\widetilde{\Sigma}_{\mathrm{IR}}^{\prime\prime}(E)$ found in Ref.~\cite{sokolovic:2025-SM} is not apparent in our case, possibly due to a slightly different hybridization with the bulk-derived bands, or a different photointensity coming from the bulk-derived bands themselves.

While Ref.~\cite{sokolovic:2025-SM} has provided a motivation for the MEC in SrTiO$_3$ based on the sum rule for the spectral function~\cite{abrikosov:1964-SM}, here we show its effect on the MDC maxima. The MDC maxima with the MEC $k_n(E)$ and without it $k_n^{\mathrm{NM}}(E)$ for the inner right (IR) and outer right (OR) branches are displayed on top of the band map $P(E, k_{[110]})$ in Fig.~\ref{fig:STO-SM}\textbf{c}. Without $|M(E,\eta)|^2\propto \sin^2(\theta)$, the spectral weight at small $\theta$ is underestimated, such that $k_n^{\mathrm{NM}}(E)$ is shifted towards higher momenta. Physically, the MDC maxima should return towards a parabolic non-interacting dispersion for $\mu-E \rightarrow E_n^{\mathrm{F}}$, with $ E_n^{\mathrm{F}}$ the Fermi energy. Fig.~\ref{fig:STO-SM}\textbf{c} shows that this prediction is obeyed for $k_{\mathrm{IR}}(E)$ and $k_{\mathrm{OR}}(E)$, but not for $k_{\mathrm{IR}}^{\mathrm{NM}}(E)$ and $k_{\mathrm{OR}}^{\mathrm{NM}}(E)$, providing further evidence that the matrix element $|M_{d_{xy}}(\eta)|^2 \propto \sin^2(\eta)$ is suitable near the $\bar{\Gamma}_{00}$-point for the $d_{xy}$-derived bands of atomically terminated SrTiO$_3$.

Finally, we explore the purpose of the MEM to encode prior knowledge on $\alpha^2F_n(\omega)$ in $m_n(\omega)$~\cite{gull:1989-SM}. In the absence of a previously extracted $\alpha^2F_n(\omega)$, the default $m_n(\omega)$ of \textsc{xARPES} provides a reasonable initial guess. However, if a previous ARPES band map is available, an existing $\alpha^2F_n(\omega)$ can be used as initial $m_n(\omega)$. As in Sec,~\ref{sec:sto} of the main text, we extract $\alpha^2F_{\mathrm{IR}}(\omega)$ because the photointensity for the inner bands is higher than that of the outer bands. Our extracted $\alpha^2F_{\mathrm{IR}}(\omega)$ and its simultaneously optimized $m_{\mathrm{IR}}(\omega)$ from Fig.~\ref{fig:Li-gr}\textbf{b} of the main text are repeated in Fig.~\ref{fig:STO-SM}\textbf{d}, and compared to setting $m_{\mathrm{OR}}^{\mathrm{MEM}}(\omega)$ equal to $\alpha^2F_{\mathrm{OR}}(\omega)$ from Ref.~\cite{sokolovic:2025-SM}, resulting in the extracted $\alpha^2F_{\mathrm{IR}}^{\mathrm{MEM}}(\omega)$. We see that when $m_{\mathrm{OR}}^{\mathrm{MEM}}(\omega)$ is used, features present in both sets of self-energies are enhanced, while features only occurring in the new data set are suppressed. Furthermore, the main spectral weight in $\alpha^2F_{\mathrm{IR}}(\omega)$ at $\omega=30$~meV has not shifted $\alpha^2F_{\mathrm{IR}}^{\mathrm{MEM}}(\omega)$, showing that the most important features are robust with respect to the model function. By contrast, the modes at $\omega_{\mathrm{LO}_3}=56$~meV, $\omega_{\mathrm{TO}_4}=69$~meV, and $\omega_{\mathrm{LO}_4}=84$~meV are less pronounced and slightly red-shifted in going from $\alpha^2F_{\mathrm{OR}}(\omega)$ of Ref.~\cite{sokolovic:2025-SM} to $\alpha^2F_{\mathrm{IR}}(\omega)$ provided here. The pronunciation is probably weaker because the phonon modes are closer to the zone center and the band bottom, where the photointensity is weaker and broadened.

\section{Supplement on quasi-freestanding graphene}\label{SI:additional_graphene}

\begin{figure*}[tb]
\includegraphics[width=\textwidth]{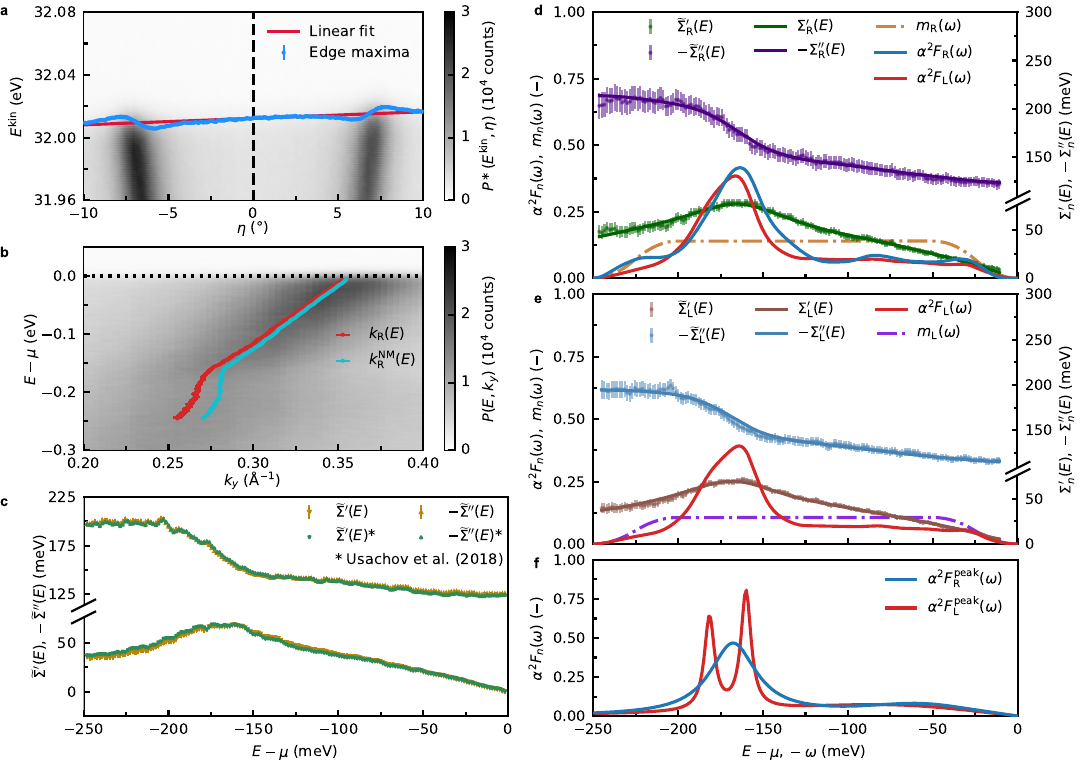}
\caption{\textbf{Supplemental analysis for Li-doped graphene}.
\textbf{a} Fermi-edge correction for the graphene band map, showing the edges of the individual Fermi-edge fits as blue bars with 95\% confidence intervals and a red line for the linear fit to these edges. The fit at the reference angle $\eta^{\mathrm{ref}}$ ($=0$ here) yields $h\widehat{\nu}-\widehat{\Phi}$. \textbf{b} The MDC maxima with (red, $k_{\mathrm{R}}(E)$) and without (teal, $k_{\mathrm{R}}^{\mathrm{NM}}(E)$) the matrix-element correction. \textbf{c} Reproduction of $\widetilde{\Sigma}'(E)$* (downward green triangle) and $-\widetilde{\Sigma}''(E)$* (upward green triangle) from Ref.~\cite{usachov:2018-SM} with the real $\widetilde{\Sigma}'(E)$ (downward gold triangle) and minus imaginary parts $-\widetilde{\Sigma}''(E)$ (upward gold triangle) with 95 \% confidence intervals using the method described in the text. \textbf{d} The results of Fig.~\ref{fig:Li-gr}\textbf{b} of the main text with the same color coding, without the Fermi-edge correction from \textbf{a}. \textbf{e} Left-hand $\alpha^2F_{\mathrm{L}}(\omega)$ (red) repeated from Fig.~6\textbf{b} of the main text, shown here with its co-optimized $m_{\mathrm{L}}(\omega)$ (orchid), based on $\widetilde{\Sigma}_{\mathrm{L}}^{\prime}(E)$ (sky blue bars) and $-\widetilde{\Sigma}_{\mathrm{L}}^{\prime\prime}(E)$ (brown bars), resulting in the reconstructions $\Sigma_{\mathrm{L}}'(E)$ (sky blue) and $-\Sigma_\mathrm{L}''(E)$ (brown). \textbf{f} Comparison between $\alpha^2F_{\mathrm{R}}^{\mathrm{peak}}(\omega)$ (blue) as a sum of three Lorentzian-based peaks based on $\widetilde{\Sigma}^{\prime}_{\mathrm{R}}(E)$ shown in Fig.~\ref{fig:Li-gr}\textbf{b} of the main text and $\alpha^2F_{\mathrm{L}}^{\mathrm{peak}}(\omega)$ (red) based on $\widetilde{\Sigma}^{\prime}_{\mathrm{L}}(E)$ shown in \textbf{e}. For $\alpha^2F_{\mathrm{L}}^{\mathrm{peak}}(\omega)$, two peaks coincide at $\omega_{\rm{L}2} =\omega_{\rm{L}3}= 168$~meV.
\label{fig:Li-gr-SM}}
\end{figure*}

In this section, we provide additional details on the extraction of the self-energy and the Eliashberg function from the Li-doped, quasi-freestanding graphene discussed in Sec.~\ref{sec:graphene} of the main text. First, we perform the Fermi-edge correction described in Sec.~\ref{sec:fermi_edge} of the main text. The resulting correction with reference angle $\eta^{\mathrm{ref}}=0$ is displayed in Fig.~\ref{fig:Li-gr-SM}\textbf{a} for the Li-doped graphene band map.

For the MEC of graphene, we use the expression for the conduction band of graphene from the supplement of Ref.~\cite{usachov:2018-SM}:
\begin{equation}\label{eq:graphene_mec}
	|M(E,\mathbf{k})|^2=1+\cos(\phi_{\mathbf{k}}),
\end{equation}
where $\phi_{\mathbf{k}}$ is the angle between the vector $\mathbf{k}-\mathbf{K}$ and the unit vector $\widehat{\mathbf{k}}_x$. Geometrically, this angle can be obtained from $\tan(\phi_{\mathbf{k}})=k_y/(k_x-K_x)$. Furthermore, we ignore the finite offset angle in the variation of $k_y$, resulting in $k_y=\sqrt{2m_{\mathrm{e}}E^{\mathrm{kin}}/\hbar^2}\sin(\eta)$~\cite{zhang:2022-SM}. Rewriting the cosine, the matrix element can finally be expressed as:
\begin{equation}\label{eq:final_mec}
	|M(E,\eta)|^2 = 1+\frac{k_x-K_x}{\sqrt{(k_x-K_x)^2
	+\frac{2m_{\mathrm{e}}E^{\mathrm{kin}}\sin^2(\eta)}{\hbar^2}}}.
\end{equation}
From Fig.~2 of Ref.~\cite{usachov:2018-SM}, we estimate that for cut~\#2, $k_x -K_x \approx -0.13~\Angstrom^{-1}$, although strictly speaking $k_x$ should slightly vary along the cut due to a finite offset angle ($\beta$ in the notation of Ref.~\cite{zhang:2022-SM}, where we reiterate that our definitions of $k_x$ and $k_y$ are interchanged). Since the first release of \textsc{xARPES} does not support PMEs that depend $E^{\mathrm{kin}}$, $E^{\mathrm{kin}}$ in Eq.~\eqref{eq:final_mec} has accordingly been substituted with $h\widehat{\nu}-\widehat{\Phi}$ in the example notebook of graphene. Conveniently, the variation in the kinetic energy is negligible across this band map, such that this substitution has little effect. The MDC fitting with and without (NM) the MEC of the Li-doped graphene band map is displayed in Fig.~\ref{fig:Li-gr-SM}\textbf{b}. As in the SrTiO$_3$ case in Fig.~\ref{fig:STO-SM}, the spectral weight without the MEC is underestimated for small $\eta$ as $|M(E,\eta)|^2$ in Eq.~\eqref{eq:final_mec} decreases towards the dark corridor, such that $k_{\mathrm{R}}^{\mathrm{NM}}(E)$ also ends up at higher momenta for this band map.

Next, we reproduce the self-energies from cut~\#2 of Ref.~\cite{usachov:2018-SM} by using an effectively identical method for the self-energy extraction. In Ref.~\cite{usachov:2018-SM}, the MDCs containing both branches are fitted for individual $E^{\mathrm{kin}}$ with an asymmetric dispersion that has two peaks of different amplitudes and positions, but identical widths. Subsequently, $-\Sigma^{\prime\prime}(E)$ is directly calculated from the width, while $\Sigma^{\prime}(E)$ is calculated as the average from the two MDC maxima following the Lorentzian approach described in Sec.~\ref{sec:mock} of the main text. We repeat the procedure while using the parabolic dispersion in \textsc{xARPES} instead of the asymmetric dispersion from the supplement of Ref.~\cite{usachov:2018-SM}. This asymmetric dispersion can be rewritten into our parabolic dispersion relation with the Taylor expansion $(1+\alpha k/\beta)^{-1}\approx 1-\alpha k/\beta$, using $\alpha$ and $\beta$ in their notation. We expect both methods to give essentially identical results, as the curvature of the non-interacting dispersion along cut \#2 is small. We follow Ref.~\cite{usachov:2018-SM} in using a quadratic background for the MDC fits, as well as assuming $k_x -K_x \approx -0.18~\Angstrom^{-1}$ for the photoemission matrix element. The self-energies $\Sigma(E)$ with the parabolic band are displayed in Fig.~\ref{fig:Li-gr-SM}\textbf{c} with 95\% confidence intervals, compared to the self-energies $\Sigma(E)^{*}$ from Ref.~\cite{usachov:2018-SM}. Very good agreement between the two methods is obtained with $m^{\mathrm{b}}=4.22~m_{\mathrm{e}}$, $k^{\mathrm{F}}=1.555~\Angstrom^{-1}$, and $k^{\mathrm{c}}=-1.2~\Angstrom^{-1}$. We attribute remaining discrepancies to different MDC fitting ranges and the different dispersion relations.

Next, we consider the effect of omitting the Fermi-edge correction on the extraction of $\alpha^2F_n(\omega)$. Fig.~\ref{fig:Li-gr-SM}\textbf{d} shows the same quantities as Fig.~\ref{fig:Li-gr}\textbf{b} of the main text, obtained here without the Fermi-edge correction. Notably, the main peak in $\alpha^2F_{\mathrm{L}}(\omega)$ is blue-shifted, while the main peak in $\alpha^2F_{\mathrm{R}}(\omega)$ is red-shifted, in accordance with the fact that $h\widehat{\nu}-\widehat{\Phi}$ is overestimated for the left-hand branch, and underestimated for the right-hand branch.

In Fig.~\ref{fig:Li-gr-SM}\textbf{e}, we show $\alpha^2F_{\mathrm{L}}(E)$ based on $\widetilde{\Sigma}_{\mathrm{L}}(E)$ with its co-optimized $m_{\mathrm{L}}(\omega)$, together with the reconstruction $\Sigma_{\mathrm{L}}(E)$. Compared to the right-hand side results, the agreement of $\Sigma_{\mathrm{L}}(E)$ with $\widetilde{\Sigma}_{\mathrm{L}}(E)$ is weaker, primarily near $E-\mu\approx-165$~meV, potentially due to the asymmetry in the experimental geometry.

Finally, we compare the stability of a MEM-based extraction of $\alpha^2F_n(\omega)$ versus fitting it as a sum of peaks. We compare Eliashberg functions as a sum of three Lorentzian peaks described by Eq.~\eqref{eq:a2f_input} of the main text, denoting the latter as $\alpha^2F_n^{\mathrm{peak}}(\omega)$. Each of these peaks consists of a positive Lorentzian labeled with $k$ at positive $\omega_k$, and a negative Lorentzian counterpart at $-\omega_k$, resulting in $\alpha^2F^{\mathrm{peak}}(\omega)$ being odd in $\omega$, resulting in a single peak for $\omega \geq 0$. We reiterate that positive semidefiniteness over a predetermined interval is the sole hard constraint on $\alpha^2F_n(\omega)$ in the MEM-based approach. Interestingly, we find that small variations in the self-energies can lead to a very different $\alpha^2F_n^{\mathrm{peak}}(\omega)$, an issue not arising with the MEM-based approach. We extract $\alpha^2F_{\mathrm{L}}^{\mathrm{peak}}(\omega)$ and $\alpha^2F_{\mathrm{R}}^{\mathrm{peak}}(\omega)$ from the respective $\widetilde{\Sigma}_{\mathrm{L}}(E)$ displayed in Fig.~\ref{fig:Li-gr}\textbf{b} of the main text and $\widetilde{\Sigma}_{\mathrm{R}}(E)$ displayed in Fig.~\ref{fig:Li-gr-SM}\textbf{e}. Fig.~\ref{fig:Li-gr}\textbf{b} of the main text shows that these sets of $\widetilde{\Sigma}_n(E)$ yield minor differences between $\alpha^2F_{\mathrm{L}}(\omega)$ and $\alpha^2F_{\mathrm{R}}(\omega)$. For the peak-based approach, we follow Ref.~\cite{usachov:2018-SM} in setting the widths $\rho_2=\rho_3$ to remove unphysically large peaks, otherwise occurring at single discretized energies with heights exceeding unity. The resulting $\alpha^2F_{\mathrm{L}}^{\mathrm{peak}}(\omega)$ and $\alpha^2F_{\mathrm{R}}^{\mathrm{peak}}(\omega)$ are displayed in Fig.~\ref{fig:Li-gr-SM}\textbf{f}. The largest peaks in $\alpha^2F_{\mathrm{R}}^{\mathrm{peak}}(\omega)$ coincide entirely at $\omega=169$~meV, contrasting a $\Delta \omega = 21$~meV splitting in $\alpha^2F_{\mathrm{L}}^{\mathrm{peak}}(\omega)$. This large difference shows that a small dissimilarity in the self-energies can result in discontinuous changes in the extracted Eliashberg functions with the peak-based approach. Instead, the MEM-based approach has no constraints beyond the model function, and small variations in the self-energies do not result in excessively large changes in the Eliashberg functions.

\section{Non-interacting band substitution}\label{SI:noninteracting band}

In this section, we explore using ARPES data to evaluate quantities calculated with theoretical methods. As an example, we investigate the $\bar{\Gamma}$-point Shockley surface state on Al(100) along the $\bar{\Gamma}-\bar{\mathrm{M}}$ high-symmetry path, using Kohn-Sham (KS) eigenvalues calculated with \textsc{Abinit}~\cite{gonze:2020-SM, romero:2020-SM, gonze:2016-SM} v10.0, and comparing them to the $T=10$~K data from Ref.~\cite{jiang:2011-SM}. Specifically, we impose a non-interacting band mass $m_{\mathrm{R}}^{\mathrm{b,KS}}$ on the right-hand (R) branch of this surface state, fitted to the KS eigenvalues, to extract the self-energies $\Sigma_{\mathrm{R}}^{\mathrm{KS}}(E)$ and the Eliashberg function $\alpha^2F_{\mathrm{R}}^{\mathrm{KS}}(\omega)$, while the other parameters $k_{\mathrm{R}}^{\mathrm{F,KS}}$, $\Gamma_{\mathrm{R}}^{\mathrm{imp,KS}}$, $\lambda_{\mathrm{R}}^{\mathrm{el,KS}}$, and $h_{\mathrm{R}}^{\mathrm{KS}}$ are determined by the Bayesian inference loop. We compare these results to the case in which all parameters are optimized with the Bayesian loop, in which case they are reported without a superscript, and comment on the differences.

For the density-functional theory (DFT) calculations, we use scalar-relativistic norm-conserving standard-accuracy pseudopotentials~\cite{hamann:2013} from the \textsc{PseudoDojo} library v0.4.1~\cite{vansetten:2018}, using the Perdew-Burke-Ernzerhof exchange-correlation functional for solids~\cite{perdew:2008} (PBEsol). We perform all calculations as spin-degenerate, with a plane wave cutoff energy of 26~Ha, and 10~mHa Gaussian smearing.

\begin{figure*}[t]
\includegraphics[width=\textwidth]{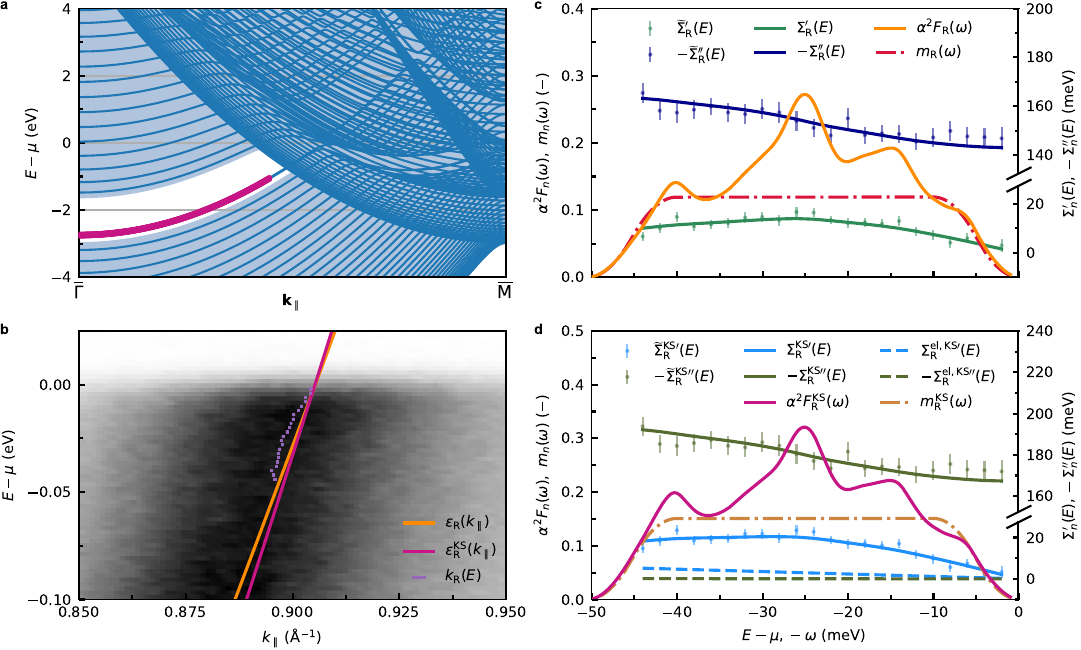}
\caption{\textbf{Many-body quantities with a DFT-based non-interacting band mass}.
\textbf{a} Surface band structure (blue lines), $k_z$-projected bulk bands (shaded area), and fits to the Kohn-Sham eigenvalues near the band bottom (magenta dots) along the $\bar{\Gamma}-\bar{\mathrm{M}}$ high-symmetry path. \textbf{b} Band map along $\bar{\Gamma}-\bar{\mathrm{M}}$, showing the MDC maxima $k_{\mathrm{R}}(E)$ (purple bars), the non-interacting dispersion from (i) the Bayesian loop $\varepsilon_{\mathrm{R}}(k_{\parallel})$ (orange), and (ii) the imposed Kohn-Sham non-interacting band mass $\varepsilon_{\mathrm{R}}^{\mathrm{KS}}(k_{\parallel})$ (magenta) shifted to match at $E=\mu$, respectively. \textbf{c} The Eliashberg function $\alpha^2F_{\mathrm{R}}(\omega)$ (orange) with all parameters optimized by the Bayesian loop, together with the model function $m_{\mathrm{R}}(\omega)$ (dash-dotted red), reconstruction of the real part of the self-energy $\Sigma_{\mathrm{R}}^{\prime}(E)$ (green) with corresponding data $\widetilde{\Sigma}_{\mathrm{R}}^{\prime}(E)$ (green), and the reconstruction of the minus imaginary part $-\Sigma_{\mathrm{R}}^{\prime\prime}(E)$ (blue) with corresponding data $-\widetilde{\Sigma}_{\mathrm{R}}^{\prime\prime}(E)$. \textbf{d} Same quantities as in \textbf{c}, but with the non-interacting mass $m_{\mathrm{R}}^{\mathrm{KS}}$ fitted to the KS eigenvalues: $\alpha^2F_{\mathrm{R}}^{\mathrm{KS}}(\omega)$ (magenta), $m_{\mathrm{R}}^{\mathrm{KS}}(\omega)$ (dash-dotted gold), $\Sigma_{\mathrm{R}}^{\mathrm{KS}\prime}(E)$ (light blue) with corresponding data $\widetilde{\Sigma}_{\mathrm{R}}^{\mathrm{KS}\prime}(E)$ (light blue bars), and $-\Sigma_{\mathrm{R}}^{\mathrm{KS}\prime\prime}(E)$ (olive) with corresponding data $-\widetilde{\Sigma}_{\mathrm{R}}^{\mathrm{KS}\prime\prime}(E)$ (olive bars). Fitting with $m_{\mathrm{R}}^{\mathrm{KS}}(\omega)$ yields a finite $\lambda_{\mathrm{R}}^{\mathrm{el,KS}}=0.117$, resulting in an appreciable $\Sigma_{\mathrm{R}}^{\mathrm{el,KS}\prime}(E)$ (dashed light blue), while $-\Sigma_{\mathrm{R}}^{\mathrm{el,KS}\prime\prime}(E)$ (dashed olive) remains small over the displayed range.
\label{fig:Al}}
\end{figure*}

Using a uniform $26\times26\times26$ $\Gamma$-centered $\mathbf{k}$-grid along the reciprocal lattice vectors for the bulk primitive unit cell, we determine the face-centered cubic (FCC) lattice constant as $a=4.0134~\Angstrom$ by converging the forces below $10^{-7}$~Ha/Bohr and the residual of the electronic potential below $10^{-14}$ Ha. We determine the surface-projected bulk band structure by filling the region between the maxima and minima of the bands obtained from calculations with 112 uniformly distributed $k_z$-values for the tetragonal representation of FCC Al (2 atoms per unit cell) sampled with a $26\times26\times18$ $\mathbf{k}$-grid. The surface-projected band structure is displayed as the shaded area in Fig.~\ref{fig:Al}\textbf{a}. Next, we construct (100)-terminated slabs using tetragonal unit cells sampled with $26\times26\times1$ $\mathbf{k}$-point grids, while relaxing the atomic coordinates along the $z$-direction. For the slabs, we converge forces below $10^{-5}$~Ha/Bohr and the potential residual below $10^{-8}$~Ha. Starting with 31 layers (${} = 31$ atoms), we first determine the vacuum thickness using the work function $\Phi$, finding $\Phi=4.30$~eV at 8 layers, which changes by less than 10 meV as the vacuum thickness is increased further. Using 8 vacuum layers, we determine the desired thickness of the slab, finding that at 41 layers, the splitting of $E^{\mathrm{F}}$ between the surface states on the sides of the slab is less than 30~meV. We non-self-consistently calculate the KS eigenvalues for 896 homogeneously distributed $\mathbf{k}$-points along the $\bar{\Gamma}-\bar{\mathrm{M}}$ path to fit the two surface-state bands, shown as the blue slab bands in Fig.~\ref{fig:Al}\textbf{a}.

Unfortunately, the surface state is only inside the so-called ``bulk (band) gap''~\cite{echenique:2001} for approximately half of this path, complicating the fitting procedure near $k_{\mathrm{R}}^{\mathrm{F}}$, where surface wavefunctions hybridize with bulk-derived bands for finite-size slabs. Therefore, we perform a least-squares parabolic fit to the first 400 values of each of the surface states shown as magenta markers in Fig.~\ref{fig:Al}\textbf{a}, finding root-mean-square errors of $1.14$~meV and $0.71$~meV for the lower and upper bands, respectively. For the two surface bands, we find the effective masses for the lower right-hand (LR) and upper right-hand (UR) bands $m_{\mathrm{LR}}^{\mathrm{KS}}=1.0936$~$m_{\mathrm{e}}$ and $m_{\mathrm{UR}}^{\mathrm{KS}}=1.0968$~$m_{\mathrm{e}}$, whose average is $m_{\mathrm{R}}^{\mathrm{KS}}=1.0952$~$m_{\mathrm{e}}$, while parabolic extrapolation to $E=\mu$ yields $k_{\mathrm{LR}}^{\mathrm{F,KS}}=0.8898~\Angstrom^{-1}$ and $k_{\mathrm{UR}}^{\mathrm{F,KS}}=0.8865~\Angstrom^{-1}$, respectively. Fits to KS eigenvalues of the surface state in the L-gap of Cu(111) were found to slightly improve by including a quartic term~\cite{berland:2012}. However, given the limited non-hybridized surface eigenvalues and the required modifications to incorporate quartic terms in \textsc{xARPES}, we proceed with $m_{\mathrm{R}}^{\mathrm{KS}}=1.0952$~$m_{\mathrm{e}}$ obtained from the quadratic fitting. The slab bands, bulk-projected bands, and parabolic fits are shown in Fig.~\ref{fig:Al}. The parabolic fits closely follow the surface-state bands, whose splitting is too small to be observed in the figure.

Turning to the experimental analysis, we determine $k_{\mathrm{R}}^{\mathrm{F}}$ by separately fitting the self-energies of both branches of the dispersion (Fig.~2(a) from Ref.~\cite{jiang:2011-SM}) followed by simultaneously imposing particle-hole symmetry ($\Sigma_{\mathrm{L,R}}^{\prime}(E=\mu)=0$) for both branches. In doing so, we realign the band map with $\phi=1.4$\textdegree and perform the Fermi-edge fitting. Following this requirement, we estimate $k_{\mathrm{L}}^{\mathrm{F}}=k_{\mathrm{R}}^{\mathrm{F}}=0.905~\Angstrom^{-1}$, in agreement with $k_{\mathrm{R}}^{\mathrm{F}}=(0.907 \pm 0.005)~\Angstrom^{-1}$ determined from a parabolic fitting in Ref.~\cite{jiang:2011-SM}. Using this $k_{\mathrm{R}}^{\mathrm{F}}$, we align the band map zoomed in on the photoemission kink (corresponding to Fig.~5a of Ref.~\cite{jiang:2011-SM}) by once again imposing that $\Sigma_{\mathrm{R}}^{\prime}(E=\mu)=0$, resulting in $\phi=15.205$\textdegree. Afterwards, we perform an MDC fitting for a parabolic dispersion, resulting in the MDC maxima $k_{\mathrm{R}}(E)$ shown on top of the band map in Fig.~\ref{fig:Al}. These maxima do not depend on any non-interacting band parameters, as we assumed that the band bottom angle $\eta_{\mathrm{R}}^{\mathrm{c}}=0$.

First, we follow the usual approach of \textsc{xARPES} to optimize the model parameters and extract $\alpha^2F_{\mathrm{R}}(\omega)$. However, instead of excluding data based on the energy resolution with $\mu - E < \Delta E$, we exclude the range $\mu - E < 0$, as the large $\Delta E=15$~meV~\cite{jiang:2011-SM} would otherwise exclude too many data points for the extraction. The calculated non-interacting dispersion $\varepsilon_{\mathrm{R}}(k_{\parallel})$ and the extracted results are displayed in of Fig.~\ref{fig:Al}\textbf{b}--\textbf{c}, respectively, while the fit parameters and associated quantities are provided in Table~\ref{tab:sup4} in the ``\textsc{xARPES}'' column.

\begin{table}[tb]
\vspace{-0.7em}\caption{Bayesian loop parameters and related quantities for the \textsc{xARPES} approach and the Kohn-Sham (KS) approach described in the text. The non-interacting band mass $m_{\mathrm{R}}^{\mathrm{b}}$ fixed during the KS procedure is written in bold.
\label{tab:sup4}}
	\renewcommand{\arraystretch}{1.2}
	\begin{tabular*}{0.8\columnwidth}{@{\extracolsep{\fill}}ccdd}
		\hline\hline
		Parameter & Unit &  \multicolumn{1}{c}{\textsc{xARPES}} & \multicolumn{1}{c}{KS}\\
		\hline
		$m_{\mathrm{R}}^{\mathrm{b}}$ & $m_{\rm{e}}$      & 1.2827  & \multicolumn{1}{r}{\textbf{1.0952}} \\ 
		$k_{\mathrm{R}}^{\mathrm{F}}$ & $\Angstrom^{-1}$  & 0.9053 & 0.9053 \\
		$\Gamma_{\mathrm{R}}^{\mathrm{imp}}$ & meV & 143.0  & 167.2 \\
		$h_{\mathrm{R}}$ & -- & 0.119 & 0.151 \\
		$\lambda_{\mathrm{R}}^{\mathrm{el}}$ & -- & 0.000 & 0.117 \\
		$\lambda_{\mathrm{R}}^{\mathrm{ph}}$ & -- & 0.744 & 0.889 \\
		$m_{\mathrm{R}}^{*}$ & $m_{\rm{e}}$  &   2.237   & 2.197 \\ 
		\hline \hline
	\end{tabular*}
\end{table}

Similarly to Eliashberg functions calculated at the $\bar{\Gamma}$- and $\bar{\mathrm{X}}$-points, the dominant spectral weight is found for $\omega \in [20, 25]$~meV, while smaller modes $\omega_j$ are found near $\omega_j \in \{15, 40\}$~meV. The Bayesian loop yields $m_{\mathrm{R}}^{\mathrm{b}}$=1.2827~$m_{\mathrm{e}}$, notably larger than $m_{\mathrm{R}}^{\mathrm{b,KS}}=1.12$~$m_{\mathrm{e}}$ estimated from a combined EDC+MDC fitting in Ref.~\cite{jiang:2011-SM}. We further optimize $k_{\mathrm{R}}^{\mathrm{F}}$ in the loop to obtain the most probable quantities, yet the obtained $k_{\mathrm{R}}^{\mathrm{F}}=0.9053~\Angstrom^{-1}$ highlights that $\phi$ was carefully determined. The coupling coefficient $\lambda_{\mathrm{R}}^{\mathrm{ph}}=0.744$ obtained here is slightly outside the confidence interval of $\lambda_{\mathrm{R}}^{\mathrm{ph}}=0.61 \pm 0.05$ from Ref.~\cite{jiang:2011-SM}, partially explained by the smaller $m_{\mathrm{b}}^{\mathrm{KS}}$ used in their approach. Using the extracted results, we calculate an effective mass of $m_{\mathrm{R}}^{*} \equiv (1+\lambda_{\mathrm{R}}^{\mathrm{ph}}+\lambda_{\mathrm{R}}^{\mathrm{el}})m_{\mathrm{R}}^{\mathrm{b}}=2.237$~$m_{\mathrm{e}}$.

Next, we fix $m_{\mathrm{R}}^{\mathrm{b,KS}}$ fitted to the Kohn-Sham eigenvalues, while optimizing the other quantities in the loop. The calculated non-interacting band $\varepsilon_{\mathrm{R}}^{\mathrm{KS}}(k_{\parallel})$ and the extracted quantities are displayed in Figs.~\ref{fig:Al}\textbf{b} and \ref{fig:Al}\textbf{d}, respectively, with relevant parameters provided in Table~\ref{tab:sup4} in the ``KS'' column. Although $\alpha^2F_{\mathrm{R}}^{\mathrm{KS}}(\omega)$ is qualitatively similar to the \textsc{xARPES} case, $\lambda_{\mathrm{R}}^{\mathrm{ph,KS}}$ is appreciably larger, while now we also find a finite $\lambda_{\mathrm{R}}^{\mathrm{el,KS}}$. These findings can be rationalized by considering the effective mass, which goes from $m_{\mathrm{R}}^{*}=2.237$~$m_{\mathrm{e}}$ to $m_{\mathrm{R}}^{*\mathrm{KS}}=2.197$~$m_{\mathrm{e}}$, a change of less than 2\%. Clearly, the loop compensates for the small imposed $m_{\mathrm{R}}^{\mathrm{b,KS}}$ with renormalizations through $\Sigma_{\mathrm{R}}^{\mathrm{el,KS}\prime}(E)$ and $\Sigma_{\mathrm{R}}^{\mathrm{ph,KS}\prime}(E)$. Given the large $E^{\mathrm{F}}$, $-\Sigma_{\mathrm{R}}^{\mathrm{el,KS}\prime\prime}(E)$ hardly deviates from zero over the range of the kink, as seen in Fig.~\ref{fig:Al}\textbf{d}, so that $\lambda_{\mathrm{R}}^{\mathrm{el}}$ changes the solution to the quasiparticle equation $E_n(\vec{k})=\varepsilon_n(\vec{k}) + \Sigma_{n}^{\prime}(E_n(\vec{k}), \vec{k})$, while barely changing MDC peak widths. Simulating the paramagnetic ground state of Al might give rise to a finite $\Sigma_{\mathrm{R}}^{\mathrm{el}}(E)$, could be captured computationally through the unfolding of a fully paramagnetic simulation~\cite{wang:2021a}, or with a DMFT simulation~\cite{abramovitch:2024}. The authors of Ref.~\cite{jiang:2011-SM} argue that $\lambda_{\mathrm{R}}^{\mathrm{el}}\approx 0.003$, in agreement with our fully optimized results. However, it is unclear whether the expression for $\Sigma_{\mathrm{R}}^{\mathrm{el}}(E)$ can be extrapolated to the band bottom, where its magnitude is determined.

We remark that $E^{\mathrm{F,KS}}=2.74$~meV fitted to the DFT results compares reasonably with $E^{\mathrm{F}}=(2.81\pm 0.03)$~eV determined with MDC+EDC fits in Ref.~\cite{jiang:2011-SM}. A larger disagreement is obtained between the fitted $m_{\mathrm{R}}^{\mathrm{b,KS}}$ and the optimized $m_{\mathrm{R}}^{\mathrm{b}}$, which questions whether $\varepsilon_{\mathrm{R}}(k_{\parallel})$ can be well-approximated by a second-order polynomial up to $E=\mu$. Unfortunately, the bulk-derived bands complicate the careful fit to the KS eigenvalues near $k_{\mathrm{R}}^{\mathrm{F}}$, while also biasing the experimental results to an unknown extent. Our results suggest that a comparison of contributions to the effective mass from first principles and experiment could be performed with surface states residing completely inside the bulk band gap, such as the Be(0001) surface state~\cite{balasubramanian:1998-SM, chien:2015-SM, polley:2024a}, while 2D systems can also be considered for this assessment. In the future, \textsc{xARPES} may be expanded to directly compare the non-interacting dispersion $\varepsilon_n(\mathbf{k})$ calculated from theoretical methods with its experimental counterpart, as well as to compare $\Sigma_n(E)$ with a calculated $\Sigma_n(E(\mathbf{k}),\mathbf{k})$.


\begin{thebibliography}{154}%
\makeatletter
\providecommand \@ifxundefined [1]{%
 \@ifx{#1\undefined}
}%
\providecommand \@ifnum [1]{%
 \ifnum #1\expandafter \@firstoftwo
 \else \expandafter \@secondoftwo
 \fi
}%
\providecommand \@ifx [1]{%
 \ifx #1\expandafter \@firstoftwo
 \else \expandafter \@secondoftwo
 \fi
}%
\providecommand \natexlab [1]{#1}%
\providecommand \enquote  [1]{``#1''}%
\providecommand \bibnamefont  [1]{#1}%
\providecommand \bibfnamefont [1]{#1}%
\providecommand \citenamefont [1]{#1}%
\providecommand \href@noop [0]{\@secondoftwo}%
\providecommand \href [0]{\begingroup \@sanitize@url \@href}%
\providecommand \@href[1]{\@@startlink{#1}\@@href}%
\providecommand \@@href[1]{\endgroup#1\@@endlink}%
\providecommand \@sanitize@url [0]{\catcode `\\12\catcode `\$12\catcode
  `\&12\catcode `\#12\catcode `\^12\catcode `\_12\catcode `\%12\relax}%
\providecommand \@@startlink[1]{}%
\providecommand \@@endlink[0]{}%
\providecommand \url  [0]{\begingroup\@sanitize@url \@url }%
\providecommand \@url [1]{\endgroup\@href {#1}{\urlprefix }}%
\providecommand \urlprefix  [0]{URL }%
\providecommand \Eprint [0]{\href }%
\providecommand \doibase [0]{https://doi.org/}%
\providecommand \selectlanguage [0]{\@gobble}%
\providecommand \bibinfo  [0]{\@secondoftwo}%
\providecommand \bibfield  [0]{\@secondoftwo}%
\providecommand \translation [1]{[#1]}%
\providecommand \BibitemOpen [0]{}%
\providecommand \bibitemStop [0]{}%
\providecommand \bibitemNoStop [0]{.\EOS\space}%
\providecommand \EOS [0]{\spacefactor3000\relax}%
\providecommand \BibitemShut  [1]{\csname bibitem#1\endcsname}%
\let\auto@bib@innerbib\@empty
\bibitem [{\citenamefont {Bruus}\ and\ \citenamefont
  {Flensberg}(2004)}]{bruus:2004}%
  \BibitemOpen
  \bibfield  {author} {\bibinfo {author} {\bibfnamefont {H.}~\bibnamefont
  {Bruus}}\ and\ \bibinfo {author} {\bibfnamefont {K.}~\bibnamefont
  {Flensberg}},\ }\href {https://doi.org/10.1093/oso/9780198566335.001.0001}
  {\emph {\bibinfo {title} {Many-Body Quantum Theory in Condensed Matter
  Physics: An Introduction}}}\ (\bibinfo  {publisher} {Oxford University Press,
  Oxford},\ \bibinfo {year} {2004})\BibitemShut {NoStop}%
\bibitem [{\citenamefont {Fr{\"o}hlich}(1950)}]{frohlich:1950}%
  \BibitemOpen
  \bibfield  {author} {\bibinfo {author} {\bibfnamefont {H.}~\bibnamefont
  {Fr{\"o}hlich}},\ }\bibfield  {title} {\bibinfo {title} {Theory of the
  superconducting state. {I}. {T}he ground state at the absolute zero of
  temperature},\ }\href {https://link.aps.org/doi/10.1103/PhysRev.79.845}
  {\bibfield  {journal} {\bibinfo  {journal} {Phys. Rev.}\ }\textbf {\bibinfo
  {volume} {79}},\ \bibinfo {pages} {845} (\bibinfo {year} {1950})}\BibitemShut
  {NoStop}%
\bibitem [{\citenamefont {Mahan}(2000)}]{mahan:2000}%
  \BibitemOpen
  \bibfield  {author} {\bibinfo {author} {\bibfnamefont {G.~D.}\ \bibnamefont
  {Mahan}},\ }\href {https://doi.org/10.1007/978-1-4757-5714-9} {\emph
  {\bibinfo {title} {Many-Particle Physics}}},\ \bibinfo {edition} {3rd}\ ed.\
  (\bibinfo  {publisher} {Kluwer Academic/Plenum Publishers},\ \bibinfo
  {address} {New York},\ \bibinfo {year} {2000})\BibitemShut {NoStop}%
\bibitem [{\citenamefont {Coleman}(2015)}]{coleman:2015}%
  \BibitemOpen
  \bibfield  {author} {\bibinfo {author} {\bibfnamefont {P.}~\bibnamefont
  {Coleman}},\ }\href {https://doi.org/10.1017/CBO9781139020916} {\emph
  {\bibinfo {title} {Introduction to Many-Body Physics}}},\ \bibinfo {edition}
  {1st}\ ed.\ (\bibinfo  {publisher} {Cambridge University Press, Cambridge},\
  \bibinfo {year} {2015})\BibitemShut {NoStop}%
\bibitem [{\citenamefont {Allen}(2015)}]{allen:2015}%
  \BibitemOpen
  \bibfield  {author} {\bibinfo {author} {\bibfnamefont {P.~B.}\ \bibnamefont
  {Allen}},\ }\bibfield  {title} {\bibinfo {title} {Electron self-energy and
  generalized drude formula for infrared conductivity of metals},\ }\href
  {https://doi.org/10.1103/PhysRevB.92.054305} {\bibfield  {journal} {\bibinfo
  {journal} {Phys. Rev. B}\ }\textbf {\bibinfo {volume} {92}},\ \bibinfo
  {pages} {054305} (\bibinfo {year} {2015})}\BibitemShut {NoStop}%
\bibitem [{\citenamefont {Berthod}(2018)}]{berthod:2018a}%
  \BibitemOpen
  \bibfield  {author} {\bibinfo {author} {\bibfnamefont {C.}~\bibnamefont
  {Berthod}},\ }\href {https://iopscience.iop.org/book/mono/978-0-7503-1741-2}
  {\emph {\bibinfo {title} {Spectroscopic Probes of Quantum Matter}}}\
  (\bibinfo  {publisher} {IOP Publishing},\ \bibinfo {address} {Bristol},\
  \bibinfo {year} {2018})\BibitemShut {NoStop}%
\bibitem [{\citenamefont {Ponc{\'e}}\ \emph {et~al.}(2016)\citenamefont
  {Ponc{\'e}}, \citenamefont {Margine}, \citenamefont {Verdi},\ and\
  \citenamefont {Giustino}}]{ponce:2016}%
  \BibitemOpen
  \bibfield  {author} {\bibinfo {author} {\bibfnamefont {S.}~\bibnamefont
  {Ponc{\'e}}}, \bibinfo {author} {\bibfnamefont {E.}~\bibnamefont {Margine}},
  \bibinfo {author} {\bibfnamefont {C.}~\bibnamefont {Verdi}},\ and\ \bibinfo
  {author} {\bibfnamefont {F.}~\bibnamefont {Giustino}},\ }\bibfield  {title}
  {\bibinfo {title} {{{EPW}}: {{Electron}}--phonon coupling, transport and
  superconducting properties using maximally localized {{Wannier}} functions},\
  }\href {https://linkinghub.elsevier.com/retrieve/pii/S0010465516302260}
  {\bibfield  {journal} {\bibinfo  {journal} {Comput. Phys. Commun.}\ }\textbf
  {\bibinfo {volume} {209}},\ \bibinfo {pages} {116} (\bibinfo {year}
  {2016})}\BibitemShut {NoStop}%
\bibitem [{\citenamefont {Allen}\ and\ \citenamefont
  {Mitrovi{\'c}}(1983)}]{allen:1983}%
  \BibitemOpen
  \bibfield  {author} {\bibinfo {author} {\bibfnamefont {P.~B.}\ \bibnamefont
  {Allen}}\ and\ \bibinfo {author} {\bibfnamefont {B.}~\bibnamefont
  {Mitrovi{\'c}}},\ }\bibfield  {title} {\bibinfo {title} {Theory of
  superconducting {$T_c$}},\ }\href
  {https://doi.org/10.1016/S0081-1947(08)60665-7} {\bibfield  {journal}
  {\bibinfo  {journal} {Solid State Phys.}\ }\textbf {\bibinfo {volume} {37}},\
  \bibinfo {pages} {1} (\bibinfo {year} {1983})}\BibitemShut {NoStop}%
\bibitem [{\citenamefont {Park}\ \emph {et~al.}(2020)\citenamefont {Park},
  \citenamefont {Zhou},\ and\ \citenamefont {Bernardi}}]{park:2020}%
  \BibitemOpen
  \bibfield  {author} {\bibinfo {author} {\bibfnamefont {J.}~\bibnamefont
  {Park}}, \bibinfo {author} {\bibfnamefont {J.-J.}\ \bibnamefont {Zhou}},\
  and\ \bibinfo {author} {\bibfnamefont {M.}~\bibnamefont {Bernardi}},\
  }\bibfield  {title} {\bibinfo {title} {Spin-phonon relaxation times in
  centrosymmetric materials from first principles},\ }\href
  {https://link.aps.org/doi/10.1103/PhysRevB.101.045202} {\bibfield  {journal}
  {\bibinfo  {journal} {Phys. Rev. B}\ }\textbf {\bibinfo {volume} {101}},\
  \bibinfo {pages} {045202} (\bibinfo {year} {2020})}\BibitemShut {NoStop}%
\bibitem [{\citenamefont {Franchini}\ \emph {et~al.}(2021)\citenamefont
  {Franchini}, \citenamefont {Reticcioli}, \citenamefont {Setv{\'i}n},\ and\
  \citenamefont {Diebold}}]{franchini:2021}%
  \BibitemOpen
  \bibfield  {author} {\bibinfo {author} {\bibfnamefont {C.}~\bibnamefont
  {Franchini}}, \bibinfo {author} {\bibfnamefont {M.}~\bibnamefont
  {Reticcioli}}, \bibinfo {author} {\bibfnamefont {M.}~\bibnamefont
  {Setv{\'i}n}},\ and\ \bibinfo {author} {\bibfnamefont {U.}~\bibnamefont
  {Diebold}},\ }\bibfield  {title} {\bibinfo {title} {Polarons in materials},\
  }\href {https://www.nature.com/articles/s41578-021-00289-w} {\bibfield
  {journal} {\bibinfo  {journal} {Nat. Rev. Mater.}\ }\textbf {\bibinfo
  {volume} {6}},\ \bibinfo {pages} {560} (\bibinfo {year} {2021})}\BibitemShut
  {NoStop}%
\bibitem [{\citenamefont {Yu}\ \emph {et~al.}(2022)\citenamefont {Yu},
  \citenamefont {Xu}, \citenamefont {Yang}, \citenamefont {Song}, \citenamefont
  {Wen}, \citenamefont {Yao}, \citenamefont {Lou}, \citenamefont {Zhang},
  \citenamefont {Li}, \citenamefont {Wei}, \citenamefont {Bao}, \citenamefont
  {Cao}, \citenamefont {Dudin}, \citenamefont {Denlinger}, \citenamefont
  {Strocov}, \citenamefont {Peng}, \citenamefont {Xu},\ and\ \citenamefont
  {Feng}}]{yu:2022a}%
  \BibitemOpen
  \bibfield  {author} {\bibinfo {author} {\bibfnamefont {T.~L.}\ \bibnamefont
  {Yu}}, \bibinfo {author} {\bibfnamefont {M.}~\bibnamefont {Xu}}, \bibinfo
  {author} {\bibfnamefont {W.~T.}\ \bibnamefont {Yang}}, \bibinfo {author}
  {\bibfnamefont {Y.~H.}\ \bibnamefont {Song}}, \bibinfo {author}
  {\bibfnamefont {C.~H.~P.}\ \bibnamefont {Wen}}, \bibinfo {author}
  {\bibfnamefont {Q.}~\bibnamefont {Yao}}, \bibinfo {author} {\bibfnamefont
  {X.}~\bibnamefont {Lou}}, \bibinfo {author} {\bibfnamefont {T.}~\bibnamefont
  {Zhang}}, \bibinfo {author} {\bibfnamefont {W.}~\bibnamefont {Li}}, \bibinfo
  {author} {\bibfnamefont {X.~Y.}\ \bibnamefont {Wei}}, \bibinfo {author}
  {\bibfnamefont {J.~K.}\ \bibnamefont {Bao}}, \bibinfo {author} {\bibfnamefont
  {G.~H.}\ \bibnamefont {Cao}}, \bibinfo {author} {\bibfnamefont
  {P.}~\bibnamefont {Dudin}}, \bibinfo {author} {\bibfnamefont {J.~D.}\
  \bibnamefont {Denlinger}}, \bibinfo {author} {\bibfnamefont {V.~N.}\
  \bibnamefont {Strocov}}, \bibinfo {author} {\bibfnamefont {R.}~\bibnamefont
  {Peng}}, \bibinfo {author} {\bibfnamefont {H.~C.}\ \bibnamefont {Xu}},\ and\
  \bibinfo {author} {\bibfnamefont {D.~L.}\ \bibnamefont {Feng}},\ }\bibfield
  {title} {\bibinfo {title} {Strong band renormalization and emergent
  ferromagnetism induced by electron-antiferromagnetic-magnon coupling},\
  }\href {https://www.nature.com/articles/s41467-022-34254-0} {\bibfield
  {journal} {\bibinfo  {journal} {Nat. Commun.}\ }\textbf {\bibinfo {volume}
  {13}},\ \bibinfo {pages} {6560} (\bibinfo {year} {2022})}\BibitemShut
  {NoStop}%
\bibitem [{\citenamefont {Mazzola}\ \emph {et~al.}(2022)\citenamefont
  {Mazzola}, \citenamefont {Yim}, \citenamefont {Sunko}, \citenamefont {Khim},
  \citenamefont {Kushwaha}, \citenamefont {Clark}, \citenamefont {Bawden},
  \citenamefont {Markovi{\'c}}, \citenamefont {Chakraborti}, \citenamefont
  {Kim}, \citenamefont {Hoesch}, \citenamefont {Mackenzie}, \citenamefont
  {Wahl},\ and\ \citenamefont {King}}]{mazzola:2022}%
  \BibitemOpen
  \bibfield  {author} {\bibinfo {author} {\bibfnamefont {F.}~\bibnamefont
  {Mazzola}}, \bibinfo {author} {\bibfnamefont {C.~M.}\ \bibnamefont {Yim}},
  \bibinfo {author} {\bibfnamefont {V.}~\bibnamefont {Sunko}}, \bibinfo
  {author} {\bibfnamefont {S.}~\bibnamefont {Khim}}, \bibinfo {author}
  {\bibfnamefont {P.}~\bibnamefont {Kushwaha}}, \bibinfo {author}
  {\bibfnamefont {O.~J.}\ \bibnamefont {Clark}}, \bibinfo {author}
  {\bibfnamefont {L.}~\bibnamefont {Bawden}}, \bibinfo {author} {\bibfnamefont
  {I.}~\bibnamefont {Markovi{\'c}}}, \bibinfo {author} {\bibfnamefont
  {D.}~\bibnamefont {Chakraborti}}, \bibinfo {author} {\bibfnamefont {T.~K.}\
  \bibnamefont {Kim}}, \bibinfo {author} {\bibfnamefont {M.}~\bibnamefont
  {Hoesch}}, \bibinfo {author} {\bibfnamefont {A.~P.}\ \bibnamefont
  {Mackenzie}}, \bibinfo {author} {\bibfnamefont {P.}~\bibnamefont {Wahl}},\
  and\ \bibinfo {author} {\bibfnamefont {P.~D.~C.}\ \bibnamefont {King}},\
  }\bibfield  {title} {\bibinfo {title} {Tuneable electron--magnon coupling of
  ferromagnetic surface states in {{PdCoO}}{\textsubscript{2}}},\ }\href
  {https://www.nature.com/articles/s41535-022-00428-8} {\bibfield  {journal}
  {\bibinfo  {journal} {npj Quantum Mater.}\ }\textbf {\bibinfo {volume} {7}},\
  \bibinfo {pages} {20} (\bibinfo {year} {2022})}\BibitemShut {NoStop}%
\bibitem [{\citenamefont {Lazzari}\ \emph {et~al.}(2018)\citenamefont
  {Lazzari}, \citenamefont {Li},\ and\ \citenamefont {Jupille}}]{lazzari:2018}%
  \BibitemOpen
  \bibfield  {author} {\bibinfo {author} {\bibfnamefont {R.}~\bibnamefont
  {Lazzari}}, \bibinfo {author} {\bibfnamefont {J.}~\bibnamefont {Li}},\ and\
  \bibinfo {author} {\bibfnamefont {J.}~\bibnamefont {Jupille}},\ }\bibfield
  {title} {\bibinfo {title} {Dielectric study of the interplay between charge
  carriers and electron energy losses in reduced titanium dioxide},\ }\href
  {https://link.aps.org/doi/10.1103/PhysRevB.98.075432} {\bibfield  {journal}
  {\bibinfo  {journal} {Phys. Rev. B}\ }\textbf {\bibinfo {volume} {98}},\
  \bibinfo {pages} {075432} (\bibinfo {year} {2018})}\BibitemShut {NoStop}%
\bibitem [{\citenamefont {Johnson}\ \emph {et~al.}(2001)\citenamefont
  {Johnson}, \citenamefont {Valla}, \citenamefont {Fedorov}, \citenamefont
  {Yusof}, \citenamefont {Wells}, \citenamefont {Li}, \citenamefont
  {Moodenbaugh}, \citenamefont {Gu}, \citenamefont {Koshizuka}, \citenamefont
  {Kendziora}, \citenamefont {Jian},\ and\ \citenamefont
  {Hinks}}]{johnson:2001}%
  \BibitemOpen
  \bibfield  {author} {\bibinfo {author} {\bibfnamefont {P.~D.}\ \bibnamefont
  {Johnson}}, \bibinfo {author} {\bibfnamefont {T.}~\bibnamefont {Valla}},
  \bibinfo {author} {\bibfnamefont {A.~V.}\ \bibnamefont {Fedorov}}, \bibinfo
  {author} {\bibfnamefont {Z.}~\bibnamefont {Yusof}}, \bibinfo {author}
  {\bibfnamefont {B.~O.}\ \bibnamefont {Wells}}, \bibinfo {author}
  {\bibfnamefont {Q.}~\bibnamefont {Li}}, \bibinfo {author} {\bibfnamefont
  {A.~R.}\ \bibnamefont {Moodenbaugh}}, \bibinfo {author} {\bibfnamefont
  {G.~D.}\ \bibnamefont {Gu}}, \bibinfo {author} {\bibfnamefont
  {N.}~\bibnamefont {Koshizuka}}, \bibinfo {author} {\bibfnamefont
  {C.}~\bibnamefont {Kendziora}}, \bibinfo {author} {\bibfnamefont
  {S.}~\bibnamefont {Jian}},\ and\ \bibinfo {author} {\bibfnamefont {D.~G.}\
  \bibnamefont {Hinks}},\ }\bibfield  {title} {\bibinfo {title} {Doping and
  {{Temperature Dependence}} of the {{Mass Enhancement Observed}} in the
  {{Cuprate
  Bi}}{\textsubscript{2}}{{Sr}}{\textsubscript{2}}{{CaCu}}{\textsubscript{2}}{{O}}{\textsubscript{8+{$\delta$}}}},\
  }\href {https://link.aps.org/doi/10.1103/PhysRevLett.87.177007} {\bibfield
  {journal} {\bibinfo  {journal} {Phys. Rev. Lett.}\ }\textbf {\bibinfo
  {volume} {87}},\ \bibinfo {pages} {177007} (\bibinfo {year}
  {2001})}\BibitemShut {NoStop}%
\bibitem [{\citenamefont {Shi}\ \emph {et~al.}(2004)\citenamefont {Shi},
  \citenamefont {Tang}, \citenamefont {Wu}, \citenamefont {Sprunger},
  \citenamefont {Yang}, \citenamefont {Brouet}, \citenamefont {Zhou},
  \citenamefont {Hussain}, \citenamefont {Shen}, \citenamefont {Zhang},\ and\
  \citenamefont {Plummer}}]{shi:2004}%
  \BibitemOpen
  \bibfield  {author} {\bibinfo {author} {\bibfnamefont {J.}~\bibnamefont
  {Shi}}, \bibinfo {author} {\bibfnamefont {S.-J.}\ \bibnamefont {Tang}},
  \bibinfo {author} {\bibfnamefont {B.}~\bibnamefont {Wu}}, \bibinfo {author}
  {\bibfnamefont {P.~T.}\ \bibnamefont {Sprunger}}, \bibinfo {author}
  {\bibfnamefont {W.~L.}\ \bibnamefont {Yang}}, \bibinfo {author}
  {\bibfnamefont {V.}~\bibnamefont {Brouet}}, \bibinfo {author} {\bibfnamefont
  {X.~J.}\ \bibnamefont {Zhou}}, \bibinfo {author} {\bibfnamefont
  {Z.}~\bibnamefont {Hussain}}, \bibinfo {author} {\bibfnamefont {Z.-X.}\
  \bibnamefont {Shen}}, \bibinfo {author} {\bibfnamefont {Z.}~\bibnamefont
  {Zhang}},\ and\ \bibinfo {author} {\bibfnamefont {E.~W.}\ \bibnamefont
  {Plummer}},\ }\bibfield  {title} {\bibinfo {title} {Direct extraction of the
  {Eliashberg} function for electron-phonon coupling: {A} case study of
  {Be(10{\=1}0)}},\ }\href
  {https://link.aps.org/doi/10.1103/PhysRevLett.92.186401} {\bibfield
  {journal} {\bibinfo  {journal} {Phys. Rev. Lett.}\ }\textbf {\bibinfo
  {volume} {92}},\ \bibinfo {pages} {186401} (\bibinfo {year}
  {2004})}\BibitemShut {NoStop}%
\bibitem [{\citenamefont {Grimvall}(1981)}]{grimvall:1981}%
  \BibitemOpen
  \bibfield  {author} {\bibinfo {author} {\bibfnamefont {G.}~\bibnamefont
  {Grimvall}},\ }\href@noop {} {\emph {\bibinfo {title} {The Electron--Phonon
  Interaction in Metals}}}\ (\bibinfo  {publisher} {North-Holland, Amsterdam},\
  \bibinfo {year} {1981})\BibitemShut {NoStop}%
\bibitem [{\citenamefont {Marsiglio}\ and\ \citenamefont
  {Carbotte}(2008)}]{marsiglio:2008}%
  \BibitemOpen
  \bibfield  {author} {\bibinfo {author} {\bibfnamefont {F.}~\bibnamefont
  {Marsiglio}}\ and\ \bibinfo {author} {\bibfnamefont {J.~P.}\ \bibnamefont
  {Carbotte}},\ }\bibfield  {title} {\bibinfo {title} {Electron-phonon
  superconductivity},\ }in\ \href {https://doi.org/10.1007/978-3-540-73253-2_3}
  {\emph {\bibinfo {booktitle} {Superconductivity: Conventional and
  Unconventional Superconductors}}},\ \bibinfo {editor} {edited by\ \bibinfo
  {editor} {\bibfnamefont {K.}~\bibnamefont {Bennemann}}\ and\ \bibinfo
  {editor} {\bibfnamefont {J.}~\bibnamefont {Kettterson}}}\ (\bibinfo
  {publisher} {Springer, Berlin},\ \bibinfo {year} {2008})\ pp.\ \bibinfo
  {pages} {73--162}\BibitemShut {NoStop}%
\bibitem [{\citenamefont {Margine}\ and\ \citenamefont
  {Giustino}(2013)}]{margine:2013}%
  \BibitemOpen
  \bibfield  {author} {\bibinfo {author} {\bibfnamefont {E.~R.}\ \bibnamefont
  {Margine}}\ and\ \bibinfo {author} {\bibfnamefont {F.}~\bibnamefont
  {Giustino}},\ }\bibfield  {title} {\bibinfo {title} {Anisotropic
  {{Migdal-Eliashberg}} theory using {{Wannier}} functions},\ }\href
  {https://link.aps.org/doi/10.1103/PhysRevB.87.024505} {\bibfield  {journal}
  {\bibinfo  {journal} {Phys. Rev. B}\ }\textbf {\bibinfo {volume} {87}},\
  \bibinfo {pages} {024505} (\bibinfo {year} {2013})}\BibitemShut {NoStop}%
\bibitem [{\citenamefont {Nagamatsu}\ \emph {et~al.}(2001)\citenamefont
  {Nagamatsu}, \citenamefont {Nakagawa}, \citenamefont {Muranaka},
  \citenamefont {Zenitani},\ and\ \citenamefont {Akimitsu}}]{nagamatsu:2001}%
  \BibitemOpen
  \bibfield  {author} {\bibinfo {author} {\bibfnamefont {J.}~\bibnamefont
  {Nagamatsu}}, \bibinfo {author} {\bibfnamefont {N.}~\bibnamefont {Nakagawa}},
  \bibinfo {author} {\bibfnamefont {T.}~\bibnamefont {Muranaka}}, \bibinfo
  {author} {\bibfnamefont {Y.}~\bibnamefont {Zenitani}},\ and\ \bibinfo
  {author} {\bibfnamefont {J.}~\bibnamefont {Akimitsu}},\ }\bibfield  {title}
  {\bibinfo {title} {Superconductivity at {39~K} in magnesium diboride},\
  }\href {https://www.nature.com/articles/35065039} {\bibfield  {journal}
  {\bibinfo  {journal} {Nature}\ }\textbf {\bibinfo {volume} {410}},\ \bibinfo
  {pages} {63} (\bibinfo {year} {2001})}\BibitemShut {NoStop}%
\bibitem [{\citenamefont {Choi}\ \emph {et~al.}(2002)\citenamefont {Choi},
  \citenamefont {Roundy}, \citenamefont {Sun}, \citenamefont {Cohen},\ and\
  \citenamefont {Louie}}]{choi:2002}%
  \BibitemOpen
  \bibfield  {author} {\bibinfo {author} {\bibfnamefont {H.~J.}\ \bibnamefont
  {Choi}}, \bibinfo {author} {\bibfnamefont {D.}~\bibnamefont {Roundy}},
  \bibinfo {author} {\bibfnamefont {H.}~\bibnamefont {Sun}}, \bibinfo {author}
  {\bibfnamefont {M.~L.}\ \bibnamefont {Cohen}},\ and\ \bibinfo {author}
  {\bibfnamefont {S.~G.}\ \bibnamefont {Louie}},\ }\bibfield  {title} {\bibinfo
  {title} {First-principles calculation of the superconducting transition in
  {{MgB}}{\textsubscript{2}} within the anisotropic {{Eliashberg}} formalism},\
  }\href {https://link.aps.org/doi/10.1103/PhysRevB.66.020513} {\bibfield
  {journal} {\bibinfo  {journal} {Phys. Rev. B}\ }\textbf {\bibinfo {volume}
  {66}},\ \bibinfo {pages} {020513} (\bibinfo {year} {2002})}\BibitemShut
  {NoStop}%
\bibitem [{\citenamefont {Marsiglio}(1999)}]{marsiglio:1999}%
  \BibitemOpen
  \bibfield  {author} {\bibinfo {author} {\bibfnamefont {F.}~\bibnamefont
  {Marsiglio}},\ }\bibfield  {title} {\bibinfo {title} {Inversion of {{Optical
  Conductivity Data}} in {{Metals}}},\ }\href
  {http://link.springer.com/10.1023/A:1007766829839} {\bibfield  {journal}
  {\bibinfo  {journal} {J. Supercond.}\ }\textbf {\bibinfo {volume} {12}},\
  \bibinfo {pages} {163} (\bibinfo {year} {1999})}\BibitemShut {NoStop}%
\bibitem [{\citenamefont {McMillan}\ and\ \citenamefont
  {Rowell}(1965)}]{mcmillan:1965}%
  \BibitemOpen
  \bibfield  {author} {\bibinfo {author} {\bibfnamefont {W.~L.}\ \bibnamefont
  {McMillan}}\ and\ \bibinfo {author} {\bibfnamefont {J.~M.}\ \bibnamefont
  {Rowell}},\ }\bibfield  {title} {\bibinfo {title} {Lead {{Phonon Spectrum
  Calculated}} from {{Superconducting Density}} of {{States}}},\ }\href
  {https://link.aps.org/doi/10.1103/PhysRevLett.14.108} {\bibfield  {journal}
  {\bibinfo  {journal} {Phys. Rev. Lett.}\ }\textbf {\bibinfo {volume} {14}},\
  \bibinfo {pages} {108} (\bibinfo {year} {1965})}\BibitemShut {NoStop}%
\bibitem [{\citenamefont {Zeljkovic}\ \emph {et~al.}(2015)\citenamefont
  {Zeljkovic}, \citenamefont {Scipioni}, \citenamefont {Walkup}, \citenamefont
  {Okada}, \citenamefont {Zhou}, \citenamefont {Sankar}, \citenamefont {Chang},
  \citenamefont {Wang}, \citenamefont {Lin}, \citenamefont {Bansil},
  \citenamefont {Chou}, \citenamefont {Wang},\ and\ \citenamefont
  {Madhavan}}]{zeljkovic:2015}%
  \BibitemOpen
  \bibfield  {author} {\bibinfo {author} {\bibfnamefont {I.}~\bibnamefont
  {Zeljkovic}}, \bibinfo {author} {\bibfnamefont {K.~L.}\ \bibnamefont
  {Scipioni}}, \bibinfo {author} {\bibfnamefont {D.}~\bibnamefont {Walkup}},
  \bibinfo {author} {\bibfnamefont {Y.}~\bibnamefont {Okada}}, \bibinfo
  {author} {\bibfnamefont {W.}~\bibnamefont {Zhou}}, \bibinfo {author}
  {\bibfnamefont {R.}~\bibnamefont {Sankar}}, \bibinfo {author} {\bibfnamefont
  {G.}~\bibnamefont {Chang}}, \bibinfo {author} {\bibfnamefont {Y.~J.}\
  \bibnamefont {Wang}}, \bibinfo {author} {\bibfnamefont {H.}~\bibnamefont
  {Lin}}, \bibinfo {author} {\bibfnamefont {A.}~\bibnamefont {Bansil}},
  \bibinfo {author} {\bibfnamefont {F.}~\bibnamefont {Chou}}, \bibinfo {author}
  {\bibfnamefont {Z.}~\bibnamefont {Wang}},\ and\ \bibinfo {author}
  {\bibfnamefont {V.}~\bibnamefont {Madhavan}},\ }\bibfield  {title} {\bibinfo
  {title} {Nanoscale determination of the mass enhancement factor in the
  lightly doped bulk insulator lead selenide},\ }\href
  {https://www.nature.com/articles/ncomms7559} {\bibfield  {journal} {\bibinfo
  {journal} {Nat. Commun.}\ }\textbf {\bibinfo {volume} {6}},\ \bibinfo {pages}
  {6559} (\bibinfo {year} {2015})}\BibitemShut {NoStop}%
\bibitem [{\citenamefont {Damascelli}\ \emph {et~al.}(2003)\citenamefont
  {Damascelli}, \citenamefont {Hussain},\ and\ \citenamefont
  {Shen}}]{damascelli:2003}%
  \BibitemOpen
  \bibfield  {author} {\bibinfo {author} {\bibfnamefont {A.}~\bibnamefont
  {Damascelli}}, \bibinfo {author} {\bibfnamefont {Z.}~\bibnamefont
  {Hussain}},\ and\ \bibinfo {author} {\bibfnamefont {Z.-X.}\ \bibnamefont
  {Shen}},\ }\bibfield  {title} {\bibinfo {title} {Angle-resolved photoemission
  studies of the cuprate superconductors},\ }\href
  {https://link.aps.org/doi/10.1103/RevModPhys.75.473} {\bibfield  {journal}
  {\bibinfo  {journal} {Rev. Mod. Phys.}\ }\textbf {\bibinfo {volume} {75}},\
  \bibinfo {pages} {473} (\bibinfo {year} {2003})}\BibitemShut {NoStop}%
\bibitem [{\citenamefont {Einstein}(1905)}]{einstein:1905}%
  \BibitemOpen
  \bibfield  {author} {\bibinfo {author} {\bibfnamefont {A.}~\bibnamefont
  {Einstein}},\ }\bibfield  {title} {\bibinfo {title} {On a heuristic point of
  view about the creation and conversion of light},\ }\href
  {https://doi.org/10.1002/andp.19053220607} {\bibfield  {journal} {\bibinfo
  {journal} {Ann. Phys.}\ }\textbf {\bibinfo {volume} {17}},\ \bibinfo {pages}
  {132} (\bibinfo {year} {1905})}\BibitemShut {NoStop}%
\bibitem [{\citenamefont {Moser}(2017)}]{moser:2017}%
  \BibitemOpen
  \bibfield  {author} {\bibinfo {author} {\bibfnamefont {S.}~\bibnamefont
  {Moser}},\ }\bibfield  {title} {\bibinfo {title} {An experimentalist's guide
  to the matrix element in angle resolved photoemission},\ }\href
  {https://linkinghub.elsevier.com/retrieve/pii/S0368204816301724} {\bibfield
  {journal} {\bibinfo  {journal} {J. Electron Spectrosc. Relat. Phenom.}\
  }\textbf {\bibinfo {volume} {214}},\ \bibinfo {pages} {29} (\bibinfo {year}
  {2017})}\BibitemShut {NoStop}%
\bibitem [{\citenamefont {Tusche}\ \emph {et~al.}(2011)\citenamefont {Tusche},
  \citenamefont {Ellguth}, \citenamefont {{\"U}nal}, \citenamefont {Chiang},
  \citenamefont {Winkelmann}, \citenamefont {Krasyuk}, \citenamefont {Hahn},
  \citenamefont {Sch{\"o}nhense},\ and\ \citenamefont
  {Kirschner}}]{tusche:2011}%
  \BibitemOpen
  \bibfield  {author} {\bibinfo {author} {\bibfnamefont {C.}~\bibnamefont
  {Tusche}}, \bibinfo {author} {\bibfnamefont {M.}~\bibnamefont {Ellguth}},
  \bibinfo {author} {\bibfnamefont {A.~A.}\ \bibnamefont {{\"U}nal}}, \bibinfo
  {author} {\bibfnamefont {C.-T.}\ \bibnamefont {Chiang}}, \bibinfo {author}
  {\bibfnamefont {A.}~\bibnamefont {Winkelmann}}, \bibinfo {author}
  {\bibfnamefont {A.}~\bibnamefont {Krasyuk}}, \bibinfo {author} {\bibfnamefont
  {M.}~\bibnamefont {Hahn}}, \bibinfo {author} {\bibfnamefont {G.}~\bibnamefont
  {Sch{\"o}nhense}},\ and\ \bibinfo {author} {\bibfnamefont {J.}~\bibnamefont
  {Kirschner}},\ }\bibfield  {title} {\bibinfo {title} {Spin resolved
  photoelectron microscopy using a two-dimensional spin-polarizing electron
  mirror},\ }\href
  {https://pubs.aip.org/apl/article/99/3/032505/341540/Spin-resolved-photoelectron-microscopy-using-a-two}
  {\bibfield  {journal} {\bibinfo  {journal} {Appl. Phys. Lett.}\ }\textbf
  {\bibinfo {volume} {99}},\ \bibinfo {pages} {032505} (\bibinfo {year}
  {2011})}\BibitemShut {NoStop}%
\bibitem [{\citenamefont {Dil}(2019)}]{dil:2019}%
  \BibitemOpen
  \bibfield  {author} {\bibinfo {author} {\bibfnamefont {J.~H.}\ \bibnamefont
  {Dil}},\ }\bibfield  {title} {\bibinfo {title} {Spin- and angle-resolved
  photoemission on topological materials},\ }\href
  {https://iopscience.iop.org/article/10.1088/2516-1075/ab168b} {\bibfield
  {journal} {\bibinfo  {journal} {Electron. Struct.}\ }\textbf {\bibinfo
  {volume} {1}},\ \bibinfo {pages} {023001} (\bibinfo {year}
  {2019})}\BibitemShut {NoStop}%
\bibitem [{\citenamefont {Giustino}(2017)}]{giustino:2017}%
  \BibitemOpen
  \bibfield  {author} {\bibinfo {author} {\bibfnamefont {F.}~\bibnamefont
  {Giustino}},\ }\bibfield  {title} {\bibinfo {title} {Electron-phonon
  interactions from first principles},\ }\href
  {https://link.aps.org/doi/10.1103/RevModPhys.89.015003} {\bibfield  {journal}
  {\bibinfo  {journal} {Rev. Mod. Phys.}\ }\textbf {\bibinfo {volume} {89}},\
  \bibinfo {pages} {015003} (\bibinfo {year} {2017})}\BibitemShut {NoStop}%
\bibitem [{\citenamefont {Abrikosov}\ \emph {et~al.}(1964)\citenamefont
  {Abrikosov}, \citenamefont {Gorkov}, \citenamefont {Dzyaloshinski},
  \citenamefont {Silverman},\ and\ \citenamefont {Weiss}}]{abrikosov:1964}%
  \BibitemOpen
  \bibfield  {author} {\bibinfo {author} {\bibfnamefont {A.~A.}\ \bibnamefont
  {Abrikosov}}, \bibinfo {author} {\bibfnamefont {L.~P.}\ \bibnamefont
  {Gorkov}}, \bibinfo {author} {\bibfnamefont {I.~E.}\ \bibnamefont
  {Dzyaloshinski}}, \bibinfo {author} {\bibfnamefont {R.~A.}\ \bibnamefont
  {Silverman}},\ and\ \bibinfo {author} {\bibfnamefont {G.~H.}\ \bibnamefont
  {Weiss}},\ }\href {https://doi.org/10.1063/1.3051555} {\emph {\bibinfo
  {title} {Methods of Quantum Field Theory in Statistical Physics}}}\ (\bibinfo
   {publisher} {Pergamon Press Ltd., Oxford},\ \bibinfo {year}
  {1964})\BibitemShut {NoStop}%
\bibitem [{\citenamefont {Veenstra}\ \emph {et~al.}(2011)\citenamefont
  {Veenstra}, \citenamefont {Goodvin}, \citenamefont {Berciu},\ and\
  \citenamefont {Damascelli}}]{veenstra:2011}%
  \BibitemOpen
  \bibfield  {author} {\bibinfo {author} {\bibfnamefont {C.~N.}\ \bibnamefont
  {Veenstra}}, \bibinfo {author} {\bibfnamefont {G.~L.}\ \bibnamefont
  {Goodvin}}, \bibinfo {author} {\bibfnamefont {M.}~\bibnamefont {Berciu}},\
  and\ \bibinfo {author} {\bibfnamefont {A.}~\bibnamefont {Damascelli}},\
  }\bibfield  {title} {\bibinfo {title} {Spectral function tour of
  electron-phonon coupling outside the {{Migdal}} limit},\ }\href
  {https://link.aps.org/doi/10.1103/PhysRevB.84.085126} {\bibfield  {journal}
  {\bibinfo  {journal} {Phys. Rev. B}\ }\textbf {\bibinfo {volume} {84}},\
  \bibinfo {pages} {085126} (\bibinfo {year} {2011})}\BibitemShut {NoStop}%
\bibitem [{\citenamefont {Li}\ \emph {et~al.}(2024)\citenamefont {Li},
  \citenamefont {Xu}, \citenamefont {Liu}, \citenamefont {Fang}, \citenamefont
  {Zheng}, \citenamefont {Dai}, \citenamefont {Li}, \citenamefont {Zhu},
  \citenamefont {Zhang}, \citenamefont {Liang}, \citenamefont {Yang},
  \citenamefont {Huang}, \citenamefont {Xi}, \citenamefont {Liu}, \citenamefont
  {Xu},\ and\ \citenamefont {Chen}}]{li:2024a}%
  \BibitemOpen
  \bibfield  {author} {\bibinfo {author} {\bibfnamefont {Y.}~\bibnamefont
  {Li}}, \bibinfo {author} {\bibfnamefont {L.}~\bibnamefont {Xu}}, \bibinfo
  {author} {\bibfnamefont {G.}~\bibnamefont {Liu}}, \bibinfo {author}
  {\bibfnamefont {Y.}~\bibnamefont {Fang}}, \bibinfo {author} {\bibfnamefont
  {H.}~\bibnamefont {Zheng}}, \bibinfo {author} {\bibfnamefont
  {S.}~\bibnamefont {Dai}}, \bibinfo {author} {\bibfnamefont {E.}~\bibnamefont
  {Li}}, \bibinfo {author} {\bibfnamefont {G.}~\bibnamefont {Zhu}}, \bibinfo
  {author} {\bibfnamefont {S.}~\bibnamefont {Zhang}}, \bibinfo {author}
  {\bibfnamefont {S.}~\bibnamefont {Liang}}, \bibinfo {author} {\bibfnamefont
  {L.}~\bibnamefont {Yang}}, \bibinfo {author} {\bibfnamefont {F.}~\bibnamefont
  {Huang}}, \bibinfo {author} {\bibfnamefont {X.}~\bibnamefont {Xi}}, \bibinfo
  {author} {\bibfnamefont {Z.}~\bibnamefont {Liu}}, \bibinfo {author}
  {\bibfnamefont {N.}~\bibnamefont {Xu}},\ and\ \bibinfo {author}
  {\bibfnamefont {Y.}~\bibnamefont {Chen}},\ }\bibfield  {title} {\bibinfo
  {title} {Evidence of strong and mode-selective electron--phonon coupling in
  the topological superconductor candidate {{2M-WS}}{\textsubscript{2}}},\
  }\href {https://www.nature.com/articles/s41467-024-50590-9} {\bibfield
  {journal} {\bibinfo  {journal} {Nat. Commun.}\ }\textbf {\bibinfo {volume}
  {15}},\ \bibinfo {pages} {6235} (\bibinfo {year} {2024})}\BibitemShut
  {NoStop}%
\bibitem [{\citenamefont {Iwasawa}(2020)}]{iwasawa:2020}%
  \BibitemOpen
  \bibfield  {author} {\bibinfo {author} {\bibfnamefont {H.}~\bibnamefont
  {Iwasawa}},\ }\bibfield  {title} {\bibinfo {title} {High-resolution
  angle-resolved photoemission spectroscopy and microscopy},\ }\href
  {https://iopscience.iop.org/article/10.1088/2516-1075/abb379} {\bibfield
  {journal} {\bibinfo  {journal} {Electron. Struct.}\ }\textbf {\bibinfo
  {volume} {2}},\ \bibinfo {pages} {043001} (\bibinfo {year}
  {2020})}\BibitemShut {NoStop}%
\bibitem [{\citenamefont {Chien}(2009)}]{chien:2009a}%
  \BibitemOpen
  \bibfield  {author} {\bibinfo {author} {\bibfnamefont {T.}~\bibnamefont
  {Chien}},\ }\emph {\bibinfo {title} {Anisotropic electron-phonon coupling on
  a two-dimensional isotropic fermi contour: $\bar{\Gamma}$ surface state of
  Be(0001)}},\ \href {https://trace.tennessee.edu/utk_graddiss/6041/} {Ph.D.
  thesis},\ \bibinfo  {school} {University of Tennessee} (\bibinfo {year}
  {2009})\BibitemShut {NoStop}%
\bibitem [{\citenamefont {Iwasawa}\ \emph {et~al.}(2012)\citenamefont
  {Iwasawa}, \citenamefont {Yoshida}, \citenamefont {Hase}, \citenamefont
  {Shimada}, \citenamefont {Namatame}, \citenamefont {Taniguchi},\ and\
  \citenamefont {Aiura}}]{iwasawa:2012}%
  \BibitemOpen
  \bibfield  {author} {\bibinfo {author} {\bibfnamefont {H.}~\bibnamefont
  {Iwasawa}}, \bibinfo {author} {\bibfnamefont {Y.}~\bibnamefont {Yoshida}},
  \bibinfo {author} {\bibfnamefont {I.}~\bibnamefont {Hase}}, \bibinfo {author}
  {\bibfnamefont {K.}~\bibnamefont {Shimada}}, \bibinfo {author} {\bibfnamefont
  {H.}~\bibnamefont {Namatame}}, \bibinfo {author} {\bibfnamefont
  {M.}~\bibnamefont {Taniguchi}},\ and\ \bibinfo {author} {\bibfnamefont
  {Y.}~\bibnamefont {Aiura}},\ }\bibfield  {title} {\bibinfo {title}
  {High-{{Energy Anomaly}} in the {{Band Dispersion}} of the {{Ruthenate
  Superconductor}}},\ }\href
  {https://link.aps.org/doi/10.1103/PhysRevLett.109.066404} {\bibfield
  {journal} {\bibinfo  {journal} {Phys. Rev. Lett.}\ }\textbf {\bibinfo
  {volume} {109}},\ \bibinfo {pages} {066404} (\bibinfo {year}
  {2012})}\BibitemShut {NoStop}%
\bibitem [{\citenamefont {Cappelli}\ \emph {et~al.}(2022)\citenamefont
  {Cappelli}, \citenamefont {Hampel}, \citenamefont {Chikina}, \citenamefont
  {Guedes}, \citenamefont {Gatti}, \citenamefont {Hunter}, \citenamefont
  {Issing}, \citenamefont {Biskup}, \citenamefont {Varela}, \citenamefont
  {Dreyer}, \citenamefont {Tamai}, \citenamefont {Georges}, \citenamefont
  {Bruno}, \citenamefont {Radovi{\'c}},\ and\ \citenamefont
  {Baumberger}}]{cappelli:2022}%
  \BibitemOpen
  \bibfield  {author} {\bibinfo {author} {\bibfnamefont {E.}~\bibnamefont
  {Cappelli}}, \bibinfo {author} {\bibfnamefont {A.}~\bibnamefont {Hampel}},
  \bibinfo {author} {\bibfnamefont {A.}~\bibnamefont {Chikina}}, \bibinfo
  {author} {\bibfnamefont {E.~B.}\ \bibnamefont {Guedes}}, \bibinfo {author}
  {\bibfnamefont {G.}~\bibnamefont {Gatti}}, \bibinfo {author} {\bibfnamefont
  {A.}~\bibnamefont {Hunter}}, \bibinfo {author} {\bibfnamefont
  {J.}~\bibnamefont {Issing}}, \bibinfo {author} {\bibfnamefont
  {N.}~\bibnamefont {Biskup}}, \bibinfo {author} {\bibfnamefont
  {M.}~\bibnamefont {Varela}}, \bibinfo {author} {\bibfnamefont {C.~E.}\
  \bibnamefont {Dreyer}}, \bibinfo {author} {\bibfnamefont {A.}~\bibnamefont
  {Tamai}}, \bibinfo {author} {\bibfnamefont {A.}~\bibnamefont {Georges}},
  \bibinfo {author} {\bibfnamefont {F.~Y.}\ \bibnamefont {Bruno}}, \bibinfo
  {author} {\bibfnamefont {M.}~\bibnamefont {Radovi{\'c}}},\ and\ \bibinfo
  {author} {\bibfnamefont {F.}~\bibnamefont {Baumberger}},\ }\bibfield  {title}
  {\bibinfo {title} {Electronic structure of the highly conductive perovskite
  oxide {{SrMoO}}{\textsubscript{3}}},\ }\href
  {https://link.aps.org/doi/10.1103/PhysRevMaterials.6.075002} {\bibfield
  {journal} {\bibinfo  {journal} {Phys. Rev. Mater.}\ }\textbf {\bibinfo
  {volume} {6}},\ \bibinfo {pages} {075002} (\bibinfo {year}
  {2022})}\BibitemShut {NoStop}%
\bibitem [{\citenamefont {Fan}(1951)}]{fan:1951}%
  \BibitemOpen
  \bibfield  {author} {\bibinfo {author} {\bibfnamefont {H.~Y.}\ \bibnamefont
  {Fan}},\ }\bibfield  {title} {\bibinfo {title} {Temperature dependence of the
  energy gap in semiconductors},\ }\href
  {https://link.aps.org/doi/10.1103/PhysRev.82.900} {\bibfield  {journal}
  {\bibinfo  {journal} {Phys. Rev.}\ }\textbf {\bibinfo {volume} {82}},\
  \bibinfo {pages} {900} (\bibinfo {year} {1951})}\BibitemShut {NoStop}%
\bibitem [{\citenamefont {Migdal}(1958)}]{migdal:1958}%
  \BibitemOpen
  \bibfield  {author} {\bibinfo {author} {\bibfnamefont {A.~B.}\ \bibnamefont
  {Migdal}},\ }\bibfield  {title} {\bibinfo {title} {Interaction between
  electrons and lattice vibrations in a normal metal},\ }\href@noop {}
  {\bibfield  {journal} {\bibinfo  {journal} {Sov. Phys. JETP}\ }\textbf
  {\bibinfo {volume} {7}},\ \bibinfo {pages} {996} (\bibinfo {year}
  {1958})}\BibitemShut {NoStop}%
\bibitem [{\citenamefont {Allen}\ and\ \citenamefont
  {Heine}(1976)}]{allen:1976}%
  \BibitemOpen
  \bibfield  {author} {\bibinfo {author} {\bibfnamefont {P.~B.}\ \bibnamefont
  {Allen}}\ and\ \bibinfo {author} {\bibfnamefont {V.}~\bibnamefont {Heine}},\
  }\bibfield  {title} {\bibinfo {title} {Theory of the temperature dependence
  of electronic band structures},\ }\href
  {https://iopscience.iop.org/article/10.1088/0022-3719/9/12/013} {\bibfield
  {journal} {\bibinfo  {journal} {J. Phys. C: Solid State Phys.}\ }\textbf
  {\bibinfo {volume} {9}},\ \bibinfo {pages} {2305} (\bibinfo {year}
  {1976})}\BibitemShut {NoStop}%
\bibitem [{\citenamefont {Allen}\ and\ \citenamefont
  {Cardona}(1981)}]{allen:1981}%
  \BibitemOpen
  \bibfield  {author} {\bibinfo {author} {\bibfnamefont {P.~B.}\ \bibnamefont
  {Allen}}\ and\ \bibinfo {author} {\bibfnamefont {M.}~\bibnamefont
  {Cardona}},\ }\bibfield  {title} {\bibinfo {title} {Theory of the temperature
  dependence of the direct gap of germanium},\ }\href
  {https://link.aps.org/doi/10.1103/PhysRevB.23.1495} {\bibfield  {journal}
  {\bibinfo  {journal} {Phys. Rev. B}\ }\textbf {\bibinfo {volume} {23}},\
  \bibinfo {pages} {1495} (\bibinfo {year} {1981})}\BibitemShut {NoStop}%
\bibitem [{\citenamefont {Marini}\ \emph {et~al.}(2015)\citenamefont {Marini},
  \citenamefont {Ponc{\'e}},\ and\ \citenamefont {Gonze}}]{marini:2015}%
  \BibitemOpen
  \bibfield  {author} {\bibinfo {author} {\bibfnamefont {A.}~\bibnamefont
  {Marini}}, \bibinfo {author} {\bibfnamefont {S.}~\bibnamefont {Ponc{\'e}}},\
  and\ \bibinfo {author} {\bibfnamefont {X.}~\bibnamefont {Gonze}},\ }\bibfield
   {title} {\bibinfo {title} {Many-body perturbation theory approach to the
  electron-phonon interaction with density-functional theory as a starting
  point},\ }\href {https://link.aps.org/doi/10.1103/PhysRevB.91.224310}
  {\bibfield  {journal} {\bibinfo  {journal} {Phys. Rev. B}\ }\textbf {\bibinfo
  {volume} {91}},\ \bibinfo {pages} {224310} (\bibinfo {year}
  {2015})}\BibitemShut {NoStop}%
\bibitem [{\citenamefont {{Lafuente-Bartolome}}\ \emph
  {et~al.}(2022)\citenamefont {{Lafuente-Bartolome}}, \citenamefont {Lian},
  \citenamefont {Sio}, \citenamefont {Gurtubay}, \citenamefont {Eiguren},\ and\
  \citenamefont {Giustino}}]{lafuente-bartolome:2022a}%
  \BibitemOpen
  \bibfield  {author} {\bibinfo {author} {\bibfnamefont {J.}~\bibnamefont
  {{Lafuente-Bartolome}}}, \bibinfo {author} {\bibfnamefont {C.}~\bibnamefont
  {Lian}}, \bibinfo {author} {\bibfnamefont {W.~H.}\ \bibnamefont {Sio}},
  \bibinfo {author} {\bibfnamefont {I.~G.}\ \bibnamefont {Gurtubay}}, \bibinfo
  {author} {\bibfnamefont {A.}~\bibnamefont {Eiguren}},\ and\ \bibinfo {author}
  {\bibfnamefont {F.}~\bibnamefont {Giustino}},\ }\bibfield  {title} {\bibinfo
  {title} {{\emph{Ab Initio}} self-consistent many-body theory of polarons at
  all couplings},\ }\href
  {https://link.aps.org/doi/10.1103/PhysRevB.106.075119} {\bibfield  {journal}
  {\bibinfo  {journal} {Phys. Rev. B}\ }\textbf {\bibinfo {volume} {106}},\
  \bibinfo {pages} {075119} (\bibinfo {year} {2022})}\BibitemShut {NoStop}%
\bibitem [{\citenamefont {Grimvall}(1976)}]{grimvall:1976}%
  \BibitemOpen
  \bibfield  {author} {\bibinfo {author} {\bibfnamefont {G.}~\bibnamefont
  {Grimvall}},\ }\bibfield  {title} {\bibinfo {title} {The electron-phonon
  interaction in normal metals},\ }\href
  {https://iopscience.iop.org/article/10.1088/0031-8949/14/1-2/013} {\bibfield
  {journal} {\bibinfo  {journal} {Phys. Scr.}\ }\textbf {\bibinfo {volume}
  {14}},\ \bibinfo {pages} {63} (\bibinfo {year} {1976})}\BibitemShut {NoStop}%
\bibitem [{\citenamefont {McDougall}\ \emph {et~al.}(1995)\citenamefont
  {McDougall}, \citenamefont {Balasubramanian},\ and\ \citenamefont
  {Jensen}}]{mcdougall:1995}%
  \BibitemOpen
  \bibfield  {author} {\bibinfo {author} {\bibfnamefont {B.~A.}\ \bibnamefont
  {McDougall}}, \bibinfo {author} {\bibfnamefont {T.}~\bibnamefont
  {Balasubramanian}},\ and\ \bibinfo {author} {\bibfnamefont {E.}~\bibnamefont
  {Jensen}},\ }\bibfield  {title} {\bibinfo {title} {Phonon contribution to
  quasiparticle lifetimes in {{Cu}} measured by angle-resolved photoemission},\
  }\href {https://link.aps.org/doi/10.1103/PhysRevB.51.13891} {\bibfield
  {journal} {\bibinfo  {journal} {Phys. Rev. B}\ }\textbf {\bibinfo {volume}
  {51}},\ \bibinfo {pages} {13891} (\bibinfo {year} {1995})}\BibitemShut
  {NoStop}%
\bibitem [{\citenamefont {Balasubramanian}\ \emph {et~al.}(1998)\citenamefont
  {Balasubramanian}, \citenamefont {Jensen}, \citenamefont {Wu},\ and\
  \citenamefont {Hulbert}}]{balasubramanian:1998}%
  \BibitemOpen
  \bibfield  {author} {\bibinfo {author} {\bibfnamefont {T.}~\bibnamefont
  {Balasubramanian}}, \bibinfo {author} {\bibfnamefont {E.}~\bibnamefont
  {Jensen}}, \bibinfo {author} {\bibfnamefont {X.~L.}\ \bibnamefont {Wu}},\
  and\ \bibinfo {author} {\bibfnamefont {S.~L.}\ \bibnamefont {Hulbert}},\
  }\bibfield  {title} {\bibinfo {title} {Large value of the electron-phonon
  coupling parameter ($\lambda$=1.15) and the possibility of surface
  superconductivity at the {Be(0001)} surface},\ }\href
  {https://link.aps.org/doi/10.1103/PhysRevB.57.R6866} {\bibfield  {journal}
  {\bibinfo  {journal} {Phys. Rev. B}\ }\textbf {\bibinfo {volume} {57}},\
  \bibinfo {pages} {R6866} (\bibinfo {year} {1998})}\BibitemShut {NoStop}%
\bibitem [{\citenamefont {Valla}\ \emph {et~al.}(1999)\citenamefont {Valla},
  \citenamefont {Fedorov}, \citenamefont {Johnson}, \citenamefont {Wells},
  \citenamefont {Hulbert}, \citenamefont {Li}, \citenamefont {Gu},\ and\
  \citenamefont {Koshizuka}}]{valla:1999}%
  \BibitemOpen
  \bibfield  {author} {\bibinfo {author} {\bibfnamefont {T.}~\bibnamefont
  {Valla}}, \bibinfo {author} {\bibfnamefont {A.~V.}\ \bibnamefont {Fedorov}},
  \bibinfo {author} {\bibfnamefont {P.~D.}\ \bibnamefont {Johnson}}, \bibinfo
  {author} {\bibfnamefont {B.~O.}\ \bibnamefont {Wells}}, \bibinfo {author}
  {\bibfnamefont {S.~L.}\ \bibnamefont {Hulbert}}, \bibinfo {author}
  {\bibfnamefont {Q.}~\bibnamefont {Li}}, \bibinfo {author} {\bibfnamefont
  {G.~D.}\ \bibnamefont {Gu}},\ and\ \bibinfo {author} {\bibfnamefont
  {N.}~\bibnamefont {Koshizuka}},\ }\bibfield  {title} {\bibinfo {title}
  {Evidence for quantum critical behavior in the optimally doped cuprate
  {Bi$_2$Sr$_2$CaCu$_2$O$_{8+\delta}$}},\ }\href
  {https://www.science.org/doi/10.1126/science.285.5436.2110} {\bibfield
  {journal} {\bibinfo  {journal} {Science}\ }\textbf {\bibinfo {volume}
  {285}},\ \bibinfo {pages} {2110} (\bibinfo {year} {1999})}\BibitemShut
  {NoStop}%
\bibitem [{\citenamefont {Jarrell}\ and\ \citenamefont
  {Gubernatis}(1996)}]{jarrell:1996}%
  \BibitemOpen
  \bibfield  {author} {\bibinfo {author} {\bibfnamefont {M.}~\bibnamefont
  {Jarrell}}\ and\ \bibinfo {author} {\bibfnamefont {J.~E.}\ \bibnamefont
  {Gubernatis}},\ }\bibfield  {title} {\bibinfo {title} {Bayesian inference and
  the analytic continuation of imaginary-time quantum {{Monte Carlo}} data},\
  }\href {https://doi.org/10.1016/0370-1573(95)00074-7} {\bibfield  {journal}
  {\bibinfo  {journal} {Phys. Rep.}\ }\textbf {\bibinfo {volume} {269}},\
  \bibinfo {pages} {63} (\bibinfo {year} {1996})}\BibitemShut {NoStop}%
\bibitem [{\citenamefont {Chien}\ \emph {et~al.}(2015)\citenamefont {Chien},
  \citenamefont {He}, \citenamefont {Mo}, \citenamefont {Hashimoto},
  \citenamefont {Hussain}, \citenamefont {Shen},\ and\ \citenamefont
  {Plummer}}]{chien:2015}%
  \BibitemOpen
  \bibfield  {author} {\bibinfo {author} {\bibfnamefont {T.}~\bibnamefont
  {Chien}}, \bibinfo {author} {\bibfnamefont {X.}~\bibnamefont {He}}, \bibinfo
  {author} {\bibfnamefont {S.-K.}\ \bibnamefont {Mo}}, \bibinfo {author}
  {\bibfnamefont {M.}~\bibnamefont {Hashimoto}}, \bibinfo {author}
  {\bibfnamefont {Z.}~\bibnamefont {Hussain}}, \bibinfo {author} {\bibfnamefont
  {Z.-X.}\ \bibnamefont {Shen}},\ and\ \bibinfo {author} {\bibfnamefont
  {E.~W.}\ \bibnamefont {Plummer}},\ }\bibfield  {title} {\bibinfo {title}
  {Electron-phonon coupling in a system with broken symmetry: {{Surface}} of
  {{Be}}(0001)},\ }\href {https://link.aps.org/doi/10.1103/PhysRevB.92.075133}
  {\bibfield  {journal} {\bibinfo  {journal} {Phys. Rev. B}\ }\textbf {\bibinfo
  {volume} {92}},\ \bibinfo {pages} {075133} (\bibinfo {year}
  {2015})}\BibitemShut {NoStop}%
\bibitem [{\citenamefont {Ludbrook}(2014)}]{ludbrook:2014}%
  \BibitemOpen
  \bibfield  {author} {\bibinfo {author} {\bibfnamefont {B.~M.}\ \bibnamefont
  {Ludbrook}},\ }\emph {\bibinfo {title} {Electron-Phonon Mediated
  Superconductivity Probed by {ARPES}: {F}rom {MgB$_2$} to Lithium-Decorated
  Graphene}},\ \href
  {https://open.library.ubc.ca/soa/cIRcle/collections/ubctheses/24/items/1.0135608}
  {Ph.D. thesis},\ \bibinfo  {school} {The University of British Columbia}
  (\bibinfo {year} {2014})\BibitemShut {NoStop}%
\bibitem [{\citenamefont {Haberer}\ \emph {et~al.}(2013)\citenamefont
  {Haberer}, \citenamefont {Petaccia}, \citenamefont {Fedorov}, \citenamefont
  {Praveen}, \citenamefont {Fabris}, \citenamefont {Piccinin}, \citenamefont
  {Vilkov}, \citenamefont {Vyalikh}, \citenamefont {Preobrajenski},
  \citenamefont {Verbitskiy}, \citenamefont {Shiozawa}, \citenamefont {Fink},
  \citenamefont {Knupfer}, \citenamefont {B{\"u}chner},\ and\ \citenamefont
  {Gr{\"u}neis}}]{haberer:2013}%
  \BibitemOpen
  \bibfield  {author} {\bibinfo {author} {\bibfnamefont {D.}~\bibnamefont
  {Haberer}}, \bibinfo {author} {\bibfnamefont {L.}~\bibnamefont {Petaccia}},
  \bibinfo {author} {\bibfnamefont {A.~V.}\ \bibnamefont {Fedorov}}, \bibinfo
  {author} {\bibfnamefont {C.~S.}\ \bibnamefont {Praveen}}, \bibinfo {author}
  {\bibfnamefont {S.}~\bibnamefont {Fabris}}, \bibinfo {author} {\bibfnamefont
  {S.}~\bibnamefont {Piccinin}}, \bibinfo {author} {\bibfnamefont
  {O.}~\bibnamefont {Vilkov}}, \bibinfo {author} {\bibfnamefont {D.~V.}\
  \bibnamefont {Vyalikh}}, \bibinfo {author} {\bibfnamefont {A.}~\bibnamefont
  {Preobrajenski}}, \bibinfo {author} {\bibfnamefont {N.~I.}\ \bibnamefont
  {Verbitskiy}}, \bibinfo {author} {\bibfnamefont {H.}~\bibnamefont
  {Shiozawa}}, \bibinfo {author} {\bibfnamefont {J.}~\bibnamefont {Fink}},
  \bibinfo {author} {\bibfnamefont {M.}~\bibnamefont {Knupfer}}, \bibinfo
  {author} {\bibfnamefont {B.}~\bibnamefont {B{\"u}chner}},\ and\ \bibinfo
  {author} {\bibfnamefont {A.}~\bibnamefont {Gr{\"u}neis}},\ }\bibfield
  {title} {\bibinfo {title} {Anisotropic {{Eliashberg}} function and
  electron-phonon coupling in doped graphene},\ }\href
  {https://link.aps.org/doi/10.1103/PhysRevB.88.081401} {\bibfield  {journal}
  {\bibinfo  {journal} {Phys. Rev. B}\ }\textbf {\bibinfo {volume} {88}},\
  \bibinfo {pages} {081401} (\bibinfo {year} {2013})}\BibitemShut {NoStop}%
\bibitem [{\citenamefont {Fedorov}\ \emph {et~al.}(2014)\citenamefont
  {Fedorov}, \citenamefont {Verbitskiy}, \citenamefont {Haberer}, \citenamefont
  {Struzzi}, \citenamefont {Petaccia}, \citenamefont {Usachov}, \citenamefont
  {Vilkov}, \citenamefont {Vyalikh}, \citenamefont {Fink}, \citenamefont
  {Knupfer}, \citenamefont {B{\"u}chner},\ and\ \citenamefont
  {Gr{\"u}neis}}]{fedorov:2014}%
  \BibitemOpen
  \bibfield  {author} {\bibinfo {author} {\bibfnamefont {A.~V.}\ \bibnamefont
  {Fedorov}}, \bibinfo {author} {\bibfnamefont {N.~I.}\ \bibnamefont
  {Verbitskiy}}, \bibinfo {author} {\bibfnamefont {D.}~\bibnamefont {Haberer}},
  \bibinfo {author} {\bibfnamefont {C.}~\bibnamefont {Struzzi}}, \bibinfo
  {author} {\bibfnamefont {L.}~\bibnamefont {Petaccia}}, \bibinfo {author}
  {\bibfnamefont {D.}~\bibnamefont {Usachov}}, \bibinfo {author} {\bibfnamefont
  {O.~Y.}\ \bibnamefont {Vilkov}}, \bibinfo {author} {\bibfnamefont {D.~V.}\
  \bibnamefont {Vyalikh}}, \bibinfo {author} {\bibfnamefont {J.}~\bibnamefont
  {Fink}}, \bibinfo {author} {\bibfnamefont {M.}~\bibnamefont {Knupfer}},
  \bibinfo {author} {\bibfnamefont {B.}~\bibnamefont {B{\"u}chner}},\ and\
  \bibinfo {author} {\bibfnamefont {A.}~\bibnamefont {Gr{\"u}neis}},\
  }\bibfield  {title} {\bibinfo {title} {Observation of a universal
  donor-dependent vibrational mode in graphene},\ }\href
  {http://www.nature.com/articles/ncomms4257} {\bibfield  {journal} {\bibinfo
  {journal} {Nat. Commun.}\ }\textbf {\bibinfo {volume} {5}},\ \bibinfo {pages}
  {3257} (\bibinfo {year} {2014})}\BibitemShut {NoStop}%
\bibitem [{\citenamefont {Ludbrook}\ \emph {et~al.}(2015)\citenamefont
  {Ludbrook}, \citenamefont {Levy}, \citenamefont {Nigge}, \citenamefont
  {Zonno}, \citenamefont {Schneider}, \citenamefont {Dvorak}, \citenamefont
  {Veenstra}, \citenamefont {Zhdanovich}, \citenamefont {Wong}, \citenamefont
  {Dosanjh}, \citenamefont {Stra{\ss}er}, \citenamefont {St{\"o}hr},
  \citenamefont {Forti}, \citenamefont {Ast}, \citenamefont {Starke},\ and\
  \citenamefont {Damascelli}}]{ludbrook:2015}%
  \BibitemOpen
  \bibfield  {author} {\bibinfo {author} {\bibfnamefont {B.~M.}\ \bibnamefont
  {Ludbrook}}, \bibinfo {author} {\bibfnamefont {G.}~\bibnamefont {Levy}},
  \bibinfo {author} {\bibfnamefont {P.}~\bibnamefont {Nigge}}, \bibinfo
  {author} {\bibfnamefont {M.}~\bibnamefont {Zonno}}, \bibinfo {author}
  {\bibfnamefont {M.}~\bibnamefont {Schneider}}, \bibinfo {author}
  {\bibfnamefont {D.~J.}\ \bibnamefont {Dvorak}}, \bibinfo {author}
  {\bibfnamefont {C.~N.}\ \bibnamefont {Veenstra}}, \bibinfo {author}
  {\bibfnamefont {S.}~\bibnamefont {Zhdanovich}}, \bibinfo {author}
  {\bibfnamefont {D.}~\bibnamefont {Wong}}, \bibinfo {author} {\bibfnamefont
  {P.}~\bibnamefont {Dosanjh}}, \bibinfo {author} {\bibfnamefont
  {C.}~\bibnamefont {Stra{\ss}er}}, \bibinfo {author} {\bibfnamefont
  {A.}~\bibnamefont {St{\"o}hr}}, \bibinfo {author} {\bibfnamefont
  {S.}~\bibnamefont {Forti}}, \bibinfo {author} {\bibfnamefont {C.~R.}\
  \bibnamefont {Ast}}, \bibinfo {author} {\bibfnamefont {U.}~\bibnamefont
  {Starke}},\ and\ \bibinfo {author} {\bibfnamefont {A.}~\bibnamefont
  {Damascelli}},\ }\bibfield  {title} {\bibinfo {title} {Evidence for
  superconductivity in {Li}-decorated monolayer graphene},\ }\href
  {http://www.pnas.org/lookup/doi/10.1073/pnas.1510435112} {\bibfield
  {journal} {\bibinfo  {journal} {Proc. Natl. Acad. Sci. U.S.A.}\ }\textbf
  {\bibinfo {volume} {112}},\ \bibinfo {pages} {11795} (\bibinfo {year}
  {2015})}\BibitemShut {NoStop}%
\bibitem [{\citenamefont {Verbitskiy}\ \emph {et~al.}(2016)\citenamefont
  {Verbitskiy}, \citenamefont {Fedorov}, \citenamefont {Tresca}, \citenamefont
  {Profeta}, \citenamefont {Petaccia}, \citenamefont {Senkovskiy},
  \citenamefont {Usachov}, \citenamefont {Vyalikh}, \citenamefont {Yashina},
  \citenamefont {Eliseev}, \citenamefont {Pichler},\ and\ \citenamefont
  {Gr{\"u}neis}}]{verbitskiy:2016}%
  \BibitemOpen
  \bibfield  {author} {\bibinfo {author} {\bibfnamefont {N.~I.}\ \bibnamefont
  {Verbitskiy}}, \bibinfo {author} {\bibfnamefont {A.~V.}\ \bibnamefont
  {Fedorov}}, \bibinfo {author} {\bibfnamefont {C.}~\bibnamefont {Tresca}},
  \bibinfo {author} {\bibfnamefont {G.}~\bibnamefont {Profeta}}, \bibinfo
  {author} {\bibfnamefont {L.}~\bibnamefont {Petaccia}}, \bibinfo {author}
  {\bibfnamefont {B.~V.}\ \bibnamefont {Senkovskiy}}, \bibinfo {author}
  {\bibfnamefont {D.~Y.}\ \bibnamefont {Usachov}}, \bibinfo {author}
  {\bibfnamefont {D.~V.}\ \bibnamefont {Vyalikh}}, \bibinfo {author}
  {\bibfnamefont {L.~V.}\ \bibnamefont {Yashina}}, \bibinfo {author}
  {\bibfnamefont {A.~A.}\ \bibnamefont {Eliseev}}, \bibinfo {author}
  {\bibfnamefont {T.}~\bibnamefont {Pichler}},\ and\ \bibinfo {author}
  {\bibfnamefont {A.}~\bibnamefont {Gr{\"u}neis}},\ }\bibfield  {title}
  {\bibinfo {title} {Environmental control of electron--phonon coupling in
  barium doped graphene},\ }\href
  {https://iopscience.iop.org/article/10.1088/2053-1583/3/4/045003} {\bibfield
  {journal} {\bibinfo  {journal} {2D Mater.}\ }\textbf {\bibinfo {volume}
  {3}},\ \bibinfo {pages} {045003} (\bibinfo {year} {2016})}\BibitemShut
  {NoStop}%
\bibitem [{\citenamefont {Usachov}\ \emph {et~al.}(2018)\citenamefont
  {Usachov}, \citenamefont {Fedorov}, \citenamefont {Vilkov}, \citenamefont
  {Ogorodnikov}, \citenamefont {Kuznetsov}, \citenamefont {Gr{\"u}neis},
  \citenamefont {Laubschat},\ and\ \citenamefont {Vyalikh}}]{usachov:2018}%
  \BibitemOpen
  \bibfield  {author} {\bibinfo {author} {\bibfnamefont {D.~Y.}\ \bibnamefont
  {Usachov}}, \bibinfo {author} {\bibfnamefont {A.~V.}\ \bibnamefont
  {Fedorov}}, \bibinfo {author} {\bibfnamefont {O.~Y.}\ \bibnamefont {Vilkov}},
  \bibinfo {author} {\bibfnamefont {I.~I.}\ \bibnamefont {Ogorodnikov}},
  \bibinfo {author} {\bibfnamefont {M.~V.}\ \bibnamefont {Kuznetsov}}, \bibinfo
  {author} {\bibfnamefont {A.}~\bibnamefont {Gr{\"u}neis}}, \bibinfo {author}
  {\bibfnamefont {C.}~\bibnamefont {Laubschat}},\ and\ \bibinfo {author}
  {\bibfnamefont {D.~V.}\ \bibnamefont {Vyalikh}},\ }\bibfield  {title}
  {\bibinfo {title} {Electron-phonon coupling in graphene placed between
  magnetic {{Li}} and {{Si}} layers on cobalt},\ }\href
  {https://link.aps.org/doi/10.1103/PhysRevB.97.085132} {\bibfield  {journal}
  {\bibinfo  {journal} {Phys. Rev. B}\ }\textbf {\bibinfo {volume} {97}},\
  \bibinfo {pages} {085132} (\bibinfo {year} {2018})}\BibitemShut {NoStop}%
\bibitem [{\citenamefont {Ehlen}\ \emph {et~al.}(2020)\citenamefont {Ehlen},
  \citenamefont {Hell}, \citenamefont {Marini}, \citenamefont {Hasdeo},
  \citenamefont {Saito}, \citenamefont {Falke}, \citenamefont {Goerbig},
  \citenamefont {Di~Santo}, \citenamefont {Petaccia}, \citenamefont {Profeta},\
  and\ \citenamefont {Gr{\"u}neis}}]{ehlen:2020}%
  \BibitemOpen
  \bibfield  {author} {\bibinfo {author} {\bibfnamefont {N.}~\bibnamefont
  {Ehlen}}, \bibinfo {author} {\bibfnamefont {M.}~\bibnamefont {Hell}},
  \bibinfo {author} {\bibfnamefont {G.}~\bibnamefont {Marini}}, \bibinfo
  {author} {\bibfnamefont {E.~H.}\ \bibnamefont {Hasdeo}}, \bibinfo {author}
  {\bibfnamefont {R.}~\bibnamefont {Saito}}, \bibinfo {author} {\bibfnamefont
  {Y.}~\bibnamefont {Falke}}, \bibinfo {author} {\bibfnamefont {M.~O.}\
  \bibnamefont {Goerbig}}, \bibinfo {author} {\bibfnamefont {G.}~\bibnamefont
  {Di~Santo}}, \bibinfo {author} {\bibfnamefont {L.}~\bibnamefont {Petaccia}},
  \bibinfo {author} {\bibfnamefont {G.}~\bibnamefont {Profeta}},\ and\ \bibinfo
  {author} {\bibfnamefont {A.}~\bibnamefont {Gr{\"u}neis}},\ }\bibfield
  {title} {\bibinfo {title} {Origin of the flat band in heavily {Cs}-doped
  graphene},\ }\href {https://pubs.acs.org/doi/10.1021/acsnano.9b08622}
  {\bibfield  {journal} {\bibinfo  {journal} {ACS Nano}\ }\textbf {\bibinfo
  {volume} {14}},\ \bibinfo {pages} {1055} (\bibinfo {year}
  {2020})}\BibitemShut {NoStop}%
\bibitem [{\citenamefont {Huempfner}\ \emph {et~al.}(2023)\citenamefont
  {Huempfner}, \citenamefont {Otto}, \citenamefont {Forker}, \citenamefont
  {M{\"u}ller},\ and\ \citenamefont {Fritz}}]{huempfner:2023}%
  \BibitemOpen
  \bibfield  {author} {\bibinfo {author} {\bibfnamefont {T.}~\bibnamefont
  {Huempfner}}, \bibinfo {author} {\bibfnamefont {F.}~\bibnamefont {Otto}},
  \bibinfo {author} {\bibfnamefont {R.}~\bibnamefont {Forker}}, \bibinfo
  {author} {\bibfnamefont {P.}~\bibnamefont {M{\"u}ller}},\ and\ \bibinfo
  {author} {\bibfnamefont {T.}~\bibnamefont {Fritz}},\ }\bibfield  {title}
  {\bibinfo {title} {Superconductivity of {{K}}-{{Intercalated Epitaxial
  Bilayer Graphene}}},\ }\href
  {https://onlinelibrary.wiley.com/doi/10.1002/admi.202300014} {\bibfield
  {journal} {\bibinfo  {journal} {Adv. Mater. Interfaces}\ }\textbf {\bibinfo
  {volume} {10}},\ \bibinfo {pages} {2300014} (\bibinfo {year}
  {2023})}\BibitemShut {NoStop}%
\bibitem [{\citenamefont {Zhong}\ \emph {et~al.}(2023)\citenamefont {Zhong},
  \citenamefont {Li}, \citenamefont {Liu}, \citenamefont {Dong}, \citenamefont
  {Aido}, \citenamefont {Arai}, \citenamefont {Li}, \citenamefont {Zhang},
  \citenamefont {Shi}, \citenamefont {Wang}, \citenamefont {Shin},
  \citenamefont {Lee}, \citenamefont {Miao}, \citenamefont {Kondo},\ and\
  \citenamefont {Okazaki}}]{zhong:2023}%
  \BibitemOpen
  \bibfield  {author} {\bibinfo {author} {\bibfnamefont {Y.}~\bibnamefont
  {Zhong}}, \bibinfo {author} {\bibfnamefont {S.}~\bibnamefont {Li}}, \bibinfo
  {author} {\bibfnamefont {H.}~\bibnamefont {Liu}}, \bibinfo {author}
  {\bibfnamefont {Y.}~\bibnamefont {Dong}}, \bibinfo {author} {\bibfnamefont
  {K.}~\bibnamefont {Aido}}, \bibinfo {author} {\bibfnamefont {Y.}~\bibnamefont
  {Arai}}, \bibinfo {author} {\bibfnamefont {H.}~\bibnamefont {Li}}, \bibinfo
  {author} {\bibfnamefont {W.}~\bibnamefont {Zhang}}, \bibinfo {author}
  {\bibfnamefont {Y.}~\bibnamefont {Shi}}, \bibinfo {author} {\bibfnamefont
  {Z.}~\bibnamefont {Wang}}, \bibinfo {author} {\bibfnamefont {S.}~\bibnamefont
  {Shin}}, \bibinfo {author} {\bibfnamefont {H.~N.}\ \bibnamefont {Lee}},
  \bibinfo {author} {\bibfnamefont {H.}~\bibnamefont {Miao}}, \bibinfo {author}
  {\bibfnamefont {T.}~\bibnamefont {Kondo}},\ and\ \bibinfo {author}
  {\bibfnamefont {K.}~\bibnamefont {Okazaki}},\ }\bibfield  {title} {\bibinfo
  {title} {Testing electron--phonon coupling for the superconductivity in
  kagome metal {{CsV}}{\textsubscript{3}}{{Sb}}{\textsubscript{5}}},\ }\href
  {https://www.nature.com/articles/s41467-023-37605-7} {\bibfield  {journal}
  {\bibinfo  {journal} {Nat. Commun.}\ }\textbf {\bibinfo {volume} {14}},\
  \bibinfo {pages} {1945} (\bibinfo {year} {2023})}\BibitemShut {NoStop}%
\bibitem [{\citenamefont {Sokolovi{\'c}}\ \emph {et~al.}(2025)\citenamefont
  {Sokolovi{\'c}}, \citenamefont {Guedes}, \citenamefont {Van~Waas},
  \citenamefont {Guo}, \citenamefont {Ponc{\'e}}, \citenamefont {Polley},
  \citenamefont {Schmid}, \citenamefont {Diebold}, \citenamefont {Radovi{\'c}},
  \citenamefont {Setv{\'i}n},\ and\ \citenamefont {Dil}}]{sokolovic:2025}%
  \BibitemOpen
  \bibfield  {author} {\bibinfo {author} {\bibfnamefont {I.}~\bibnamefont
  {Sokolovi{\'c}}}, \bibinfo {author} {\bibfnamefont {E.~B.}\ \bibnamefont
  {Guedes}}, \bibinfo {author} {\bibfnamefont {T.~P.}\ \bibnamefont
  {Van~Waas}}, \bibinfo {author} {\bibfnamefont {F.}~\bibnamefont {Guo}},
  \bibinfo {author} {\bibfnamefont {S.}~\bibnamefont {Ponc{\'e}}}, \bibinfo
  {author} {\bibfnamefont {C.}~\bibnamefont {Polley}}, \bibinfo {author}
  {\bibfnamefont {M.}~\bibnamefont {Schmid}}, \bibinfo {author} {\bibfnamefont
  {U.}~\bibnamefont {Diebold}}, \bibinfo {author} {\bibfnamefont
  {M.}~\bibnamefont {Radovi{\'c}}}, \bibinfo {author} {\bibfnamefont
  {M.}~\bibnamefont {Setv{\'i}n}},\ and\ \bibinfo {author} {\bibfnamefont
  {J.~H.}\ \bibnamefont {Dil}},\ }\bibfield  {title} {\bibinfo {title} {Duality
  and degeneracy lifting in two-dimensional electron liquids on
  {{SrTiO}}{\textsubscript{3}}(001)},\ }\href
  {https://www.nature.com/articles/s41467-025-59258-4} {\bibfield  {journal}
  {\bibinfo  {journal} {Nat. Commun.}\ }\textbf {\bibinfo {volume} {16}},\
  \bibinfo {pages} {4594} (\bibinfo {year} {2025})}\BibitemShut {NoStop}%
\bibitem [{\citenamefont {R{\o}st}\ \emph {et~al.}(2024)\citenamefont
  {R{\o}st}, \citenamefont {Mazzola}, \citenamefont {Bakkelund}, \citenamefont
  {{\AA}sland}, \citenamefont {Hu}, \citenamefont {Cooil}, \citenamefont
  {Polley},\ and\ \citenamefont {Wells}}]{rost:2024}%
  \BibitemOpen
  \bibfield  {author} {\bibinfo {author} {\bibfnamefont {H.~I.}\ \bibnamefont
  {R{\o}st}}, \bibinfo {author} {\bibfnamefont {F.}~\bibnamefont {Mazzola}},
  \bibinfo {author} {\bibfnamefont {J.}~\bibnamefont {Bakkelund}}, \bibinfo
  {author} {\bibfnamefont {A.~C.}\ \bibnamefont {{\AA}sland}}, \bibinfo
  {author} {\bibfnamefont {J.}~\bibnamefont {Hu}}, \bibinfo {author}
  {\bibfnamefont {S.~P.}\ \bibnamefont {Cooil}}, \bibinfo {author}
  {\bibfnamefont {C.~M.}\ \bibnamefont {Polley}},\ and\ \bibinfo {author}
  {\bibfnamefont {J.~W.}\ \bibnamefont {Wells}},\ }\bibfield  {title} {\bibinfo
  {title} {Disentangling electron-boson interactions on the surface of a
  familiar ferromagnet},\ }\href
  {https://link.aps.org/doi/10.1103/PhysRevB.109.035137} {\bibfield  {journal}
  {\bibinfo  {journal} {Phys. Rev. B}\ }\textbf {\bibinfo {volume} {109}},\
  \bibinfo {pages} {035137} (\bibinfo {year} {2024})}\BibitemShut {NoStop}%
\bibitem [{\citenamefont {Bok}\ \emph {et~al.}(2010)\citenamefont {Bok},
  \citenamefont {Yun}, \citenamefont {Choi}, \citenamefont {Zhang},
  \citenamefont {Zhou},\ and\ \citenamefont {Varma}}]{bok:2010}%
  \BibitemOpen
  \bibfield  {author} {\bibinfo {author} {\bibfnamefont {J.~M.}\ \bibnamefont
  {Bok}}, \bibinfo {author} {\bibfnamefont {J.~H.}\ \bibnamefont {Yun}},
  \bibinfo {author} {\bibfnamefont {H.-Y.}\ \bibnamefont {Choi}}, \bibinfo
  {author} {\bibfnamefont {W.}~\bibnamefont {Zhang}}, \bibinfo {author}
  {\bibfnamefont {X.~J.}\ \bibnamefont {Zhou}},\ and\ \bibinfo {author}
  {\bibfnamefont {C.~M.}\ \bibnamefont {Varma}},\ }\bibfield  {title} {\bibinfo
  {title} {Momentum dependence of the single-particle self-energy and
  fluctuation spectrum of slightly underdoped
  {Bi$_2$Sr$_2$CaCu$_2$O$_{8+\delta}$} from high-resolution laser
  angle-resolved photoemission},\ }\href
  {https://link.aps.org/doi/10.1103/PhysRevB.81.174516} {\bibfield  {journal}
  {\bibinfo  {journal} {Phys. Rev. B}\ }\textbf {\bibinfo {volume} {81}},\
  \bibinfo {pages} {174516} (\bibinfo {year} {2010})}\BibitemShut {NoStop}%
\bibitem [{\citenamefont {Zhao}\ \emph {et~al.}(2011)\citenamefont {Zhao},
  \citenamefont {Wang}, \citenamefont {Shi}, \citenamefont {Zhang},
  \citenamefont {Liu}, \citenamefont {Meng}, \citenamefont {Liu}, \citenamefont
  {Dong}, \citenamefont {Zhang}, \citenamefont {Lu}, \citenamefont {Wang},
  \citenamefont {Zhu}, \citenamefont {Wang}, \citenamefont {Peng},
  \citenamefont {Wang}, \citenamefont {Zhang}, \citenamefont {Yang},
  \citenamefont {Chen}, \citenamefont {Xu},\ and\ \citenamefont
  {Zhou}}]{zhao:2011}%
  \BibitemOpen
  \bibfield  {author} {\bibinfo {author} {\bibfnamefont {L.}~\bibnamefont
  {Zhao}}, \bibinfo {author} {\bibfnamefont {J.}~\bibnamefont {Wang}}, \bibinfo
  {author} {\bibfnamefont {J.}~\bibnamefont {Shi}}, \bibinfo {author}
  {\bibfnamefont {W.}~\bibnamefont {Zhang}}, \bibinfo {author} {\bibfnamefont
  {H.}~\bibnamefont {Liu}}, \bibinfo {author} {\bibfnamefont {J.}~\bibnamefont
  {Meng}}, \bibinfo {author} {\bibfnamefont {G.}~\bibnamefont {Liu}}, \bibinfo
  {author} {\bibfnamefont {X.}~\bibnamefont {Dong}}, \bibinfo {author}
  {\bibfnamefont {J.}~\bibnamefont {Zhang}}, \bibinfo {author} {\bibfnamefont
  {W.}~\bibnamefont {Lu}}, \bibinfo {author} {\bibfnamefont {G.}~\bibnamefont
  {Wang}}, \bibinfo {author} {\bibfnamefont {Y.}~\bibnamefont {Zhu}}, \bibinfo
  {author} {\bibfnamefont {X.}~\bibnamefont {Wang}}, \bibinfo {author}
  {\bibfnamefont {Q.}~\bibnamefont {Peng}}, \bibinfo {author} {\bibfnamefont
  {Z.}~\bibnamefont {Wang}}, \bibinfo {author} {\bibfnamefont {S.}~\bibnamefont
  {Zhang}}, \bibinfo {author} {\bibfnamefont {F.}~\bibnamefont {Yang}},
  \bibinfo {author} {\bibfnamefont {C.}~\bibnamefont {Chen}}, \bibinfo {author}
  {\bibfnamefont {Z.}~\bibnamefont {Xu}},\ and\ \bibinfo {author}
  {\bibfnamefont {X.~J.}\ \bibnamefont {Zhou}},\ }\bibfield  {title} {\bibinfo
  {title} {Quantitative determination of {{Eliashberg}} function and evidence
  of strong electron coupling with multiple phonon modes in heavily overdoped
  ({{Bi}},{{Pb}}){\textsubscript{2}}{{Sr}}{\textsubscript{2}}{{CuO}}{\textsubscript{6+{$\delta$}}}},\
  }\href {https://link.aps.org/doi/10.1103/PhysRevB.83.184515} {\bibfield
  {journal} {\bibinfo  {journal} {Phys. Rev. B}\ }\textbf {\bibinfo {volume}
  {83}},\ \bibinfo {pages} {184515} (\bibinfo {year} {2011})}\BibitemShut
  {NoStop}%
\bibitem [{\citenamefont {Yun}\ \emph {et~al.}(2011)\citenamefont {Yun},
  \citenamefont {Bok}, \citenamefont {Choi}, \citenamefont {Zhang},
  \citenamefont {Zhou},\ and\ \citenamefont {Varma}}]{yun:2011}%
  \BibitemOpen
  \bibfield  {author} {\bibinfo {author} {\bibfnamefont {J.~H.}\ \bibnamefont
  {Yun}}, \bibinfo {author} {\bibfnamefont {J.~M.}\ \bibnamefont {Bok}},
  \bibinfo {author} {\bibfnamefont {H.-Y.}\ \bibnamefont {Choi}}, \bibinfo
  {author} {\bibfnamefont {W.}~\bibnamefont {Zhang}}, \bibinfo {author}
  {\bibfnamefont {X.~J.}\ \bibnamefont {Zhou}},\ and\ \bibinfo {author}
  {\bibfnamefont {C.~M.}\ \bibnamefont {Varma}},\ }\bibfield  {title} {\bibinfo
  {title} {Analysis of laser angle-resolved photoemission spectra of
  {{Ba}}{\textsubscript{2}}{{Sr}}{\textsubscript{2}}{{CaCu}}{\textsubscript{2}}{{O}}{\textsubscript{8+{$\delta$}}}
  in the superconducting state: {{Angle-resolved}} self-energy and the
  fluctuation spectrum},\ }\href
  {https://link.aps.org/doi/10.1103/PhysRevB.84.104521} {\bibfield  {journal}
  {\bibinfo  {journal} {Phys. Rev. B}\ }\textbf {\bibinfo {volume} {84}},\
  \bibinfo {pages} {104521} (\bibinfo {year} {2011})}\BibitemShut {NoStop}%
\bibitem [{\citenamefont {Marzari}\ \emph {et~al.}(2021)\citenamefont
  {Marzari}, \citenamefont {Ferretti},\ and\ \citenamefont
  {Wolverton}}]{marzari:2021}%
  \BibitemOpen
  \bibfield  {author} {\bibinfo {author} {\bibfnamefont {N.}~\bibnamefont
  {Marzari}}, \bibinfo {author} {\bibfnamefont {A.}~\bibnamefont {Ferretti}},\
  and\ \bibinfo {author} {\bibfnamefont {C.}~\bibnamefont {Wolverton}},\
  }\bibfield  {title} {\bibinfo {title} {Electronic-structure methods for
  materials design},\ }\href
  {https://www.nature.com/articles/s41563-021-01013-3} {\bibfield  {journal}
  {\bibinfo  {journal} {Nat. Mater.}\ }\textbf {\bibinfo {volume} {20}},\
  \bibinfo {pages} {736} (\bibinfo {year} {2021})}\BibitemShut {NoStop}%
\bibitem [{\citenamefont {Gonze}(1997)}]{gonze:1997b}%
  \BibitemOpen
  \bibfield  {author} {\bibinfo {author} {\bibfnamefont {X.}~\bibnamefont
  {Gonze}},\ }\bibfield  {title} {\bibinfo {title} {First-principles responses
  of solids to atomic displacements and homogeneous electric fields:
  {{Implementation}} of a conjugate-gradient algorithm},\ }\href
  {https://link.aps.org/doi/10.1103/PhysRevB.55.10337} {\bibfield  {journal}
  {\bibinfo  {journal} {Phys. Rev. B}\ }\textbf {\bibinfo {volume} {55}},\
  \bibinfo {pages} {10337} (\bibinfo {year} {1997})}\BibitemShut {NoStop}%
\bibitem [{\citenamefont {Gonze}\ and\ \citenamefont {Lee}(1997)}]{gonze:1997}%
  \BibitemOpen
  \bibfield  {author} {\bibinfo {author} {\bibfnamefont {X.}~\bibnamefont
  {Gonze}}\ and\ \bibinfo {author} {\bibfnamefont {C.}~\bibnamefont {Lee}},\
  }\bibfield  {title} {\bibinfo {title} {Dynamical matrices, {{Born}} effective
  charges, dielectric permittivity tensors, and interatomic force constants
  from density-functional perturbation theory},\ }\href
  {https://link.aps.org/doi/10.1103/PhysRevB.55.10355} {\bibfield  {journal}
  {\bibinfo  {journal} {Phys. Rev. B}\ }\textbf {\bibinfo {volume} {55}},\
  \bibinfo {pages} {10355} (\bibinfo {year} {1997})}\BibitemShut {NoStop}%
\bibitem [{\citenamefont {Baroni}\ \emph {et~al.}(2001)\citenamefont {Baroni},
  \citenamefont {{de Gironcoli}}, \citenamefont {Dal~Corso},\ and\
  \citenamefont {Giannozzi}}]{baroni:2001}%
  \BibitemOpen
  \bibfield  {author} {\bibinfo {author} {\bibfnamefont {S.}~\bibnamefont
  {Baroni}}, \bibinfo {author} {\bibfnamefont {S.}~\bibnamefont {{de
  Gironcoli}}}, \bibinfo {author} {\bibfnamefont {A.}~\bibnamefont
  {Dal~Corso}},\ and\ \bibinfo {author} {\bibfnamefont {P.}~\bibnamefont
  {Giannozzi}},\ }\bibfield  {title} {\bibinfo {title} {Phonons and related
  crystal properties from density-functional perturbation theory},\ }\href
  {https://link.aps.org/doi/10.1103/RevModPhys.73.515} {\bibfield  {journal}
  {\bibinfo  {journal} {Rev. Mod. Phys.}\ }\textbf {\bibinfo {volume} {73}},\
  \bibinfo {pages} {515} (\bibinfo {year} {2001})}\BibitemShut {NoStop}%
\bibitem [{\citenamefont {Ponc{\'e}}\ \emph {et~al.}(2015)\citenamefont
  {Ponc{\'e}}, \citenamefont {Gillet}, \citenamefont {Laflamme~Janssen},
  \citenamefont {Marini}, \citenamefont {Verstraete},\ and\ \citenamefont
  {Gonze}}]{ponce:2015}%
  \BibitemOpen
  \bibfield  {author} {\bibinfo {author} {\bibfnamefont {S.}~\bibnamefont
  {Ponc{\'e}}}, \bibinfo {author} {\bibfnamefont {Y.}~\bibnamefont {Gillet}},
  \bibinfo {author} {\bibfnamefont {J.}~\bibnamefont {Laflamme~Janssen}},
  \bibinfo {author} {\bibfnamefont {A.}~\bibnamefont {Marini}}, \bibinfo
  {author} {\bibfnamefont {M.}~\bibnamefont {Verstraete}},\ and\ \bibinfo
  {author} {\bibfnamefont {X.}~\bibnamefont {Gonze}},\ }\bibfield  {title}
  {\bibinfo {title} {Temperature dependence of the electronic structure of
  semiconductors and insulators},\ }\href
  {https://pubs.aip.org/jcp/article/143/10/102813/898920/Temperature-dependence-of-the-electronic-structure}
  {\bibfield  {journal} {\bibinfo  {journal} {J. Chem. Phys.}\ }\textbf
  {\bibinfo {volume} {143}},\ \bibinfo {pages} {102813} (\bibinfo {year}
  {2015})}\BibitemShut {NoStop}%
\bibitem [{\citenamefont {Eiguren}\ and\ \citenamefont
  {Ambrosch-Draxl}(2008)}]{eiguren:2008}%
  \BibitemOpen
  \bibfield  {author} {\bibinfo {author} {\bibfnamefont {A.}~\bibnamefont
  {Eiguren}}\ and\ \bibinfo {author} {\bibfnamefont {C.}~\bibnamefont
  {Ambrosch-Draxl}},\ }\bibfield  {title} {\bibinfo {title} {Wannier
  interpolation scheme for phonon-induced potentials: Application to bulk
  {{MgB}}$_2$, {{W}}, and the (1$\times$ 1) {{H}}-covered {{W}}(110) surface},\
  }\href {https://link.aps.org/doi/10.1103/PhysRevB.78.045124} {\bibfield
  {journal} {\bibinfo  {journal} {Phys. Rev. B}\ }\textbf {\bibinfo {volume}
  {78}},\ \bibinfo {pages} {045124} (\bibinfo {year} {2008})}\BibitemShut
  {NoStop}%
\bibitem [{\citenamefont {Gonze}\ \emph {et~al.}(2020)\citenamefont {Gonze},
  \citenamefont {Amadon}, \citenamefont {Antonius}, \citenamefont {Arnardi},
  \citenamefont {Baguet}, \citenamefont {Beuken}, \citenamefont {Bieder},
  \citenamefont {Bottin}, \citenamefont {Bouchet}, \citenamefont {Bousquet},
  \citenamefont {Brouwer}, \citenamefont {Bruneval}, \citenamefont {Brunin},
  \citenamefont {Cavignac}, \citenamefont {Charraud}, \citenamefont {Chen},
  \citenamefont {Côté}, \citenamefont {Cottenier}, \citenamefont {Denier},
  \citenamefont {Geneste}, \citenamefont {Ghosez}, \citenamefont {Giantomassi},
  \citenamefont {Gillet}, \citenamefont {Gingras}, \citenamefont {Hamann},
  \citenamefont {Hautier}, \citenamefont {He}, \citenamefont {Helbig},
  \citenamefont {Holzwarth}, \citenamefont {Jia}, \citenamefont {Jollet},
  \citenamefont {Lafargue-Dit-Hauret}, \citenamefont {Lejaeghere},
  \citenamefont {Marques}, \citenamefont {Martin}, \citenamefont {Martins},
  \citenamefont {Miranda}, \citenamefont {Naccarato}, \citenamefont {Persson},
  \citenamefont {Petretto}, \citenamefont {Planes}, \citenamefont {Pouillon},
  \citenamefont {Prokhorenko}, \citenamefont {Ricci}, \citenamefont
  {Rignanese}, \citenamefont {Romero}, \citenamefont {Schmitt}, \citenamefont
  {Torrent}, \citenamefont {van Setten}, \citenamefont {Van~Troeye},
  \citenamefont {Verstraete}, \citenamefont {Zérah},\ and\ \citenamefont
  {Zwanziger}}]{gonze:2020}%
  \BibitemOpen
  \bibfield  {author} {\bibinfo {author} {\bibfnamefont {X.}~\bibnamefont
  {Gonze}}, \bibinfo {author} {\bibfnamefont {B.}~\bibnamefont {Amadon}},
  \bibinfo {author} {\bibfnamefont {G.}~\bibnamefont {Antonius}}, \bibinfo
  {author} {\bibfnamefont {F.}~\bibnamefont {Arnardi}}, \bibinfo {author}
  {\bibfnamefont {L.}~\bibnamefont {Baguet}}, \bibinfo {author} {\bibfnamefont
  {J.-M.}\ \bibnamefont {Beuken}}, \bibinfo {author} {\bibfnamefont
  {J.}~\bibnamefont {Bieder}}, \bibinfo {author} {\bibfnamefont
  {F.}~\bibnamefont {Bottin}}, \bibinfo {author} {\bibfnamefont
  {J.}~\bibnamefont {Bouchet}}, \bibinfo {author} {\bibfnamefont
  {E.}~\bibnamefont {Bousquet}}, \bibinfo {author} {\bibfnamefont
  {N.}~\bibnamefont {Brouwer}}, \bibinfo {author} {\bibfnamefont
  {F.}~\bibnamefont {Bruneval}}, \bibinfo {author} {\bibfnamefont
  {G.}~\bibnamefont {Brunin}}, \bibinfo {author} {\bibfnamefont
  {T.}~\bibnamefont {Cavignac}}, \bibinfo {author} {\bibfnamefont {J.-B.}\
  \bibnamefont {Charraud}}, \bibinfo {author} {\bibfnamefont {W.}~\bibnamefont
  {Chen}}, \bibinfo {author} {\bibfnamefont {M.}~\bibnamefont {Côté}},
  \bibinfo {author} {\bibfnamefont {S.}~\bibnamefont {Cottenier}}, \bibinfo
  {author} {\bibfnamefont {J.}~\bibnamefont {Denier}}, \bibinfo {author}
  {\bibfnamefont {G.}~\bibnamefont {Geneste}}, \bibinfo {author} {\bibfnamefont
  {P.}~\bibnamefont {Ghosez}}, \bibinfo {author} {\bibfnamefont
  {M.}~\bibnamefont {Giantomassi}}, \bibinfo {author} {\bibfnamefont
  {Y.}~\bibnamefont {Gillet}}, \bibinfo {author} {\bibfnamefont
  {O.}~\bibnamefont {Gingras}}, \bibinfo {author} {\bibfnamefont {D.~R.}\
  \bibnamefont {Hamann}}, \bibinfo {author} {\bibfnamefont {G.}~\bibnamefont
  {Hautier}}, \bibinfo {author} {\bibfnamefont {X.}~\bibnamefont {He}},
  \bibinfo {author} {\bibfnamefont {N.}~\bibnamefont {Helbig}}, \bibinfo
  {author} {\bibfnamefont {N.}~\bibnamefont {Holzwarth}}, \bibinfo {author}
  {\bibfnamefont {Y.}~\bibnamefont {Jia}}, \bibinfo {author} {\bibfnamefont
  {F.}~\bibnamefont {Jollet}}, \bibinfo {author} {\bibfnamefont
  {W.}~\bibnamefont {Lafargue-Dit-Hauret}}, \bibinfo {author} {\bibfnamefont
  {K.}~\bibnamefont {Lejaeghere}}, \bibinfo {author} {\bibfnamefont {M.~A.}\
  \bibnamefont {Marques}}, \bibinfo {author} {\bibfnamefont {A.}~\bibnamefont
  {Martin}}, \bibinfo {author} {\bibfnamefont {C.}~\bibnamefont {Martins}},
  \bibinfo {author} {\bibfnamefont {H.~P.}\ \bibnamefont {Miranda}}, \bibinfo
  {author} {\bibfnamefont {F.}~\bibnamefont {Naccarato}}, \bibinfo {author}
  {\bibfnamefont {K.}~\bibnamefont {Persson}}, \bibinfo {author} {\bibfnamefont
  {G.}~\bibnamefont {Petretto}}, \bibinfo {author} {\bibfnamefont
  {V.}~\bibnamefont {Planes}}, \bibinfo {author} {\bibfnamefont
  {Y.}~\bibnamefont {Pouillon}}, \bibinfo {author} {\bibfnamefont
  {S.}~\bibnamefont {Prokhorenko}}, \bibinfo {author} {\bibfnamefont
  {F.}~\bibnamefont {Ricci}}, \bibinfo {author} {\bibfnamefont {G.-M.}\
  \bibnamefont {Rignanese}}, \bibinfo {author} {\bibfnamefont {A.~H.}\
  \bibnamefont {Romero}}, \bibinfo {author} {\bibfnamefont {M.~M.}\
  \bibnamefont {Schmitt}}, \bibinfo {author} {\bibfnamefont {M.}~\bibnamefont
  {Torrent}}, \bibinfo {author} {\bibfnamefont {M.~J.}\ \bibnamefont {van
  Setten}}, \bibinfo {author} {\bibfnamefont {B.}~\bibnamefont {Van~Troeye}},
  \bibinfo {author} {\bibfnamefont {M.~J.}\ \bibnamefont {Verstraete}},
  \bibinfo {author} {\bibfnamefont {G.}~\bibnamefont {Zérah}},\ and\ \bibinfo
  {author} {\bibfnamefont {J.~W.}\ \bibnamefont {Zwanziger}},\ }\bibfield
  {title} {\bibinfo {title} {The \textsc{Abinit} project: Impact, environment
  and recent developments},\ }\href
  {https://linkinghub.elsevier.com/retrieve/pii/S0010465519303741} {\bibfield
  {journal} {\bibinfo  {journal} {Comput. Phys. Commun.}\ }\textbf {\bibinfo
  {volume} {248}},\ \bibinfo {pages} {107042} (\bibinfo {year}
  {2020})}\BibitemShut {NoStop}%
\bibitem [{\citenamefont {Marzari}\ \emph {et~al.}(2012)\citenamefont
  {Marzari}, \citenamefont {Mostofi}, \citenamefont {Yates}, \citenamefont
  {Souza},\ and\ \citenamefont {Vanderbilt}}]{marzari:2012}%
  \BibitemOpen
  \bibfield  {author} {\bibinfo {author} {\bibfnamefont {N.}~\bibnamefont
  {Marzari}}, \bibinfo {author} {\bibfnamefont {A.~A.}\ \bibnamefont
  {Mostofi}}, \bibinfo {author} {\bibfnamefont {J.~R.}\ \bibnamefont {Yates}},
  \bibinfo {author} {\bibfnamefont {I.}~\bibnamefont {Souza}},\ and\ \bibinfo
  {author} {\bibfnamefont {D.}~\bibnamefont {Vanderbilt}},\ }\bibfield  {title}
  {\bibinfo {title} {Maximally localized {{Wannier}} functions: {{Theory}} and
  applications},\ }\href {https://link.aps.org/doi/10.1103/RevModPhys.84.1419}
  {\bibfield  {journal} {\bibinfo  {journal} {Rev. Mod. Phys.}\ }\textbf
  {\bibinfo {volume} {84}},\ \bibinfo {pages} {1419} (\bibinfo {year}
  {2012})}\BibitemShut {NoStop}%
\bibitem [{\citenamefont {Pizzi}\ \emph {et~al.}(2020)\citenamefont {Pizzi},
  \citenamefont {Vitale}, \citenamefont {Arita}, \citenamefont {Bl{\"u}gel},
  \citenamefont {Freimuth}, \citenamefont {G{\'e}ranton}, \citenamefont
  {Gibertini}, \citenamefont {Gresch}, \citenamefont {Johnson}, \citenamefont
  {Koretsune}, \citenamefont {{Iba{\~n}ez-Azpiroz}}, \citenamefont {Lee},
  \citenamefont {Lihm}, \citenamefont {Marchand}, \citenamefont {Marrazzo},
  \citenamefont {Mokrousov}, \citenamefont {Mustafa}, \citenamefont {Nohara},
  \citenamefont {Nomura}, \citenamefont {Paulatto}, \citenamefont {Ponc{\'e}},
  \citenamefont {Ponweiser}, \citenamefont {Qiao}, \citenamefont {Th{\"o}le},
  \citenamefont {Tsirkin}, \citenamefont {Wierzbowska}, \citenamefont
  {Marzari}, \citenamefont {Vanderbilt}, \citenamefont {Souza}, \citenamefont
  {Mostofi},\ and\ \citenamefont {Yates}}]{pizzi:2020}%
  \BibitemOpen
  \bibfield  {author} {\bibinfo {author} {\bibfnamefont {G.}~\bibnamefont
  {Pizzi}}, \bibinfo {author} {\bibfnamefont {V.}~\bibnamefont {Vitale}},
  \bibinfo {author} {\bibfnamefont {R.}~\bibnamefont {Arita}}, \bibinfo
  {author} {\bibfnamefont {S.}~\bibnamefont {Bl{\"u}gel}}, \bibinfo {author}
  {\bibfnamefont {F.}~\bibnamefont {Freimuth}}, \bibinfo {author}
  {\bibfnamefont {G.}~\bibnamefont {G{\'e}ranton}}, \bibinfo {author}
  {\bibfnamefont {M.}~\bibnamefont {Gibertini}}, \bibinfo {author}
  {\bibfnamefont {D.}~\bibnamefont {Gresch}}, \bibinfo {author} {\bibfnamefont
  {C.}~\bibnamefont {Johnson}}, \bibinfo {author} {\bibfnamefont
  {T.}~\bibnamefont {Koretsune}}, \bibinfo {author} {\bibfnamefont
  {J.}~\bibnamefont {{Iba{\~n}ez-Azpiroz}}}, \bibinfo {author} {\bibfnamefont
  {H.}~\bibnamefont {Lee}}, \bibinfo {author} {\bibfnamefont {J.-M.}\
  \bibnamefont {Lihm}}, \bibinfo {author} {\bibfnamefont {D.}~\bibnamefont
  {Marchand}}, \bibinfo {author} {\bibfnamefont {A.}~\bibnamefont {Marrazzo}},
  \bibinfo {author} {\bibfnamefont {Y.}~\bibnamefont {Mokrousov}}, \bibinfo
  {author} {\bibfnamefont {J.~I.}\ \bibnamefont {Mustafa}}, \bibinfo {author}
  {\bibfnamefont {Y.}~\bibnamefont {Nohara}}, \bibinfo {author} {\bibfnamefont
  {Y.}~\bibnamefont {Nomura}}, \bibinfo {author} {\bibfnamefont
  {L.}~\bibnamefont {Paulatto}}, \bibinfo {author} {\bibfnamefont
  {S.}~\bibnamefont {Ponc{\'e}}}, \bibinfo {author} {\bibfnamefont
  {T.}~\bibnamefont {Ponweiser}}, \bibinfo {author} {\bibfnamefont
  {J.}~\bibnamefont {Qiao}}, \bibinfo {author} {\bibfnamefont {F.}~\bibnamefont
  {Th{\"o}le}}, \bibinfo {author} {\bibfnamefont {S.~S.}\ \bibnamefont
  {Tsirkin}}, \bibinfo {author} {\bibfnamefont {M.}~\bibnamefont
  {Wierzbowska}}, \bibinfo {author} {\bibfnamefont {N.}~\bibnamefont
  {Marzari}}, \bibinfo {author} {\bibfnamefont {D.}~\bibnamefont {Vanderbilt}},
  \bibinfo {author} {\bibfnamefont {I.}~\bibnamefont {Souza}}, \bibinfo
  {author} {\bibfnamefont {A.~A.}\ \bibnamefont {Mostofi}},\ and\ \bibinfo
  {author} {\bibfnamefont {J.~R.}\ \bibnamefont {Yates}},\ }\bibfield  {title}
  {\bibinfo {title} {Wannier90 as a community code: New features and
  applications},\ }\href
  {https://iopscience.iop.org/article/10.1088/1361-648X/ab51ff} {\bibfield
  {journal} {\bibinfo  {journal} {J. Phys.: Condens. Matter}\ }\textbf
  {\bibinfo {volume} {32}},\ \bibinfo {pages} {165902} (\bibinfo {year}
  {2020})}\BibitemShut {NoStop}%
\bibitem [{\citenamefont {Giustino}\ \emph {et~al.}(2007)\citenamefont
  {Giustino}, \citenamefont {Cohen},\ and\ \citenamefont
  {Louie}}]{giustino:2007}%
  \BibitemOpen
  \bibfield  {author} {\bibinfo {author} {\bibfnamefont {F.}~\bibnamefont
  {Giustino}}, \bibinfo {author} {\bibfnamefont {M.~L.}\ \bibnamefont
  {Cohen}},\ and\ \bibinfo {author} {\bibfnamefont {S.~G.}\ \bibnamefont
  {Louie}},\ }\bibfield  {title} {\bibinfo {title} {Electron-phonon interaction
  using {{Wannier}} functions},\ }\href
  {https://link.aps.org/doi/10.1103/PhysRevB.76.165108} {\bibfield  {journal}
  {\bibinfo  {journal} {Phys. Rev. B}\ }\textbf {\bibinfo {volume} {76}},\
  \bibinfo {pages} {165108} (\bibinfo {year} {2007})}\BibitemShut {NoStop}%
\bibitem [{\citenamefont {Lee}\ \emph {et~al.}(2023)\citenamefont {Lee},
  \citenamefont {Ponc{\'e}}, \citenamefont {Bushick}, \citenamefont
  {Hajinazar}, \citenamefont {{Lafuente-Bartolome}}, \citenamefont {Leveillee},
  \citenamefont {Lian}, \citenamefont {Lihm}, \citenamefont {Macheda},
  \citenamefont {Mori}, \citenamefont {Paudyal}, \citenamefont {Sio},
  \citenamefont {Tiwari}, \citenamefont {Zacharias}, \citenamefont {Zhang},
  \citenamefont {Bonini}, \citenamefont {Kioupakis}, \citenamefont {Margine},\
  and\ \citenamefont {Giustino}}]{lee:2023}%
  \BibitemOpen
  \bibfield  {author} {\bibinfo {author} {\bibfnamefont {H.}~\bibnamefont
  {Lee}}, \bibinfo {author} {\bibfnamefont {S.}~\bibnamefont {Ponc{\'e}}},
  \bibinfo {author} {\bibfnamefont {K.}~\bibnamefont {Bushick}}, \bibinfo
  {author} {\bibfnamefont {S.}~\bibnamefont {Hajinazar}}, \bibinfo {author}
  {\bibfnamefont {J.}~\bibnamefont {{Lafuente-Bartolome}}}, \bibinfo {author}
  {\bibfnamefont {J.}~\bibnamefont {Leveillee}}, \bibinfo {author}
  {\bibfnamefont {C.}~\bibnamefont {Lian}}, \bibinfo {author} {\bibfnamefont
  {J.-M.}\ \bibnamefont {Lihm}}, \bibinfo {author} {\bibfnamefont
  {F.}~\bibnamefont {Macheda}}, \bibinfo {author} {\bibfnamefont
  {H.}~\bibnamefont {Mori}}, \bibinfo {author} {\bibfnamefont {H.}~\bibnamefont
  {Paudyal}}, \bibinfo {author} {\bibfnamefont {W.~H.}\ \bibnamefont {Sio}},
  \bibinfo {author} {\bibfnamefont {S.}~\bibnamefont {Tiwari}}, \bibinfo
  {author} {\bibfnamefont {M.}~\bibnamefont {Zacharias}}, \bibinfo {author}
  {\bibfnamefont {X.}~\bibnamefont {Zhang}}, \bibinfo {author} {\bibfnamefont
  {N.}~\bibnamefont {Bonini}}, \bibinfo {author} {\bibfnamefont
  {E.}~\bibnamefont {Kioupakis}}, \bibinfo {author} {\bibfnamefont {E.~R.}\
  \bibnamefont {Margine}},\ and\ \bibinfo {author} {\bibfnamefont
  {F.}~\bibnamefont {Giustino}},\ }\bibfield  {title} {\bibinfo {title}
  {Electron--phonon physics from first principles using the {{EPW}} code},\
  }\href {https://www.nature.com/articles/s41524-023-01107-3} {\bibfield
  {journal} {\bibinfo  {journal} {npj Comput. Mater.}\ }\textbf {\bibinfo
  {volume} {9}},\ \bibinfo {pages} {156} (\bibinfo {year} {2023})}\BibitemShut
  {NoStop}%
\bibitem [{\citenamefont {Gonze}\ \emph {et~al.}(2016)\citenamefont {Gonze},
  \citenamefont {Jollet}, \citenamefont {Abreu~Araujo}, \citenamefont {Adams},
  \citenamefont {Amadon}, \citenamefont {Applencourt}, \citenamefont {Audouze},
  \citenamefont {Beuken}, \citenamefont {Bieder}, \citenamefont {Bokhanchuk},
  \citenamefont {Bousquet}, \citenamefont {Bruneval}, \citenamefont {Caliste},
  \citenamefont {C{\^o}t{\'e}}, \citenamefont {Dahm}, \citenamefont {Da~Pieve},
  \citenamefont {Delaveau}, \citenamefont {Di~Gennaro}, \citenamefont {Dorado},
  \citenamefont {Espejo}, \citenamefont {Geneste}, \citenamefont {Genovese},
  \citenamefont {Gerossier}, \citenamefont {Giantomassi}, \citenamefont
  {Gillet}, \citenamefont {Hamann}, \citenamefont {He}, \citenamefont {Jomard},
  \citenamefont {Laflamme~Janssen}, \citenamefont {Le~Roux}, \citenamefont
  {Levitt}, \citenamefont {Lherbier}, \citenamefont {Liu}, \citenamefont
  {Luka{\v c}evi{\'c}}, \citenamefont {Martin}, \citenamefont {Martins},
  \citenamefont {Oliveira}, \citenamefont {Ponc{\'e}}, \citenamefont
  {Pouillon}, \citenamefont {Rangel}, \citenamefont {Rignanese}, \citenamefont
  {Romero}, \citenamefont {Rousseau}, \citenamefont {Rubel}, \citenamefont
  {Shukri}, \citenamefont {Stankovski}, \citenamefont {Torrent}, \citenamefont
  {Van~Setten}, \citenamefont {Van~Troeye}, \citenamefont {Verstraete},
  \citenamefont {Waroquiers}, \citenamefont {Wiktor}, \citenamefont {Xu},
  \citenamefont {Zhou},\ and\ \citenamefont {Zwanziger}}]{gonze:2016}%
  \BibitemOpen
  \bibfield  {author} {\bibinfo {author} {\bibfnamefont {X.}~\bibnamefont
  {Gonze}}, \bibinfo {author} {\bibfnamefont {F.}~\bibnamefont {Jollet}},
  \bibinfo {author} {\bibfnamefont {F.}~\bibnamefont {Abreu~Araujo}}, \bibinfo
  {author} {\bibfnamefont {D.}~\bibnamefont {Adams}}, \bibinfo {author}
  {\bibfnamefont {B.}~\bibnamefont {Amadon}}, \bibinfo {author} {\bibfnamefont
  {T.}~\bibnamefont {Applencourt}}, \bibinfo {author} {\bibfnamefont
  {C.}~\bibnamefont {Audouze}}, \bibinfo {author} {\bibfnamefont {J.-M.}\
  \bibnamefont {Beuken}}, \bibinfo {author} {\bibfnamefont {J.}~\bibnamefont
  {Bieder}}, \bibinfo {author} {\bibfnamefont {A.}~\bibnamefont {Bokhanchuk}},
  \bibinfo {author} {\bibfnamefont {E.}~\bibnamefont {Bousquet}}, \bibinfo
  {author} {\bibfnamefont {F.}~\bibnamefont {Bruneval}}, \bibinfo {author}
  {\bibfnamefont {D.}~\bibnamefont {Caliste}}, \bibinfo {author} {\bibfnamefont
  {M.}~\bibnamefont {C{\^o}t{\'e}}}, \bibinfo {author} {\bibfnamefont
  {F.}~\bibnamefont {Dahm}}, \bibinfo {author} {\bibfnamefont {F.}~\bibnamefont
  {Da~Pieve}}, \bibinfo {author} {\bibfnamefont {M.}~\bibnamefont {Delaveau}},
  \bibinfo {author} {\bibfnamefont {M.}~\bibnamefont {Di~Gennaro}}, \bibinfo
  {author} {\bibfnamefont {B.}~\bibnamefont {Dorado}}, \bibinfo {author}
  {\bibfnamefont {C.}~\bibnamefont {Espejo}}, \bibinfo {author} {\bibfnamefont
  {G.}~\bibnamefont {Geneste}}, \bibinfo {author} {\bibfnamefont
  {L.}~\bibnamefont {Genovese}}, \bibinfo {author} {\bibfnamefont
  {A.}~\bibnamefont {Gerossier}}, \bibinfo {author} {\bibfnamefont
  {M.}~\bibnamefont {Giantomassi}}, \bibinfo {author} {\bibfnamefont
  {Y.}~\bibnamefont {Gillet}}, \bibinfo {author} {\bibfnamefont
  {D.}~\bibnamefont {Hamann}}, \bibinfo {author} {\bibfnamefont
  {L.}~\bibnamefont {He}}, \bibinfo {author} {\bibfnamefont {G.}~\bibnamefont
  {Jomard}}, \bibinfo {author} {\bibfnamefont {J.}~\bibnamefont
  {Laflamme~Janssen}}, \bibinfo {author} {\bibfnamefont {S.}~\bibnamefont
  {Le~Roux}}, \bibinfo {author} {\bibfnamefont {A.}~\bibnamefont {Levitt}},
  \bibinfo {author} {\bibfnamefont {A.}~\bibnamefont {Lherbier}}, \bibinfo
  {author} {\bibfnamefont {F.}~\bibnamefont {Liu}}, \bibinfo {author}
  {\bibfnamefont {I.}~\bibnamefont {Luka{\v c}evi{\'c}}}, \bibinfo {author}
  {\bibfnamefont {A.}~\bibnamefont {Martin}}, \bibinfo {author} {\bibfnamefont
  {C.}~\bibnamefont {Martins}}, \bibinfo {author} {\bibfnamefont
  {M.}~\bibnamefont {Oliveira}}, \bibinfo {author} {\bibfnamefont
  {S.}~\bibnamefont {Ponc{\'e}}}, \bibinfo {author} {\bibfnamefont
  {Y.}~\bibnamefont {Pouillon}}, \bibinfo {author} {\bibfnamefont
  {T.}~\bibnamefont {Rangel}}, \bibinfo {author} {\bibfnamefont {G.-M.}\
  \bibnamefont {Rignanese}}, \bibinfo {author} {\bibfnamefont {A.}~\bibnamefont
  {Romero}}, \bibinfo {author} {\bibfnamefont {B.}~\bibnamefont {Rousseau}},
  \bibinfo {author} {\bibfnamefont {O.}~\bibnamefont {Rubel}}, \bibinfo
  {author} {\bibfnamefont {A.}~\bibnamefont {Shukri}}, \bibinfo {author}
  {\bibfnamefont {M.}~\bibnamefont {Stankovski}}, \bibinfo {author}
  {\bibfnamefont {M.}~\bibnamefont {Torrent}}, \bibinfo {author} {\bibfnamefont
  {M.}~\bibnamefont {Van~Setten}}, \bibinfo {author} {\bibfnamefont
  {B.}~\bibnamefont {Van~Troeye}}, \bibinfo {author} {\bibfnamefont
  {M.}~\bibnamefont {Verstraete}}, \bibinfo {author} {\bibfnamefont
  {D.}~\bibnamefont {Waroquiers}}, \bibinfo {author} {\bibfnamefont
  {J.}~\bibnamefont {Wiktor}}, \bibinfo {author} {\bibfnamefont
  {B.}~\bibnamefont {Xu}}, \bibinfo {author} {\bibfnamefont {A.}~\bibnamefont
  {Zhou}},\ and\ \bibinfo {author} {\bibfnamefont {J.}~\bibnamefont
  {Zwanziger}},\ }\bibfield  {title} {\bibinfo {title} {Recent developments in
  the {{ABINIT}} software package},\ }\href
  {https://linkinghub.elsevier.com/retrieve/pii/S0010465516300923} {\bibfield
  {journal} {\bibinfo  {journal} {Comput. Phys. Commun.}\ }\textbf {\bibinfo
  {volume} {205}},\ \bibinfo {pages} {106} (\bibinfo {year}
  {2016})}\BibitemShut {NoStop}%
\bibitem [{\citenamefont {Cepellotti}\ \emph {et~al.}(2022)\citenamefont
  {Cepellotti}, \citenamefont {Coulter}, \citenamefont {Johansson},
  \citenamefont {Fedorova},\ and\ \citenamefont {Kozinsky}}]{cepellotti:2022}%
  \BibitemOpen
  \bibfield  {author} {\bibinfo {author} {\bibfnamefont {A.}~\bibnamefont
  {Cepellotti}}, \bibinfo {author} {\bibfnamefont {J.}~\bibnamefont {Coulter}},
  \bibinfo {author} {\bibfnamefont {A.}~\bibnamefont {Johansson}}, \bibinfo
  {author} {\bibfnamefont {N.~S.}\ \bibnamefont {Fedorova}},\ and\ \bibinfo
  {author} {\bibfnamefont {B.}~\bibnamefont {Kozinsky}},\ }\bibfield  {title}
  {\bibinfo {title} {Phoebe: A high-performance framework for solving phonon
  and electron {{Boltzmann}} transport equations},\ }\href
  {https://iopscience.iop.org/article/10.1088/2515-7639/ac86f6} {\bibfield
  {journal} {\bibinfo  {journal} {J. Phys. Mater.}\ }\textbf {\bibinfo {volume}
  {5}},\ \bibinfo {pages} {035003} (\bibinfo {year} {2022})}\BibitemShut
  {NoStop}%
\bibitem [{\citenamefont {Zhou}\ \emph {et~al.}(2021)\citenamefont {Zhou},
  \citenamefont {Park}, \citenamefont {Lu}, \citenamefont {Maliyov},
  \citenamefont {Tong},\ and\ \citenamefont {Bernardi}}]{zhou:2021a}%
  \BibitemOpen
  \bibfield  {author} {\bibinfo {author} {\bibfnamefont {J.-J.}\ \bibnamefont
  {Zhou}}, \bibinfo {author} {\bibfnamefont {J.}~\bibnamefont {Park}}, \bibinfo
  {author} {\bibfnamefont {I.-T.}\ \bibnamefont {Lu}}, \bibinfo {author}
  {\bibfnamefont {I.}~\bibnamefont {Maliyov}}, \bibinfo {author} {\bibfnamefont
  {X.}~\bibnamefont {Tong}},\ and\ \bibinfo {author} {\bibfnamefont
  {M.}~\bibnamefont {Bernardi}},\ }\bibfield  {title} {\bibinfo {title}
  {\textsc{Perturbo}: A software package for ab initio electron–phonon
  interactions, charge transport and ultrafast dynamics},\ }\href
  {https://linkinghub.elsevier.com/retrieve/pii/S0010465521000837} {\bibfield
  {journal} {\bibinfo  {journal} {Comput. Phys. Commun.}\ }\textbf {\bibinfo
  {volume} {264}},\ \bibinfo {pages} {107970} (\bibinfo {year}
  {2021})}\BibitemShut {NoStop}%
\bibitem [{\citenamefont {Protik}\ \emph {et~al.}(2022)\citenamefont {Protik},
  \citenamefont {Li}, \citenamefont {Pruneda}, \citenamefont {Broido},\ and\
  \citenamefont {Ordejón}}]{protik:2022}%
  \BibitemOpen
  \bibfield  {author} {\bibinfo {author} {\bibfnamefont {N.~H.}\ \bibnamefont
  {Protik}}, \bibinfo {author} {\bibfnamefont {C.}~\bibnamefont {Li}}, \bibinfo
  {author} {\bibfnamefont {M.}~\bibnamefont {Pruneda}}, \bibinfo {author}
  {\bibfnamefont {D.}~\bibnamefont {Broido}},\ and\ \bibinfo {author}
  {\bibfnamefont {P.}~\bibnamefont {Ordejón}},\ }\bibfield  {title} {\bibinfo
  {title} {The \texttt{elphbolt} ab initio solver for the coupled
  electron-phonon boltzmann transport equations},\ }\href
  {https://www.nature.com/articles/s41524-022-00710-0} {\bibfield  {journal}
  {\bibinfo  {journal} {npj Comput. Mater.}\ }\textbf {\bibinfo {volume} {8}},\
  \bibinfo {pages} {28} (\bibinfo {year} {2022})}\BibitemShut {NoStop}%
\bibitem [{\citenamefont {Marini}\ \emph {et~al.}(2024)\citenamefont {Marini},
  \citenamefont {Marchese}, \citenamefont {Profeta}, \citenamefont {Sjakste},
  \citenamefont {Macheda}, \citenamefont {Vast}, \citenamefont {Mauri},\ and\
  \citenamefont {Calandra}}]{marini:2024}%
  \BibitemOpen
  \bibfield  {author} {\bibinfo {author} {\bibfnamefont {G.}~\bibnamefont
  {Marini}}, \bibinfo {author} {\bibfnamefont {G.}~\bibnamefont {Marchese}},
  \bibinfo {author} {\bibfnamefont {G.}~\bibnamefont {Profeta}}, \bibinfo
  {author} {\bibfnamefont {J.}~\bibnamefont {Sjakste}}, \bibinfo {author}
  {\bibfnamefont {F.}~\bibnamefont {Macheda}}, \bibinfo {author} {\bibfnamefont
  {N.}~\bibnamefont {Vast}}, \bibinfo {author} {\bibfnamefont {F.}~\bibnamefont
  {Mauri}},\ and\ \bibinfo {author} {\bibfnamefont {M.}~\bibnamefont
  {Calandra}},\ }\bibfield  {title} {\bibinfo {title} {\textsc{epi}$q$: An
  open-source software for the calculation of electron-phonon interaction
  related properties},\ }\href
  {https://linkinghub.elsevier.com/retrieve/pii/S0010465523002953} {\bibfield
  {journal} {\bibinfo  {journal} {Comput. Phys. Commun.}\ }\textbf {\bibinfo
  {volume} {295}},\ \bibinfo {pages} {108950} (\bibinfo {year}
  {2024})}\BibitemShut {NoStop}%
\bibitem [{\citenamefont {Li}\ \emph {et~al.}(2019)\citenamefont {Li},
  \citenamefont {Antonius}, \citenamefont {Wu}, \citenamefont {{da Jornada}},\
  and\ \citenamefont {Louie}}]{li:2019a}%
  \BibitemOpen
  \bibfield  {author} {\bibinfo {author} {\bibfnamefont {Z.}~\bibnamefont
  {Li}}, \bibinfo {author} {\bibfnamefont {G.}~\bibnamefont {Antonius}},
  \bibinfo {author} {\bibfnamefont {M.}~\bibnamefont {Wu}}, \bibinfo {author}
  {\bibfnamefont {F.~H.}\ \bibnamefont {{da Jornada}}},\ and\ \bibinfo {author}
  {\bibfnamefont {S.~G.}\ \bibnamefont {Louie}},\ }\bibfield  {title} {\bibinfo
  {title} {Electron-phonon coupling from \textit{Ab Initio} linear-response
  theory within the \textit{GW} method: {C}orrelation-enhanced interactions and
  superconductivity in {Ba$_{1-x}$K$_x$BiO$_3$}},\ }\href
  {https://link.aps.org/doi/10.1103/PhysRevLett.122.186402} {\bibfield
  {journal} {\bibinfo  {journal} {Phys. Rev. Lett.}\ }\textbf {\bibinfo
  {volume} {122}},\ \bibinfo {pages} {186402} (\bibinfo {year}
  {2019})}\BibitemShut {NoStop}%
\bibitem [{\citenamefont {Abramovitch}\ \emph {et~al.}(2023)\citenamefont
  {Abramovitch}, \citenamefont {Zhou}, \citenamefont {Mravlje}, \citenamefont
  {Georges},\ and\ \citenamefont {Bernardi}}]{abramovitch:2023}%
  \BibitemOpen
  \bibfield  {author} {\bibinfo {author} {\bibfnamefont {D.~J.}\ \bibnamefont
  {Abramovitch}}, \bibinfo {author} {\bibfnamefont {J.-J.}\ \bibnamefont
  {Zhou}}, \bibinfo {author} {\bibfnamefont {J.}~\bibnamefont {Mravlje}},
  \bibinfo {author} {\bibfnamefont {A.}~\bibnamefont {Georges}},\ and\ \bibinfo
  {author} {\bibfnamefont {M.}~\bibnamefont {Bernardi}},\ }\bibfield  {title}
  {\bibinfo {title} {Combining electron-phonon and dynamical mean-field theory
  calculations of correlated materials: {T}ransport in the correlated metal
  {Sr$_2$RuO$_4$}},\ }\href
  {https://link.aps.org/doi/10.1103/PhysRevMaterials.7.093801} {\bibfield
  {journal} {\bibinfo  {journal} {Phys. Rev. Mater.}\ }\textbf {\bibinfo
  {volume} {7}},\ \bibinfo {pages} {093801} (\bibinfo {year}
  {2023})}\BibitemShut {NoStop}%
\bibitem [{\citenamefont {Matthiessen}\ and\ \citenamefont
  {Vogt}(1864)}]{matthiessen:1864}%
  \BibitemOpen
  \bibfield  {author} {\bibinfo {author} {\bibfnamefont {A.}~\bibnamefont
  {Matthiessen}}\ and\ \bibinfo {author} {\bibfnamefont {A.~C.}\ \bibnamefont
  {Vogt}},\ }\bibfield  {title} {\bibinfo {title} {{{IV}}. {{On}} the influence
  of temperature on the electric conducting-power of alloys},\ }\href
  {https://royalsocietypublishing.org/doi/10.1098/rstl.1864.0004} {\bibfield
  {journal} {\bibinfo  {journal} {Philos. Trans. R. Soc. Lond.}\ }\textbf
  {\bibinfo {volume} {154}},\ \bibinfo {pages} {167} (\bibinfo {year}
  {1864})}\BibitemShut {NoStop}%
\bibitem [{\citenamefont {Kemper}\ \emph {et~al.}(2018)\citenamefont {Kemper},
  \citenamefont {Abdurazakov},\ and\ \citenamefont {Freericks}}]{kemper:2018}%
  \BibitemOpen
  \bibfield  {author} {\bibinfo {author} {\bibfnamefont {A.~F.}\ \bibnamefont
  {Kemper}}, \bibinfo {author} {\bibfnamefont {O.}~\bibnamefont
  {Abdurazakov}},\ and\ \bibinfo {author} {\bibfnamefont {J.~K.}\ \bibnamefont
  {Freericks}},\ }\bibfield  {title} {\bibinfo {title} {General principles for
  the nonequilibrium relaxation of populations in quantum materials},\ }\href
  {https://link.aps.org/doi/10.1103/PhysRevX.8.041009} {\bibfield  {journal}
  {\bibinfo  {journal} {Phys. Rev. X}\ }\textbf {\bibinfo {volume} {8}},\
  \bibinfo {pages} {041009} (\bibinfo {year} {2018})}\BibitemShut {NoStop}%
\bibitem [{\citenamefont {Kordyuk}\ \emph {et~al.}(2005)\citenamefont
  {Kordyuk}, \citenamefont {Borisenko}, \citenamefont {Koitzsch}, \citenamefont
  {Fink}, \citenamefont {Knupfer},\ and\ \citenamefont
  {Berger}}]{kordyuk:2005}%
  \BibitemOpen
  \bibfield  {author} {\bibinfo {author} {\bibfnamefont {A.~A.}\ \bibnamefont
  {Kordyuk}}, \bibinfo {author} {\bibfnamefont {S.~V.}\ \bibnamefont
  {Borisenko}}, \bibinfo {author} {\bibfnamefont {A.}~\bibnamefont {Koitzsch}},
  \bibinfo {author} {\bibfnamefont {J.}~\bibnamefont {Fink}}, \bibinfo {author}
  {\bibfnamefont {M.}~\bibnamefont {Knupfer}},\ and\ \bibinfo {author}
  {\bibfnamefont {H.}~\bibnamefont {Berger}},\ }\bibfield  {title} {\bibinfo
  {title} {Bare electron dispersion from experiment: {{Self-consistent}}
  self-energy analysis of photoemission data},\ }\href
  {https://link.aps.org/doi/10.1103/PhysRevB.71.214513} {\bibfield  {journal}
  {\bibinfo  {journal} {Phys. Rev. B}\ }\textbf {\bibinfo {volume} {71}},\
  \bibinfo {pages} {214513} (\bibinfo {year} {2005})}\BibitemShut {NoStop}%
\bibitem [{\citenamefont {Pletikosi{\'c}}\ \emph {et~al.}(2012)\citenamefont
  {Pletikosi{\'c}}, \citenamefont {Kralj}, \citenamefont {Milun},\ and\
  \citenamefont {Pervan}}]{pletikosic:2012}%
  \BibitemOpen
  \bibfield  {author} {\bibinfo {author} {\bibfnamefont {I.}~\bibnamefont
  {Pletikosi{\'c}}}, \bibinfo {author} {\bibfnamefont {M.}~\bibnamefont
  {Kralj}}, \bibinfo {author} {\bibfnamefont {M.}~\bibnamefont {Milun}},\ and\
  \bibinfo {author} {\bibfnamefont {P.}~\bibnamefont {Pervan}},\ }\bibfield
  {title} {\bibinfo {title} {Finding the bare band: {{Electron}} coupling to
  two phonon modes in potassium-doped graphene on {{Ir}}(111)},\ }\href
  {https://link.aps.org/doi/10.1103/PhysRevB.85.155447} {\bibfield  {journal}
  {\bibinfo  {journal} {Phys. Rev. B}\ }\textbf {\bibinfo {volume} {85}},\
  \bibinfo {pages} {155447} (\bibinfo {year} {2012})}\BibitemShut {NoStop}%
\bibitem [{\citenamefont {Stansbury}\ and\ \citenamefont
  {Lanzara}(2020)}]{stansbury:2020}%
  \BibitemOpen
  \bibfield  {author} {\bibinfo {author} {\bibfnamefont {C.}~\bibnamefont
  {Stansbury}}\ and\ \bibinfo {author} {\bibfnamefont {A.}~\bibnamefont
  {Lanzara}},\ }\bibfield  {title} {\bibinfo {title} {{{PyARPES}}: {{An}}
  analysis framework for multimodal angle-resolved photoemission
  spectroscopies},\ }\href
  {https://linkinghub.elsevier.com/retrieve/pii/S2352711019301633} {\bibfield
  {journal} {\bibinfo  {journal} {SoftwareX}\ }\textbf {\bibinfo {volume}
  {11}},\ \bibinfo {pages} {100472} (\bibinfo {year} {2020})}\BibitemShut
  {NoStop}%
\bibitem [{\citenamefont {Polley}\ \emph {et~al.}(2024)\citenamefont {Polley},
  \citenamefont {Leandersson}, \citenamefont {Adell}, \citenamefont {Osiecki},
  \citenamefont {Carbone}, \citenamefont {Ali}, \citenamefont {Fedderwitz},\
  and\ \citenamefont {Balasubramanian}}]{polley:2024}%
  \BibitemOpen
  \bibfield  {author} {\bibinfo {author} {\bibfnamefont {C.~M.}\ \bibnamefont
  {Polley}}, \bibinfo {author} {\bibfnamefont {M.}~\bibnamefont {Leandersson}},
  \bibinfo {author} {\bibfnamefont {J.}~\bibnamefont {Adell}}, \bibinfo
  {author} {\bibfnamefont {J.}~\bibnamefont {Osiecki}}, \bibinfo {author}
  {\bibfnamefont {D.}~\bibnamefont {Carbone}}, \bibinfo {author} {\bibfnamefont
  {K.}~\bibnamefont {Ali}}, \bibinfo {author} {\bibfnamefont {H.}~\bibnamefont
  {Fedderwitz}},\ and\ \bibinfo {author} {\bibfnamefont {T.}~\bibnamefont
  {Balasubramanian}},\ }\bibfield  {title} {\bibinfo {title} {The {Bloch}
  beamline at {MAX IV}: Micro-spot {ARPES} from a conventional, full-featured
  beamline},\ }\href
  {https://www.tandfonline.com/doi/full/10.1080/08940886.2024.2391252}
  {\bibfield  {journal} {\bibinfo  {journal} {Synchrotron Radiat. News}\
  }\textbf {\bibinfo {volume} {37}},\ \bibinfo {pages} {18} (\bibinfo {year}
  {2024})}\BibitemShut {NoStop}%
\bibitem [{\citenamefont {Bisti}(2025)}]{navarp:website}%
  \BibitemOpen
  \bibfield  {author} {\bibinfo {author} {\bibfnamefont {F.}~\bibnamefont
  {Bisti}},\ }\href {https://fbisti.gitlab.io/navarp/index.html} {\bibinfo
  {title} {{{NavARP}}}} (\bibinfo {year} {Last accessed: 5 July
  2025})\BibitemShut {NoStop}%
\bibitem [{\citenamefont {Han}(2025)}]{erlabpy:website}%
  \BibitemOpen
  \bibfield  {author} {\bibinfo {author} {\bibfnamefont {K.}~\bibnamefont
  {Han}},\ }\href {https://erlabpy.readthedocs.io} {\bibinfo {title}
  {{{ERLabPy}}}} (\bibinfo {year} {Last accessed: 21 May 2025})\BibitemShut
  {NoStop}%
\bibitem [{\citenamefont {Bryan}(1990)}]{bryan:1990}%
  \BibitemOpen
  \bibfield  {author} {\bibinfo {author} {\bibfnamefont {R.~K.}\ \bibnamefont
  {Bryan}},\ }\bibfield  {title} {\bibinfo {title} {Maximum entropy analysis of
  oversampled data problems},\ }\href
  {http://link.springer.com/10.1007/BF02427376} {\bibfield  {journal} {\bibinfo
   {journal} {Eur. Biophys. J.}\ }\textbf {\bibinfo {volume} {18}},\ \bibinfo
  {pages} {165} (\bibinfo {year} {1990})}\BibitemShut {NoStop}%
\bibitem [{\citenamefont {Fero}\ \emph {et~al.}(2014)\citenamefont {Fero},
  \citenamefont {Smallwood}, \citenamefont {Affeldt},\ and\ \citenamefont
  {Lanzara}}]{fero:2014}%
  \BibitemOpen
  \bibfield  {author} {\bibinfo {author} {\bibfnamefont {A.}~\bibnamefont
  {Fero}}, \bibinfo {author} {\bibfnamefont {C.}~\bibnamefont {Smallwood}},
  \bibinfo {author} {\bibfnamefont {G.}~\bibnamefont {Affeldt}},\ and\ \bibinfo
  {author} {\bibfnamefont {A.}~\bibnamefont {Lanzara}},\ }\bibfield  {title}
  {\bibinfo {title} {Impact of work function induced electric fields on
  laser-based angle-resolved photoemission spectroscopy},\ }\href
  {https://linkinghub.elsevier.com/retrieve/pii/S0368204814000267} {\bibfield
  {journal} {\bibinfo  {journal} {J. Electron. Spectrosc. Relat. Phenom.}\
  }\textbf {\bibinfo {volume} {195}},\ \bibinfo {pages} {237} (\bibinfo {year}
  {2014})}\BibitemShut {NoStop}%
\bibitem [{\citenamefont {Ishida}\ and\ \citenamefont
  {Shin}(2018)}]{ishida:2018}%
  \BibitemOpen
  \bibfield  {author} {\bibinfo {author} {\bibfnamefont {Y.}~\bibnamefont
  {Ishida}}\ and\ \bibinfo {author} {\bibfnamefont {S.}~\bibnamefont {Shin}},\
  }\bibfield  {title} {\bibinfo {title} {Functions to map photoelectron
  distributions in a variety of setups in angle-resolved photoemission
  spectroscopy},\ }\href
  {https://pubs.aip.org/rsi/article/89/4/043903/362369/Functions-to-map-photoelectron-distributions-in-a}
  {\bibfield  {journal} {\bibinfo  {journal} {Rev. Sci. Instrum.}\ }\textbf
  {\bibinfo {volume} {89}},\ \bibinfo {pages} {043903} (\bibinfo {year}
  {2018})}\BibitemShut {NoStop}%
\bibitem [{\citenamefont {Berglund}\ and\ \citenamefont
  {Spicer}(1964)}]{berglund:1964}%
  \BibitemOpen
  \bibfield  {author} {\bibinfo {author} {\bibfnamefont {C.~N.}\ \bibnamefont
  {Berglund}}\ and\ \bibinfo {author} {\bibfnamefont {W.~E.}\ \bibnamefont
  {Spicer}},\ }\bibfield  {title} {\bibinfo {title} {Photoemission studies of
  copper and silver: {T}heory},\ }\href
  {https://link.aps.org/doi/10.1103/PhysRev.136.A1030} {\bibfield  {journal}
  {\bibinfo  {journal} {Phys. Rev.}\ }\textbf {\bibinfo {volume} {136}},\
  \bibinfo {pages} {A1030} (\bibinfo {year} {1964})}\BibitemShut {NoStop}%
\bibitem [{\citenamefont {Schaich}\ and\ \citenamefont
  {Ashcroft}(1971)}]{schaich:1971}%
  \BibitemOpen
  \bibfield  {author} {\bibinfo {author} {\bibfnamefont {W.~L.}\ \bibnamefont
  {Schaich}}\ and\ \bibinfo {author} {\bibfnamefont {N.~W.}\ \bibnamefont
  {Ashcroft}},\ }\bibfield  {title} {\bibinfo {title} {Model calculations in
  the theory of photoemission},\ }\href
  {https://link.aps.org/doi/10.1103/PhysRevB.3.2452} {\bibfield  {journal}
  {\bibinfo  {journal} {Phys. Rev. B}\ }\textbf {\bibinfo {volume} {3}},\
  \bibinfo {pages} {2452} (\bibinfo {year} {1971})}\BibitemShut {NoStop}%
\bibitem [{\citenamefont {Mahan}(1970)}]{mahan:1970}%
  \BibitemOpen
  \bibfield  {author} {\bibinfo {author} {\bibfnamefont {G.~D.}\ \bibnamefont
  {Mahan}},\ }\bibfield  {title} {\bibinfo {title} {Theory of photoemission in
  simple metals},\ }\href {https://link.aps.org/doi/10.1103/PhysRevB.2.4334}
  {\bibfield  {journal} {\bibinfo  {journal} {Phys. Rev. B}\ }\textbf {\bibinfo
  {volume} {2}},\ \bibinfo {pages} {4334} (\bibinfo {year} {1970})}\BibitemShut
  {NoStop}%
\bibitem [{\citenamefont {Keiter}(1978)}]{keiter:1978}%
  \BibitemOpen
  \bibfield  {author} {\bibinfo {author} {\bibfnamefont {H.}~\bibnamefont
  {Keiter}},\ }\bibfield  {title} {\bibinfo {title} {Comment on many-body
  aspects of external photoemission},\ }\href
  {http://link.springer.com/10.1007/BF01320982} {\bibfield  {journal} {\bibinfo
   {journal} {Z. Phys. B.}\ }\textbf {\bibinfo {volume} {30}},\ \bibinfo
  {pages} {167} (\bibinfo {year} {1978})}\BibitemShut {NoStop}%
\bibitem [{\citenamefont {Almbladh}(2006)}]{almbladh:2006}%
  \BibitemOpen
  \bibfield  {author} {\bibinfo {author} {\bibfnamefont {C.-O.}\ \bibnamefont
  {Almbladh}},\ }\bibfield  {title} {\bibinfo {title} {Photoemission beyond the
  sudden approximation},\ }\href
  {https://iopscience.iop.org/article/10.1088/1742-6596/35/1/011} {\bibfield
  {journal} {\bibinfo  {journal} {J. Phys.: Conf. Ser.}\ }\textbf {\bibinfo
  {volume} {35}},\ \bibinfo {pages} {127} (\bibinfo {year} {2006})}\BibitemShut
  {NoStop}%
\bibitem [{\citenamefont {Hermeking}\ and\ \citenamefont
  {Wehrum}(1975)}]{hermeking:1975}%
  \BibitemOpen
  \bibfield  {author} {\bibinfo {author} {\bibfnamefont {H.}~\bibnamefont
  {Hermeking}}\ and\ \bibinfo {author} {\bibfnamefont {R.~P.}\ \bibnamefont
  {Wehrum}},\ }\bibfield  {title} {\bibinfo {title} {On the derivation of a
  golden rule formula for photoemission in the quadratic response formalism},\
  }\href {https://iopscience.iop.org/article/10.1088/0022-3719/8/20/024}
  {\bibfield  {journal} {\bibinfo  {journal} {J. Phys. C: Solid State Phys.}\
  }\textbf {\bibinfo {volume} {8}},\ \bibinfo {pages} {3468} (\bibinfo {year}
  {1975})}\BibitemShut {NoStop}%
\bibitem [{\citenamefont {H{\"u}fner}(1995)}]{hufner:1995}%
  \BibitemOpen
  \bibfield  {author} {\bibinfo {author} {\bibfnamefont {S.}~\bibnamefont
  {H{\"u}fner}},\ }\href {http://link.springer.com/10.1007/978-3-662-03150-6}
  {\emph {\bibinfo {title} {Photoelectron {{Spectroscopy}}}}},\ \bibinfo
  {series} {Springer {{Series}} in {{Solid-State Sciences}}}, Vol.~\bibinfo
  {volume} {82}\ (\bibinfo  {publisher} {Springer},\ \bibinfo {address}
  {Berlin, Heidelberg},\ \bibinfo {year} {1995})\BibitemShut {NoStop}%
\bibitem [{\citenamefont {Smith}\ \emph {et~al.}(1993)\citenamefont {Smith},
  \citenamefont {Thiry},\ and\ \citenamefont {Petroff}}]{smith:1993}%
  \BibitemOpen
  \bibfield  {author} {\bibinfo {author} {\bibfnamefont {N.~V.}\ \bibnamefont
  {Smith}}, \bibinfo {author} {\bibfnamefont {P.}~\bibnamefont {Thiry}},\ and\
  \bibinfo {author} {\bibfnamefont {Y.}~\bibnamefont {Petroff}},\ }\bibfield
  {title} {\bibinfo {title} {Photoemission linewidths and quasiparticle
  lifetimes},\ }\href {https://link.aps.org/doi/10.1103/PhysRevB.47.15476}
  {\bibfield  {journal} {\bibinfo  {journal} {Phys. Rev. B}\ }\textbf {\bibinfo
  {volume} {47}},\ \bibinfo {pages} {15476} (\bibinfo {year}
  {1993})}\BibitemShut {NoStop}%
\bibitem [{\citenamefont {Strocov}\ \emph {et~al.}(2023)\citenamefont
  {Strocov}, \citenamefont {Lev}, \citenamefont {Alarab}, \citenamefont
  {Constantinou}, \citenamefont {Wang}, \citenamefont {Schmitt}, \citenamefont
  {Stock}, \citenamefont {Nicola{\"i}}, \citenamefont {O{\v c}en{\'a}{\v
  s}ek},\ and\ \citenamefont {Min{\'a}r}}]{strocov:2023}%
  \BibitemOpen
  \bibfield  {author} {\bibinfo {author} {\bibfnamefont {V.~N.}\ \bibnamefont
  {Strocov}}, \bibinfo {author} {\bibfnamefont {L.~L.}\ \bibnamefont {Lev}},
  \bibinfo {author} {\bibfnamefont {F.}~\bibnamefont {Alarab}}, \bibinfo
  {author} {\bibfnamefont {P.}~\bibnamefont {Constantinou}}, \bibinfo {author}
  {\bibfnamefont {X.}~\bibnamefont {Wang}}, \bibinfo {author} {\bibfnamefont
  {T.}~\bibnamefont {Schmitt}}, \bibinfo {author} {\bibfnamefont {T.~J.~Z.}\
  \bibnamefont {Stock}}, \bibinfo {author} {\bibfnamefont {L.}~\bibnamefont
  {Nicola{\"i}}}, \bibinfo {author} {\bibfnamefont {J.}~\bibnamefont {O{\v
  c}en{\'a}{\v s}ek}},\ and\ \bibinfo {author} {\bibfnamefont {J.}~\bibnamefont
  {Min{\'a}r}},\ }\bibfield  {title} {\bibinfo {title} {High-energy
  photoemission final states beyond the free-electron approximation},\ }\href
  {https://www.nature.com/articles/s41467-023-40432-5} {\bibfield  {journal}
  {\bibinfo  {journal} {Nat. Commun.}\ }\textbf {\bibinfo {volume} {14}},\
  \bibinfo {pages} {4827} (\bibinfo {year} {2023})}\BibitemShut {NoStop}%
\bibitem [{\citenamefont {Iwasawa}\ \emph {et~al.}(2017)\citenamefont
  {Iwasawa}, \citenamefont {Schwier}, \citenamefont {Arita}, \citenamefont
  {Ino}, \citenamefont {Namatame}, \citenamefont {Taniguchi}, \citenamefont
  {Aiura},\ and\ \citenamefont {Shimada}}]{iwasawa:2017}%
  \BibitemOpen
  \bibfield  {author} {\bibinfo {author} {\bibfnamefont {H.}~\bibnamefont
  {Iwasawa}}, \bibinfo {author} {\bibfnamefont {E.~F.}\ \bibnamefont
  {Schwier}}, \bibinfo {author} {\bibfnamefont {M.}~\bibnamefont {Arita}},
  \bibinfo {author} {\bibfnamefont {A.}~\bibnamefont {Ino}}, \bibinfo {author}
  {\bibfnamefont {H.}~\bibnamefont {Namatame}}, \bibinfo {author}
  {\bibfnamefont {M.}~\bibnamefont {Taniguchi}}, \bibinfo {author}
  {\bibfnamefont {Y.}~\bibnamefont {Aiura}},\ and\ \bibinfo {author}
  {\bibfnamefont {K.}~\bibnamefont {Shimada}},\ }\bibfield  {title} {\bibinfo
  {title} {Development of laser-based scanning $\mu$-{{ARPES}} system with
  ultimate energy and momentum resolutions},\ }\href
  {https://linkinghub.elsevier.com/retrieve/pii/S0304399117301365} {\bibfield
  {journal} {\bibinfo  {journal} {Ultramicroscopy}\ }\textbf {\bibinfo {volume}
  {182}},\ \bibinfo {pages} {85} (\bibinfo {year} {2017})}\BibitemShut
  {NoStop}%
\bibitem [{\citenamefont {F{\"o}rster}\ \emph {et~al.}(2016)\citenamefont
  {F{\"o}rster}, \citenamefont {Kr{\"u}ger},\ and\ \citenamefont
  {Rohlfing}}]{forster:2016}%
  \BibitemOpen
  \bibfield  {author} {\bibinfo {author} {\bibfnamefont {T.}~\bibnamefont
  {F{\"o}rster}}, \bibinfo {author} {\bibfnamefont {P.}~\bibnamefont
  {Kr{\"u}ger}},\ and\ \bibinfo {author} {\bibfnamefont {M.}~\bibnamefont
  {Rohlfing}},\ }\bibfield  {title} {\bibinfo {title} {{{{\emph{GW}}}}
  calculations for {{Bi}}{\textsubscript{2}}{{Te}}{\textsubscript{3}} and
  {{Sb}}{\textsubscript{2}}{{Te}}{\textsubscript{3}} thin films: {{Electronic}}
  and topological properties},\ }\href
  {https://link.aps.org/doi/10.1103/PhysRevB.93.205442} {\bibfield  {journal}
  {\bibinfo  {journal} {Phys. Rev. B}\ }\textbf {\bibinfo {volume} {93}},\
  \bibinfo {pages} {205442} (\bibinfo {year} {2016})}\BibitemShut {NoStop}%
\bibitem [{\citenamefont {Verga}\ \emph {et~al.}(2003)\citenamefont {Verga},
  \citenamefont {Knigavko},\ and\ \citenamefont {Marsiglio}}]{verga:2003}%
  \BibitemOpen
  \bibfield  {author} {\bibinfo {author} {\bibfnamefont {S.}~\bibnamefont
  {Verga}}, \bibinfo {author} {\bibfnamefont {A.}~\bibnamefont {Knigavko}},\
  and\ \bibinfo {author} {\bibfnamefont {F.}~\bibnamefont {Marsiglio}},\
  }\bibfield  {title} {\bibinfo {title} {Inversion of angle-resolved
  photoemission measurements in high-{{{\emph{T}}}}{\emph{{\textsubscript{c}}}}
  cuprates},\ }\href {https://link.aps.org/doi/10.1103/PhysRevB.67.054503}
  {\bibfield  {journal} {\bibinfo  {journal} {Phys. Rev. B}\ }\textbf {\bibinfo
  {volume} {67}},\ \bibinfo {pages} {054503} (\bibinfo {year}
  {2003})}\BibitemShut {NoStop}%
\bibitem [{\citenamefont {Li}\ \emph {et~al.}(2021)\citenamefont {Li},
  \citenamefont {Wu}, \citenamefont {Chan},\ and\ \citenamefont
  {Louie}}]{li:2021b}%
  \BibitemOpen
  \bibfield  {author} {\bibinfo {author} {\bibfnamefont {Z.}~\bibnamefont
  {Li}}, \bibinfo {author} {\bibfnamefont {M.}~\bibnamefont {Wu}}, \bibinfo
  {author} {\bibfnamefont {Y.-H.}\ \bibnamefont {Chan}},\ and\ \bibinfo
  {author} {\bibfnamefont {S.~G.}\ \bibnamefont {Louie}},\ }\bibfield  {title}
  {\bibinfo {title} {Unmasking the origin of kinks in the photoemission spectra
  of cuprate superconductors},\ }\href
  {https://link.aps.org/doi/10.1103/PhysRevLett.126.146401} {\bibfield
  {journal} {\bibinfo  {journal} {Phys. Rev. Lett.}\ }\textbf {\bibinfo
  {volume} {126}},\ \bibinfo {pages} {146401} (\bibinfo {year}
  {2021})}\BibitemShut {NoStop}%
\bibitem [{\citenamefont {Chubukov}\ and\ \citenamefont
  {Maslov}(2012)}]{chubukov:2012}%
  \BibitemOpen
  \bibfield  {author} {\bibinfo {author} {\bibfnamefont {A.~V.}\ \bibnamefont
  {Chubukov}}\ and\ \bibinfo {author} {\bibfnamefont {D.~L.}\ \bibnamefont
  {Maslov}},\ }\bibfield  {title} {\bibinfo {title}
  {First-{Matsubara}-frequency rule in a {Fermi} liquid. {I}. {F}ermionic
  self-energy},\ }\href {https://link.aps.org/doi/10.1103/PhysRevB.86.155136}
  {\bibfield  {journal} {\bibinfo  {journal} {Phys. Rev. B}\ }\textbf {\bibinfo
  {volume} {86}},\ \bibinfo {pages} {155136} (\bibinfo {year}
  {2012})}\BibitemShut {NoStop}%
\bibitem [{\citenamefont {Allen}(1978)}]{allen:1978}%
  \BibitemOpen
  \bibfield  {author} {\bibinfo {author} {\bibfnamefont {P.~B.}\ \bibnamefont
  {Allen}},\ }\bibfield  {title} {\bibinfo {title} {Solids with thermal or
  static disorder. {I}. {O}ne-electron properties},\ }\href
  {https://link.aps.org/doi/10.1103/PhysRevB.18.5217} {\bibfield  {journal}
  {\bibinfo  {journal} {Phys. Rev. B}\ }\textbf {\bibinfo {volume} {18}},\
  \bibinfo {pages} {5217} (\bibinfo {year} {1978})}\BibitemShut {NoStop}%
\bibitem [{\citenamefont {Allen}\ and\ \citenamefont
  {Silberglitt}(1974)}]{allen:1974a}%
  \BibitemOpen
  \bibfield  {author} {\bibinfo {author} {\bibfnamefont {P.~B.}\ \bibnamefont
  {Allen}}\ and\ \bibinfo {author} {\bibfnamefont {R.}~\bibnamefont
  {Silberglitt}},\ }\bibfield  {title} {\bibinfo {title} {Some effects of
  phonon dynamics on electron lifetime, mass renormalization, and
  superconducting transition temperature},\ }\href
  {https://link.aps.org/doi/10.1103/PhysRevB.9.4733} {\bibfield  {journal}
  {\bibinfo  {journal} {Phys. Rev. B}\ }\textbf {\bibinfo {volume} {9}},\
  \bibinfo {pages} {4733} (\bibinfo {year} {1974})}\BibitemShut {NoStop}%
\bibitem [{\citenamefont {Eiguren}\ \emph {et~al.}(2009)\citenamefont
  {Eiguren}, \citenamefont {{Ambrosch-Draxl}},\ and\ \citenamefont
  {Echenique}}]{eiguren:2009}%
  \BibitemOpen
  \bibfield  {author} {\bibinfo {author} {\bibfnamefont {A.}~\bibnamefont
  {Eiguren}}, \bibinfo {author} {\bibfnamefont {C.}~\bibnamefont
  {{Ambrosch-Draxl}}},\ and\ \bibinfo {author} {\bibfnamefont {P.~M.}\
  \bibnamefont {Echenique}},\ }\bibfield  {title} {\bibinfo {title}
  {Self-consistently renormalized quasiparticles under the electron-phonon
  interaction},\ }\href {https://link.aps.org/doi/10.1103/PhysRevB.79.245103}
  {\bibfield  {journal} {\bibinfo  {journal} {Phys. Rev. B}\ }\textbf {\bibinfo
  {volume} {79}},\ \bibinfo {pages} {245103} (\bibinfo {year}
  {2009})}\BibitemShut {NoStop}%
\bibitem [{\citenamefont {Wen}\ \emph {et~al.}(2018)\citenamefont {Wen},
  \citenamefont {Xu}, \citenamefont {Yao}, \citenamefont {Peng}, \citenamefont
  {Niu}, \citenamefont {Chen}, \citenamefont {Liu}, \citenamefont {Shen},
  \citenamefont {Song}, \citenamefont {Lou}, \citenamefont {Fang},
  \citenamefont {Liu}, \citenamefont {Song}, \citenamefont {Jiao},
  \citenamefont {Duan}, \citenamefont {Wen}, \citenamefont {Dudin},
  \citenamefont {Kotliar}, \citenamefont {Yin},\ and\ \citenamefont
  {Feng}}]{wen:2018}%
  \BibitemOpen
  \bibfield  {author} {\bibinfo {author} {\bibfnamefont {C.~H.~P.}\
  \bibnamefont {Wen}}, \bibinfo {author} {\bibfnamefont {H.~C.}\ \bibnamefont
  {Xu}}, \bibinfo {author} {\bibfnamefont {Q.}~\bibnamefont {Yao}}, \bibinfo
  {author} {\bibfnamefont {R.}~\bibnamefont {Peng}}, \bibinfo {author}
  {\bibfnamefont {X.~H.}\ \bibnamefont {Niu}}, \bibinfo {author} {\bibfnamefont
  {Q.~Y.}\ \bibnamefont {Chen}}, \bibinfo {author} {\bibfnamefont {Z.~T.}\
  \bibnamefont {Liu}}, \bibinfo {author} {\bibfnamefont {D.~W.}\ \bibnamefont
  {Shen}}, \bibinfo {author} {\bibfnamefont {Q.}~\bibnamefont {Song}}, \bibinfo
  {author} {\bibfnamefont {X.}~\bibnamefont {Lou}}, \bibinfo {author}
  {\bibfnamefont {Y.~F.}\ \bibnamefont {Fang}}, \bibinfo {author}
  {\bibfnamefont {X.~S.}\ \bibnamefont {Liu}}, \bibinfo {author} {\bibfnamefont
  {Y.~H.}\ \bibnamefont {Song}}, \bibinfo {author} {\bibfnamefont {Y.~J.}\
  \bibnamefont {Jiao}}, \bibinfo {author} {\bibfnamefont {T.~F.}\ \bibnamefont
  {Duan}}, \bibinfo {author} {\bibfnamefont {H.~H.}\ \bibnamefont {Wen}},
  \bibinfo {author} {\bibfnamefont {P.}~\bibnamefont {Dudin}}, \bibinfo
  {author} {\bibfnamefont {G.}~\bibnamefont {Kotliar}}, \bibinfo {author}
  {\bibfnamefont {Z.~P.}\ \bibnamefont {Yin}},\ and\ \bibinfo {author}
  {\bibfnamefont {D.~L.}\ \bibnamefont {Feng}},\ }\bibfield  {title} {\bibinfo
  {title} {Unveiling the superconducting mechanism of
  {Ba$_{0.51}$K$_{0.49}$BiO$_3$}},\ }\href
  {https://link.aps.org/doi/10.1103/PhysRevLett.121.117002} {\bibfield
  {journal} {\bibinfo  {journal} {Phys. Rev. Lett.}\ }\textbf {\bibinfo
  {volume} {121}},\ \bibinfo {pages} {117002} (\bibinfo {year}
  {2018})}\BibitemShut {NoStop}%
\bibitem [{\citenamefont {Rudin}\ \emph {et~al.}(1998)\citenamefont {Rudin},
  \citenamefont {Bauer}, \citenamefont {Liu},\ and\ \citenamefont
  {Freericks}}]{rudin:1998}%
  \BibitemOpen
  \bibfield  {author} {\bibinfo {author} {\bibfnamefont {S.~P.}\ \bibnamefont
  {Rudin}}, \bibinfo {author} {\bibfnamefont {R.}~\bibnamefont {Bauer}},
  \bibinfo {author} {\bibfnamefont {A.~Y.}\ \bibnamefont {Liu}},\ and\ \bibinfo
  {author} {\bibfnamefont {J.~K.}\ \bibnamefont {Freericks}},\ }\bibfield
  {title} {\bibinfo {title} {Reevaluating electron-phonon coupling strengths:
  {{Indium}} as a test case for {\emph{ab initio}} and many-body theory
  methods},\ }\href {https://link.aps.org/doi/10.1103/PhysRevB.58.14511}
  {\bibfield  {journal} {\bibinfo  {journal} {Phys. Rev. B}\ }\textbf {\bibinfo
  {volume} {58}},\ \bibinfo {pages} {14511} (\bibinfo {year}
  {1998})}\BibitemShut {NoStop}%
\bibitem [{\citenamefont {Gull}(1989)}]{gull:1989}%
  \BibitemOpen
  \bibfield  {author} {\bibinfo {author} {\bibfnamefont {S.~F.}\ \bibnamefont
  {Gull}},\ }\bibfield  {title} {\bibinfo {title} {Developments in maximum
  entropy data analysis},\ }in\ \href
  {https://link.springer.com/chapter/10.1007/978-94-015-7860-8_4} {\emph
  {\bibinfo {booktitle} {Developments in Maximum Entropy Data Analysis}}},\
  \bibinfo {series and number} {Maximum {{Entropy}} and {{Bayesian Methods}}},\
  \bibinfo {editor} {edited by\ \bibinfo {editor} {\bibfnamefont
  {J.}~\bibnamefont {Skilling}}}\ (\bibinfo  {publisher} {Springer,
  Dordrecht},\ \bibinfo {year} {1989})\ pp.\ \bibinfo {pages}
  {53--71}\BibitemShut {NoStop}%
\bibitem [{\citenamefont {Kaufmann}\ and\ \citenamefont
  {Held}(2023)}]{kaufmann:2023}%
  \BibitemOpen
  \bibfield  {author} {\bibinfo {author} {\bibfnamefont {J.}~\bibnamefont
  {Kaufmann}}\ and\ \bibinfo {author} {\bibfnamefont {K.}~\bibnamefont
  {Held}},\ }\bibfield  {title} {\bibinfo {title} {Ana\_cont: {{Python}}
  package for analytic continuation},\ }\href
  {https://linkinghub.elsevier.com/retrieve/pii/S0010465522002387} {\bibfield
  {journal} {\bibinfo  {journal} {Comput. Phys. Commun.}\ }\textbf {\bibinfo
  {volume} {282}},\ \bibinfo {pages} {108519} (\bibinfo {year}
  {2023})}\BibitemShut {NoStop}%
\bibitem [{\citenamefont {Levy}\ \emph {et~al.}(2014)\citenamefont {Levy},
  \citenamefont {Nettke}, \citenamefont {Ludbrook}, \citenamefont {Veenstra},\
  and\ \citenamefont {Damascelli}}]{levy:2014}%
  \BibitemOpen
  \bibfield  {author} {\bibinfo {author} {\bibfnamefont {G.}~\bibnamefont
  {Levy}}, \bibinfo {author} {\bibfnamefont {W.}~\bibnamefont {Nettke}},
  \bibinfo {author} {\bibfnamefont {B.~M.}\ \bibnamefont {Ludbrook}}, \bibinfo
  {author} {\bibfnamefont {C.~N.}\ \bibnamefont {Veenstra}},\ and\ \bibinfo
  {author} {\bibfnamefont {A.}~\bibnamefont {Damascelli}},\ }\bibfield  {title}
  {\bibinfo {title} {Deconstruction of resolution effects in angle-resolved
  photoemission},\ }\href {https://link.aps.org/doi/10.1103/PhysRevB.90.045150}
  {\bibfield  {journal} {\bibinfo  {journal} {Phys. Rev. B}\ }\textbf {\bibinfo
  {volume} {90}},\ \bibinfo {pages} {045150} (\bibinfo {year}
  {2014})}\BibitemShut {NoStop}%
\bibitem [{\citenamefont {Plumb}\ \emph {et~al.}(2010)\citenamefont {Plumb},
  \citenamefont {Reber}, \citenamefont {Koralek}, \citenamefont {Sun},
  \citenamefont {Douglas}, \citenamefont {Aiura}, \citenamefont {Oka},
  \citenamefont {Eisaki},\ and\ \citenamefont {Dessau}}]{plumb:2010}%
  \BibitemOpen
  \bibfield  {author} {\bibinfo {author} {\bibfnamefont {N.~C.}\ \bibnamefont
  {Plumb}}, \bibinfo {author} {\bibfnamefont {T.~J.}\ \bibnamefont {Reber}},
  \bibinfo {author} {\bibfnamefont {J.~D.}\ \bibnamefont {Koralek}}, \bibinfo
  {author} {\bibfnamefont {Z.}~\bibnamefont {Sun}}, \bibinfo {author}
  {\bibfnamefont {J.~F.}\ \bibnamefont {Douglas}}, \bibinfo {author}
  {\bibfnamefont {Y.}~\bibnamefont {Aiura}}, \bibinfo {author} {\bibfnamefont
  {K.}~\bibnamefont {Oka}}, \bibinfo {author} {\bibfnamefont {H.}~\bibnamefont
  {Eisaki}},\ and\ \bibinfo {author} {\bibfnamefont {D.~S.}\ \bibnamefont
  {Dessau}},\ }\bibfield  {title} {\bibinfo {title} {Low-energy ($<10$~mev)
  feature in the nodal electron self-energy and strong temperature dependence
  of the {Fermi} velocity in {Bi$_2$Sr$_2$CaCu$_2$O$_{8+\delta}$}},\ }\href
  {https://link.aps.org/doi/10.1103/PhysRevLett.105.046402} {\bibfield
  {journal} {\bibinfo  {journal} {Phys. Rev. Lett.}\ }\textbf {\bibinfo
  {volume} {105}},\ \bibinfo {pages} {046402} (\bibinfo {year}
  {2010})}\BibitemShut {NoStop}%
\bibitem [{\citenamefont {Jiang}\ \emph {et~al.}(2009)\citenamefont {Jiang},
  \citenamefont {Higashiguchi}, \citenamefont {Tobida}, \citenamefont {Tanaka},
  \citenamefont {Fukuda}, \citenamefont {Hayashi}, \citenamefont {Shimada},
  \citenamefont {Namatame},\ and\ \citenamefont {Taniguchi}}]{jiang:2009}%
  \BibitemOpen
  \bibfield  {author} {\bibinfo {author} {\bibfnamefont {J.}~\bibnamefont
  {Jiang}}, \bibinfo {author} {\bibfnamefont {M.}~\bibnamefont {Higashiguchi}},
  \bibinfo {author} {\bibfnamefont {N.}~\bibnamefont {Tobida}}, \bibinfo
  {author} {\bibfnamefont {K.}~\bibnamefont {Tanaka}}, \bibinfo {author}
  {\bibfnamefont {S.}~\bibnamefont {Fukuda}}, \bibinfo {author} {\bibfnamefont
  {H.}~\bibnamefont {Hayashi}}, \bibinfo {author} {\bibfnamefont
  {K.}~\bibnamefont {Shimada}}, \bibinfo {author} {\bibfnamefont
  {H.}~\bibnamefont {Namatame}},\ and\ \bibinfo {author} {\bibfnamefont
  {M.}~\bibnamefont {Taniguchi}},\ }\bibfield  {title} {\bibinfo {title}
  {High-resolution angle-resolved photoemission study of the {Al(100)} single
  crystal},\ }\href
  {http://www.jstage.jst.go.jp/article/ejssnt/7/0/7_0_57/_article} {\bibfield
  {journal} {\bibinfo  {journal} {e-J. Surf. Sci. Nanotechnol.}\ }\textbf
  {\bibinfo {volume} {7}},\ \bibinfo {pages} {57} (\bibinfo {year}
  {2009})}\BibitemShut {NoStop}%
\bibitem [{\citenamefont {Zhou}\ \emph {et~al.}(2005)\citenamefont {Zhou},
  \citenamefont {Shi}, \citenamefont {Yoshida}, \citenamefont {Cuk},
  \citenamefont {Yang}, \citenamefont {Brouet}, \citenamefont {Nakamura},
  \citenamefont {Mannella}, \citenamefont {Komiya}, \citenamefont {Ando},
  \citenamefont {Zhou}, \citenamefont {Ti}, \citenamefont {Xiong},
  \citenamefont {Zhao}, \citenamefont {Sasagawa}, \citenamefont {Kakeshita},
  \citenamefont {Eisaki}, \citenamefont {Uchida}, \citenamefont {Fujimori},
  \citenamefont {Zhang}, \citenamefont {Plummer}, \citenamefont {Laughlin},
  \citenamefont {Hussain},\ and\ \citenamefont {Shen}}]{zhou:2005}%
  \BibitemOpen
  \bibfield  {author} {\bibinfo {author} {\bibfnamefont {X.~J.}\ \bibnamefont
  {Zhou}}, \bibinfo {author} {\bibfnamefont {J.}~\bibnamefont {Shi}}, \bibinfo
  {author} {\bibfnamefont {T.}~\bibnamefont {Yoshida}}, \bibinfo {author}
  {\bibfnamefont {T.}~\bibnamefont {Cuk}}, \bibinfo {author} {\bibfnamefont
  {W.~L.}\ \bibnamefont {Yang}}, \bibinfo {author} {\bibfnamefont
  {V.}~\bibnamefont {Brouet}}, \bibinfo {author} {\bibfnamefont
  {J.}~\bibnamefont {Nakamura}}, \bibinfo {author} {\bibfnamefont
  {N.}~\bibnamefont {Mannella}}, \bibinfo {author} {\bibfnamefont
  {S.}~\bibnamefont {Komiya}}, \bibinfo {author} {\bibfnamefont
  {Y.}~\bibnamefont {Ando}}, \bibinfo {author} {\bibfnamefont {F.}~\bibnamefont
  {Zhou}}, \bibinfo {author} {\bibfnamefont {W.~X.}\ \bibnamefont {Ti}},
  \bibinfo {author} {\bibfnamefont {J.~W.}\ \bibnamefont {Xiong}}, \bibinfo
  {author} {\bibfnamefont {Z.~X.}\ \bibnamefont {Zhao}}, \bibinfo {author}
  {\bibfnamefont {T.}~\bibnamefont {Sasagawa}}, \bibinfo {author}
  {\bibfnamefont {T.}~\bibnamefont {Kakeshita}}, \bibinfo {author}
  {\bibfnamefont {H.}~\bibnamefont {Eisaki}}, \bibinfo {author} {\bibfnamefont
  {S.}~\bibnamefont {Uchida}}, \bibinfo {author} {\bibfnamefont
  {A.}~\bibnamefont {Fujimori}}, \bibinfo {author} {\bibfnamefont
  {Z.}~\bibnamefont {Zhang}}, \bibinfo {author} {\bibfnamefont {E.~W.}\
  \bibnamefont {Plummer}}, \bibinfo {author} {\bibfnamefont {R.~B.}\
  \bibnamefont {Laughlin}}, \bibinfo {author} {\bibfnamefont {Z.}~\bibnamefont
  {Hussain}},\ and\ \bibinfo {author} {\bibfnamefont {Z.-X.}\ \bibnamefont
  {Shen}},\ }\bibfield  {title} {\bibinfo {title} {Multiple bosonic mode
  coupling in the electron self-energy of {(La$_{2-x}$Sr$_x$)CuO$_4$}},\ }\href
  {https://link.aps.org/doi/10.1103/PhysRevLett.95.117001} {\bibfield
  {journal} {\bibinfo  {journal} {Phys. Rev. Lett.}\ }\textbf {\bibinfo
  {volume} {95}},\ \bibinfo {pages} {117001} (\bibinfo {year}
  {2005})}\BibitemShut {NoStop}%
\bibitem [{\citenamefont {King}\ \emph {et~al.}(2014)\citenamefont {King},
  \citenamefont {McKeown~Walker}, \citenamefont {Tamai}, \citenamefont {{de la
  Torre}}, \citenamefont {Eknapakul}, \citenamefont {Buaphet}, \citenamefont
  {Mo}, \citenamefont {Meevasana}, \citenamefont {Bahramy},\ and\ \citenamefont
  {Baumberger}}]{king:2014}%
  \BibitemOpen
  \bibfield  {author} {\bibinfo {author} {\bibfnamefont {P.~D.~C.}\
  \bibnamefont {King}}, \bibinfo {author} {\bibfnamefont {S.}~\bibnamefont
  {McKeown~Walker}}, \bibinfo {author} {\bibfnamefont {A.}~\bibnamefont
  {Tamai}}, \bibinfo {author} {\bibfnamefont {A.}~\bibnamefont {{de la
  Torre}}}, \bibinfo {author} {\bibfnamefont {T.}~\bibnamefont {Eknapakul}},
  \bibinfo {author} {\bibfnamefont {P.}~\bibnamefont {Buaphet}}, \bibinfo
  {author} {\bibfnamefont {S.-K.}\ \bibnamefont {Mo}}, \bibinfo {author}
  {\bibfnamefont {W.}~\bibnamefont {Meevasana}}, \bibinfo {author}
  {\bibfnamefont {M.~S.}\ \bibnamefont {Bahramy}},\ and\ \bibinfo {author}
  {\bibfnamefont {F.}~\bibnamefont {Baumberger}},\ }\bibfield  {title}
  {\bibinfo {title} {Quasiparticle dynamics and spin--orbital texture of the
  {{SrTiO}}{\textsubscript{3}} two-dimensional electron gas},\ }\href
  {http://www.nature.com/articles/ncomms4414} {\bibfield  {journal} {\bibinfo
  {journal} {Nat. Commun.}\ }\textbf {\bibinfo {volume} {5}},\ \bibinfo {pages}
  {3414} (\bibinfo {year} {2014})}\BibitemShut {NoStop}%
\bibitem [{\citenamefont {Day}\ \emph {et~al.}(2019)\citenamefont {Day},
  \citenamefont {Zwartsenberg}, \citenamefont {Elfimov},\ and\ \citenamefont
  {Damascelli}}]{day:2019}%
  \BibitemOpen
  \bibfield  {author} {\bibinfo {author} {\bibfnamefont {R.~P.}\ \bibnamefont
  {Day}}, \bibinfo {author} {\bibfnamefont {B.}~\bibnamefont {Zwartsenberg}},
  \bibinfo {author} {\bibfnamefont {I.~S.}\ \bibnamefont {Elfimov}},\ and\
  \bibinfo {author} {\bibfnamefont {A.}~\bibnamefont {Damascelli}},\ }\bibfield
   {title} {\bibinfo {title} {Computational framework chinook for
  angle-resolved photoemission spectroscopy},\ }\href
  {http://www.nature.com/articles/s41535-019-0194-8} {\bibfield  {journal}
  {\bibinfo  {journal} {npj Quantum Mater.}\ }\textbf {\bibinfo {volume} {4}},\
  \bibinfo {pages} {54} (\bibinfo {year} {2019})}\BibitemShut {NoStop}%
\bibitem [{\citenamefont {Sch{\"u}ler}\ \emph {et~al.}(2022)\citenamefont
  {Sch{\"u}ler}, \citenamefont {Pincelli}, \citenamefont {Dong}, \citenamefont
  {Devereaux}, \citenamefont {Wolf}, \citenamefont {Rettig}, \citenamefont
  {Ernstorfer},\ and\ \citenamefont {Beaulieu}}]{schuler:2022}%
  \BibitemOpen
  \bibfield  {author} {\bibinfo {author} {\bibfnamefont {M.}~\bibnamefont
  {Sch{\"u}ler}}, \bibinfo {author} {\bibfnamefont {T.}~\bibnamefont
  {Pincelli}}, \bibinfo {author} {\bibfnamefont {S.}~\bibnamefont {Dong}},
  \bibinfo {author} {\bibfnamefont {T.~P.}\ \bibnamefont {Devereaux}}, \bibinfo
  {author} {\bibfnamefont {M.}~\bibnamefont {Wolf}}, \bibinfo {author}
  {\bibfnamefont {L.}~\bibnamefont {Rettig}}, \bibinfo {author} {\bibfnamefont
  {R.}~\bibnamefont {Ernstorfer}},\ and\ \bibinfo {author} {\bibfnamefont
  {S.}~\bibnamefont {Beaulieu}},\ }\bibfield  {title} {\bibinfo {title}
  {Polarization-modulated angle-resolved photoemission spectroscopy: {T}oward
  circular dichroism without circular photons and {Bloch} wave-function
  reconstruction},\ }\href
  {https://link.aps.org/doi/10.1103/PhysRevX.12.011019} {\bibfield  {journal}
  {\bibinfo  {journal} {Phys. Rev. X}\ }\textbf {\bibinfo {volume} {12}},\
  \bibinfo {pages} {011019} (\bibinfo {year} {2022})}\BibitemShut {NoStop}%
\bibitem [{\citenamefont {Kern}\ \emph {et~al.}(2023)\citenamefont {Kern},
  \citenamefont {Haags}, \citenamefont {Egger}, \citenamefont {Yang},
  \citenamefont {Kirschner}, \citenamefont {Wolff}, \citenamefont {Seyller},
  \citenamefont {Gottwald}, \citenamefont {Richter}, \citenamefont
  {De~Giovannini}, \citenamefont {Rubio}, \citenamefont {Ramsey}, \citenamefont
  {Bocquet}, \citenamefont {Soubatch}, \citenamefont {Tautz}, \citenamefont
  {Puschnig},\ and\ \citenamefont {Moser}}]{kern:2023}%
  \BibitemOpen
  \bibfield  {author} {\bibinfo {author} {\bibfnamefont {C.~S.}\ \bibnamefont
  {Kern}}, \bibinfo {author} {\bibfnamefont {A.}~\bibnamefont {Haags}},
  \bibinfo {author} {\bibfnamefont {L.}~\bibnamefont {Egger}}, \bibinfo
  {author} {\bibfnamefont {X.}~\bibnamefont {Yang}}, \bibinfo {author}
  {\bibfnamefont {H.}~\bibnamefont {Kirschner}}, \bibinfo {author}
  {\bibfnamefont {S.}~\bibnamefont {Wolff}}, \bibinfo {author} {\bibfnamefont
  {T.}~\bibnamefont {Seyller}}, \bibinfo {author} {\bibfnamefont
  {A.}~\bibnamefont {Gottwald}}, \bibinfo {author} {\bibfnamefont
  {M.}~\bibnamefont {Richter}}, \bibinfo {author} {\bibfnamefont
  {U.}~\bibnamefont {De~Giovannini}}, \bibinfo {author} {\bibfnamefont
  {A.}~\bibnamefont {Rubio}}, \bibinfo {author} {\bibfnamefont {M.~G.}\
  \bibnamefont {Ramsey}}, \bibinfo {author} {\bibfnamefont {F.~C.}\
  \bibnamefont {Bocquet}}, \bibinfo {author} {\bibfnamefont {S.}~\bibnamefont
  {Soubatch}}, \bibinfo {author} {\bibfnamefont {F.~S.}\ \bibnamefont {Tautz}},
  \bibinfo {author} {\bibfnamefont {P.}~\bibnamefont {Puschnig}},\ and\
  \bibinfo {author} {\bibfnamefont {S.}~\bibnamefont {Moser}},\ }\bibfield
  {title} {\bibinfo {title} {Simple extension of the plane-wave final state in
  photoemission: {{Bringing}} understanding to the photon-energy dependence of
  two-dimensional materials},\ }\href
  {https://link.aps.org/doi/10.1103/PhysRevResearch.5.033075} {\bibfield
  {journal} {\bibinfo  {journal} {Phys. Rev. Res.}\ }\textbf {\bibinfo {volume}
  {5}},\ \bibinfo {pages} {033075} (\bibinfo {year} {2023})}\BibitemShut
  {NoStop}%
\bibitem [{\citenamefont {Ryoo}\ and\ \citenamefont {Park}(2025)}]{ryoo:2025}%
  \BibitemOpen
  \bibfield  {author} {\bibinfo {author} {\bibfnamefont {J.~H.}\ \bibnamefont
  {Ryoo}}\ and\ \bibinfo {author} {\bibfnamefont {C.-H.}\ \bibnamefont
  {Park}},\ }\bibfield  {title} {\bibinfo {title} {{Lippmann-Schwinger}
  approach for accurate photoelectron wave functions and angle-resolved
  photoemission spectra from first principles},\ }\href
  {https://link.aps.org/doi/10.1103/gwmm-6l57} {\bibfield  {journal} {\bibinfo
  {journal} {Phys. Rev. Lett.}\ }\textbf {\bibinfo {volume} {135}},\ \bibinfo
  {pages} {056403} (\bibinfo {year} {2025})}\BibitemShut {NoStop}%
\bibitem [{\citenamefont {Ebert}\ \emph {et~al.}(2011)\citenamefont {Ebert},
  \citenamefont {K{\"o}dderitzsch},\ and\ \citenamefont
  {Min{\'a}r}}]{ebert:2011}%
  \BibitemOpen
  \bibfield  {author} {\bibinfo {author} {\bibfnamefont {H.}~\bibnamefont
  {Ebert}}, \bibinfo {author} {\bibfnamefont {D.}~\bibnamefont
  {K{\"o}dderitzsch}},\ and\ \bibinfo {author} {\bibfnamefont {J.}~\bibnamefont
  {Min{\'a}r}},\ }\bibfield  {title} {\bibinfo {title} {Calculating condensed
  matter properties using the {{KKR-Green}}'s function method---recent
  developments and applications},\ }\href
  {https://iopscience.iop.org/article/10.1088/0034-4885/74/9/096501} {\bibfield
   {journal} {\bibinfo  {journal} {Rep. Prog. Phys.}\ }\textbf {\bibinfo
  {volume} {74}},\ \bibinfo {pages} {096501} (\bibinfo {year}
  {2011})}\BibitemShut {NoStop}%
\bibitem [{\citenamefont {Zhou}\ \emph
  {et~al.}(2018{\natexlab{a}})\citenamefont {Zhou}, \citenamefont {Hellman},\
  and\ \citenamefont {Bernardi}}]{zhou:2018a}%
  \BibitemOpen
  \bibfield  {author} {\bibinfo {author} {\bibfnamefont {J.-J.}\ \bibnamefont
  {Zhou}}, \bibinfo {author} {\bibfnamefont {O.}~\bibnamefont {Hellman}},\ and\
  \bibinfo {author} {\bibfnamefont {M.}~\bibnamefont {Bernardi}},\ }\bibfield
  {title} {\bibinfo {title} {Electron-phonon scattering in the presence of soft
  modes and electron mobility in {SrTiO$_3$} perovskite from first
  principles},\ }\href
  {https://link.aps.org/doi/10.1103/PhysRevLett.121.226603} {\bibfield
  {journal} {\bibinfo  {journal} {Phys. Rev. Lett.}\ }\textbf {\bibinfo
  {volume} {121}},\ \bibinfo {pages} {226603} (\bibinfo {year}
  {2018}{\natexlab{a}})}\BibitemShut {NoStop}%
\bibitem [{\citenamefont {Zhou}\ and\ \citenamefont
  {Bernardi}(2019)}]{zhou:2019}%
  \BibitemOpen
  \bibfield  {author} {\bibinfo {author} {\bibfnamefont {J.-J.}\ \bibnamefont
  {Zhou}}\ and\ \bibinfo {author} {\bibfnamefont {M.}~\bibnamefont
  {Bernardi}},\ }\bibfield  {title} {\bibinfo {title} {Predicting charge
  transport in the presence of polarons: {T}he beyond-quasiparticle regime in
  {SrTiO$_3$}},\ }\href
  {https://link.aps.org/doi/10.1103/PhysRevResearch.1.033138} {\bibfield
  {journal} {\bibinfo  {journal} {Phys. Rev. Res.}\ }\textbf {\bibinfo {volume}
  {1}},\ \bibinfo {pages} {033138} (\bibinfo {year} {2019})}\BibitemShut
  {NoStop}%
\bibitem [{\citenamefont {Chen}\ \emph {et~al.}(2015)\citenamefont {Chen},
  \citenamefont {Avila}, \citenamefont {Frantzeskakis}, \citenamefont {Levy},\
  and\ \citenamefont {Asensio}}]{chen:2015}%
  \BibitemOpen
  \bibfield  {author} {\bibinfo {author} {\bibfnamefont {C.}~\bibnamefont
  {Chen}}, \bibinfo {author} {\bibfnamefont {J.}~\bibnamefont {Avila}},
  \bibinfo {author} {\bibfnamefont {E.}~\bibnamefont {Frantzeskakis}}, \bibinfo
  {author} {\bibfnamefont {A.}~\bibnamefont {Levy}},\ and\ \bibinfo {author}
  {\bibfnamefont {M.~C.}\ \bibnamefont {Asensio}},\ }\bibfield  {title}
  {\bibinfo {title} {Observation of a two-dimensional liquid of
  {{Fr{\"o}hlich}} polarons at the bare {{SrTiO}}{\textsubscript{3}} surface},\
  }\href {https://www.nature.com/articles/ncomms9585} {\bibfield  {journal}
  {\bibinfo  {journal} {Nat. Commun.}\ }\textbf {\bibinfo {volume} {6}},\
  \bibinfo {pages} {8585} (\bibinfo {year} {2015})}\BibitemShut {NoStop}%
\bibitem [{\citenamefont {Wang}\ \emph {et~al.}(2016)\citenamefont {Wang},
  \citenamefont {Walker}, \citenamefont {Tamai}, \citenamefont {Wang},
  \citenamefont {Ristic}, \citenamefont {Bruno},\ and\ \citenamefont
  {Baumberger}}]{wang:2016}%
  \BibitemOpen
  \bibfield  {author} {\bibinfo {author} {\bibfnamefont {Z.}~\bibnamefont
  {Wang}}, \bibinfo {author} {\bibfnamefont {S.~M.}\ \bibnamefont {Walker}},
  \bibinfo {author} {\bibfnamefont {A.}~\bibnamefont {Tamai}}, \bibinfo
  {author} {\bibfnamefont {Y.}~\bibnamefont {Wang}}, \bibinfo {author}
  {\bibfnamefont {Z.}~\bibnamefont {Ristic}}, \bibinfo {author} {\bibfnamefont
  {F.~Y.}\ \bibnamefont {Bruno}},\ and\ \bibinfo {author} {\bibfnamefont
  {F.}~\bibnamefont {Baumberger}},\ }\bibfield  {title} {\bibinfo {title}
  {Tailoring the nature and strength of electron--phonon interactions in the
  {{SrTiO}}{\textsubscript{3}}(001) {{2D}} electron liquid},\ }\href
  {https://www.nature.com/articles/nmat4623} {\bibfield  {journal} {\bibinfo
  {journal} {Nat. Mater.}\ }\textbf {\bibinfo {volume} {15}},\ \bibinfo {pages}
  {6} (\bibinfo {year} {2016})}\BibitemShut {NoStop}%
\bibitem [{\citenamefont {Guedes}\ \emph {et~al.}(2020)\citenamefont {Guedes},
  \citenamefont {Muff}, \citenamefont {Fanciulli}, \citenamefont {Weber},
  \citenamefont {Caputo}, \citenamefont {Wang}, \citenamefont {Plumb},
  \citenamefont {Radovi{\'c}},\ and\ \citenamefont {Dil}}]{guedes:2020}%
  \BibitemOpen
  \bibfield  {author} {\bibinfo {author} {\bibfnamefont {E.~B.}\ \bibnamefont
  {Guedes}}, \bibinfo {author} {\bibfnamefont {S.}~\bibnamefont {Muff}},
  \bibinfo {author} {\bibfnamefont {M.}~\bibnamefont {Fanciulli}}, \bibinfo
  {author} {\bibfnamefont {A.~P.}\ \bibnamefont {Weber}}, \bibinfo {author}
  {\bibfnamefont {M.}~\bibnamefont {Caputo}}, \bibinfo {author} {\bibfnamefont
  {Z.}~\bibnamefont {Wang}}, \bibinfo {author} {\bibfnamefont {N.~C.}\
  \bibnamefont {Plumb}}, \bibinfo {author} {\bibfnamefont {M.}~\bibnamefont
  {Radovi{\'c}}},\ and\ \bibinfo {author} {\bibfnamefont {J.~H.}\ \bibnamefont
  {Dil}},\ }\bibfield  {title} {\bibinfo {title} {Single spin-polarized
  {{Fermi}} surface in {{SrTiO}}{\textsubscript{3}} thin films},\ }\href
  {https://link.aps.org/doi/10.1103/PhysRevResearch.2.033173} {\bibfield
  {journal} {\bibinfo  {journal} {Phys. Rev. Res.}\ }\textbf {\bibinfo {volume}
  {2}},\ \bibinfo {pages} {033173} (\bibinfo {year} {2020})}\BibitemShut
  {NoStop}%
\bibitem [{\citenamefont {Faeth}\ \emph {et~al.}(2021)\citenamefont {Faeth},
  \citenamefont {Xie}, \citenamefont {Yang}, \citenamefont {Kawasaki},
  \citenamefont {Nelson}, \citenamefont {Zhang}, \citenamefont {Parzyck},
  \citenamefont {Mishra}, \citenamefont {Li}, \citenamefont {Jozwiak},
  \citenamefont {Bostwick}, \citenamefont {Rotenberg}, \citenamefont {Schlom},\
  and\ \citenamefont {Shen}}]{faeth:2021}%
  \BibitemOpen
  \bibfield  {author} {\bibinfo {author} {\bibfnamefont {B.~D.}\ \bibnamefont
  {Faeth}}, \bibinfo {author} {\bibfnamefont {S.}~\bibnamefont {Xie}}, \bibinfo
  {author} {\bibfnamefont {S.}~\bibnamefont {Yang}}, \bibinfo {author}
  {\bibfnamefont {J.~K.}\ \bibnamefont {Kawasaki}}, \bibinfo {author}
  {\bibfnamefont {J.~N.}\ \bibnamefont {Nelson}}, \bibinfo {author}
  {\bibfnamefont {S.}~\bibnamefont {Zhang}}, \bibinfo {author} {\bibfnamefont
  {C.}~\bibnamefont {Parzyck}}, \bibinfo {author} {\bibfnamefont
  {P.}~\bibnamefont {Mishra}}, \bibinfo {author} {\bibfnamefont
  {C.}~\bibnamefont {Li}}, \bibinfo {author} {\bibfnamefont {C.}~\bibnamefont
  {Jozwiak}}, \bibinfo {author} {\bibfnamefont {A.}~\bibnamefont {Bostwick}},
  \bibinfo {author} {\bibfnamefont {E.}~\bibnamefont {Rotenberg}}, \bibinfo
  {author} {\bibfnamefont {D.~G.}\ \bibnamefont {Schlom}},\ and\ \bibinfo
  {author} {\bibfnamefont {K.~M.}\ \bibnamefont {Shen}},\ }\bibfield  {title}
  {\bibinfo {title} {Interfacial electron-phonon coupling constants extracted
  from intrinsic replica bands in monolayer {FeSe/SrTiO$_3$}},\ }\href
  {https://link.aps.org/doi/10.1103/PhysRevLett.127.016803} {\bibfield
  {journal} {\bibinfo  {journal} {Phys. Rev. Lett.}\ }\textbf {\bibinfo
  {volume} {127}},\ \bibinfo {pages} {016803} (\bibinfo {year}
  {2021})}\BibitemShut {NoStop}%
\bibitem [{\citenamefont {Goldberg}\ \emph {et~al.}(1981)\citenamefont
  {Goldberg}, \citenamefont {Fadley},\ and\ \citenamefont
  {Kono}}]{goldberg:1981}%
  \BibitemOpen
  \bibfield  {author} {\bibinfo {author} {\bibfnamefont {S.}~\bibnamefont
  {Goldberg}}, \bibinfo {author} {\bibfnamefont {C.}~\bibnamefont {Fadley}},\
  and\ \bibinfo {author} {\bibfnamefont {S.}~\bibnamefont {Kono}},\ }\bibfield
  {title} {\bibinfo {title} {Photoionization cross-sections for atomic orbitals
  with random and fixed spatial orientation},\ }\href
  {https://linkinghub.elsevier.com/retrieve/pii/0368204881850670} {\bibfield
  {journal} {\bibinfo  {journal} {J. Electron Spectrosc. Relat. Phenom.}\
  }\textbf {\bibinfo {volume} {21}},\ \bibinfo {pages} {285} (\bibinfo {year}
  {1981})}\BibitemShut {NoStop}%
\bibitem [{\citenamefont {Vogt}(1988)}]{vogt:1988}%
  \BibitemOpen
  \bibfield  {author} {\bibinfo {author} {\bibfnamefont {H.}~\bibnamefont
  {Vogt}},\ }\bibfield  {title} {\bibinfo {title} {Hyper-{{Raman}} tensors of
  the zone-center optical phonons in {{SrTiO}}{\textsubscript{3}} and
  {{KTaO}}{\textsubscript{3}}},\ }\href
  {https://link.aps.org/doi/10.1103/PhysRevB.38.5699} {\bibfield  {journal}
  {\bibinfo  {journal} {Phys. Rev. B}\ }\textbf {\bibinfo {volume} {38}},\
  \bibinfo {pages} {5699} (\bibinfo {year} {1988})}\BibitemShut {NoStop}%
\bibitem [{\citenamefont {Zhou}\ \emph
  {et~al.}(2018{\natexlab{b}})\citenamefont {Zhou}, \citenamefont {He},
  \citenamefont {Liu}, \citenamefont {Zhao}, \citenamefont {Yu},\ and\
  \citenamefont {Zhang}}]{zhou:2018}%
  \BibitemOpen
  \bibfield  {author} {\bibinfo {author} {\bibfnamefont {X.}~\bibnamefont
  {Zhou}}, \bibinfo {author} {\bibfnamefont {S.}~\bibnamefont {He}}, \bibinfo
  {author} {\bibfnamefont {G.}~\bibnamefont {Liu}}, \bibinfo {author}
  {\bibfnamefont {L.}~\bibnamefont {Zhao}}, \bibinfo {author} {\bibfnamefont
  {L.}~\bibnamefont {Yu}},\ and\ \bibinfo {author} {\bibfnamefont
  {W.}~\bibnamefont {Zhang}},\ }\bibfield  {title} {\bibinfo {title} {New
  developments in laser-based photoemission spectroscopy and its scientific
  applications: A key issues review},\ }\href
  {https://iopscience.iop.org/article/10.1088/1361-6633/aab0cc} {\bibfield
  {journal} {\bibinfo  {journal} {Rep. Prog. Phys.}\ }\textbf {\bibinfo
  {volume} {81}},\ \bibinfo {pages} {062101} (\bibinfo {year}
  {2018}{\natexlab{b}})}\BibitemShut {NoStop}%
\bibitem [{\citenamefont {Cancellieri}\ \emph {et~al.}(2016)\citenamefont
  {Cancellieri}, \citenamefont {Mishchenko}, \citenamefont {Aschauer},
  \citenamefont {Filippetti}, \citenamefont {Faber}, \citenamefont {Bari{\v
  s}i{\'c}}, \citenamefont {Rogalev}, \citenamefont {Schmitt}, \citenamefont
  {Nagaosa},\ and\ \citenamefont {Strocov}}]{cancellieri:2016}%
  \BibitemOpen
  \bibfield  {author} {\bibinfo {author} {\bibfnamefont {C.}~\bibnamefont
  {Cancellieri}}, \bibinfo {author} {\bibfnamefont {A.~S.}\ \bibnamefont
  {Mishchenko}}, \bibinfo {author} {\bibfnamefont {U.}~\bibnamefont
  {Aschauer}}, \bibinfo {author} {\bibfnamefont {A.}~\bibnamefont
  {Filippetti}}, \bibinfo {author} {\bibfnamefont {C.}~\bibnamefont {Faber}},
  \bibinfo {author} {\bibfnamefont {O.~S.}\ \bibnamefont {Bari{\v s}i{\'c}}},
  \bibinfo {author} {\bibfnamefont {V.~A.}\ \bibnamefont {Rogalev}}, \bibinfo
  {author} {\bibfnamefont {T.}~\bibnamefont {Schmitt}}, \bibinfo {author}
  {\bibfnamefont {N.}~\bibnamefont {Nagaosa}},\ and\ \bibinfo {author}
  {\bibfnamefont {V.~N.}\ \bibnamefont {Strocov}},\ }\bibfield  {title}
  {\bibinfo {title} {Polaronic metal state at the
  {{LaAlO}}{\textsubscript{3}}/{{SrTiO}}{\textsubscript{3}} interface},\ }\href
  {http://www.nature.com/articles/ncomms10386} {\bibfield  {journal} {\bibinfo
  {journal} {Nat. Commun.}\ }\textbf {\bibinfo {volume} {7}},\ \bibinfo {pages}
  {10386} (\bibinfo {year} {2016})}\BibitemShut {NoStop}%
\bibitem [{\citenamefont {Alarab}\ \emph {et~al.}(2024)\citenamefont {Alarab},
  \citenamefont {Hricovini}, \citenamefont {Leikert}, \citenamefont {Richter},
  \citenamefont {Schmitt}, \citenamefont {Sing}, \citenamefont {Claessen},
  \citenamefont {Min{\'a}r},\ and\ \citenamefont {Strocov}}]{alarab:2024}%
  \BibitemOpen
  \bibfield  {author} {\bibinfo {author} {\bibfnamefont {F.}~\bibnamefont
  {Alarab}}, \bibinfo {author} {\bibfnamefont {K.}~\bibnamefont {Hricovini}},
  \bibinfo {author} {\bibfnamefont {B.}~\bibnamefont {Leikert}}, \bibinfo
  {author} {\bibfnamefont {C.}~\bibnamefont {Richter}}, \bibinfo {author}
  {\bibfnamefont {T.}~\bibnamefont {Schmitt}}, \bibinfo {author} {\bibfnamefont
  {M.}~\bibnamefont {Sing}}, \bibinfo {author} {\bibfnamefont {R.}~\bibnamefont
  {Claessen}}, \bibinfo {author} {\bibfnamefont {J.}~\bibnamefont
  {Min{\'a}r}},\ and\ \bibinfo {author} {\bibfnamefont {V.~N.}\ \bibnamefont
  {Strocov}},\ }\bibfield  {title} {\bibinfo {title} {Nature of the metallic
  and in-gap states in {Ni}-doped {SrTiO$_3$}},\ }\href
  {https://pubs.aip.org/apm/article/12/1/011118/3061433/Nature-of-the-metallic-and-in-gap-states-in-Ni}
  {\bibfield  {journal} {\bibinfo  {journal} {APL Mater.}\ }\textbf {\bibinfo
  {volume} {12}},\ \bibinfo {pages} {011118} (\bibinfo {year}
  {2024})}\BibitemShut {NoStop}%
\bibitem [{\citenamefont {{Santander-Syro}}\ \emph {et~al.}(2014)\citenamefont
  {{Santander-Syro}}, \citenamefont {Fortuna}, \citenamefont {Bareille},
  \citenamefont {R{\"o}del}, \citenamefont {Landolt}, \citenamefont {Plumb},
  \citenamefont {Dil},\ and\ \citenamefont
  {Radovi{\'c}}}]{santander-syro:2014}%
  \BibitemOpen
  \bibfield  {author} {\bibinfo {author} {\bibfnamefont {A.~F.}\ \bibnamefont
  {{Santander-Syro}}}, \bibinfo {author} {\bibfnamefont {F.}~\bibnamefont
  {Fortuna}}, \bibinfo {author} {\bibfnamefont {C.}~\bibnamefont {Bareille}},
  \bibinfo {author} {\bibfnamefont {T.~C.}\ \bibnamefont {R{\"o}del}}, \bibinfo
  {author} {\bibfnamefont {G.}~\bibnamefont {Landolt}}, \bibinfo {author}
  {\bibfnamefont {N.~C.}\ \bibnamefont {Plumb}}, \bibinfo {author}
  {\bibfnamefont {J.~H.}\ \bibnamefont {Dil}},\ and\ \bibinfo {author}
  {\bibfnamefont {M.}~\bibnamefont {Radovi{\'c}}},\ }\bibfield  {title}
  {\bibinfo {title} {Giant spin splitting of the two-dimensional electron gas
  at the surface of {{SrTiO}}{\textsubscript{3}}},\ }\href
  {https://www.nature.com/articles/nmat4107} {\bibfield  {journal} {\bibinfo
  {journal} {Nat. Mater.}\ }\textbf {\bibinfo {volume} {13}},\ \bibinfo {pages}
  {1085} (\bibinfo {year} {2014})}\BibitemShut {NoStop}%
\bibitem [{\citenamefont {Fabian}\ and\ \citenamefont
  {Das~Sarma}(1999)}]{fabian:1999}%
  \BibitemOpen
  \bibfield  {author} {\bibinfo {author} {\bibfnamefont {J.}~\bibnamefont
  {Fabian}}\ and\ \bibinfo {author} {\bibfnamefont {S.}~\bibnamefont
  {Das~Sarma}},\ }\bibfield  {title} {\bibinfo {title} {Phonon-induced spin
  relaxation of conduction electrons in aluminum},\ }\href
  {https://link.aps.org/doi/10.1103/PhysRevLett.83.1211} {\bibfield  {journal}
  {\bibinfo  {journal} {Phys. Rev. Lett.}\ }\textbf {\bibinfo {volume} {83}},\
  \bibinfo {pages} {1211} (\bibinfo {year} {1999})}\BibitemShut {NoStop}%
\bibitem [{\citenamefont {Hwang}\ \emph {et~al.}(2007)\citenamefont {Hwang},
  \citenamefont {Timusk}, \citenamefont {Schachinger},\ and\ \citenamefont
  {Carbotte}}]{hwang:2007}%
  \BibitemOpen
  \bibfield  {author} {\bibinfo {author} {\bibfnamefont {J.}~\bibnamefont
  {Hwang}}, \bibinfo {author} {\bibfnamefont {T.}~\bibnamefont {Timusk}},
  \bibinfo {author} {\bibfnamefont {E.}~\bibnamefont {Schachinger}},\ and\
  \bibinfo {author} {\bibfnamefont {J.~P.}\ \bibnamefont {Carbotte}},\
  }\bibfield  {title} {\bibinfo {title} {Evolution of the bosonic spectral
  density of the high-temperature superconductor
  {{Bi}}{\textsubscript{2}}{{Sr}}{\textsubscript{2}}{{CaCu}}{\textsubscript{2}}{{O}}{\textsubscript{8+{$\delta$}}}},\
  }\href {https://link.aps.org/doi/10.1103/PhysRevB.75.144508} {\bibfield
  {journal} {\bibinfo  {journal} {Phys. Rev. B}\ }\textbf {\bibinfo {volume}
  {75}},\ \bibinfo {pages} {144508} (\bibinfo {year} {2007})}\BibitemShut
  {NoStop}%
\bibitem [{\citenamefont {Zhang}\ \emph {et~al.}(2022)\citenamefont {Zhang},
  \citenamefont {Pincelli}, \citenamefont {Jozwiak}, \citenamefont {Kondo},
  \citenamefont {Ernstorfer}, \citenamefont {Sato},\ and\ \citenamefont
  {Zhou}}]{zhang:2022}%
  \BibitemOpen
  \bibfield  {author} {\bibinfo {author} {\bibfnamefont {H.}~\bibnamefont
  {Zhang}}, \bibinfo {author} {\bibfnamefont {T.}~\bibnamefont {Pincelli}},
  \bibinfo {author} {\bibfnamefont {C.}~\bibnamefont {Jozwiak}}, \bibinfo
  {author} {\bibfnamefont {T.}~\bibnamefont {Kondo}}, \bibinfo {author}
  {\bibfnamefont {R.}~\bibnamefont {Ernstorfer}}, \bibinfo {author}
  {\bibfnamefont {T.}~\bibnamefont {Sato}},\ and\ \bibinfo {author}
  {\bibfnamefont {S.}~\bibnamefont {Zhou}},\ }\bibfield  {title} {\bibinfo
  {title} {Angle-resolved photoemission spectroscopy},\ }\href
  {https://www.nature.com/articles/s43586-022-00133-7} {\bibfield  {journal}
  {\bibinfo  {journal} {Nat. Rev. Methods Primers}\ }\textbf {\bibinfo {volume}
  {2}},\ \bibinfo {pages} {54} (\bibinfo {year} {2022})}\BibitemShut {NoStop}%
\bibitem [{\citenamefont {Ponc{\'e}}\ \emph {et~al.}(2023)\citenamefont
  {Ponc{\'e}}, \citenamefont {Royo}, \citenamefont {Stengel}, \citenamefont
  {Marzari},\ and\ \citenamefont {Gibertini}}]{ponce:2023}%
  \BibitemOpen
  \bibfield  {author} {\bibinfo {author} {\bibfnamefont {S.}~\bibnamefont
  {Ponc{\'e}}}, \bibinfo {author} {\bibfnamefont {M.}~\bibnamefont {Royo}},
  \bibinfo {author} {\bibfnamefont {M.}~\bibnamefont {Stengel}}, \bibinfo
  {author} {\bibfnamefont {N.}~\bibnamefont {Marzari}},\ and\ \bibinfo {author}
  {\bibfnamefont {M.}~\bibnamefont {Gibertini}},\ }\bibfield  {title} {\bibinfo
  {title} {Long-range electrostatic contribution to electron-phonon couplings
  and mobilities of two-dimensional and bulk materials},\ }\href
  {https://link.aps.org/doi/10.1103/PhysRevB.107.155424} {\bibfield  {journal}
  {\bibinfo  {journal} {Phys. Rev. B}\ }\textbf {\bibinfo {volume} {107}},\
  \bibinfo {pages} {155424} (\bibinfo {year} {2023})}\BibitemShut {NoStop}%
\bibitem [{\citenamefont {Margine}\ and\ \citenamefont
  {Giustino}(2014)}]{margine:2014}%
  \BibitemOpen
  \bibfield  {author} {\bibinfo {author} {\bibfnamefont {E.~R.}\ \bibnamefont
  {Margine}}\ and\ \bibinfo {author} {\bibfnamefont {F.}~\bibnamefont
  {Giustino}},\ }\bibfield  {title} {\bibinfo {title} {Two-gap
  superconductivity in heavily {\emph{n}}-doped graphene:
  {{{\emph{Ab}}}}{\emph{ initio}} {{Migdal-Eliashberg}} theory},\ }\href
  {https://link.aps.org/doi/10.1103/PhysRevB.90.014518} {\bibfield  {journal}
  {\bibinfo  {journal} {Phys. Rev. B}\ }\textbf {\bibinfo {volume} {90}},\
  \bibinfo {pages} {014518} (\bibinfo {year} {2014})}\BibitemShut {NoStop}%
\bibitem [{\citenamefont {Novko}(2017)}]{novko:2017}%
  \BibitemOpen
  \bibfield  {author} {\bibinfo {author} {\bibfnamefont {D.}~\bibnamefont
  {Novko}},\ }\bibfield  {title} {\bibinfo {title} {Dopant-induced plasmon
  decay in graphene},\ }\href
  {https://pubs.acs.org/doi/10.1021/acs.nanolett.7b03553} {\bibfield  {journal}
  {\bibinfo  {journal} {Nano Lett.}\ }\textbf {\bibinfo {volume} {17}},\
  \bibinfo {pages} {6991} (\bibinfo {year} {2017})}\BibitemShut {NoStop}%
\bibitem [{\citenamefont {Hwang}\ and\ \citenamefont
  {Das~Sarma}(2008)}]{hwang:2008}%
  \BibitemOpen
  \bibfield  {author} {\bibinfo {author} {\bibfnamefont {E.~H.}\ \bibnamefont
  {Hwang}}\ and\ \bibinfo {author} {\bibfnamefont {S.}~\bibnamefont
  {Das~Sarma}},\ }\bibfield  {title} {\bibinfo {title} {Quasiparticle spectral
  function in doped graphene: {{Electron-electron}} interaction effects in
  {{ARPES}}},\ }\href {https://link.aps.org/doi/10.1103/PhysRevB.77.081412}
  {\bibfield  {journal} {\bibinfo  {journal} {Phys. Rev. B}\ }\textbf {\bibinfo
  {volume} {77}},\ \bibinfo {pages} {081412} (\bibinfo {year}
  {2008})}\BibitemShut {NoStop}%
\bibitem [{\citenamefont {Park}\ \emph {et~al.}(2009)\citenamefont {Park},
  \citenamefont {Giustino}, \citenamefont {Spataru}, \citenamefont {Cohen},\
  and\ \citenamefont {Louie}}]{park:2009}%
  \BibitemOpen
  \bibfield  {author} {\bibinfo {author} {\bibfnamefont {C.-H.}\ \bibnamefont
  {Park}}, \bibinfo {author} {\bibfnamefont {F.}~\bibnamefont {Giustino}},
  \bibinfo {author} {\bibfnamefont {C.~D.}\ \bibnamefont {Spataru}}, \bibinfo
  {author} {\bibfnamefont {M.~L.}\ \bibnamefont {Cohen}},\ and\ \bibinfo
  {author} {\bibfnamefont {S.~G.}\ \bibnamefont {Louie}},\ }\bibfield  {title}
  {\bibinfo {title} {First-principles study of electron linewidths in
  graphene},\ }\href {https://link.aps.org/doi/10.1103/PhysRevLett.102.076803}
  {\bibfield  {journal} {\bibinfo  {journal} {Phys. Rev. Lett.}\ }\textbf
  {\bibinfo {volume} {102}},\ \bibinfo {pages} {076803} (\bibinfo {year}
  {2009})}\BibitemShut {NoStop}%
\bibitem [{\citenamefont {Gr{\"u}neis}\ \emph {et~al.}(2009)\citenamefont
  {Gr{\"u}neis}, \citenamefont {Attaccalite}, \citenamefont {Rubio},
  \citenamefont {Vyalikh}, \citenamefont {Molodtsov}, \citenamefont {Fink},
  \citenamefont {Follath}, \citenamefont {Eberhardt}, \citenamefont
  {B{\"u}chner},\ and\ \citenamefont {Pichler}}]{gruneis:2009a}%
  \BibitemOpen
  \bibfield  {author} {\bibinfo {author} {\bibfnamefont {A.}~\bibnamefont
  {Gr{\"u}neis}}, \bibinfo {author} {\bibfnamefont {C.}~\bibnamefont
  {Attaccalite}}, \bibinfo {author} {\bibfnamefont {A.}~\bibnamefont {Rubio}},
  \bibinfo {author} {\bibfnamefont {D.~V.}\ \bibnamefont {Vyalikh}}, \bibinfo
  {author} {\bibfnamefont {S.~L.}\ \bibnamefont {Molodtsov}}, \bibinfo {author}
  {\bibfnamefont {J.}~\bibnamefont {Fink}}, \bibinfo {author} {\bibfnamefont
  {R.}~\bibnamefont {Follath}}, \bibinfo {author} {\bibfnamefont
  {W.}~\bibnamefont {Eberhardt}}, \bibinfo {author} {\bibfnamefont
  {B.}~\bibnamefont {B{\"u}chner}},\ and\ \bibinfo {author} {\bibfnamefont
  {T.}~\bibnamefont {Pichler}},\ }\bibfield  {title} {\bibinfo {title}
  {Electronic structure and electron-phonon coupling of doped graphene layers
  in {{KC}}{\textsubscript{8}}},\ }\href
  {https://link.aps.org/doi/10.1103/PhysRevB.79.205106} {\bibfield  {journal}
  {\bibinfo  {journal} {Phys. Rev. B}\ }\textbf {\bibinfo {volume} {79}},\
  \bibinfo {pages} {205106} (\bibinfo {year} {2009})}\BibitemShut {NoStop}%
\bibitem [{\citenamefont {Tang}\ \emph {et~al.}(2004)\citenamefont {Tang},
  \citenamefont {Shi}, \citenamefont {Wu}, \citenamefont {Sprunger},
  \citenamefont {Yang}, \citenamefont {Brouet}, \citenamefont {Zhou},
  \citenamefont {Hussain}, \citenamefont {Shen}, \citenamefont {Zhang},\ and\
  \citenamefont {Plummer}}]{tang:2004}%
  \BibitemOpen
  \bibfield  {author} {\bibinfo {author} {\bibfnamefont {S.-J.}\ \bibnamefont
  {Tang}}, \bibinfo {author} {\bibfnamefont {J.}~\bibnamefont {Shi}}, \bibinfo
  {author} {\bibfnamefont {B.}~\bibnamefont {Wu}}, \bibinfo {author}
  {\bibfnamefont {P.~T.}\ \bibnamefont {Sprunger}}, \bibinfo {author}
  {\bibfnamefont {W.~L.}\ \bibnamefont {Yang}}, \bibinfo {author}
  {\bibfnamefont {V.}~\bibnamefont {Brouet}}, \bibinfo {author} {\bibfnamefont
  {X.~J.}\ \bibnamefont {Zhou}}, \bibinfo {author} {\bibfnamefont
  {Z.}~\bibnamefont {Hussain}}, \bibinfo {author} {\bibfnamefont {Z.-X.}\
  \bibnamefont {Shen}}, \bibinfo {author} {\bibfnamefont {Z.}~\bibnamefont
  {Zhang}},\ and\ \bibinfo {author} {\bibfnamefont {E.~W.}\ \bibnamefont
  {Plummer}},\ }\bibfield  {title} {\bibinfo {title} {A spectroscopic view of
  electron--phonon coupling at metal surfaces},\ }\href
  {https://onlinelibrary.wiley.com/doi/10.1002/pssb.200404890} {\bibfield
  {journal} {\bibinfo  {journal} {Phys. Status Solidi (b)}\ }\textbf {\bibinfo
  {volume} {241}},\ \bibinfo {pages} {2345} (\bibinfo {year}
  {2004})}\BibitemShut {NoStop}%
\bibitem [{\citenamefont {Fajardo}\ and\ \citenamefont
  {Winkler}(2019)}]{fajardo:2019}%
  \BibitemOpen
  \bibfield  {author} {\bibinfo {author} {\bibfnamefont {E.~A.}\ \bibnamefont
  {Fajardo}}\ and\ \bibinfo {author} {\bibfnamefont {R.}~\bibnamefont
  {Winkler}},\ }\bibfield  {title} {\bibinfo {title} {Effective dynamics of
  two-dimensional {{Bloch}} electrons in external fields derived from
  symmetry},\ }\href {https://link.aps.org/doi/10.1103/PhysRevB.100.125301}
  {\bibfield  {journal} {\bibinfo  {journal} {Phys. Rev. B}\ }\textbf {\bibinfo
  {volume} {100}},\ \bibinfo {pages} {125301} (\bibinfo {year}
  {2019})}\BibitemShut {NoStop}%
\bibitem [{\citenamefont {Romero}\ \emph {et~al.}(2020)\citenamefont {Romero},
  \citenamefont {Allan}, \citenamefont {Amadon}, \citenamefont {Antonius},
  \citenamefont {Applencourt}, \citenamefont {Baguet}, \citenamefont {Bieder},
  \citenamefont {Bottin}, \citenamefont {Bouchet}, \citenamefont {Bousquet},
  \citenamefont {Bruneval}, \citenamefont {Brunin}, \citenamefont {Caliste},
  \citenamefont {Côté}, \citenamefont {Denier}, \citenamefont {Dreyer},
  \citenamefont {Ghosez}, \citenamefont {Giantomassi}, \citenamefont {Gillet},
  \citenamefont {Gingras}, \citenamefont {Hamann}, \citenamefont {Hautier},
  \citenamefont {Jollet}, \citenamefont {Jomard}, \citenamefont {Martin},
  \citenamefont {Miranda}, \citenamefont {Naccarato}, \citenamefont {Petretto},
  \citenamefont {Pike}, \citenamefont {Planes}, \citenamefont {Prokhorenko},
  \citenamefont {Rangel}, \citenamefont {Ricci}, \citenamefont {Rignanese},
  \citenamefont {Royo}, \citenamefont {Stengel}, \citenamefont {Torrent},
  \citenamefont {Van~Setten}, \citenamefont {Van~Troeye}, \citenamefont
  {Verstraete}, \citenamefont {Wiktor}, \citenamefont {Zwanziger},\ and\
  \citenamefont {Gonze}}]{romero:2020}%
  \BibitemOpen
  \bibfield  {author} {\bibinfo {author} {\bibfnamefont {A.~H.}\ \bibnamefont
  {Romero}}, \bibinfo {author} {\bibfnamefont {D.~C.}\ \bibnamefont {Allan}},
  \bibinfo {author} {\bibfnamefont {B.}~\bibnamefont {Amadon}}, \bibinfo
  {author} {\bibfnamefont {G.}~\bibnamefont {Antonius}}, \bibinfo {author}
  {\bibfnamefont {T.}~\bibnamefont {Applencourt}}, \bibinfo {author}
  {\bibfnamefont {L.}~\bibnamefont {Baguet}}, \bibinfo {author} {\bibfnamefont
  {J.}~\bibnamefont {Bieder}}, \bibinfo {author} {\bibfnamefont
  {F.}~\bibnamefont {Bottin}}, \bibinfo {author} {\bibfnamefont
  {J.}~\bibnamefont {Bouchet}}, \bibinfo {author} {\bibfnamefont
  {E.}~\bibnamefont {Bousquet}}, \bibinfo {author} {\bibfnamefont
  {F.}~\bibnamefont {Bruneval}}, \bibinfo {author} {\bibfnamefont
  {G.}~\bibnamefont {Brunin}}, \bibinfo {author} {\bibfnamefont
  {D.}~\bibnamefont {Caliste}}, \bibinfo {author} {\bibfnamefont
  {M.}~\bibnamefont {Côté}}, \bibinfo {author} {\bibfnamefont
  {J.}~\bibnamefont {Denier}}, \bibinfo {author} {\bibfnamefont
  {C.}~\bibnamefont {Dreyer}}, \bibinfo {author} {\bibfnamefont
  {P.}~\bibnamefont {Ghosez}}, \bibinfo {author} {\bibfnamefont
  {M.}~\bibnamefont {Giantomassi}}, \bibinfo {author} {\bibfnamefont
  {Y.}~\bibnamefont {Gillet}}, \bibinfo {author} {\bibfnamefont
  {O.}~\bibnamefont {Gingras}}, \bibinfo {author} {\bibfnamefont {D.~R.}\
  \bibnamefont {Hamann}}, \bibinfo {author} {\bibfnamefont {G.}~\bibnamefont
  {Hautier}}, \bibinfo {author} {\bibfnamefont {F.}~\bibnamefont {Jollet}},
  \bibinfo {author} {\bibfnamefont {G.}~\bibnamefont {Jomard}}, \bibinfo
  {author} {\bibfnamefont {A.}~\bibnamefont {Martin}}, \bibinfo {author}
  {\bibfnamefont {H.~P.~C.}\ \bibnamefont {Miranda}}, \bibinfo {author}
  {\bibfnamefont {F.}~\bibnamefont {Naccarato}}, \bibinfo {author}
  {\bibfnamefont {G.}~\bibnamefont {Petretto}}, \bibinfo {author}
  {\bibfnamefont {N.~A.}\ \bibnamefont {Pike}}, \bibinfo {author}
  {\bibfnamefont {V.}~\bibnamefont {Planes}}, \bibinfo {author} {\bibfnamefont
  {S.}~\bibnamefont {Prokhorenko}}, \bibinfo {author} {\bibfnamefont
  {T.}~\bibnamefont {Rangel}}, \bibinfo {author} {\bibfnamefont
  {F.}~\bibnamefont {Ricci}}, \bibinfo {author} {\bibfnamefont {G.-M.}\
  \bibnamefont {Rignanese}}, \bibinfo {author} {\bibfnamefont {M.}~\bibnamefont
  {Royo}}, \bibinfo {author} {\bibfnamefont {M.}~\bibnamefont {Stengel}},
  \bibinfo {author} {\bibfnamefont {M.}~\bibnamefont {Torrent}}, \bibinfo
  {author} {\bibfnamefont {M.~J.}\ \bibnamefont {Van~Setten}}, \bibinfo
  {author} {\bibfnamefont {B.}~\bibnamefont {Van~Troeye}}, \bibinfo {author}
  {\bibfnamefont {M.~J.}\ \bibnamefont {Verstraete}}, \bibinfo {author}
  {\bibfnamefont {J.}~\bibnamefont {Wiktor}}, \bibinfo {author} {\bibfnamefont
  {J.~W.}\ \bibnamefont {Zwanziger}},\ and\ \bibinfo {author} {\bibfnamefont
  {X.}~\bibnamefont {Gonze}},\ }\bibfield  {title} {\bibinfo {title}
  {\textsc{abinit}: Overview and focus on selected capabilities},\ }\href
  {https://pubs.aip.org/jcp/article/152/12/124102/953753/ABINIT-Overview-and-focus-on-selected-capabilities}
  {\bibfield  {journal} {\bibinfo  {journal} {J. Chem. Phys.}\ }\textbf
  {\bibinfo {volume} {152}},\ \bibinfo {pages} {124102} (\bibinfo {year}
  {2020})}\BibitemShut {NoStop}%
\bibitem [{\citenamefont {Jiang}\ \emph {et~al.}(2011)\citenamefont {Jiang},
  \citenamefont {Shimada}, \citenamefont {Hayashi}, \citenamefont {Iwasawa},
  \citenamefont {Aiura}, \citenamefont {Namatame},\ and\ \citenamefont
  {Taniguchi}}]{jiang:2011}%
  \BibitemOpen
  \bibfield  {author} {\bibinfo {author} {\bibfnamefont {J.}~\bibnamefont
  {Jiang}}, \bibinfo {author} {\bibfnamefont {K.}~\bibnamefont {Shimada}},
  \bibinfo {author} {\bibfnamefont {H.}~\bibnamefont {Hayashi}}, \bibinfo
  {author} {\bibfnamefont {H.}~\bibnamefont {Iwasawa}}, \bibinfo {author}
  {\bibfnamefont {Y.}~\bibnamefont {Aiura}}, \bibinfo {author} {\bibfnamefont
  {H.}~\bibnamefont {Namatame}},\ and\ \bibinfo {author} {\bibfnamefont
  {M.}~\bibnamefont {Taniguchi}},\ }\bibfield  {title} {\bibinfo {title}
  {Coupling parameters of many-body interactions for the {{Al}}(100) surface
  state: {{A}} high-resolution angle-resolved photoemission spectroscopy
  study},\ }\href {https://link.aps.org/doi/10.1103/PhysRevB.84.155124}
  {\bibfield  {journal} {\bibinfo  {journal} {Phys. Rev. B}\ }\textbf {\bibinfo
  {volume} {84}},\ \bibinfo {pages} {155124} (\bibinfo {year}
  {2011})}\BibitemShut {NoStop}%
\bibitem [{\citenamefont {Lihm}\ and\ \citenamefont
  {Ponc{\'e}}(2025)}]{lihm:2025}%
  \BibitemOpen
  \bibfield  {author} {\bibinfo {author} {\bibfnamefont {J.-M.}\ \bibnamefont
  {Lihm}}\ and\ \bibinfo {author} {\bibfnamefont {S.}~\bibnamefont
  {Ponc{\'e}}},\ }\bibfield  {title} {\bibinfo {title} {Nonperturbative
  self-consistent electron-phonon spectral functions and transport},\ }\href
  {https://link.aps.org/doi/10.1103/PhysRevLett.134.186401} {\bibfield
  {journal} {\bibinfo  {journal} {Phys. Rev. Lett.}\ }\textbf {\bibinfo
  {volume} {134}},\ \bibinfo {pages} {186401} (\bibinfo {year}
  {2025})}\BibitemShut {NoStop}%
\bibitem [{\citenamefont {Aryasetiawan}\ \emph {et~al.}(1996)\citenamefont
  {Aryasetiawan}, \citenamefont {Hedin},\ and\ \citenamefont
  {Karlsson}}]{aryasetiawan:1996}%
  \BibitemOpen
  \bibfield  {author} {\bibinfo {author} {\bibfnamefont {F.}~\bibnamefont
  {Aryasetiawan}}, \bibinfo {author} {\bibfnamefont {L.}~\bibnamefont
  {Hedin}},\ and\ \bibinfo {author} {\bibfnamefont {K.}~\bibnamefont
  {Karlsson}},\ }\bibfield  {title} {\bibinfo {title} {Multiple plasmon
  satellites in {Na} and {Al} spectral functions from \textit{Ab Initio}
  cumulant expansion},\ }\href
  {https://link.aps.org/doi/10.1103/PhysRevLett.77.2268} {\bibfield  {journal}
  {\bibinfo  {journal} {Phys. Rev. Lett.}\ }\textbf {\bibinfo {volume} {77}},\
  \bibinfo {pages} {2268} (\bibinfo {year} {1996})}\BibitemShut {NoStop}%
\bibitem [{\citenamefont {Varma}\ \emph {et~al.}(1989)\citenamefont {Varma},
  \citenamefont {Littlewood}, \citenamefont {{Schmitt-Rink}}, \citenamefont
  {Abrahams},\ and\ \citenamefont {Ruckenstein}}]{varma:1989}%
  \BibitemOpen
  \bibfield  {author} {\bibinfo {author} {\bibfnamefont {C.~M.}\ \bibnamefont
  {Varma}}, \bibinfo {author} {\bibfnamefont {P.~B.}\ \bibnamefont
  {Littlewood}}, \bibinfo {author} {\bibfnamefont {S.}~\bibnamefont
  {{Schmitt-Rink}}}, \bibinfo {author} {\bibfnamefont {E.}~\bibnamefont
  {Abrahams}},\ and\ \bibinfo {author} {\bibfnamefont {A.~E.}\ \bibnamefont
  {Ruckenstein}},\ }\bibfield  {title} {\bibinfo {title} {Phenomenology of the
  normal state of {Cu-O} high-temperature superconductors},\ }\href
  {https://link.aps.org/doi/10.1103/PhysRevLett.63.1996} {\bibfield  {journal}
  {\bibinfo  {journal} {Phys. Rev. Lett.}\ }\textbf {\bibinfo {volume} {63}},\
  \bibinfo {pages} {1996} (\bibinfo {year} {1989})}\BibitemShut {NoStop}%
\bibitem [{\citenamefont {Bostwick}\ \emph {et~al.}(2007)\citenamefont
  {Bostwick}, \citenamefont {Ohta}, \citenamefont {Seyller}, \citenamefont
  {Horn},\ and\ \citenamefont {Rotenberg}}]{bostwick:2007a}%
  \BibitemOpen
  \bibfield  {author} {\bibinfo {author} {\bibfnamefont {A.}~\bibnamefont
  {Bostwick}}, \bibinfo {author} {\bibfnamefont {T.}~\bibnamefont {Ohta}},
  \bibinfo {author} {\bibfnamefont {T.}~\bibnamefont {Seyller}}, \bibinfo
  {author} {\bibfnamefont {K.}~\bibnamefont {Horn}},\ and\ \bibinfo {author}
  {\bibfnamefont {E.}~\bibnamefont {Rotenberg}},\ }\bibfield  {title} {\bibinfo
  {title} {Quasiparticle dynamics in graphene},\ }\href
  {https://www.nature.com/articles/nphys477} {\bibfield  {journal} {\bibinfo
  {journal} {Nat. Phys.}\ }\textbf {\bibinfo {volume} {3}},\ \bibinfo {pages}
  {36} (\bibinfo {year} {2007})}\BibitemShut {NoStop}%
\bibitem [{\citenamefont {{LMFIT developers}}(2025)}]{lmfit:website}%
  \BibitemOpen
  \bibfield  {author} {\bibinfo {author} {\bibnamefont {{LMFIT developers}}},\
  }\href {https://lmfit.github.io/lmfit-py/index.html} {\bibinfo {title}
  {{{LMFIT}}}} (\bibinfo {year} {Last accessed: 5 July 2025})\BibitemShut
  {NoStop}%
\bibitem [{\citenamefont {Virtanen}\ \emph {et~al.}(2020)\citenamefont
  {Virtanen}, \citenamefont {Gommers}, \citenamefont {Oliphant}, \citenamefont
  {Haberland}, \citenamefont {Reddy}, \citenamefont {Cournapeau}, \citenamefont
  {Burovski}, \citenamefont {Peterson}, \citenamefont {Weckesser},
  \citenamefont {Bright}, \citenamefont {{van der Walt}}, \citenamefont
  {Brett}, \citenamefont {Wilson}, \citenamefont {Millman}, \citenamefont
  {Mayorov}, \citenamefont {Nelson}, \citenamefont {Jones}, \citenamefont
  {Kern}, \citenamefont {Larson}, \citenamefont {Carey}, \citenamefont {Polat},
  \citenamefont {Feng}, \citenamefont {Moore}, \citenamefont {VanderPlas},
  \citenamefont {Laxalde}, \citenamefont {Perktold}, \citenamefont {Cimrman},
  \citenamefont {Henriksen}, \citenamefont {Quintero}, \citenamefont {Harris},
  \citenamefont {Archibald}, \citenamefont {Ribeiro}, \citenamefont
  {Pedregosa}, \citenamefont {{van Mulbregt}}, \citenamefont {{SciPy 1.0
  Contributors}}, \citenamefont {Vijaykumar}, \citenamefont {Bardelli},
  \citenamefont {Rothberg}, \citenamefont {Hilboll}, \citenamefont {Kloeckner},
  \citenamefont {Scopatz}, \citenamefont {Lee}, \citenamefont {Rokem},
  \citenamefont {Woods}, \citenamefont {Fulton}, \citenamefont {Masson},
  \citenamefont {H{\"a}ggstr{\"o}m}, \citenamefont {Fitzgerald}, \citenamefont
  {Nicholson}, \citenamefont {Hagen}, \citenamefont {Pasechnik}, \citenamefont
  {Olivetti}, \citenamefont {Martin}, \citenamefont {Wieser}, \citenamefont
  {Silva}, \citenamefont {Lenders}, \citenamefont {Wilhelm}, \citenamefont
  {Young}, \citenamefont {Price}, \citenamefont {Ingold}, \citenamefont
  {Allen}, \citenamefont {Lee}, \citenamefont {Audren}, \citenamefont {Probst},
  \citenamefont {Dietrich}, \citenamefont {Silterra}, \citenamefont {Webber},
  \citenamefont {Slavi{\v c}}, \citenamefont {Nothman}, \citenamefont
  {Buchner}, \citenamefont {Kulick}, \citenamefont {Sch{\"o}nberger},
  \citenamefont {{de Miranda Cardoso}}, \citenamefont {Reimer}, \citenamefont
  {Harrington}, \citenamefont {Rodr{\'i}guez}, \citenamefont
  {{Nunez-Iglesias}}, \citenamefont {Kuczynski}, \citenamefont {Tritz},
  \citenamefont {Thoma}, \citenamefont {Newville}, \citenamefont
  {K{\"u}mmerer}, \citenamefont {Bolingbroke}, \citenamefont {Tartre},
  \citenamefont {Pak}, \citenamefont {Smith}, \citenamefont {Nowaczyk},
  \citenamefont {Shebanov}, \citenamefont {Pavlyk}, \citenamefont {Brodtkorb},
  \citenamefont {Lee}, \citenamefont {McGibbon}, \citenamefont {Feldbauer},
  \citenamefont {Lewis}, \citenamefont {Tygier}, \citenamefont {Sievert},
  \citenamefont {Vigna}, \citenamefont {Peterson}, \citenamefont {More},
  \citenamefont {Pudlik}, \citenamefont {Oshima}, \citenamefont {Pingel},
  \citenamefont {Robitaille}, \citenamefont {Spura}, \citenamefont {Jones},
  \citenamefont {Cera}, \citenamefont {Leslie}, \citenamefont {Zito},
  \citenamefont {Krauss}, \citenamefont {Upadhyay}, \citenamefont {Halchenko},\
  and\ \citenamefont {{V{\'a}zquez-Baeza}}}]{virtanen:2020}%
  \BibitemOpen
  \bibfield  {author} {\bibinfo {author} {\bibfnamefont {P.}~\bibnamefont
  {Virtanen}}, \bibinfo {author} {\bibfnamefont {R.}~\bibnamefont {Gommers}},
  \bibinfo {author} {\bibfnamefont {T.~E.}\ \bibnamefont {Oliphant}}, \bibinfo
  {author} {\bibfnamefont {M.}~\bibnamefont {Haberland}}, \bibinfo {author}
  {\bibfnamefont {T.}~\bibnamefont {Reddy}}, \bibinfo {author} {\bibfnamefont
  {D.}~\bibnamefont {Cournapeau}}, \bibinfo {author} {\bibfnamefont
  {E.}~\bibnamefont {Burovski}}, \bibinfo {author} {\bibfnamefont
  {P.}~\bibnamefont {Peterson}}, \bibinfo {author} {\bibfnamefont
  {W.}~\bibnamefont {Weckesser}}, \bibinfo {author} {\bibfnamefont
  {J.}~\bibnamefont {Bright}}, \bibinfo {author} {\bibfnamefont {S.~J.}\
  \bibnamefont {{van der Walt}}}, \bibinfo {author} {\bibfnamefont
  {M.}~\bibnamefont {Brett}}, \bibinfo {author} {\bibfnamefont
  {J.}~\bibnamefont {Wilson}}, \bibinfo {author} {\bibfnamefont {K.~J.}\
  \bibnamefont {Millman}}, \bibinfo {author} {\bibfnamefont {N.}~\bibnamefont
  {Mayorov}}, \bibinfo {author} {\bibfnamefont {A.~R.~J.}\ \bibnamefont
  {Nelson}}, \bibinfo {author} {\bibfnamefont {E.}~\bibnamefont {Jones}},
  \bibinfo {author} {\bibfnamefont {R.}~\bibnamefont {Kern}}, \bibinfo {author}
  {\bibfnamefont {E.}~\bibnamefont {Larson}}, \bibinfo {author} {\bibfnamefont
  {C.~J.}\ \bibnamefont {Carey}}, \bibinfo {author} {\bibfnamefont
  {{\.I}.}~\bibnamefont {Polat}}, \bibinfo {author} {\bibfnamefont
  {Y.}~\bibnamefont {Feng}}, \bibinfo {author} {\bibfnamefont {E.~W.}\
  \bibnamefont {Moore}}, \bibinfo {author} {\bibfnamefont {J.}~\bibnamefont
  {VanderPlas}}, \bibinfo {author} {\bibfnamefont {D.}~\bibnamefont {Laxalde}},
  \bibinfo {author} {\bibfnamefont {J.}~\bibnamefont {Perktold}}, \bibinfo
  {author} {\bibfnamefont {R.}~\bibnamefont {Cimrman}}, \bibinfo {author}
  {\bibfnamefont {I.}~\bibnamefont {Henriksen}}, \bibinfo {author}
  {\bibfnamefont {E.~A.}\ \bibnamefont {Quintero}}, \bibinfo {author}
  {\bibfnamefont {C.~R.}\ \bibnamefont {Harris}}, \bibinfo {author}
  {\bibfnamefont {A.~M.}\ \bibnamefont {Archibald}}, \bibinfo {author}
  {\bibfnamefont {A.~H.}\ \bibnamefont {Ribeiro}}, \bibinfo {author}
  {\bibfnamefont {F.}~\bibnamefont {Pedregosa}}, \bibinfo {author}
  {\bibfnamefont {P.}~\bibnamefont {{van Mulbregt}}}, \bibinfo {author}
  {\bibnamefont {{SciPy 1.0 Contributors}}}, \bibinfo {author} {\bibfnamefont
  {A.}~\bibnamefont {Vijaykumar}}, \bibinfo {author} {\bibfnamefont {A.~P.}\
  \bibnamefont {Bardelli}}, \bibinfo {author} {\bibfnamefont {A.}~\bibnamefont
  {Rothberg}}, \bibinfo {author} {\bibfnamefont {A.}~\bibnamefont {Hilboll}},
  \bibinfo {author} {\bibfnamefont {A.}~\bibnamefont {Kloeckner}}, \bibinfo
  {author} {\bibfnamefont {A.}~\bibnamefont {Scopatz}}, \bibinfo {author}
  {\bibfnamefont {A.}~\bibnamefont {Lee}}, \bibinfo {author} {\bibfnamefont
  {A.}~\bibnamefont {Rokem}}, \bibinfo {author} {\bibfnamefont {C.~N.}\
  \bibnamefont {Woods}}, \bibinfo {author} {\bibfnamefont {C.}~\bibnamefont
  {Fulton}}, \bibinfo {author} {\bibfnamefont {C.}~\bibnamefont {Masson}},
  \bibinfo {author} {\bibfnamefont {C.}~\bibnamefont {H{\"a}ggstr{\"o}m}},
  \bibinfo {author} {\bibfnamefont {C.}~\bibnamefont {Fitzgerald}}, \bibinfo
  {author} {\bibfnamefont {D.~A.}\ \bibnamefont {Nicholson}}, \bibinfo {author}
  {\bibfnamefont {D.~R.}\ \bibnamefont {Hagen}}, \bibinfo {author}
  {\bibfnamefont {D.~V.}\ \bibnamefont {Pasechnik}}, \bibinfo {author}
  {\bibfnamefont {E.}~\bibnamefont {Olivetti}}, \bibinfo {author}
  {\bibfnamefont {E.}~\bibnamefont {Martin}}, \bibinfo {author} {\bibfnamefont
  {E.}~\bibnamefont {Wieser}}, \bibinfo {author} {\bibfnamefont
  {F.}~\bibnamefont {Silva}}, \bibinfo {author} {\bibfnamefont
  {F.}~\bibnamefont {Lenders}}, \bibinfo {author} {\bibfnamefont
  {F.}~\bibnamefont {Wilhelm}}, \bibinfo {author} {\bibfnamefont
  {G.}~\bibnamefont {Young}}, \bibinfo {author} {\bibfnamefont {G.~A.}\
  \bibnamefont {Price}}, \bibinfo {author} {\bibfnamefont {G.-L.}\ \bibnamefont
  {Ingold}}, \bibinfo {author} {\bibfnamefont {G.~E.}\ \bibnamefont {Allen}},
  \bibinfo {author} {\bibfnamefont {G.~R.}\ \bibnamefont {Lee}}, \bibinfo
  {author} {\bibfnamefont {H.}~\bibnamefont {Audren}}, \bibinfo {author}
  {\bibfnamefont {I.}~\bibnamefont {Probst}}, \bibinfo {author} {\bibfnamefont
  {J.~P.}\ \bibnamefont {Dietrich}}, \bibinfo {author} {\bibfnamefont
  {J.}~\bibnamefont {Silterra}}, \bibinfo {author} {\bibfnamefont {J.~T.}\
  \bibnamefont {Webber}}, \bibinfo {author} {\bibfnamefont {J.}~\bibnamefont
  {Slavi{\v c}}}, \bibinfo {author} {\bibfnamefont {J.}~\bibnamefont
  {Nothman}}, \bibinfo {author} {\bibfnamefont {J.}~\bibnamefont {Buchner}},
  \bibinfo {author} {\bibfnamefont {J.}~\bibnamefont {Kulick}}, \bibinfo
  {author} {\bibfnamefont {J.~L.}\ \bibnamefont {Sch{\"o}nberger}}, \bibinfo
  {author} {\bibfnamefont {J.~V.}\ \bibnamefont {{de Miranda Cardoso}}},
  \bibinfo {author} {\bibfnamefont {J.}~\bibnamefont {Reimer}}, \bibinfo
  {author} {\bibfnamefont {J.}~\bibnamefont {Harrington}}, \bibinfo {author}
  {\bibfnamefont {J.~L.~C.}\ \bibnamefont {Rodr{\'i}guez}}, \bibinfo {author}
  {\bibfnamefont {J.}~\bibnamefont {{Nunez-Iglesias}}}, \bibinfo {author}
  {\bibfnamefont {J.}~\bibnamefont {Kuczynski}}, \bibinfo {author}
  {\bibfnamefont {K.}~\bibnamefont {Tritz}}, \bibinfo {author} {\bibfnamefont
  {M.}~\bibnamefont {Thoma}}, \bibinfo {author} {\bibfnamefont
  {M.}~\bibnamefont {Newville}}, \bibinfo {author} {\bibfnamefont
  {M.}~\bibnamefont {K{\"u}mmerer}}, \bibinfo {author} {\bibfnamefont
  {M.}~\bibnamefont {Bolingbroke}}, \bibinfo {author} {\bibfnamefont
  {M.}~\bibnamefont {Tartre}}, \bibinfo {author} {\bibfnamefont
  {M.}~\bibnamefont {Pak}}, \bibinfo {author} {\bibfnamefont {N.~J.}\
  \bibnamefont {Smith}}, \bibinfo {author} {\bibfnamefont {N.}~\bibnamefont
  {Nowaczyk}}, \bibinfo {author} {\bibfnamefont {N.}~\bibnamefont {Shebanov}},
  \bibinfo {author} {\bibfnamefont {O.}~\bibnamefont {Pavlyk}}, \bibinfo
  {author} {\bibfnamefont {P.~A.}\ \bibnamefont {Brodtkorb}}, \bibinfo {author}
  {\bibfnamefont {P.}~\bibnamefont {Lee}}, \bibinfo {author} {\bibfnamefont
  {R.~T.}\ \bibnamefont {McGibbon}}, \bibinfo {author} {\bibfnamefont
  {R.}~\bibnamefont {Feldbauer}}, \bibinfo {author} {\bibfnamefont
  {S.}~\bibnamefont {Lewis}}, \bibinfo {author} {\bibfnamefont
  {S.}~\bibnamefont {Tygier}}, \bibinfo {author} {\bibfnamefont
  {S.}~\bibnamefont {Sievert}}, \bibinfo {author} {\bibfnamefont
  {S.}~\bibnamefont {Vigna}}, \bibinfo {author} {\bibfnamefont
  {S.}~\bibnamefont {Peterson}}, \bibinfo {author} {\bibfnamefont
  {S.}~\bibnamefont {More}}, \bibinfo {author} {\bibfnamefont {T.}~\bibnamefont
  {Pudlik}}, \bibinfo {author} {\bibfnamefont {T.}~\bibnamefont {Oshima}},
  \bibinfo {author} {\bibfnamefont {T.~J.}\ \bibnamefont {Pingel}}, \bibinfo
  {author} {\bibfnamefont {T.~P.}\ \bibnamefont {Robitaille}}, \bibinfo
  {author} {\bibfnamefont {T.}~\bibnamefont {Spura}}, \bibinfo {author}
  {\bibfnamefont {T.~R.}\ \bibnamefont {Jones}}, \bibinfo {author}
  {\bibfnamefont {T.}~\bibnamefont {Cera}}, \bibinfo {author} {\bibfnamefont
  {T.}~\bibnamefont {Leslie}}, \bibinfo {author} {\bibfnamefont
  {T.}~\bibnamefont {Zito}}, \bibinfo {author} {\bibfnamefont {T.}~\bibnamefont
  {Krauss}}, \bibinfo {author} {\bibfnamefont {U.}~\bibnamefont {Upadhyay}},
  \bibinfo {author} {\bibfnamefont {Y.~O.}\ \bibnamefont {Halchenko}},\ and\
  \bibinfo {author} {\bibfnamefont {Y.}~\bibnamefont {{V{\'a}zquez-Baeza}}},\
  }\bibfield  {title} {\bibinfo {title} {{{SciPy}} 1.0: Fundamental algorithms
  for scientific computing in {{Python}}},\ }\href
  {http://www.nature.com/articles/s41592-019-0686-2} {\bibfield  {journal}
  {\bibinfo  {journal} {Nat. Methods}\ }\textbf {\bibinfo {volume} {17}},\
  \bibinfo {pages} {261} (\bibinfo {year} {2020})}\BibitemShut {NoStop}%
\bibitem [{\citenamefont {Hellbr{\"u}ck}\ \emph {et~al.}(2024)\citenamefont
  {Hellbr{\"u}ck}, \citenamefont {Puppin}, \citenamefont {Guo}, \citenamefont
  {Hickstein}, \citenamefont {Benhabib}, \citenamefont {Grioni}, \citenamefont
  {Dil}, \citenamefont {LaGrange}, \citenamefont {R{\o}nnow},\ and\
  \citenamefont {Carbone}}]{hellbruck:2024}%
  \BibitemOpen
  \bibfield  {author} {\bibinfo {author} {\bibfnamefont {L.}~\bibnamefont
  {Hellbr{\"u}ck}}, \bibinfo {author} {\bibfnamefont {M.}~\bibnamefont
  {Puppin}}, \bibinfo {author} {\bibfnamefont {F.}~\bibnamefont {Guo}},
  \bibinfo {author} {\bibfnamefont {D.~D.}\ \bibnamefont {Hickstein}}, \bibinfo
  {author} {\bibfnamefont {S.}~\bibnamefont {Benhabib}}, \bibinfo {author}
  {\bibfnamefont {M.}~\bibnamefont {Grioni}}, \bibinfo {author} {\bibfnamefont
  {J.~H.}\ \bibnamefont {Dil}}, \bibinfo {author} {\bibfnamefont
  {T.}~\bibnamefont {LaGrange}}, \bibinfo {author} {\bibfnamefont {H.~M.}\
  \bibnamefont {R{\o}nnow}},\ and\ \bibinfo {author} {\bibfnamefont
  {F.}~\bibnamefont {Carbone}},\ }\bibfield  {title} {\bibinfo {title}
  {High-resolution {{MHz}} time- and angle-resolved photoemission spectroscopy
  based on a tunable vacuum ultraviolet source},\ }\href
  {https://pubs.aip.org/rsi/article/95/3/033007/3278843/High-resolution-MHz-time-and-angle-resolved}
  {\bibfield  {journal} {\bibinfo  {journal} {Rev. Sci. Instrum.}\ }\textbf
  {\bibinfo {volume} {95}},\ \bibinfo {pages} {033007} (\bibinfo {year}
  {2024})}\BibitemShut {NoStop}%
\end{thebibliography}

\begin{thebibliography}{34}%
\makeatletter
\providecommand \@ifxundefined [1]{%
 \@ifx{#1\undefined}
}%
\providecommand \@ifnum [1]{%
 \ifnum #1\expandafter \@firstoftwo
 \else \expandafter \@secondoftwo
 \fi
}%
\providecommand \@ifx [1]{%
 \ifx #1\expandafter \@firstoftwo
 \else \expandafter \@secondoftwo
 \fi
}%
\providecommand \natexlab [1]{#1}%
\providecommand \enquote  [1]{``#1''}%
\providecommand \bibnamefont  [1]{#1}%
\providecommand \bibfnamefont [1]{#1}%
\providecommand \citenamefont [1]{#1}%
\providecommand \href@noop [0]{\@secondoftwo}%
\providecommand \href [0]{\begingroup \@sanitize@url \@href}%
\providecommand \@href[1]{\@@startlink{#1}\@@href}%
\providecommand \@@href[1]{\endgroup#1\@@endlink}%
\providecommand \@sanitize@url [0]{\catcode `\\12\catcode `\$12\catcode
  `\&12\catcode `\#12\catcode `\^12\catcode `\_12\catcode `\%12\relax}%
\providecommand \@@startlink[1]{}%
\providecommand \@@endlink[0]{}%
\providecommand \url  [0]{\begingroup\@sanitize@url \@url }%
\providecommand \@url [1]{\endgroup\@href {#1}{\urlprefix }}%
\providecommand \urlprefix  [0]{URL }%
\providecommand \Eprint [0]{\href }%
\providecommand \doibase [0]{https://doi.org/}%
\providecommand \selectlanguage [0]{\@gobble}%
\providecommand \bibinfo  [0]{\@secondoftwo}%
\providecommand \bibfield  [0]{\@secondoftwo}%
\providecommand \translation [1]{[#1]}%
\providecommand \BibitemOpen [0]{}%
\providecommand \bibitemStop [0]{}%
\providecommand \bibitemNoStop [0]{.\EOS\space}%
\providecommand \EOS [0]{\spacefactor3000\relax}%
\providecommand \BibitemShut  [1]{\csname bibitem#1\endcsname}%
\let\auto@bib@innerbib\@empty
\bibitem [{\citenamefont {Luttinger}(1961)}]{luttinger:1961}%
  \BibitemOpen
  \bibfield  {author} {\bibinfo {author} {\bibfnamefont {J.~M.}\ \bibnamefont
  {Luttinger}},\ }\bibfield  {title} {\bibinfo {title} {Analytic properties of
  single-particle propagators for many-fermion systems},\ }\href
  {https://link.aps.org/doi/10.1103/PhysRev.121.942} {\bibfield  {journal}
  {\bibinfo  {journal} {Phys. Rev.}\ }\textbf {\bibinfo {volume} {121}},\
  \bibinfo {pages} {942} (\bibinfo {year} {1961})}\BibitemShut {NoStop}%
\bibitem [{\citenamefont {Georges}\ and\ \citenamefont
  {Kotliar}(1992)}]{georges:1992}%
  \BibitemOpen
  \bibfield  {author} {\bibinfo {author} {\bibfnamefont {A.}~\bibnamefont
  {Georges}}\ and\ \bibinfo {author} {\bibfnamefont {G.}~\bibnamefont
  {Kotliar}},\ }\bibfield  {title} {\bibinfo {title} {Hubbard model in infinite
  dimensions},\ }\href {https://link.aps.org/doi/10.1103/PhysRevB.45.6479}
  {\bibfield  {journal} {\bibinfo  {journal} {Phys. Rev. B}\ }\textbf {\bibinfo
  {volume} {45}},\ \bibinfo {pages} {6479} (\bibinfo {year}
  {1992})}\BibitemShut {NoStop}%
\bibitem [{\citenamefont {Berthod}(2018)}]{berthod:2018a-SM}%
  \BibitemOpen
  \bibfield  {author} {\bibinfo {author} {\bibfnamefont {C.}~\bibnamefont
  {Berthod}},\ }\href {https://iopscience.iop.org/book/mono/978-0-7503-1741-2}
  {\emph {\bibinfo {title} {Spectroscopic Probes of Quantum Matter}}}\
  (\bibinfo  {publisher} {IOP Publishing},\ \bibinfo {address} {Bristol},\
  \bibinfo {year} {2018})\BibitemShut {NoStop}%
\bibitem [{\citenamefont {Bloom}(1975)}]{bloom:1975}%
  \BibitemOpen
  \bibfield  {author} {\bibinfo {author} {\bibfnamefont {P.}~\bibnamefont
  {Bloom}},\ }\bibfield  {title} {\bibinfo {title} {Two-dimensional {Fermi}
  gas},\ }\href {https://link.aps.org/doi/10.1103/PhysRevB.12.125} {\bibfield
  {journal} {\bibinfo  {journal} {Phys. Rev. B}\ }\textbf {\bibinfo {volume}
  {12}},\ \bibinfo {pages} {125} (\bibinfo {year} {1975})}\BibitemShut
  {NoStop}%
\bibitem [{\citenamefont {Jungwirth}\ and\ \citenamefont
  {MacDonald}(1996)}]{jungwirth:1996}%
  \BibitemOpen
  \bibfield  {author} {\bibinfo {author} {\bibfnamefont {T.}~\bibnamefont
  {Jungwirth}}\ and\ \bibinfo {author} {\bibfnamefont {A.~H.}\ \bibnamefont
  {MacDonald}},\ }\bibfield  {title} {\bibinfo {title} {Electron-electron
  interactions and two-dimensional--two-dimensional tunneling},\ }\href
  {https://link.aps.org/doi/10.1103/PhysRevB.53.7403} {\bibfield  {journal}
  {\bibinfo  {journal} {Phys. Rev. B}\ }\textbf {\bibinfo {volume} {53}},\
  \bibinfo {pages} {7403} (\bibinfo {year} {1996})}\BibitemShut {NoStop}%
\bibitem [{\citenamefont {Abramovitch}\ \emph {et~al.}(2024)\citenamefont
  {Abramovitch}, \citenamefont {Mravlje}, \citenamefont {Zhou}, \citenamefont
  {Georges},\ and\ \citenamefont {Bernardi}}]{abramovitch:2024}%
  \BibitemOpen
  \bibfield  {author} {\bibinfo {author} {\bibfnamefont {D.~J.}\ \bibnamefont
  {Abramovitch}}, \bibinfo {author} {\bibfnamefont {J.}~\bibnamefont
  {Mravlje}}, \bibinfo {author} {\bibfnamefont {J.-J.}\ \bibnamefont {Zhou}},
  \bibinfo {author} {\bibfnamefont {A.}~\bibnamefont {Georges}},\ and\ \bibinfo
  {author} {\bibfnamefont {M.}~\bibnamefont {Bernardi}},\ }\bibfield  {title}
  {\bibinfo {title} {Respective roles of electron-phonon and electron-electron
  interactions in the transport and quasiparticle properties of {SrVO$_3$}},\
  }\href {https://link.aps.org/doi/10.1103/PhysRevLett.133.186501} {\bibfield
  {journal} {\bibinfo  {journal} {Phys. Rev. Lett.}\ }\textbf {\bibinfo
  {volume} {133}},\ \bibinfo {pages} {186501} (\bibinfo {year}
  {2024})}\BibitemShut {NoStop}%
\bibitem [{\citenamefont {Cappelli}\ \emph {et~al.}(2022)\citenamefont
  {Cappelli}, \citenamefont {Hampel}, \citenamefont {Chikina}, \citenamefont
  {Guedes}, \citenamefont {Gatti}, \citenamefont {Hunter}, \citenamefont
  {Issing}, \citenamefont {Biskup}, \citenamefont {Varela}, \citenamefont
  {Dreyer}, \citenamefont {Tamai}, \citenamefont {Georges}, \citenamefont
  {Bruno}, \citenamefont {Radovi{\'c}},\ and\ \citenamefont
  {Baumberger}}]{cappelli:2022-SM}%
  \BibitemOpen
  \bibfield  {author} {\bibinfo {author} {\bibfnamefont {E.}~\bibnamefont
  {Cappelli}}, \bibinfo {author} {\bibfnamefont {A.}~\bibnamefont {Hampel}},
  \bibinfo {author} {\bibfnamefont {A.}~\bibnamefont {Chikina}}, \bibinfo
  {author} {\bibfnamefont {E.~B.}\ \bibnamefont {Guedes}}, \bibinfo {author}
  {\bibfnamefont {G.}~\bibnamefont {Gatti}}, \bibinfo {author} {\bibfnamefont
  {A.}~\bibnamefont {Hunter}}, \bibinfo {author} {\bibfnamefont
  {J.}~\bibnamefont {Issing}}, \bibinfo {author} {\bibfnamefont
  {N.}~\bibnamefont {Biskup}}, \bibinfo {author} {\bibfnamefont
  {M.}~\bibnamefont {Varela}}, \bibinfo {author} {\bibfnamefont {C.~E.}\
  \bibnamefont {Dreyer}}, \bibinfo {author} {\bibfnamefont {A.}~\bibnamefont
  {Tamai}}, \bibinfo {author} {\bibfnamefont {A.}~\bibnamefont {Georges}},
  \bibinfo {author} {\bibfnamefont {F.~Y.}\ \bibnamefont {Bruno}}, \bibinfo
  {author} {\bibfnamefont {M.}~\bibnamefont {Radovi{\'c}}},\ and\ \bibinfo
  {author} {\bibfnamefont {F.}~\bibnamefont {Baumberger}},\ }\bibfield  {title}
  {\bibinfo {title} {Electronic structure of the highly conductive perovskite
  oxide {{SrMoO}}{\textsubscript{3}}},\ }\href
  {https://link.aps.org/doi/10.1103/PhysRevMaterials.6.075002} {\bibfield
  {journal} {\bibinfo  {journal} {Phys. Rev. Mater.}\ }\textbf {\bibinfo
  {volume} {6}},\ \bibinfo {pages} {075002} (\bibinfo {year}
  {2022})}\BibitemShut {NoStop}%
\bibitem [{\citenamefont {Behnia}(2022)}]{behnia:2022}%
  \BibitemOpen
  \bibfield  {author} {\bibinfo {author} {\bibfnamefont {K.}~\bibnamefont
  {Behnia}},\ }\bibfield  {title} {\bibinfo {title} {On the origin and the
  amplitude of {$T$}-square resistivity in {Fermi} liquids},\ }\href
  {https://onlinelibrary.wiley.com/doi/10.1002/andp.202100588} {\bibfield
  {journal} {\bibinfo  {journal} {Ann. Phys.}\ }\textbf {\bibinfo {volume}
  {534}},\ \bibinfo {pages} {2100588} (\bibinfo {year} {2022})}\BibitemShut
  {NoStop}%
\bibitem [{\citenamefont {Stricker}\ \emph {et~al.}(2014)\citenamefont
  {Stricker}, \citenamefont {Mravlje}, \citenamefont {Berthod}, \citenamefont
  {Fittipaldi}, \citenamefont {Vecchione}, \citenamefont {Georges},\ and\
  \citenamefont {{van der Marel}}}]{stricker:2014}%
  \BibitemOpen
  \bibfield  {author} {\bibinfo {author} {\bibfnamefont {D.}~\bibnamefont
  {Stricker}}, \bibinfo {author} {\bibfnamefont {J.}~\bibnamefont {Mravlje}},
  \bibinfo {author} {\bibfnamefont {C.}~\bibnamefont {Berthod}}, \bibinfo
  {author} {\bibfnamefont {R.}~\bibnamefont {Fittipaldi}}, \bibinfo {author}
  {\bibfnamefont {A.}~\bibnamefont {Vecchione}}, \bibinfo {author}
  {\bibfnamefont {A.}~\bibnamefont {Georges}},\ and\ \bibinfo {author}
  {\bibfnamefont {D.}~\bibnamefont {{van der Marel}}},\ }\bibfield  {title}
  {\bibinfo {title} {Optical response of {Sr$_2$RuO$_4$} reveals universal
  {Fermi}-liquid scaling and quasiparticles beyond {Landau} theory},\ }\href
  {https://link.aps.org/doi/10.1103/PhysRevLett.113.087404} {\bibfield
  {journal} {\bibinfo  {journal} {Phys. Rev. Lett.}\ }\textbf {\bibinfo
  {volume} {113}},\ \bibinfo {pages} {087404} (\bibinfo {year}
  {2014})}\BibitemShut {NoStop}%
\bibitem [{\citenamefont {Jacko}\ \emph {et~al.}(2009)\citenamefont {Jacko},
  \citenamefont {Fj{\ae}restad},\ and\ \citenamefont {Powell}}]{jacko:2009}%
  \BibitemOpen
  \bibfield  {author} {\bibinfo {author} {\bibfnamefont {A.~C.}\ \bibnamefont
  {Jacko}}, \bibinfo {author} {\bibfnamefont {J.~O.}\ \bibnamefont
  {Fj{\ae}restad}},\ and\ \bibinfo {author} {\bibfnamefont {B.~J.}\
  \bibnamefont {Powell}},\ }\bibfield  {title} {\bibinfo {title} {A unified
  explanation of the {{Kadowaki}}--{{Woods}} ratio in strongly correlated
  metals},\ }\href {https://www.nature.com/articles/nphys1249} {\bibfield
  {journal} {\bibinfo  {journal} {Nat. Phys.}\ }\textbf {\bibinfo {volume}
  {5}},\ \bibinfo {pages} {422} (\bibinfo {year} {2009})}\BibitemShut {NoStop}%
\bibitem [{\citenamefont {Chubukov}\ and\ \citenamefont
  {Maslov}(2012)}]{chubukov:2012-SM}%
  \BibitemOpen
  \bibfield  {author} {\bibinfo {author} {\bibfnamefont {A.~V.}\ \bibnamefont
  {Chubukov}}\ and\ \bibinfo {author} {\bibfnamefont {D.~L.}\ \bibnamefont
  {Maslov}},\ }\bibfield  {title} {\bibinfo {title}
  {First-{Matsubara}-frequency rule in a {Fermi} liquid. {I}. {F}ermionic
  self-energy},\ }\href {https://link.aps.org/doi/10.1103/PhysRevB.86.155136}
  {\bibfield  {journal} {\bibinfo  {journal} {Phys. Rev. B}\ }\textbf {\bibinfo
  {volume} {86}},\ \bibinfo {pages} {155136} (\bibinfo {year}
  {2012})}\BibitemShut {NoStop}%
\bibitem [{\citenamefont {Bryan}(1990)}]{bryan:1990-SM}%
  \BibitemOpen
  \bibfield  {author} {\bibinfo {author} {\bibfnamefont {R.~K.}\ \bibnamefont
  {Bryan}},\ }\bibfield  {title} {\bibinfo {title} {Maximum entropy analysis of
  oversampled data problems},\ }\href
  {http://link.springer.com/10.1007/BF02427376} {\bibfield  {journal} {\bibinfo
   {journal} {Eur. Biophys. J.}\ }\textbf {\bibinfo {volume} {18}},\ \bibinfo
  {pages} {165} (\bibinfo {year} {1990})}\BibitemShut {NoStop}%
\bibitem [{\citenamefont {Shi}\ \emph {et~al.}(2004)\citenamefont {Shi},
  \citenamefont {Tang}, \citenamefont {Wu}, \citenamefont {Sprunger},
  \citenamefont {Yang}, \citenamefont {Brouet}, \citenamefont {Zhou},
  \citenamefont {Hussain}, \citenamefont {Shen}, \citenamefont {Zhang},\ and\
  \citenamefont {Plummer}}]{shi:2004-SM}%
  \BibitemOpen
  \bibfield  {author} {\bibinfo {author} {\bibfnamefont {J.}~\bibnamefont
  {Shi}}, \bibinfo {author} {\bibfnamefont {S.-J.}\ \bibnamefont {Tang}},
  \bibinfo {author} {\bibfnamefont {B.}~\bibnamefont {Wu}}, \bibinfo {author}
  {\bibfnamefont {P.~T.}\ \bibnamefont {Sprunger}}, \bibinfo {author}
  {\bibfnamefont {W.~L.}\ \bibnamefont {Yang}}, \bibinfo {author}
  {\bibfnamefont {V.}~\bibnamefont {Brouet}}, \bibinfo {author} {\bibfnamefont
  {X.~J.}\ \bibnamefont {Zhou}}, \bibinfo {author} {\bibfnamefont
  {Z.}~\bibnamefont {Hussain}}, \bibinfo {author} {\bibfnamefont {Z.-X.}\
  \bibnamefont {Shen}}, \bibinfo {author} {\bibfnamefont {Z.}~\bibnamefont
  {Zhang}},\ and\ \bibinfo {author} {\bibfnamefont {E.~W.}\ \bibnamefont
  {Plummer}},\ }\bibfield  {title} {\bibinfo {title} {Direct extraction of the
  {Eliashberg} function for electron-phonon coupling: {A} case study of
  {Be(10{\=1}0)}},\ }\href
  {https://link.aps.org/doi/10.1103/PhysRevLett.92.186401} {\bibfield
  {journal} {\bibinfo  {journal} {Phys. Rev. Lett.}\ }\textbf {\bibinfo
  {volume} {92}},\ \bibinfo {pages} {186401} (\bibinfo {year}
  {2004})}\BibitemShut {NoStop}%
\bibitem [{\citenamefont {Bok}\ \emph {et~al.}(2010)\citenamefont {Bok},
  \citenamefont {Yun}, \citenamefont {Choi}, \citenamefont {Zhang},
  \citenamefont {Zhou},\ and\ \citenamefont {Varma}}]{bok:2010-SM}%
  \BibitemOpen
  \bibfield  {author} {\bibinfo {author} {\bibfnamefont {J.~M.}\ \bibnamefont
  {Bok}}, \bibinfo {author} {\bibfnamefont {J.~H.}\ \bibnamefont {Yun}},
  \bibinfo {author} {\bibfnamefont {H.-Y.}\ \bibnamefont {Choi}}, \bibinfo
  {author} {\bibfnamefont {W.}~\bibnamefont {Zhang}}, \bibinfo {author}
  {\bibfnamefont {X.~J.}\ \bibnamefont {Zhou}},\ and\ \bibinfo {author}
  {\bibfnamefont {C.~M.}\ \bibnamefont {Varma}},\ }\bibfield  {title} {\bibinfo
  {title} {Momentum dependence of the single-particle self-energy and
  fluctuation spectrum of slightly underdoped
  {Bi$_2$Sr$_2$CaCu$_2$O$_{8+\delta}$} from high-resolution laser
  angle-resolved photoemission},\ }\href
  {https://link.aps.org/doi/10.1103/PhysRevB.81.174516} {\bibfield  {journal}
  {\bibinfo  {journal} {Phys. Rev. B}\ }\textbf {\bibinfo {volume} {81}},\
  \bibinfo {pages} {174516} (\bibinfo {year} {2010})}\BibitemShut {NoStop}%
\bibitem [{\citenamefont {Grimvall}(1981)}]{grimvall:1981-SM}%
  \BibitemOpen
  \bibfield  {author} {\bibinfo {author} {\bibfnamefont {G.}~\bibnamefont
  {Grimvall}},\ }\href@noop {} {\emph {\bibinfo {title} {The Electron--Phonon
  Interaction in Metals}}}\ (\bibinfo  {publisher} {North-Holland, Amsterdam},\
  \bibinfo {year} {1981})\BibitemShut {NoStop}%
\bibitem [{\citenamefont {Nelder}\ and\ \citenamefont
  {Mead}(1965)}]{nelder:1965}%
  \BibitemOpen
  \bibfield  {author} {\bibinfo {author} {\bibfnamefont {J.~A.}\ \bibnamefont
  {Nelder}}\ and\ \bibinfo {author} {\bibfnamefont {R.}~\bibnamefont {Mead}},\
  }\bibfield  {title} {\bibinfo {title} {A simplex method for function
  minimization},\ }\href
  {https://academic.oup.com/comjnl/article-lookup/doi/10.1093/comjnl/7.4.308}
  {\bibfield  {journal} {\bibinfo  {journal} {Comput. J.}\ }\textbf {\bibinfo
  {volume} {7}},\ \bibinfo {pages} {308} (\bibinfo {year} {1965})}\BibitemShut
  {NoStop}%
\bibitem [{\citenamefont {Sokolovi{\'c}}\ \emph {et~al.}(2025)\citenamefont
  {Sokolovi{\'c}}, \citenamefont {Guedes}, \citenamefont {Van~Waas},
  \citenamefont {Guo}, \citenamefont {Ponc{\'e}}, \citenamefont {Polley},
  \citenamefont {Schmid}, \citenamefont {Diebold}, \citenamefont {Radovi{\'c}},
  \citenamefont {Setv{\'i}n},\ and\ \citenamefont {Dil}}]{sokolovic:2025-SM}%
  \BibitemOpen
  \bibfield  {author} {\bibinfo {author} {\bibfnamefont {I.}~\bibnamefont
  {Sokolovi{\'c}}}, \bibinfo {author} {\bibfnamefont {E.~B.}\ \bibnamefont
  {Guedes}}, \bibinfo {author} {\bibfnamefont {T.~P.}\ \bibnamefont
  {Van~Waas}}, \bibinfo {author} {\bibfnamefont {F.}~\bibnamefont {Guo}},
  \bibinfo {author} {\bibfnamefont {S.}~\bibnamefont {Ponc{\'e}}}, \bibinfo
  {author} {\bibfnamefont {C.}~\bibnamefont {Polley}}, \bibinfo {author}
  {\bibfnamefont {M.}~\bibnamefont {Schmid}}, \bibinfo {author} {\bibfnamefont
  {U.}~\bibnamefont {Diebold}}, \bibinfo {author} {\bibfnamefont
  {M.}~\bibnamefont {Radovi{\'c}}}, \bibinfo {author} {\bibfnamefont
  {M.}~\bibnamefont {Setv{\'i}n}},\ and\ \bibinfo {author} {\bibfnamefont
  {J.~H.}\ \bibnamefont {Dil}},\ }\bibfield  {title} {\bibinfo {title} {Duality
  and degeneracy lifting in two-dimensional electron liquids on
  {{SrTiO}}{\textsubscript{3}}(001)},\ }\href
  {https://www.nature.com/articles/s41467-025-59258-4} {\bibfield  {journal}
  {\bibinfo  {journal} {Nat. Commun.}\ }\textbf {\bibinfo {volume} {16}},\
  \bibinfo {pages} {4594} (\bibinfo {year} {2025})}\BibitemShut {NoStop}%
\bibitem [{\citenamefont {Abrikosov}\ \emph {et~al.}(1964)\citenamefont
  {Abrikosov}, \citenamefont {Gorkov}, \citenamefont {Dzyaloshinski},
  \citenamefont {Silverman},\ and\ \citenamefont {Weiss}}]{abrikosov:1964-SM}%
  \BibitemOpen
  \bibfield  {author} {\bibinfo {author} {\bibfnamefont {A.~A.}\ \bibnamefont
  {Abrikosov}}, \bibinfo {author} {\bibfnamefont {L.~P.}\ \bibnamefont
  {Gorkov}}, \bibinfo {author} {\bibfnamefont {I.~E.}\ \bibnamefont
  {Dzyaloshinski}}, \bibinfo {author} {\bibfnamefont {R.~A.}\ \bibnamefont
  {Silverman}},\ and\ \bibinfo {author} {\bibfnamefont {G.~H.}\ \bibnamefont
  {Weiss}},\ }\href {https://doi.org/10.1063/1.3051555} {\emph {\bibinfo
  {title} {Methods of Quantum Field Theory in Statistical Physics}}}\ (\bibinfo
   {publisher} {Pergamon Press Ltd., Oxford},\ \bibinfo {year}
  {1964})\BibitemShut {NoStop}%
\bibitem [{\citenamefont {Gull}(1989)}]{gull:1989-SM}%
  \BibitemOpen
  \bibfield  {author} {\bibinfo {author} {\bibfnamefont {S.~F.}\ \bibnamefont
  {Gull}},\ }\bibfield  {title} {\bibinfo {title} {Developments in maximum
  entropy data analysis},\ }in\ \href
  {https://link.springer.com/chapter/10.1007/978-94-015-7860-8_4} {\emph
  {\bibinfo {booktitle} {Developments in Maximum Entropy Data Analysis}}},\
  \bibinfo {series and number} {Maximum {{Entropy}} and {{Bayesian Methods}}},\
  \bibinfo {editor} {edited by\ \bibinfo {editor} {\bibfnamefont
  {J.}~\bibnamefont {Skilling}}}\ (\bibinfo  {publisher} {Springer,
  Dordrecht},\ \bibinfo {year} {1989})\ pp.\ \bibinfo {pages}
  {53--71}\BibitemShut {NoStop}%
\bibitem [{\citenamefont {Usachov}\ \emph {et~al.}(2018)\citenamefont
  {Usachov}, \citenamefont {Fedorov}, \citenamefont {Vilkov}, \citenamefont
  {Ogorodnikov}, \citenamefont {Kuznetsov}, \citenamefont {Gr{\"u}neis},
  \citenamefont {Laubschat},\ and\ \citenamefont {Vyalikh}}]{usachov:2018-SM}%
  \BibitemOpen
  \bibfield  {author} {\bibinfo {author} {\bibfnamefont {D.~Y.}\ \bibnamefont
  {Usachov}}, \bibinfo {author} {\bibfnamefont {A.~V.}\ \bibnamefont
  {Fedorov}}, \bibinfo {author} {\bibfnamefont {O.~Y.}\ \bibnamefont {Vilkov}},
  \bibinfo {author} {\bibfnamefont {I.~I.}\ \bibnamefont {Ogorodnikov}},
  \bibinfo {author} {\bibfnamefont {M.~V.}\ \bibnamefont {Kuznetsov}}, \bibinfo
  {author} {\bibfnamefont {A.}~\bibnamefont {Gr{\"u}neis}}, \bibinfo {author}
  {\bibfnamefont {C.}~\bibnamefont {Laubschat}},\ and\ \bibinfo {author}
  {\bibfnamefont {D.~V.}\ \bibnamefont {Vyalikh}},\ }\bibfield  {title}
  {\bibinfo {title} {Electron-phonon coupling in graphene placed between
  magnetic {{Li}} and {{Si}} layers on cobalt},\ }\href
  {https://link.aps.org/doi/10.1103/PhysRevB.97.085132} {\bibfield  {journal}
  {\bibinfo  {journal} {Phys. Rev. B}\ }\textbf {\bibinfo {volume} {97}},\
  \bibinfo {pages} {085132} (\bibinfo {year} {2018})}\BibitemShut {NoStop}%
\bibitem [{\citenamefont {Zhang}\ \emph {et~al.}(2022)\citenamefont {Zhang},
  \citenamefont {Pincelli}, \citenamefont {Jozwiak}, \citenamefont {Kondo},
  \citenamefont {Ernstorfer}, \citenamefont {Sato},\ and\ \citenamefont
  {Zhou}}]{zhang:2022-SM}%
  \BibitemOpen
  \bibfield  {author} {\bibinfo {author} {\bibfnamefont {H.}~\bibnamefont
  {Zhang}}, \bibinfo {author} {\bibfnamefont {T.}~\bibnamefont {Pincelli}},
  \bibinfo {author} {\bibfnamefont {C.}~\bibnamefont {Jozwiak}}, \bibinfo
  {author} {\bibfnamefont {T.}~\bibnamefont {Kondo}}, \bibinfo {author}
  {\bibfnamefont {R.}~\bibnamefont {Ernstorfer}}, \bibinfo {author}
  {\bibfnamefont {T.}~\bibnamefont {Sato}},\ and\ \bibinfo {author}
  {\bibfnamefont {S.}~\bibnamefont {Zhou}},\ }\bibfield  {title} {\bibinfo
  {title} {Angle-resolved photoemission spectroscopy},\ }\href
  {https://www.nature.com/articles/s43586-022-00133-7} {\bibfield  {journal}
  {\bibinfo  {journal} {Nat. Rev. Methods Primers}\ }\textbf {\bibinfo {volume}
  {2}},\ \bibinfo {pages} {54} (\bibinfo {year} {2022})}\BibitemShut {NoStop}%
\bibitem [{\citenamefont {Gonze}\ \emph {et~al.}(2020)\citenamefont {Gonze},
  \citenamefont {Amadon}, \citenamefont {Antonius}, \citenamefont {Arnardi},
  \citenamefont {Baguet}, \citenamefont {Beuken}, \citenamefont {Bieder},
  \citenamefont {Bottin}, \citenamefont {Bouchet}, \citenamefont {Bousquet},
  \citenamefont {Brouwer}, \citenamefont {Bruneval}, \citenamefont {Brunin},
  \citenamefont {Cavignac}, \citenamefont {Charraud}, \citenamefont {Chen},
  \citenamefont {Côté}, \citenamefont {Cottenier}, \citenamefont {Denier},
  \citenamefont {Geneste}, \citenamefont {Ghosez}, \citenamefont {Giantomassi},
  \citenamefont {Gillet}, \citenamefont {Gingras}, \citenamefont {Hamann},
  \citenamefont {Hautier}, \citenamefont {He}, \citenamefont {Helbig},
  \citenamefont {Holzwarth}, \citenamefont {Jia}, \citenamefont {Jollet},
  \citenamefont {Lafargue-Dit-Hauret}, \citenamefont {Lejaeghere},
  \citenamefont {Marques}, \citenamefont {Martin}, \citenamefont {Martins},
  \citenamefont {Miranda}, \citenamefont {Naccarato}, \citenamefont {Persson},
  \citenamefont {Petretto}, \citenamefont {Planes}, \citenamefont {Pouillon},
  \citenamefont {Prokhorenko}, \citenamefont {Ricci}, \citenamefont
  {Rignanese}, \citenamefont {Romero}, \citenamefont {Schmitt}, \citenamefont
  {Torrent}, \citenamefont {van Setten}, \citenamefont {Van~Troeye},
  \citenamefont {Verstraete}, \citenamefont {Zérah},\ and\ \citenamefont
  {Zwanziger}}]{gonze:2020-SM}%
  \BibitemOpen
  \bibfield  {author} {\bibinfo {author} {\bibfnamefont {X.}~\bibnamefont
  {Gonze}}, \bibinfo {author} {\bibfnamefont {B.}~\bibnamefont {Amadon}},
  \bibinfo {author} {\bibfnamefont {G.}~\bibnamefont {Antonius}}, \bibinfo
  {author} {\bibfnamefont {F.}~\bibnamefont {Arnardi}}, \bibinfo {author}
  {\bibfnamefont {L.}~\bibnamefont {Baguet}}, \bibinfo {author} {\bibfnamefont
  {J.-M.}\ \bibnamefont {Beuken}}, \bibinfo {author} {\bibfnamefont
  {J.}~\bibnamefont {Bieder}}, \bibinfo {author} {\bibfnamefont
  {F.}~\bibnamefont {Bottin}}, \bibinfo {author} {\bibfnamefont
  {J.}~\bibnamefont {Bouchet}}, \bibinfo {author} {\bibfnamefont
  {E.}~\bibnamefont {Bousquet}}, \bibinfo {author} {\bibfnamefont
  {N.}~\bibnamefont {Brouwer}}, \bibinfo {author} {\bibfnamefont
  {F.}~\bibnamefont {Bruneval}}, \bibinfo {author} {\bibfnamefont
  {G.}~\bibnamefont {Brunin}}, \bibinfo {author} {\bibfnamefont
  {T.}~\bibnamefont {Cavignac}}, \bibinfo {author} {\bibfnamefont {J.-B.}\
  \bibnamefont {Charraud}}, \bibinfo {author} {\bibfnamefont {W.}~\bibnamefont
  {Chen}}, \bibinfo {author} {\bibfnamefont {M.}~\bibnamefont {Côté}},
  \bibinfo {author} {\bibfnamefont {S.}~\bibnamefont {Cottenier}}, \bibinfo
  {author} {\bibfnamefont {J.}~\bibnamefont {Denier}}, \bibinfo {author}
  {\bibfnamefont {G.}~\bibnamefont {Geneste}}, \bibinfo {author} {\bibfnamefont
  {P.}~\bibnamefont {Ghosez}}, \bibinfo {author} {\bibfnamefont
  {M.}~\bibnamefont {Giantomassi}}, \bibinfo {author} {\bibfnamefont
  {Y.}~\bibnamefont {Gillet}}, \bibinfo {author} {\bibfnamefont
  {O.}~\bibnamefont {Gingras}}, \bibinfo {author} {\bibfnamefont {D.~R.}\
  \bibnamefont {Hamann}}, \bibinfo {author} {\bibfnamefont {G.}~\bibnamefont
  {Hautier}}, \bibinfo {author} {\bibfnamefont {X.}~\bibnamefont {He}},
  \bibinfo {author} {\bibfnamefont {N.}~\bibnamefont {Helbig}}, \bibinfo
  {author} {\bibfnamefont {N.}~\bibnamefont {Holzwarth}}, \bibinfo {author}
  {\bibfnamefont {Y.}~\bibnamefont {Jia}}, \bibinfo {author} {\bibfnamefont
  {F.}~\bibnamefont {Jollet}}, \bibinfo {author} {\bibfnamefont
  {W.}~\bibnamefont {Lafargue-Dit-Hauret}}, \bibinfo {author} {\bibfnamefont
  {K.}~\bibnamefont {Lejaeghere}}, \bibinfo {author} {\bibfnamefont {M.~A.}\
  \bibnamefont {Marques}}, \bibinfo {author} {\bibfnamefont {A.}~\bibnamefont
  {Martin}}, \bibinfo {author} {\bibfnamefont {C.}~\bibnamefont {Martins}},
  \bibinfo {author} {\bibfnamefont {H.~P.}\ \bibnamefont {Miranda}}, \bibinfo
  {author} {\bibfnamefont {F.}~\bibnamefont {Naccarato}}, \bibinfo {author}
  {\bibfnamefont {K.}~\bibnamefont {Persson}}, \bibinfo {author} {\bibfnamefont
  {G.}~\bibnamefont {Petretto}}, \bibinfo {author} {\bibfnamefont
  {V.}~\bibnamefont {Planes}}, \bibinfo {author} {\bibfnamefont
  {Y.}~\bibnamefont {Pouillon}}, \bibinfo {author} {\bibfnamefont
  {S.}~\bibnamefont {Prokhorenko}}, \bibinfo {author} {\bibfnamefont
  {F.}~\bibnamefont {Ricci}}, \bibinfo {author} {\bibfnamefont {G.-M.}\
  \bibnamefont {Rignanese}}, \bibinfo {author} {\bibfnamefont {A.~H.}\
  \bibnamefont {Romero}}, \bibinfo {author} {\bibfnamefont {M.~M.}\
  \bibnamefont {Schmitt}}, \bibinfo {author} {\bibfnamefont {M.}~\bibnamefont
  {Torrent}}, \bibinfo {author} {\bibfnamefont {M.~J.}\ \bibnamefont {van
  Setten}}, \bibinfo {author} {\bibfnamefont {B.}~\bibnamefont {Van~Troeye}},
  \bibinfo {author} {\bibfnamefont {M.~J.}\ \bibnamefont {Verstraete}},
  \bibinfo {author} {\bibfnamefont {G.}~\bibnamefont {Zérah}},\ and\ \bibinfo
  {author} {\bibfnamefont {J.~W.}\ \bibnamefont {Zwanziger}},\ }\bibfield
  {title} {\bibinfo {title} {The \textsc{Abinit} project: Impact, environment
  and recent developments},\ }\href
  {https://linkinghub.elsevier.com/retrieve/pii/S0010465519303741} {\bibfield
  {journal} {\bibinfo  {journal} {Comput. Phys. Commun.}\ }\textbf {\bibinfo
  {volume} {248}},\ \bibinfo {pages} {107042} (\bibinfo {year}
  {2020})}\BibitemShut {NoStop}%
\bibitem [{\citenamefont {Romero}\ \emph {et~al.}(2020)\citenamefont {Romero},
  \citenamefont {Allan}, \citenamefont {Amadon}, \citenamefont {Antonius},
  \citenamefont {Applencourt}, \citenamefont {Baguet}, \citenamefont {Bieder},
  \citenamefont {Bottin}, \citenamefont {Bouchet}, \citenamefont {Bousquet},
  \citenamefont {Bruneval}, \citenamefont {Brunin}, \citenamefont {Caliste},
  \citenamefont {Côté}, \citenamefont {Denier}, \citenamefont {Dreyer},
  \citenamefont {Ghosez}, \citenamefont {Giantomassi}, \citenamefont {Gillet},
  \citenamefont {Gingras}, \citenamefont {Hamann}, \citenamefont {Hautier},
  \citenamefont {Jollet}, \citenamefont {Jomard}, \citenamefont {Martin},
  \citenamefont {Miranda}, \citenamefont {Naccarato}, \citenamefont {Petretto},
  \citenamefont {Pike}, \citenamefont {Planes}, \citenamefont {Prokhorenko},
  \citenamefont {Rangel}, \citenamefont {Ricci}, \citenamefont {Rignanese},
  \citenamefont {Royo}, \citenamefont {Stengel}, \citenamefont {Torrent},
  \citenamefont {Van~Setten}, \citenamefont {Van~Troeye}, \citenamefont
  {Verstraete}, \citenamefont {Wiktor}, \citenamefont {Zwanziger},\ and\
  \citenamefont {Gonze}}]{romero:2020-SM}%
  \BibitemOpen
  \bibfield  {author} {\bibinfo {author} {\bibfnamefont {A.~H.}\ \bibnamefont
  {Romero}}, \bibinfo {author} {\bibfnamefont {D.~C.}\ \bibnamefont {Allan}},
  \bibinfo {author} {\bibfnamefont {B.}~\bibnamefont {Amadon}}, \bibinfo
  {author} {\bibfnamefont {G.}~\bibnamefont {Antonius}}, \bibinfo {author}
  {\bibfnamefont {T.}~\bibnamefont {Applencourt}}, \bibinfo {author}
  {\bibfnamefont {L.}~\bibnamefont {Baguet}}, \bibinfo {author} {\bibfnamefont
  {J.}~\bibnamefont {Bieder}}, \bibinfo {author} {\bibfnamefont
  {F.}~\bibnamefont {Bottin}}, \bibinfo {author} {\bibfnamefont
  {J.}~\bibnamefont {Bouchet}}, \bibinfo {author} {\bibfnamefont
  {E.}~\bibnamefont {Bousquet}}, \bibinfo {author} {\bibfnamefont
  {F.}~\bibnamefont {Bruneval}}, \bibinfo {author} {\bibfnamefont
  {G.}~\bibnamefont {Brunin}}, \bibinfo {author} {\bibfnamefont
  {D.}~\bibnamefont {Caliste}}, \bibinfo {author} {\bibfnamefont
  {M.}~\bibnamefont {Côté}}, \bibinfo {author} {\bibfnamefont
  {J.}~\bibnamefont {Denier}}, \bibinfo {author} {\bibfnamefont
  {C.}~\bibnamefont {Dreyer}}, \bibinfo {author} {\bibfnamefont
  {P.}~\bibnamefont {Ghosez}}, \bibinfo {author} {\bibfnamefont
  {M.}~\bibnamefont {Giantomassi}}, \bibinfo {author} {\bibfnamefont
  {Y.}~\bibnamefont {Gillet}}, \bibinfo {author} {\bibfnamefont
  {O.}~\bibnamefont {Gingras}}, \bibinfo {author} {\bibfnamefont {D.~R.}\
  \bibnamefont {Hamann}}, \bibinfo {author} {\bibfnamefont {G.}~\bibnamefont
  {Hautier}}, \bibinfo {author} {\bibfnamefont {F.}~\bibnamefont {Jollet}},
  \bibinfo {author} {\bibfnamefont {G.}~\bibnamefont {Jomard}}, \bibinfo
  {author} {\bibfnamefont {A.}~\bibnamefont {Martin}}, \bibinfo {author}
  {\bibfnamefont {H.~P.~C.}\ \bibnamefont {Miranda}}, \bibinfo {author}
  {\bibfnamefont {F.}~\bibnamefont {Naccarato}}, \bibinfo {author}
  {\bibfnamefont {G.}~\bibnamefont {Petretto}}, \bibinfo {author}
  {\bibfnamefont {N.~A.}\ \bibnamefont {Pike}}, \bibinfo {author}
  {\bibfnamefont {V.}~\bibnamefont {Planes}}, \bibinfo {author} {\bibfnamefont
  {S.}~\bibnamefont {Prokhorenko}}, \bibinfo {author} {\bibfnamefont
  {T.}~\bibnamefont {Rangel}}, \bibinfo {author} {\bibfnamefont
  {F.}~\bibnamefont {Ricci}}, \bibinfo {author} {\bibfnamefont {G.-M.}\
  \bibnamefont {Rignanese}}, \bibinfo {author} {\bibfnamefont {M.}~\bibnamefont
  {Royo}}, \bibinfo {author} {\bibfnamefont {M.}~\bibnamefont {Stengel}},
  \bibinfo {author} {\bibfnamefont {M.}~\bibnamefont {Torrent}}, \bibinfo
  {author} {\bibfnamefont {M.~J.}\ \bibnamefont {Van~Setten}}, \bibinfo
  {author} {\bibfnamefont {B.}~\bibnamefont {Van~Troeye}}, \bibinfo {author}
  {\bibfnamefont {M.~J.}\ \bibnamefont {Verstraete}}, \bibinfo {author}
  {\bibfnamefont {J.}~\bibnamefont {Wiktor}}, \bibinfo {author} {\bibfnamefont
  {J.~W.}\ \bibnamefont {Zwanziger}},\ and\ \bibinfo {author} {\bibfnamefont
  {X.}~\bibnamefont {Gonze}},\ }\bibfield  {title} {\bibinfo {title}
  {\textsc{abinit}: Overview and focus on selected capabilities},\ }\href
  {https://pubs.aip.org/jcp/article/152/12/124102/953753/ABINIT-Overview-and-focus-on-selected-capabilities}
  {\bibfield  {journal} {\bibinfo  {journal} {J. Chem. Phys.}\ }\textbf
  {\bibinfo {volume} {152}},\ \bibinfo {pages} {124102} (\bibinfo {year}
  {2020})}\BibitemShut {NoStop}%
\bibitem [{\citenamefont {Gonze}\ \emph {et~al.}(2016)\citenamefont {Gonze},
  \citenamefont {Jollet}, \citenamefont {Abreu~Araujo}, \citenamefont {Adams},
  \citenamefont {Amadon}, \citenamefont {Applencourt}, \citenamefont {Audouze},
  \citenamefont {Beuken}, \citenamefont {Bieder}, \citenamefont {Bokhanchuk},
  \citenamefont {Bousquet}, \citenamefont {Bruneval}, \citenamefont {Caliste},
  \citenamefont {C{\^o}t{\'e}}, \citenamefont {Dahm}, \citenamefont {Da~Pieve},
  \citenamefont {Delaveau}, \citenamefont {Di~Gennaro}, \citenamefont {Dorado},
  \citenamefont {Espejo}, \citenamefont {Geneste}, \citenamefont {Genovese},
  \citenamefont {Gerossier}, \citenamefont {Giantomassi}, \citenamefont
  {Gillet}, \citenamefont {Hamann}, \citenamefont {He}, \citenamefont {Jomard},
  \citenamefont {Laflamme~Janssen}, \citenamefont {Le~Roux}, \citenamefont
  {Levitt}, \citenamefont {Lherbier}, \citenamefont {Liu}, \citenamefont
  {Luka{\v c}evi{\'c}}, \citenamefont {Martin}, \citenamefont {Martins},
  \citenamefont {Oliveira}, \citenamefont {Ponc{\'e}}, \citenamefont
  {Pouillon}, \citenamefont {Rangel}, \citenamefont {Rignanese}, \citenamefont
  {Romero}, \citenamefont {Rousseau}, \citenamefont {Rubel}, \citenamefont
  {Shukri}, \citenamefont {Stankovski}, \citenamefont {Torrent}, \citenamefont
  {Van~Setten}, \citenamefont {Van~Troeye}, \citenamefont {Verstraete},
  \citenamefont {Waroquiers}, \citenamefont {Wiktor}, \citenamefont {Xu},
  \citenamefont {Zhou},\ and\ \citenamefont {Zwanziger}}]{gonze:2016-SM}%
  \BibitemOpen
  \bibfield  {author} {\bibinfo {author} {\bibfnamefont {X.}~\bibnamefont
  {Gonze}}, \bibinfo {author} {\bibfnamefont {F.}~\bibnamefont {Jollet}},
  \bibinfo {author} {\bibfnamefont {F.}~\bibnamefont {Abreu~Araujo}}, \bibinfo
  {author} {\bibfnamefont {D.}~\bibnamefont {Adams}}, \bibinfo {author}
  {\bibfnamefont {B.}~\bibnamefont {Amadon}}, \bibinfo {author} {\bibfnamefont
  {T.}~\bibnamefont {Applencourt}}, \bibinfo {author} {\bibfnamefont
  {C.}~\bibnamefont {Audouze}}, \bibinfo {author} {\bibfnamefont {J.-M.}\
  \bibnamefont {Beuken}}, \bibinfo {author} {\bibfnamefont {J.}~\bibnamefont
  {Bieder}}, \bibinfo {author} {\bibfnamefont {A.}~\bibnamefont {Bokhanchuk}},
  \bibinfo {author} {\bibfnamefont {E.}~\bibnamefont {Bousquet}}, \bibinfo
  {author} {\bibfnamefont {F.}~\bibnamefont {Bruneval}}, \bibinfo {author}
  {\bibfnamefont {D.}~\bibnamefont {Caliste}}, \bibinfo {author} {\bibfnamefont
  {M.}~\bibnamefont {C{\^o}t{\'e}}}, \bibinfo {author} {\bibfnamefont
  {F.}~\bibnamefont {Dahm}}, \bibinfo {author} {\bibfnamefont {F.}~\bibnamefont
  {Da~Pieve}}, \bibinfo {author} {\bibfnamefont {M.}~\bibnamefont {Delaveau}},
  \bibinfo {author} {\bibfnamefont {M.}~\bibnamefont {Di~Gennaro}}, \bibinfo
  {author} {\bibfnamefont {B.}~\bibnamefont {Dorado}}, \bibinfo {author}
  {\bibfnamefont {C.}~\bibnamefont {Espejo}}, \bibinfo {author} {\bibfnamefont
  {G.}~\bibnamefont {Geneste}}, \bibinfo {author} {\bibfnamefont
  {L.}~\bibnamefont {Genovese}}, \bibinfo {author} {\bibfnamefont
  {A.}~\bibnamefont {Gerossier}}, \bibinfo {author} {\bibfnamefont
  {M.}~\bibnamefont {Giantomassi}}, \bibinfo {author} {\bibfnamefont
  {Y.}~\bibnamefont {Gillet}}, \bibinfo {author} {\bibfnamefont
  {D.}~\bibnamefont {Hamann}}, \bibinfo {author} {\bibfnamefont
  {L.}~\bibnamefont {He}}, \bibinfo {author} {\bibfnamefont {G.}~\bibnamefont
  {Jomard}}, \bibinfo {author} {\bibfnamefont {J.}~\bibnamefont
  {Laflamme~Janssen}}, \bibinfo {author} {\bibfnamefont {S.}~\bibnamefont
  {Le~Roux}}, \bibinfo {author} {\bibfnamefont {A.}~\bibnamefont {Levitt}},
  \bibinfo {author} {\bibfnamefont {A.}~\bibnamefont {Lherbier}}, \bibinfo
  {author} {\bibfnamefont {F.}~\bibnamefont {Liu}}, \bibinfo {author}
  {\bibfnamefont {I.}~\bibnamefont {Luka{\v c}evi{\'c}}}, \bibinfo {author}
  {\bibfnamefont {A.}~\bibnamefont {Martin}}, \bibinfo {author} {\bibfnamefont
  {C.}~\bibnamefont {Martins}}, \bibinfo {author} {\bibfnamefont
  {M.}~\bibnamefont {Oliveira}}, \bibinfo {author} {\bibfnamefont
  {S.}~\bibnamefont {Ponc{\'e}}}, \bibinfo {author} {\bibfnamefont
  {Y.}~\bibnamefont {Pouillon}}, \bibinfo {author} {\bibfnamefont
  {T.}~\bibnamefont {Rangel}}, \bibinfo {author} {\bibfnamefont {G.-M.}\
  \bibnamefont {Rignanese}}, \bibinfo {author} {\bibfnamefont {A.}~\bibnamefont
  {Romero}}, \bibinfo {author} {\bibfnamefont {B.}~\bibnamefont {Rousseau}},
  \bibinfo {author} {\bibfnamefont {O.}~\bibnamefont {Rubel}}, \bibinfo
  {author} {\bibfnamefont {A.}~\bibnamefont {Shukri}}, \bibinfo {author}
  {\bibfnamefont {M.}~\bibnamefont {Stankovski}}, \bibinfo {author}
  {\bibfnamefont {M.}~\bibnamefont {Torrent}}, \bibinfo {author} {\bibfnamefont
  {M.}~\bibnamefont {Van~Setten}}, \bibinfo {author} {\bibfnamefont
  {B.}~\bibnamefont {Van~Troeye}}, \bibinfo {author} {\bibfnamefont
  {M.}~\bibnamefont {Verstraete}}, \bibinfo {author} {\bibfnamefont
  {D.}~\bibnamefont {Waroquiers}}, \bibinfo {author} {\bibfnamefont
  {J.}~\bibnamefont {Wiktor}}, \bibinfo {author} {\bibfnamefont
  {B.}~\bibnamefont {Xu}}, \bibinfo {author} {\bibfnamefont {A.}~\bibnamefont
  {Zhou}},\ and\ \bibinfo {author} {\bibfnamefont {J.}~\bibnamefont
  {Zwanziger}},\ }\bibfield  {title} {\bibinfo {title} {Recent developments in
  the {{ABINIT}} software package},\ }\href
  {https://linkinghub.elsevier.com/retrieve/pii/S0010465516300923} {\bibfield
  {journal} {\bibinfo  {journal} {Comput. Phys. Commun.}\ }\textbf {\bibinfo
  {volume} {205}},\ \bibinfo {pages} {106} (\bibinfo {year}
  {2016})}\BibitemShut {NoStop}%
\bibitem [{\citenamefont {Jiang}\ \emph {et~al.}(2011)\citenamefont {Jiang},
  \citenamefont {Shimada}, \citenamefont {Hayashi}, \citenamefont {Iwasawa},
  \citenamefont {Aiura}, \citenamefont {Namatame},\ and\ \citenamefont
  {Taniguchi}}]{jiang:2011-SM}%
  \BibitemOpen
  \bibfield  {author} {\bibinfo {author} {\bibfnamefont {J.}~\bibnamefont
  {Jiang}}, \bibinfo {author} {\bibfnamefont {K.}~\bibnamefont {Shimada}},
  \bibinfo {author} {\bibfnamefont {H.}~\bibnamefont {Hayashi}}, \bibinfo
  {author} {\bibfnamefont {H.}~\bibnamefont {Iwasawa}}, \bibinfo {author}
  {\bibfnamefont {Y.}~\bibnamefont {Aiura}}, \bibinfo {author} {\bibfnamefont
  {H.}~\bibnamefont {Namatame}},\ and\ \bibinfo {author} {\bibfnamefont
  {M.}~\bibnamefont {Taniguchi}},\ }\bibfield  {title} {\bibinfo {title}
  {Coupling parameters of many-body interactions for the {{Al}}(100) surface
  state: {{A}} high-resolution angle-resolved photoemission spectroscopy
  study},\ }\href {https://link.aps.org/doi/10.1103/PhysRevB.84.155124}
  {\bibfield  {journal} {\bibinfo  {journal} {Phys. Rev. B}\ }\textbf {\bibinfo
  {volume} {84}},\ \bibinfo {pages} {155124} (\bibinfo {year}
  {2011})}\BibitemShut {NoStop}%
\bibitem [{\citenamefont {Hamann}(2013)}]{hamann:2013}%
  \BibitemOpen
  \bibfield  {author} {\bibinfo {author} {\bibfnamefont {D.~R.}\ \bibnamefont
  {Hamann}},\ }\bibfield  {title} {\bibinfo {title} {Optimized norm-conserving
  {{Vanderbilt}} pseudopotentials},\ }\href
  {https://link.aps.org/doi/10.1103/PhysRevB.88.085117} {\bibfield  {journal}
  {\bibinfo  {journal} {Phys. Rev. B}\ }\textbf {\bibinfo {volume} {88}},\
  \bibinfo {pages} {085117} (\bibinfo {year} {2013})}\BibitemShut {NoStop}%
\bibitem [{\citenamefont {Van~Setten}\ \emph {et~al.}(2018)\citenamefont
  {Van~Setten}, \citenamefont {Giantomassi}, \citenamefont {Bousquet},
  \citenamefont {Verstraete}, \citenamefont {Hamann}, \citenamefont {Gonze},\
  and\ \citenamefont {Rignanese}}]{vansetten:2018}%
  \BibitemOpen
  \bibfield  {author} {\bibinfo {author} {\bibfnamefont {M.}~\bibnamefont
  {Van~Setten}}, \bibinfo {author} {\bibfnamefont {M.}~\bibnamefont
  {Giantomassi}}, \bibinfo {author} {\bibfnamefont {E.}~\bibnamefont
  {Bousquet}}, \bibinfo {author} {\bibfnamefont {M.}~\bibnamefont
  {Verstraete}}, \bibinfo {author} {\bibfnamefont {D.}~\bibnamefont {Hamann}},
  \bibinfo {author} {\bibfnamefont {X.}~\bibnamefont {Gonze}},\ and\ \bibinfo
  {author} {\bibfnamefont {G.-M.}\ \bibnamefont {Rignanese}},\ }\bibfield
  {title} {\bibinfo {title} {The \textsc{PseudoDojo}: Training and grading a 85
  element optimized norm-conserving pseudopotential table},\ }\href
  {https://linkinghub.elsevier.com/retrieve/pii/S0010465518300250} {\bibfield
  {journal} {\bibinfo  {journal} {Comput. Phys. Commun.}\ }\textbf {\bibinfo
  {volume} {226}},\ \bibinfo {pages} {39} (\bibinfo {year} {2018})}\BibitemShut
  {NoStop}%
\bibitem [{\citenamefont {Perdew}\ \emph {et~al.}(2008)\citenamefont {Perdew},
  \citenamefont {Ruzsinszky}, \citenamefont {Csonka}, \citenamefont {Vydrov},
  \citenamefont {Scuseria}, \citenamefont {Constantin}, \citenamefont {Zhou},\
  and\ \citenamefont {Burke}}]{perdew:2008}%
  \BibitemOpen
  \bibfield  {author} {\bibinfo {author} {\bibfnamefont {J.~P.}\ \bibnamefont
  {Perdew}}, \bibinfo {author} {\bibfnamefont {A.}~\bibnamefont {Ruzsinszky}},
  \bibinfo {author} {\bibfnamefont {G.~I.}\ \bibnamefont {Csonka}}, \bibinfo
  {author} {\bibfnamefont {O.~A.}\ \bibnamefont {Vydrov}}, \bibinfo {author}
  {\bibfnamefont {G.~E.}\ \bibnamefont {Scuseria}}, \bibinfo {author}
  {\bibfnamefont {L.~A.}\ \bibnamefont {Constantin}}, \bibinfo {author}
  {\bibfnamefont {X.}~\bibnamefont {Zhou}},\ and\ \bibinfo {author}
  {\bibfnamefont {K.}~\bibnamefont {Burke}},\ }\bibfield  {title} {\bibinfo
  {title} {Restoring the density-gradient expansion for exchange in solids and
  surfaces},\ }\href {https://link.aps.org/doi/10.1103/PhysRevLett.100.136406}
  {\bibfield  {journal} {\bibinfo  {journal} {Phys. Rev. Lett.}\ }\textbf
  {\bibinfo {volume} {100}},\ \bibinfo {pages} {136406} (\bibinfo {year}
  {2008})}\BibitemShut {NoStop}%
\bibitem [{\citenamefont {Echenique}(2001)}]{echenique:2001}%
  \BibitemOpen
  \bibfield  {author} {\bibinfo {author} {\bibfnamefont {P.}~\bibnamefont
  {Echenique}},\ }\bibfield  {title} {\bibinfo {title} {Surface-state electron
  dynamics in noble metals},\ }\href
  {https://linkinghub.elsevier.com/retrieve/pii/S0079681601000296} {\bibfield
  {journal} {\bibinfo  {journal} {Prog. Surf. Sci.}\ }\textbf {\bibinfo
  {volume} {67}},\ \bibinfo {pages} {271} (\bibinfo {year} {2001})}\BibitemShut
  {NoStop}%
\bibitem [{\citenamefont {Berland}\ \emph {et~al.}(2012)\citenamefont
  {Berland}, \citenamefont {Einstein},\ and\ \citenamefont
  {Hyldgaard}}]{berland:2012}%
  \BibitemOpen
  \bibfield  {author} {\bibinfo {author} {\bibfnamefont {K.}~\bibnamefont
  {Berland}}, \bibinfo {author} {\bibfnamefont {T.~L.}\ \bibnamefont
  {Einstein}},\ and\ \bibinfo {author} {\bibfnamefont {P.}~\bibnamefont
  {Hyldgaard}},\ }\bibfield  {title} {\bibinfo {title} {Response of the
  {Shockley} surface state to an external electrical field: {A}
  density-functional theory study of {Cu}(111)},\ }\href
  {https://link.aps.org/doi/10.1103/PhysRevB.85.035427} {\bibfield  {journal}
  {\bibinfo  {journal} {Phys. Rev. B}\ }\textbf {\bibinfo {volume} {85}},\
  \bibinfo {pages} {035427} (\bibinfo {year} {2012})}\BibitemShut {NoStop}%
\bibitem [{\citenamefont {Wang}\ \emph {et~al.}(2021)\citenamefont {Wang},
  \citenamefont {Malyi}, \citenamefont {Zhao},\ and\ \citenamefont
  {Zunger}}]{wang:2021a}%
  \BibitemOpen
  \bibfield  {author} {\bibinfo {author} {\bibfnamefont {Z.}~\bibnamefont
  {Wang}}, \bibinfo {author} {\bibfnamefont {O.~I.}\ \bibnamefont {Malyi}},
  \bibinfo {author} {\bibfnamefont {X.}~\bibnamefont {Zhao}},\ and\ \bibinfo
  {author} {\bibfnamefont {A.}~\bibnamefont {Zunger}},\ }\bibfield  {title}
  {\bibinfo {title} {Mass enhancement in 3{$d$} and {$s$}-{$p$} perovskites
  from symmetry breaking},\ }\href
  {https://link.aps.org/doi/10.1103/PhysRevB.103.165110} {\bibfield  {journal}
  {\bibinfo  {journal} {Phys. Rev. B}\ }\textbf {\bibinfo {volume} {103}},\
  \bibinfo {pages} {165110} (\bibinfo {year} {2021})}\BibitemShut {NoStop}%
\bibitem [{\citenamefont {Balasubramanian}\ \emph {et~al.}(1998)\citenamefont
  {Balasubramanian}, \citenamefont {Jensen}, \citenamefont {Wu},\ and\
  \citenamefont {Hulbert}}]{balasubramanian:1998-SM}%
  \BibitemOpen
  \bibfield  {author} {\bibinfo {author} {\bibfnamefont {T.}~\bibnamefont
  {Balasubramanian}}, \bibinfo {author} {\bibfnamefont {E.}~\bibnamefont
  {Jensen}}, \bibinfo {author} {\bibfnamefont {X.~L.}\ \bibnamefont {Wu}},\
  and\ \bibinfo {author} {\bibfnamefont {S.~L.}\ \bibnamefont {Hulbert}},\
  }\bibfield  {title} {\bibinfo {title} {Large value of the electron-phonon
  coupling parameter ($\lambda$=1.15) and the possibility of surface
  superconductivity at the {Be(0001)} surface},\ }\href
  {https://link.aps.org/doi/10.1103/PhysRevB.57.R6866} {\bibfield  {journal}
  {\bibinfo  {journal} {Phys. Rev. B}\ }\textbf {\bibinfo {volume} {57}},\
  \bibinfo {pages} {R6866} (\bibinfo {year} {1998})}\BibitemShut {NoStop}%
\bibitem [{\citenamefont {Chien}\ \emph {et~al.}(2015)\citenamefont {Chien},
  \citenamefont {He}, \citenamefont {Mo}, \citenamefont {Hashimoto},
  \citenamefont {Hussain}, \citenamefont {Shen},\ and\ \citenamefont
  {Plummer}}]{chien:2015-SM}%
  \BibitemOpen
  \bibfield  {author} {\bibinfo {author} {\bibfnamefont {T.}~\bibnamefont
  {Chien}}, \bibinfo {author} {\bibfnamefont {X.}~\bibnamefont {He}}, \bibinfo
  {author} {\bibfnamefont {S.-K.}\ \bibnamefont {Mo}}, \bibinfo {author}
  {\bibfnamefont {M.}~\bibnamefont {Hashimoto}}, \bibinfo {author}
  {\bibfnamefont {Z.}~\bibnamefont {Hussain}}, \bibinfo {author} {\bibfnamefont
  {Z.-X.}\ \bibnamefont {Shen}},\ and\ \bibinfo {author} {\bibfnamefont
  {E.~W.}\ \bibnamefont {Plummer}},\ }\bibfield  {title} {\bibinfo {title}
  {Electron-phonon coupling in a system with broken symmetry: {{Surface}} of
  {{Be}}(0001)},\ }\href {https://link.aps.org/doi/10.1103/PhysRevB.92.075133}
  {\bibfield  {journal} {\bibinfo  {journal} {Phys. Rev. B}\ }\textbf {\bibinfo
  {volume} {92}},\ \bibinfo {pages} {075133} (\bibinfo {year}
  {2015})}\BibitemShut {NoStop}%
\bibitem [{\citenamefont {Polley}\ and\ \citenamefont
  {Balasubramanian}(2024)}]{polley:2024a}%
  \BibitemOpen
  \bibfield  {author} {\bibinfo {author} {\bibfnamefont {C.}~\bibnamefont
  {Polley}}\ and\ \bibinfo {author} {\bibfnamefont {T.}~\bibnamefont
  {Balasubramanian}},\ }\bibfield  {title} {\bibinfo {title} {Electronic
  structure of thin epitaxial {{Be}} films on {{Si}}(111)},\ }\href
  {https://linkinghub.elsevier.com/retrieve/pii/S0039602823001899} {\bibfield
  {journal} {\bibinfo  {journal} {Surf. Sci.}\ }\textbf {\bibinfo {volume}
  {741}},\ \bibinfo {pages} {122436} (\bibinfo {year} {2024})}\BibitemShut
  {NoStop}%
\end{thebibliography}
\end{document}